\documentclass[11pt,fleqn]{book} 
\usepackage[braket, qm]{qcircuit}
\usepackage{multirow}
\usepackage{array}
\usepackage{booktabs}
\usepackage[symbol]{footmisc}
\usepackage{subfigure}
\usepackage{mathrsfs}
\usepackage{diagbox}
\usepackage{gensymb}
\usepackage{cancel}
\usepackage{soul}
\usepackage[titletoc]{appendix}
\usepackage{tabstackengine}

\usepackage{cite}

\newcommand{\sx}{\sigma^{x}}
\newcommand{\sy}{\sigma^{y}}
\newcommand{\sz}{\sigma^{z}}
\newcommand{\sM}{\sigma^{+}}
\newcommand{\sm}{\sigma^{-}}
\newcommand{\hdet}{\mathrm{HDet}}

\newcolumntype{C}[1]{>{\centering\arraybackslash$}p{#1}<{$}}

\newcommand{\overbar}[1]{\mkern 1.5mu\overline{\mkern-1.5mu#1\mkern-1.5mu}\mkern 1.5mu}
\newcommand{\floor}[1]{\left\lfloor #1 \right\rfloor}
\newcommand{\qwxo}[2][-1]{\ar @{-} [#1,0]_{#2}}

\let\OLDthebibliography\thebibliography
\renewcommand\thebibliography[1]{
  \OLDthebibliography{#1}
  \setlength{\parskip}{2.5pt}
  \setlength{\itemsep}{2.5pt plus 0.5ex}
}

\usepackage{graphicx} 

\usepackage{tikz} 

\usepackage[english]{babel} 

\usepackage{enumitem} 
\setlist{nolistsep} 

\usepackage{booktabs} 

\usepackage{xcolor} 
\definecolor{clr}{RGB}{65,90,227}

\usepackage{geometry} 

\usepackage{avant} 
\usepackage{lmodern}

\usepackage{microtype} 
\usepackage[utf8]{inputenc} 
\usepackage[T1]{fontenc} 

\usepackage{titletoc} 

\contentsmargin{0cm} 

\titlecontents{part}
	[0cm] 
	{\addvspace{20pt}\bfseries} 
	{}
	{}
	{}

\titlecontents{chapter}
	[1.25cm] 
	{\addvspace{12pt}\large\sffamily\bfseries} 
	{\color{clr!60}\contentslabel[\Large\thecontentslabel]{1.25cm}\color{clr}} 
	{\color{clr}} 
	{\color{clr!60}\normalsize\;\titlerule*[.5pc]{.}\;\thecontentspage} 

\titlecontents{section}
	[1.5cm] 
	{\addvspace{3pt}\sffamily\bfseries} 
	{\contentslabel[\thecontentslabel]{1.25cm}} 
	{} 
	{\hfill\color{black}\thecontentspage} 

\titlecontents{subsection}
	[1.5cm] 
	{\addvspace{1pt}\sffamily\small} 
	{\contentslabel[\thecontentslabel]{1.25cm}} 
	{} 
	{\ \titlerule*[.5pc]{.}\;\thecontentspage} 

\usepackage{caption}
\titlecontents{figure}
	[1.25cm] 
	{\addvspace{1pt}\sffamily\small} 
	{\thecontentslabel\hspace*{1em}} 
	{} 
	{\ \titlerule*[.5pc]{.}\;\thecontentspage} 

\titlecontents{table}
	[1.25cm] 
	{\addvspace{1pt}\sffamily\small} 
	{\thecontentslabel\hspace*{1em}} 
	{} 
	{\ \titlerule*[.5pc]{.}\;\thecontentspage} 

\titlecontents{lchapter}
	[0em] 
	{\addvspace{15pt}\large\sffamily\bfseries} 
	{\color{clr}\contentslabel[\Large\thecontentslabel]{1.25cm}\color{clr}} 
	{}  
	{\color{clr}\normalsize\sffamily\bfseries\;\titlerule*[.5pc]{.}\;\thecontentspage} 

\titlecontents{lsection}
	[0em] 
	{\sffamily\small} 
	{\contentslabel[\thecontentslabel]{1.25cm}} 
	{}
	{}

\titlecontents{lsubsection}
	[.5em] 
	{\sffamily\footnotesize} 
	{\contentslabel[\thecontentslabel]{1.25cm}}
	{}
	{}

\usepackage{fancyhdr} 

\fancypagestyle{plain}{
	\fancyhead{}%
}

\makeatletter
\renewcommand{\cleardoublepage}{
\clearpage\ifodd\c@page\else
\hbox{}
\vspace*{\fill}
\thispagestyle{empty}
\newpage
\fi}

\usepackage{amsmath,amsfonts,amssymb,amsthm} 

\newtheoremstyle{clrnumbox}
{0pt}
{0pt}
{\normalfont}
{}
{\small\bf\sffamily\color{clr}}
{\;}
{0.25em}
{\small\sffamily\color{clr}\thmname{#1}\nobreakspace\thmnumber{\@ifnotempty{#1}{}\@upn{#2}}
\thmnote{\nobreakspace\the\thm@notefont\sffamily\bfseries\color{black}---\nobreakspace#3.}} 

\newtheoremstyle{blacknumex}
{5pt}
{5pt}
{\normalfont}
{} 
{\small\bf\sffamily}
{\;}
{0.25em}
{\small\sffamily{\tiny\ensuremath{\blacksquare}}\nobreakspace\thmname{#1}\nobreakspace\thmnumber{\@ifnotempty{#1}{}\@upn{#2}}
\thmnote{\nobreakspace\the\thm@notefont\sffamily\bfseries---\nobreakspace#3.}}

\newtheoremstyle{blacknumbox} 
{0pt}
{0pt}
{\normalfont}
{}
{\small\bf\sffamily}
{\;}
{0.25em}
{\small\sffamily\thmname{#1}\nobreakspace\thmnumber{\@ifnotempty{#1}{}\@upn{#2}}
\thmnote{\nobreakspace\the\thm@notefont\sffamily\bfseries---\nobreakspace#3.}}

\newtheoremstyle{clrnum}
{5pt}
{5pt}
{\normalfont}
{}
{\small\bf\sffamily\color{clr}}
{\;}
{0.25em}
{\small\sffamily\color{clr}\thmname{#1}\nobreakspace\thmnumber{\@ifnotempty{#1}{}\@upn{#2}}
\thmnote{\nobreakspace\the\thm@notefont\sffamily\bfseries\color{black}---\nobreakspace#3.}} 
\makeatother

\newcounter{dummy} 
\numberwithin{dummy}{section}
\theoremstyle{clrnumbox}
\newtheorem{theoremeT}[dummy]{Theorem}

\newtheorem{exerciseT}{Exercise}[chapter]
\theoremstyle{blacknumex}
\newtheorem{conjectureT}{Conjecture}[chapter]
\theoremstyle{blacknumex}
\newtheorem{exampleT}{Example}[chapter]
\theoremstyle{blacknumbox}

\newtheorem{definitionT}{Definition}[section]
\newtheorem{corollaryT}[dummy]{Corollary}
\newtheorem{lemmaT}[dummy]{Lemma}
\theoremstyle{clrnum}

\RequirePackage[framemethod=default]{mdframed} 

\newmdenv[skipabove=7pt,
skipbelow=7pt,
backgroundcolor=black!5,
linecolor=clr,
innerleftmargin=5pt,
innerrightmargin=5pt,
innertopmargin=5pt,
leftmargin=0cm,
rightmargin=0cm,
innerbottommargin=5pt]{tBox}

\newmdenv[skipabove=7pt,
skipbelow=7pt,
rightline=false,
leftline=true,
topline=false,
bottomline=false,
backgroundcolor=clr!10,
linecolor=clr,
innerleftmargin=5pt,
innerrightmargin=5pt,
innertopmargin=5pt,
innerbottommargin=5pt,
leftmargin=0cm,
rightmargin=0cm,
linewidth=4pt]{eBox}	

\newmdenv[skipabove=7pt,
skipbelow=7pt,
rightline=false,
leftline=true,
topline=false,
bottomline=false,
linecolor=clr,
innerleftmargin=5pt,
innerrightmargin=5pt,
innertopmargin=0pt,
leftmargin=0cm,
rightmargin=0cm,
linewidth=4pt,
innerbottommargin=0pt]{dBox}	

\newmdenv[skipabove=7pt,
skipbelow=7pt,
rightline=false,
leftline=true,
topline=false,
bottomline=false,
linecolor=gray,
backgroundcolor=black!5,
innerleftmargin=5pt,
innerrightmargin=5pt,
innertopmargin=5pt,
leftmargin=0cm,
rightmargin=0cm,
linewidth=4pt,
innerbottommargin=5pt]{cBox}

\newenvironment{theorem}{\begin{tBox}\begin{theoremeT}}{\end{theoremeT}\end{tBox}}

\newenvironment{conjecture}{\begin{eBox}\begin{conjectureT}}{\hfill{\color{clr}\tiny\ensuremath{\blacksquare}}\end{conjectureT}\end{eBox}}				  
\newenvironment{definition}{\begin{cBox}\begin{definitionT}}{\end{definitionT}\end{cBox}}	
		
\newenvironment{corollary}{\begin{cBox}\begin{corollaryT}}{\end{corollaryT}\end{cBox}}
\newenvironment{lemma}{\begin{dBox}\begin{lemmaT}}{\end{lemmaT}\end{dBox}}		

\makeatletter
\renewcommand{\@seccntformat}[1]{\llap{\textcolor{clr}{\csname the#1\endcsname}\hspace{1em}}}                    
\renewcommand{\section}{\@startsection{section}{1}{\z@}
{-4ex \@plus -1ex \@minus -.4ex}
{1ex \@plus.2ex }
{\normalfont\large\sffamily\bfseries}}
\renewcommand{\subsection}{\@startsection {subsection}{2}{\z@}
{-3ex \@plus -0.1ex \@minus -.4ex}
{0.5ex \@plus.2ex }
{\normalfont\sffamily\bfseries}}
\renewcommand{\subsubsection}{\@startsection {subsubsection}{3}{\z@}
{-2ex \@plus -0.1ex \@minus -.2ex}
{.2ex \@plus.2ex }
{\normalfont\small\sffamily\bfseries}}                        
\renewcommand\paragraph{\@startsection{paragraph}{4}{\z@}
{-2ex \@plus-.2ex \@minus .2ex}
{.1ex}
{\normalfont\small\sffamily\bfseries}}

\newcommand{\@mypartnumtocformat}[2]{%
	\setlength\fboxsep{0pt}%
	\noindent\colorbox{clr!20}{\strut\parbox[c][.7cm]{\ecart}{\color{clr!70}\Large\sffamily\bfseries\centering#1}}\hskip\esp\colorbox{clr!40}{\strut\parbox[c][.7cm]{\linewidth-\ecart-\esp}{\Large\sffamily\centering#2}}%
}

\newcommand{\@myparttocformat}[1]{%
	\setlength\fboxsep{0pt}%
	\noindent\colorbox{clr!40}{\strut\parbox[c][.7cm]{\linewidth}{\Large\sffamily\centering#1}}%
}

\newlength\esp
\newlength\ecart
\def\@part[#1]#2{%
\ifnum \c@secnumdepth >-2\relax%
\refstepcounter{part}%
\addcontentsline{toc}{part}{\texorpdfstring{\protect\@mypartnumtocformat{\thepart}{#1}}{\partname~\thepart\ ---\ #1}}
\else%
\addcontentsline{toc}{part}{\texorpdfstring{\protect\@myparttocformat{#1}}{#1}}%
\fi%
\startcontents%
\markboth{}{}%
{\thispagestyle{empty}%
\begin{tikzpicture}[remember picture,overlay]%
\node at (current page.north west){\begin{tikzpicture}[remember picture,overlay]%
\fill[clr!20](0cm,0cm) rectangle (\paperwidth,-\paperheight);
\node[anchor=north] at (4cm,-3.25cm){\color{clr!40}\fontsize{220}{100}\sffamily\bfseries\thepart}; 
\node[anchor=south east] at (\paperwidth-1cm,-\paperheight+1cm){\parbox[t][][t]{8.5cm}{
\printcontents{l}{0}{\setcounter{tocdepth}{1}}
}};
\node[anchor=north east] at (\paperwidth-1.5cm,-3.25cm){\parbox[t][][t]{15cm}{\strut\raggedleft\color{clr}\fontsize{30}{30}\sffamily\bfseries#2}};
\end{tikzpicture}};
\end{tikzpicture}}%
\@endpart}
\def\@spart#1{%
\startcontents%
\phantomsection
{\thispagestyle{empty}%
\begin{tikzpicture}[remember picture,overlay]%
\node at (current page.north west){\begin{tikzpicture}[remember picture,overlay]%
\fill[clr!20](0cm,0cm) rectangle (\paperwidth,-\paperheight);
\node[anchor=north east] at (\paperwidth-1.5cm,-3.25cm){\parbox[t][][t]{15cm}{\strut\raggedleft\color{clr}\fontsize{30}{30}\sffamily\bfseries#1}};
\end{tikzpicture}};
\end{tikzpicture}}
\addcontentsline{toc}{part}{\texorpdfstring{%
\setlength\fboxsep{0pt}%
\noindent\protect\colorbox{clr!40}{\strut\protect\parbox[c][.7cm]{\linewidth}{\Large\sffamily\protect\centering #1\quad\mbox{}}}}{#1}}%
\@endpart}
\def\@endpart{\vfil\newpage
\if@twoside
\if@openright
\null
\thispagestyle{empty}%
\newpage
\fi
\fi
\if@tempswa
\twocolumn
\fi}

\newif\ifusechapterimage
\usechapterimagetrue
\newcommand{\thechapterimage}{}%
\newcommand{\chapterimage}[1]{\ifusechapterimage\renewcommand{\thechapterimage}{#1}\fi}%
\newcommand{\autodot}{.}
\newlength\chaptertitleheight
\newsavebox\chaptertitlebox
\def\@makechapterhead#1{%
{\parindent \z@ \raggedright \normalfont
\ifnum \c@secnumdepth >\m@ne
\setbox\chaptertitlebox=\hbox{%
\parbox{15cm}{\huge\sffamily\bfseries\color{black}\thechapter\autodot~#1\strut}}
\setlength\chaptertitleheight{\dimexpr\ht\chaptertitlebox+\dp\chaptertitlebox}
\if@mainmatter
\begin{tikzpicture}[remember picture,overlay]
\node at (current page.north west)
{\begin{tikzpicture}[remember picture,overlay]
\node[anchor=north west,inner sep=0pt] at (0,0) {\ifusechapterimage\includegraphics[width=\paperwidth]{\thechapterimage}\fi};
\draw[anchor=west] (\Gm@lmargin,-9cm) node [line width=2pt,rounded corners=15pt,draw=clr,fill=white,fill opacity=0.9,inner sep=15pt,minimum height=1.2\chaptertitleheight]{\strut\makebox[22cm]{}};
\draw[anchor=west] (\Gm@lmargin+.3cm,-9cm) node [text width=15cm] {\huge\sffamily\bfseries\color{black}\thechapter\autodot~#1\strut};
\end{tikzpicture}};
\end{tikzpicture}
\else
\begin{tikzpicture}[remember picture,overlay]
\node at (current page.north west)
{\begin{tikzpicture}[remember picture,overlay]
\node[anchor=north west,inner sep=0pt] at (0,0) {\ifusechapterimage\includegraphics[width=\paperwidth]{\thechapterimage}\fi};
\draw[anchor=west] (\Gm@lmargin,-9cm) node [line width=2pt,rounded corners=15pt,draw=clr,fill=white,fill opacity=0.9,inner sep=15pt,minimum height=1.2\chaptertitleheight]{\strut\makebox[22cm]{}};
\draw[anchor=west] (\Gm@lmargin+.3cm,-9cm) node [text width=15cm] {\huge\sffamily\bfseries\color{black}#1\strut};
\end{tikzpicture}};
\end{tikzpicture}
\fi\fi\par\vspace*{270\p@}}}

\def\@makeschapterhead#1{%
\begin{tikzpicture}[remember picture,overlay]
\node at (current page.north west)
{\begin{tikzpicture}[remember picture,overlay]
\node[anchor=north west,inner sep=0pt] at (0,0) {\ifusechapterimage\includegraphics[width=\paperwidth]{\thechapterimage}\fi};
\draw[anchor=west] (\Gm@lmargin,-9cm) node [line width=2pt,rounded corners=15pt,draw=clr,fill=white,fill opacity=0.9,inner sep=15pt]{\strut\makebox[22cm]{}};
\draw[anchor=west] (\Gm@lmargin+.3cm,-9cm) node {\huge\sffamily\bfseries\color{black}#1\strut};
\end{tikzpicture}};
\end{tikzpicture}
\par\vspace*{270\p@}}
\makeatother

\usepackage[linkcolor=clr,citecolor=clr,pdfencoding=auto, psdextra]{hyperref}
\hypersetup{hidelinks,
colorlinks=true ,breaklinks=true,urlcolor=clr, bookmarks=false,
bookmarksopen=false}

\usepackage{bookmark}
\bookmarksetup{
open,
numbered,
addtohook={%
\ifnum\bookmarkget{level}=0 
\bookmarksetup{bold}%
\fi
\ifnum\bookmarkget{level}=-1 
\bookmarksetup{color=clr,bold}%
\fi
}
}

\begin{document}


\frontmatter

\pagestyle{empty}
\begin{center}

\vfill
\centerline{\mbox{\includegraphics[width=60mm]{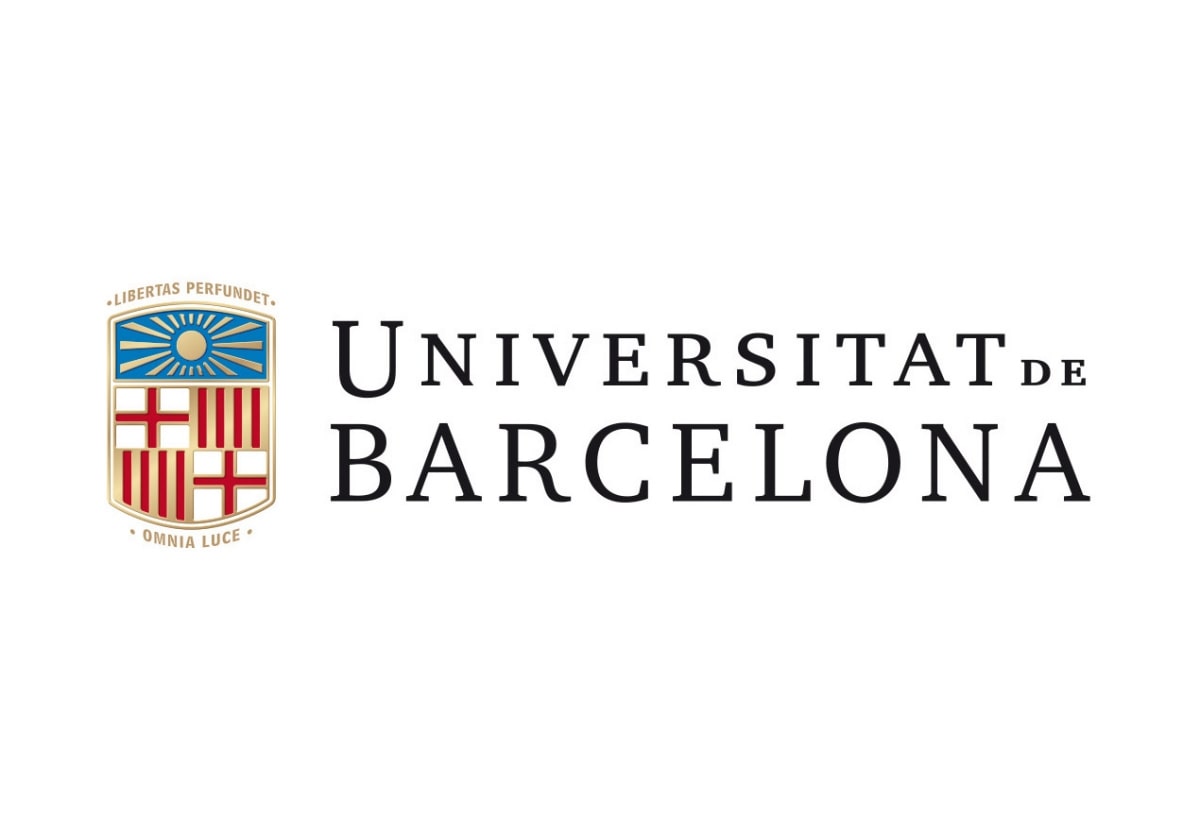}}}

\medskip
{\large \textbf{TESI DOCTORAL}}

\vfill
\vspace{5mm}

{\LARGE Alba Cervera Lierta}

\vspace{15mm}

{\Huge \textbf{Maximal Entanglement}}\\
\vspace{0.5 cm}
{\Large \textbf{Applications in Quantum Information and Particle Physics}}

\vfill

{\large Departament de Física Quàntica i Astrofísica}

\vfill

{\large Director de tesi: \large Dr. José Ignacio Latorre Sentís} 

\vfill

{\large Barcelona, Abril de 2019}

\end{center}


\newpage
~\vfill
\thispagestyle{empty}

\noindent Copyright \copyright\ 2019 Alba Cervera Lierta (CC-\textit{by})\\ 

\noindent Cover art and design by Carlos Villafranca.\\
\noindent Chapter 2 image art and design by Carlos Villafranca. \\

\noindent Quantum circuits have been designed with \emph{Q-circuit} package. Source can be found at \url{https://cquic.unm.edu/resources/}. \\

\noindent Feynman diagrams have been designed with \emph{TikZ-Feynman} package. Documentation can be found in Joshua P. Ellis, \emph{Computer Physics Communications} \textbf{210}, 103-123 (2017). \\

\noindent \LaTeX \hspace{0.05cm} template ``The Legrand Orange Book'' v. 2.4 by Mathias Legrand with modifications. \\

\noindent Unless explicitly stated in the caption, I have produced all the figures that appear in this work, which will often closely follow those in the corresponding publications.\\ 

\noindent \textit{First printing, April 2019.} 


\newpage
\thispagestyle{empty}

\begin{flushright}
\begin{minipage}{0.7\textwidth}

\end{minipage}%
\begin{minipage}{0.3\textwidth}
\vspace{5cm}
\textit{A les meves \`avies i avis.}\\
\end{minipage}
\end{flushright}


\frontmatter

\chapterimage{gracias2}
\addcontentsline{toc}{chapter}{Agra\"iments}
\chapter*{Agra\"iments}


Aquesta tesi no hauria estat possible sense el suport i l'ajut de moltes persones. Algunes d'aquestes han format part de la meva vida abans de començar aquest camí, altres les he conegut a punt d'acabar-lo, però totes i cada una d'elles m'han influenciat tant acadèmicament com personalment.

La primera d'elles és en José Ignacio, a qui considero més que un director, un mentor. Gràcies per les teves lliçons. Crec que tant jo com qualsevol dels teus estudiants recordarem sempre les llargues i inspiradores converses davant la pissarra de guix del teu despatx. I mil gràcies per donar-me l'oportunitat d'embarcar-me en aquesta aventura científica i de treballar tot el possible perquè no l'abandonés.

En segon lloc, vull agrair al que va ser també el meu director durant uns mesos. Gràcies Juan per convidar-me a treballar amb tu a Oxford. Van ser uns mesos clau on vaig prendre la decisió de continuar amb aquest doctorat com sigui. Estic segura que sense aquella experiència avui no seria on sóc. Aprofito per agrair també a totes aquelles persones que vaig conèixer durant la meva estància al Rudolph Pierls Institute: Marco, Nathan, Stefano, Emanuele, Valerio i, per descomptat, Luca, amb qui he treballat, sofert i gaudit amb un dels projectes d'aquesta tesi. \emph{Grazie mille}. També vull agrair als meus coautors, Karol, Dardo, Germán i Albert, per les seves aportacions i interessants discussions.

Durant aquests quatre anys, he tingut companys al meu voltant amb els que he compartit alegries i frustracions, i que s'han convertit en veritables amics. Gràcies Dani per compartir la teva experiència i feina amb mi. Les llargues tardes al despatx parlant de política i física i les lliçons d'escacs al Petronilla fan que et perdoni per no poder deixar sense vigilància els meus Kinder. Gràcies també a Javier, Adrià, Clara, Marc, Isma i Elis per compartir aquests anys de doctorat i els anteriors de la carrera. Les converses i cafés a mig matí, mitja tarda i/o al mig dia han fet suportables els dies més pesats. Gracias Ivan por los mismos motivos y por nuestras discusiones sobre física de partículas: aprendí mucho contigo. 

També voldria agrair a totes aquelles persones que han format part de la meva vida fora de la Universitat, en especial les que sempre han estat disponibles per fer unes \emph{birres} o uns vermuts per desconnectar i gaudir de la vida. Gràcies Sergi, Irene, Fran, Marc, Josep, Toni, Gemma, Joan, Marc, Alba i molts més.

Aquesta tesi ha sigut el reflex de significants canvis al voltant de la investigació en informació quàntica a Barcelona. El motiu pel qual he dedicat els últims dos anys al camp de la computació quàntica ha sigut la creació d'un grup d'investigació fruit de la col·laboració entre la Universitat de Barcelona i el Barcelona Supercomputing Center (BSC). Voldria agrair al BSC per donar-me suport econòmic en aquests darrers anys i per apostar per aquest camp. En especial vull donar les gràcies a tots els que formen part del grup Quantic o hi col·laboren: Pol, Artur, Luca, Sofyan, Chris, Carlos, David, Adrian, Diego, Sergi, Elies i Josep. També voldria agrair a l'empresa IBM per atorgar-me el premi ``Teach Me QISKit'' i donar-me així visibilitat en aquest camp. 

Gracias Carlos por plantar la semilla de todo esto. Fuiste una fuente de inspiración para mi y muchos otros. Gracias por ser el primero en mostrarme la belleza de la física. 

Gràcies també als meus pares, Jordi i Gemma, i al meu germà, Gerard, per animar-me sempre a estudiar allò que més m'agradés, per animar-me en els moments més baixos i per compartir amb mi les alegries que han anat sorgint. Sense el seu suport res del que he fet hagués estat possible. Gràcies també als meus avis, Roser, Llorenç, Cleo i Luis, per tota una vida de sacrificis que han fet possible que jo ara hagi arribat on hagi volgut. Sé que algun d'ells hagués triat aquest mateix camí si hagués pogut, però les circumstàncies de l'època no ho van fer possible. Per aquest motiu els hi dedico especialment aquesta tesi, que no és més que el fruit del seu esforç.

Por último, a la persona que mejor conoce los entresijos no escritos de esta tesis. Gracias Miguel, por tu apoyo, por tu optimismo, por tus cuidados, por soportar los malos momentos, por celebrar los buenos, por tu infinita paciencia, por escucharme, por tu dedicación y, en definitiva, por quererme. Esta tesis también va dedicada a ti.

\chapterimage{Publications}
\addcontentsline{toc}{chapter}{List of publications}
\chapter*{List of publications}

\begin{itemize}
\item D. Alsina, A. Cervera, D. Goyeneche, J. I. Latorre and K. \.{Z}yczkowski, \\
\textit{Operational Approach to Bell Inequalities: Application to qutrits}, \\
Physical Review A \textbf{94}, 032102 (2016).
\vspace{0.3cm}
\item A. Cervera-Lierta, J. I. Latorre, J. Rojo and L. Rottoli,\\ \textit{Maximal Entanglement in High Energy Physics},\\ SciPost Physics \textbf{3}, 036 (2017).
\vspace{0.3cm}
\item A. Cervera-Lierta, A. Gasull, J. I. Latorre and G. Sierra, \\ 
\textit{Multipartite entanglement in spin chains and the Hyperdeterminant},\\
 Journal of Physical A: Mathematical and Theoretical \textbf{51}, 505301 (2018).
\vspace{0.3cm}
\item A. Cervera-Lierta,\\
\textit{Exact Ising model simulation on a quantum computer},\\
Quantum \textbf{2}, 114 (2018).
\vspace{0.3cm}
\item A. Cervera-Lierta, J. I. Latorre, D. Goyeneche, \\
\textit{Quantum circuits for maximally entangled states}, \\
arXiv:1904.07955 [quant-ph].

\end{itemize}

\chapterimage{resum2}
\addcontentsline{toc}{chapter}{Resum}
\chapter*{Resum}

L'entrella\c cament \'es una de les principals caracter\'istiques de la mec\`anica qu\`antica. \'Es probablement un dels fen\`omens qu\`antics m\'es debatuts i estudiants degut,  en part, a la seva naturalesa antiintu\"itiva i, m\'es recentment, a les seves aplicacions en el camp de la informaci\'o qu\`antica. \'Es precisament una propietat que va m\'es enll\`a del que la f\'isica cl\`assica pot explicar. La motivaci\'o d'aquesta tesi \'es estudiar l'entrella\c cament en general i sota quines circumst\`ancies \'es m\`axim en particular.

Primerament, analitzem el paper de l'entrella\c cament en la construcci\'o de la frontera entre la f\'isica cl\`assica i la f\'isica qu\`antica. Els experiments de Bell ens permeten calcular una s\`erie de correladors que ens ajudaran a distingir si les part\'icules implicades en l'experiment obeeixen el que es coneix com a \emph{realisme local}. La violaci\'o de les desigualtats de Bell demostra que no hi ha una teoria cl\`assica de variables ocultes que expliqui els resultats de l'experiment, \'es a dir, la f\'isica qu\`antica subjacent no pot ser explicada des de la f\'isica cl\`assica.

La caracteritzaci\'o de desigualtats de Bell per a qualsevol nombre de part\'icules i dimensions locals \'es un problema obert en informaci\'o qu\`antica. En aquesta tesi, estudiem i dedu\"im noves desigualtats de Bell en termes d'operadors, focalitzant-nos especialment en aquelles que involucren qutrits. Les desigualtats per qubits són violades m\`aximament pels estats altament entrella\c cats coneguts com a GHZ. Les desigualtats de qutrits, o altres dimensions m\'es grans, són violades m\`aximament per estats que són una deformaci\'o dels GHZ. Aquest resultat mostra l'estreta relaci\'o, per\`o no equival\`encia, entre no-localitat i m\`axim entrella\c cament.

Seguidament, estudiem l'entrella\c cament multipartit i la seva aplicaci\'o en la detecci\'o de transicions de fase qu\`antiques. En particular, estudiem l'entrella\c cament entre quatre part\'icules de dimensi\'o dos, \'es a dir, entre quatre qubits. Com a figura de m\`erit, utilitzem l'hiperdeterminant i els dos invariants polin\`omics que el formen, anomenats $S$ i $T$. Analitzem uns quants estats qu\`antics rellevants per acabar concloent que aquesta figura de m\`erit capta un tipus concret d'entrella\c cament multipartit. Quan calculem l'hiperdeterminant en cadenes d'espins $1/2$, obtenim un pic pronunciat al voltant de la transici\'o de fase en el cas del model d'Ising. En el cas del model $XXZ$, l'hiperdeterminant \'es sempre zero. En aquest cas, utilitzem els invariants $S$ i $T$. El resultat \'es que el valor dels invariants canvia bruscament en els punts on hi ha transici\'o de fase. Finalment, estudiem la funci\'o d'ona de Haldane-Shastry i obtenim resultats similars als del model $XXZ$.

En una segona part de la tesi, ens centrem en el camp de la computaci\'o qu\`antica. Gr\`acies als aven\c cos que s'han dut a terme en els darrers anys en relaci\'o al control dels \`atoms, fotons i processos qu\`antics en general, la computaci\'o qu\`antica ha esdevingut una realitat. Actualment, diverses empreses estan en proc\`es de construcci\'o dels seus propis ordinadors qu\`antics. Per aquest motiu, es fa necess\`aria la recerca de m\`etodes per testejar i comparar aquests primers prototips d'ordinadors. 

Per una banda, proposem i testegem un m\`etode que consisteix en la simulaci\'o exacta del model d'Ising. Aquest model es pot resoldre anal\'iticament, per tant, els resultats obtinguts d'un ordinador qu\`antic els podrem comparar amb el seu valor te\`oric. Proposem un circuit qu\`antic que diagonalitza l'Hamiltoni\`a d'Ising i que, per tant, fa possible la simulaci\'o en el temps i la preparaci\'o d'estats t\`ermics. Testegem aquest circuit pel cas d'una cadena formada per quatre espins en els ordinadors que ofereix la multinacional IBM i l'empresa Rigetti Computing. Els resultats difereixen notablement del valor te\`oric esperat tot i que els valors dels temps de decoher\`encia i la fidelitat de les portes haurien de fer possible uns resultats millors. Aix\`o fa pensar que hi ha altres fonts d'errors que no es tenen en compte en general i que clarament cobren rellev\`ancia fins i tot en circuits tan petits.

Per altra banda, proposem un test tant dur com necessari per a un ordinador qu\`antic: la simulaci\'o d'estats altament entrella\c cats. S'ha demostrat que l'avantatge dels algorismes qu\`antics respecte als cl\`assics recau principalment en la generaci\'o alta d'entrella\c cament en algun moment de l'algorisme. A m\'es, estats de baix entrella\c cament poden ser simulats de forma eficient amb t\`ecniques cl\`assiques. Si volem construir ordinadors qu\`antics i que ens siguin \'utils per realitzar aquelles tasques que els cl\`assics no poden dur a terme, necessitarem que aquests dispositius puguin generar i suportar estats altament entrella\c cats. La nostra proposta \'es simular estats absolutament m\`aximament entrella\c cats, \'es a dir, estats on totes les seves biparticions estan m\`aximament entrella\c cades. Presentem els circuits expl\'icits per realitzar aquestes simulacions tant per qubits com per qudits de dimensi\'o m\'es gran que dos. A m\'es, tamb\'e analitzem com l'entropia de cada bipartici\'o sempre augmenta o es mant\'e, mai disminueix, i utilitzem aquesta propietat per trobar els circuits m\'es \`optims, \'es a dir, amb un menor nombre de portes qu\`antiques.

Finalment, ens centrem en l'origen m\'es fonamental de l'entrella\c cament: els processos de part\'icules elementals. Estudiem quina ha de ser l'estructura de la interacci\'o de QED per tal de poder generar estats m\`aximament entrella\c cats en termes de les helicitats de les part\'icules sortints. El resultat demostra que, a primer ordre en teoria de pertorbacions, la interacci\'o de QED es recupera imposant m\`axim entrella\c cament. Tamb\'e estudiem quines implicacions t\'e aquesta imposici\'o en processos que involucrin corrents d\`ebils neutres. El resultat a primer ordre \'es que el valor de l'angle de Weinberg ha de ser de $\pi/6$, molt proper al valor experimental. Per \'ultim, estudiem un exemple d'interacci\'o forta: la interacci\'o glu\'o-glu\'o. El resultat \'es que els gluons es poden entrella\c car m\`aximament independentment dels valors de les constants d'estructura, per tant, no podem obtenir m\'es informaci\'o sobre la interacci\'o mitjan\c cant aquesta conjectura de m\`axim entrella\c cament.

\mainmatter


\chapterimage{drops2} 

\pagestyle{empty} 

\tableofcontents 

\cleardoublepage 

\pagestyle{fancy} 


\chapterimage{Content}
\chapter{Introduction}

\vspace{-1.5cm}
\begin{flushright}
\begin{minipage}{0.6\textwidth}
\textit{When two systems [...] enter into temporary physical interaction due to known forces between them, and when after times of mutual influence the systems separate again, then they can no longer be described in the same way as before [...]. I would not call that \emph{one} but rather \emph{the} characteristic trait of quantum mechanics, the one that enforces its entire departure from classical lines of thought. By the interaction the two representatives (or $\psi$-functions) have become entangled.}
\begin{flushright}
--Erwin Schr\"odinger, \\
``Discussion of probability relations between separated systems'', 1935.
\end{flushright}
\end{minipage}
\end{flushright}
\vspace{1cm}

Entanglement is a quantum phenomenon that occurs when two or more quantum systems cannot be described independently from the others. In a sense, entanglement is an example of a quantum correlation, where once we have collapsed the wave function of one part of the system, the state of the other is determined by the result on the first one. Indeed, even if the two or more systems are separated by a spacelike distance, the result after the collapse of one of them determines the result of the other. This is probably one of the most striking traits of quantum mechanics that has generated a huge amount of both interest and discussion.

One can be tempted to believe that it is possible to use this apparent instant action -- or, in Einstein's words, ``spooky action at a distance'' -- to communicate information faster than light. However, there is no way for an observer that performs one of the measurements to elucidate if her result has been obtained randomly, according to wave function probability amplitudes, or is a consequence of the collapse of the other observer. They have to communicate with each other and ask which one has performed the measurement first, that is, using a classical communication channel which, of course, obeys causality. Then, it is not possible to use entanglement to communicate faster than light and the explanation of what has really happened before and after the measurements are performed remains in the field of interpretation of quantum mechanics. In the end, the quantum wave function, entangled or not, is a mathematical object that we use to describe the information of a system. 

Several experiments have highlighted the fact that entanglement is a \emph{genuine} quantum property that goes beyond any classical description. The violation of Bell inequalities shows that there is no hidden variable theory that allows explaining the correlations that entanglement predicts \cite{Aspect82}. It is not possible to describe quantum mechanics using classical laws. Entanglement is also phenomena that involve the whole system itself; local operations on each subsystem does not change the amount of entanglement. This fact cannot be remedied by using classical protocols; it has been proved that Local Operations and Classical Communication (LOCC) methods cannot change the entanglement of a system \cite{Nielsen99}.

One may expect that such a distinctive trait of quantum mechanics have several physical applications. Indeed, entanglement can also be understood as the resource that enables genuine quantum protocols such as cryptography based on Bell inequalities \cite{Ekert91} and teleportation \cite{Bennett93}. In addition, large entanglement is expected to be present in quantum registers when a quantum algorithm produces a relevant advantage in performance over a classical computer such as Shor's algorithm \cite{Shor97}.

The quantification of entanglement for any number of parties is an open problem in quantum physics. The natural growth of complexity in the study of multipartite entanglement is illustrated in the example of four-party entanglement by the existence of 9 Stochastic LOCC classes of pure 4-qubit states \cite{Verstraete02}. Then, it is not surprising the existence of multiple non-equivalent figures of merit to quantify multipartite entanglement \cite{Okovi09,Osterloh08,Osterloh16}. 
Among all of them, the most well-known is probably the Von Neumann entropy. For a bipartite system, maximal entropy and maximal entanglement are equivalent and are usually used indistinctly. This is the ideal case, where the maximum value of the figure of merit corresponds with the maximal entanglement. However, this fact is not reproduced in the multipartite case, where different figures of merit have different values for the same state. For that reason, there is a seek for a formal definition of a ``maximal entangled'' state. A proposal is the Absolutely Maximally Entangled states, those states that are maximally entangled in all their bipartitions.

Entanglement is a key property of quantum mechanics and, consequently, it plays an important role in Nature. In order to observe quantum phenomena of this kind, Nature should be able to generate entangled states. It is then natural to ask ourselves how entanglement is generated at its most fundamental level, i.e. at the level of fundamental interactions. Violation of Bell inequalities has been proved with entangled photons that have been generated, at its most fundamental level, by a matter-light interaction.

The aim of the present thesis is to address several studies where entanglement is present and plays a central role. This thesis deals with examples of applications and open problems described in the above paragraphs. In Chapter \ref{Ch:Bell_Ineq}, we present Bell inequalities for multipartite systems of local dimension 2 (qubits) and also dimension 3 (qutrits). In Chapter \ref{Ch:HDet}, we analyse four-partite entanglement in spin chains using as a figure of merit the hyperdeterminant. In Chapter \ref{Ch:Ising}, we perform an experiment in a real quantum computer consisting of the exact simulation of the $XY$ model. In Chapter \ref{Ch:AME}, we propose quantum circuits that generate Absolutely Maximally Entangled states of any dimension. Finally, in Chapter \ref{Ch:MaxEnt}, we analyse the generation of maximal entanglement in particle physics and its implications in the determination of the interaction structure. The conclusions of this thesis are exposed in Chapter \ref{Ch:Conclusions}. Moreover, supplementary material can be found in the appendices. Appendix \ref{app:quantum_gates} provides a summary of quantum gates and quantum circuits. Appendix \ref{app:Feynman} summarizes the conventions used in Chapter \ref{Ch:MaxEnt} and the set of Feynman rules used to do the amplitudes computation. In appendix \ref{app:QED} it can be found an exhaustive analysis of entanglement generation in all tree-level QED processes. Some extra material resulting from the other chapters is written in the appendix \ref{app:OddsEnds}.


\chapterimage{Bell_image} 

\chapter{Novel Bell Inequalities \label{Ch:Bell_Ineq}}


\vspace{-1.5cm}
\begin{flushright}
\begin{minipage}{0.6\textwidth}
\textit{
...what is proved by impossibility proofs is lack of imagination.}
\begin{flushright}
--John S. Bell, \\
``On the impossible pilot wave'', 1982.
\end{flushright}
\end{minipage}
\end{flushright}
\vspace{1cm}

On 1935, Albert Einstein, Boris Podolsky and Nathan Rosen published an article that directly defied the young theory of quantum mechanics \cite{EPR}. On their paper, ``Can quantum-mechanical description of physical reality be considered complete?'', they proposed a \emph{gedankenexperiment} which demonstrated that quantum physical theory is not complete by identifying elements of reality that were not included in the main theory. According to their definition
\begin{quote}
\textit{If, without in any way disturbing a system, we can predict with certainty (i.e., with probability equal to unity) the value of a physical quantity, then there exists an element of physical reality corresponding to this physical quantity.}
\end{quote}

Many physicists discussed this statement and its implications \cite{VonNeumann,Bohm1,Bohm2}. In order to explain the results obtained from quantum mechanical  experiments, it was necessary to assume a local hidden variables theory, that is, to introduce unknown variables that assure the local realism behavior that should underlie these results. It was almost thirty years later when a new article appeared and became the milestone of this discussion. John S. Bell's paper ``On the Einstein Podolsky Rosen paradox'' proposed a real experiment to test whether or not Nature behaves as expected from the EPR point of view \cite{Bell64}. 

Bell's original experiment predicts an upper bound for a linear combination of correlators between some measurements performed by three observers. This upper bound is computed according to the laws of classical physics; to be precise, according to a \textit{local realistic} theory. If we perform this experiment and we find a violation of this inequality, then Nature, in particular quantum mechanics, can not be described by the laws of classical physics.

After Bell, many other scientists proposed Bell-type inequalities. In particular, Clauser and Horne -- and, afterward, together with Shimony and Holt -- proposed a more experimentally realizable Bell experiment \cite{CH,CHSH}. Almost fifty years after EPR's paper, experimentalists obtained strong evidence that local hidden variables theories were ruled out \cite{Aspect82}. However, there are still open issues related with how these experiments are performed and that could invalidate partially the results obtained. These list of open issues are called \emph{loopholes}, but as the technology is improving, more of them are being closed. In fact, there are already experiments that claim that have closed all ``closeable'' loopholes \cite{Hensen15}.

Both Bell original and CHSH inequalities involve two observers that can perform a measurement with two different settings obtaining two possible outputs. However, it is necessary to test other more sophisticated systems, namely those involving more parties or with more possible outcomes. There have been numerous attempts to go beyond the CHSH inequalities. Mermin introduced a set of inequalities for an arbitrary number of qubits that were maximally violated by the GHZ state \cite{Mermin90,GHZ90}. A systematic mathematical treatment of these inequalities was carried out a decade later \cite{Werner01,Zukowski02}. It was also at that time that inequality for two $d$-dimensional particles was discovered \cite{CGLMP} and with it came the first realization that maximally entangled particles did not always maximally violate a Bell inequality \cite{Acin02}. This fact showed that entanglement was not in a one-to-one correspondence with non-locality. Progress in generalization to a larger number of $d$-dimensional particles has been more modest \cite{Acin04,Qutrits}. For a general review of Bell nonlocality see Ref. \cite{ReviewBI}.

In this chapter, we construct Bell inequalities for systems composed of several subsystems with more than two levels each. In particular, we focus our attention on quantum systems consisting of qutrits. Inequalities for three outcomes have been written in terms of probabilities, although they can also be treated with expectation values \cite{Chen02,Arnault12}. We have extended this formalism in order to build new inequalities for three outcomes and a different number of parties and find its classical and quantum bounds for qutrits in a semi-systematic way. We have found some regular patterns for the coefficients of the inequalities and for the settings and states that maximally violate these inequalities. This mechanism is potentially generalizable to other dimensions.

After this introduction, we start with a short summary of what is understood as a \emph{Bell experiment}. In Sec. \ref{sec:BI_qubits}, \ref{sec:BIqutrits} and \ref{sec:CGLMPd} we review some well-known Bell inequalities and deduce them with a different approach. We use this new formalism to extend these inequalities to a larger number of parties and to find which are the optimal settings that violate maximally these inequalities. In Sec. \ref{sec:mapping} we propose a novel method to obtain Bell inequalities from maximally entangled states and show three examples. Finally, the conclusions and some open questions are written in Sec. \ref{sec:conclusionsBI}.

\section{Bell experiment}

A typical Bell experiment involves two or more systems ($A$, $B$, $C$, ...) that have interacted in the past -- for instance, they have a common origin -- have been separated a large distance and are measured by independent observers. These observers can perform a measurement of some physical quantity, for example, particles spin or photon polarization, in different ways. For instance, they can choose between two or more settings that project in two different directions. As a result of each measurement, the observers obtain an output labelled with a macroscopic value. Observer of system $A$, let's call her Alice, obtains $a$ outcome if she measures with one of the settings and $a'$ output if she measures with another setting. Similarly, observer $B$, called Bob, obtains $b$ or $b'$ outputs, observer $C$, Charlie, obtains $c$ and $c'$ outputs, etc. After repeating the experiment many times, Alice, Bob and their possible colleagues compare their results and compute the expected value of all pair of measurements, i.e. $\langle ab\rangle$, $\langle ab'\rangle$, etc. Here we label the setting using the same letter as the corresponding output in an abuse of language. A Bell inequality is a linear relation between these expected values that predicts an upper bound if the system follows the laws of \emph{local realism}.

In the above experiment, the notion of \emph{locality} refers to the fact that the outcomes obtained do not depend on the measurement settings performed by the other observers. For instance, the result if Alice obtains +1 when she measures with the first setting is independent of the setting Bob has chosen to measure his subsystem. On the other hand, \emph{realism} is included in the assumption that measurement outcomes depend only on the setting used and on \textit{hidden variables} $\lambda$. These hidden variables could be a list of values or stochastic variables.

A violation of a Bell inequality implies that one or both assumptions, locality or realism, are false, and this is actually what happens when Bell experiments are performed on quantum mechanical systems. 

Let's formalize the description of a Bell experiment. When Alice and Bob measure with the first setting, the outputs obtained could depend on some probability distribution. Moreover, we can expect that the results of two measurements are in general dependent, i.e.  
\begin{equation}
p(a,b)\neq p(a) p(b),
\end{equation}
since both systems have interacted in the past. However, if we assume local realism, we should be able to find a description of these events in terms of some variables $\lambda$ that give an explanation of the results obtained. These hidden variables represent the \textit{elements of reality} that EPR mentioned in their article. Then, under this assumption,
\begin{equation}
p(a,b;\lambda)= p(a;\lambda) p(b;\lambda).
\end{equation}
These hidden variables can follow a probabilistic distribution $q(\lambda)$ which should be independent on the settings used if we assume locality, i.e. $q(\lambda;a,a',b,b'...)=q(\lambda)$. Thus we can compute the result of each experiment as
\begin{equation}
p(a,b)=\int d\lambda q(\lambda)p(a;\lambda)p(b;\lambda).
\label{eq:joint_prob}
\end{equation}

Notice that the very entanglement definition contradicts Eq. \eqref{eq:joint_prob} since it is not possible to factorize an entangled state into its subsystems. So, although non-locality and entanglement are not equivalent definitions \cite{Acin1,Acin2,ReviewBI}, entanglement will be closely related to the violation of a Bell inequality.

In the following sections, we will study Bell inequalities involving multiple parties, settings and dimensions. In particular, we are interested in the operational formulation of these inequalities and the maximal values that they can achieve from a local realism point of view (LR) or a quantum mechanical point of view (QM). For the last one, each setting will be represented by a quantum mechanical operator such that
\begin{equation}
\hat{a}=\sum_{i=0}^{d-1}a_{i}|a_{i}\rangle\langle a_{i}|,
\end{equation}
where $d$ is the local dimension ($d=2$ for qubits, $d=3$ for qutrits, ...) and $a_{i}$ are the possible outcomes. Similarly, we can write the operators for the other settings: $\hat{a'}$, $\hat{b}$, $\hat{b'}$, etc. To simplify the notation, we will consider indistinctly $a$, $a'$, $b$, etc., labels as operators and as outcomes as well so, for now on, we will avoid the use of hats to write the quantum operators. In this formulation, the Bell inequality becomes an operator and its upper bound corresponds with its maximum expected value.
\begin{definition}[Bell operator]
A Bell operator of $n$ parties, $s$ settings and $d$ possible outputs per setting will be denoted as $\mathcal{B}_{nsd}$. The maximum expected value of this operator according to a local realism theory is $\langle\mathcal{B}\rangle_{\mathrm{LR}}$ and according to quantum mechanics $\langle\mathcal{B}\rangle_{\mathrm{QM}}$.
\end{definition}
The maximum value of the Bell inequality corresponds with the larger eigenvalue of this operator which, for a fixed $n$ and $d$, depends on the settings choice. Thus, we will look for the optimal settings that lead to this maximum value.

\section{Bell inequalities for two outcomes \label{sec:BI_qubits}}

Let's start with Bell inequalities with two possible outcomes, that is those inequalities applied to qubit states (see App. \ref{app:quantum_gates}). We will also consider two settings for each party, i.e. we will study Bell inequalities of the form $\mathcal{B}_{n22}$.

\subsection{Two parties}

The most relevant Bell inequality for two outcomes is the one proposed by Clauser, Horne, Shimony and Holt \cite{CHSH}:
\begin{equation}
E(\mathcal{B}_{CHSH})=E(a,b)+E(a,b')+E(a',b)-E(a',b'),
\end{equation}
where the two possible outcomes measured with the two settings are $a,a'=\pm 1$ for subsystem $A$ and and $b,b'=\pm 1$ for subsystem $B$. The function $E(a,b)$ represents the correlation, classical or quantum, between $a$ and $b$ measurements and it will be estimated after $A$ and $B$ observers have repeated many times the experiment. 

The local realism bounds of this inequality are
\begin{equation}
-2\leq E(\mathcal{B}_{CHSH}) \leq 2 \ .
\end{equation}
We can write the above inequality in the following way:
\begin{equation}
\mathcal{B}_{CHSH}=a\left(b+b'\right)+a'\left(b-b'\right),
\label{eq:CHSH}
\end{equation}
where we have removed the correlation function for simplicity. One can easily notice that when $b+b'$ are maximum, i.e. have a value of 2, then $b-b'$ vanish, obtaining the expected LR bound for the inequality.

In quantum mechanics, the variables $a$, $a'$, $b$ and $b'$ are represented by Hermitian operators acting on the Hilbert spaces $\mathscr{H}_{A}$ and $\mathscr{H}_{B}$. The expected value of these operators is $E(a,b)=\langle\psi|a\otimes b|\psi\rangle=\langle a\otimes b\rangle$, where $|\psi\rangle$ is the quantum state of the whole system. For the properties of the expected values of quantum systems, $\langle a\otimes b\rangle=\langle a\rangle\otimes\langle b\rangle$ allowing us to write the CHSH inequality in the same form as in Eq. \eqref{eq:CHSH}. In addition, for simplicity, we will not write the Kronecker product between quantum operators nor the bracket notation for expected values.
 
It was proven by Cirel'son that the maximum quantum value for $\langle\mathcal{B}_{CHSH}\rangle$ is $2\sqrt{2}$, so CHSH inequality is violated by quantum mechanics.

An enlightening computation of classical an quantum bounds of this inequality was given in Ref. \cite{Landau87}. The squared of the Bell operator $\mathcal{B}_{CHSH}$ can be written as
\begin{equation}
\mathcal{B}_{CHSH}^2= 4\mathbb{I} - [a,a'][b,b'] \ ,
\end{equation}
where $\mathbb{I}=\mathbb{I}_{A}\otimes\mathbb{I}_{B}$ and $a^2=a'^{2}=b^2=b'^{2}=\mathbb{I}$ due to these operators represent dichotomic observables. For LR, i.e. classical physics, observables commute, so the classical bound obtained is $\langle\mathcal{B}_{CHSH}\rangle_{\mathrm{LR}}=\sqrt{4}=2$. On the contrary, the largest absolute value of all possible eigenvalues for commutators of Hermitian operators of dimension $2^2$ is 2 and it is achieved by considering the Pauli matrices, the generators of $SU(2)$. Pauli matrices have the property $[\sigma_{i},\sigma_{j}]=2\epsilon_{ijk}\sigma_{k}$, so taking
\begin{equation}
[a,a'][b,b']=[\sigma_{i},\sigma_{j}][\sigma_{j}\sigma_{i}]=(2\epsilon_{ijk}\sigma_{k})(2\epsilon_{jik}\sigma_{k})=2\sigma_{k}(-2\sigma_{k})= -4 \ ,
\end{equation}
the expected value of Bell operator become $\langle\mathcal{B}_{CHSH}\rangle_{\mathrm{QM}}=\sqrt{8}=2\sqrt{2}$, as expected from Cirel'son result.

To compare the LR and the QM bound we will define the following ratio:

\begin{definition}[Bell ratio]
Given the maximum expected values $\langle\mathcal{B}\rangle_{\mathrm{QM}}$ and $\langle\mathcal{B}\rangle_{\mathrm{LR}}$, the Bell ratio is defined as
\begin{equation}
R(\mathcal{B})\equiv\frac{\langle\mathcal{B}\rangle_{\mathrm{QM}}}{\langle\mathcal{B}\rangle_{\mathrm{LR}}}.
\end{equation}
\end{definition}

This ratio quantifies the strength of the inequality generated by the Bell operator $\mathcal{B}$. Note that a Bell inequality is characterized by the ratio $R(\mathcal{B})>1$. For example, for the CHSH inequality we have $R(\mathcal{B})=\sqrt{2}$. Although we will focus on the study of this ratio, there exist other measures to compare classical and quantum values of a Bell inequality. Other works analyze the $p$ value \cite{Hensen15} or the Kullback-Leibler relative entropy \cite{Dam05}.

\subsection{Three parties}

For three parties, we first construct the most general symmetric Bell operator
\begin{equation}
\mathcal{B}_{322}=z_{0}(abc)+z_1(abc'+ab'c+a'bc)+z_2(ab'c'+a'bc'+a'b'c)+z_3(a'b'c'),
\end{equation}
where $\vec{z}=(z_{0},z_1,z_2,z_3)\in\mathbb{R}$. For $\vec{z}=(0,1,0,-1)$, the above inequality becomes the three-qubit Mermin operator \cite{Moradi09}
\begin{equation}
M_{3}=abc'+ab'c+a'bc-a'b'c'.
\label{eq:M3}
\end{equation}
Taking the square of $M_{3}$,
\begin{equation}
M_{3}^2=4-([a,a'][b,b']+[a,a'][c,c']+[b,b'][c,c']),
\end{equation}
allows us to obtain the classical value $\langle M_{3}\rangle_{\mathrm{LR}}=2$ and the quantum value $\langle M_{3}\rangle_{\mathrm{QM}}=4$ since each commutator can achieve a maximum absolute value of 2. Remember that we avoid to write identities and Kronecker products; the above expression involve three parties, so when one or more parties do not appear in the expression, an identity operator should be assumed, meaning the corresponding observer do not perform any measurement.

A different set of coefficients was proposed by Svetlichny \cite{Svetlichny87}. The choice $\vec{z}=(1,1,-1,1)$ leads to the operator
\begin{equation}
S_{3}=abc + abc'+ab'c+a'bc - (ab'c'+a'bc'+a'b'c) + a'b'c',
\label{eq:S3}
\end{equation}
which square form becomes
\begin{equation}
S_{3}^2=8-2([a,a'][b,b']+[a,a'][c,c']+[b,b'][c,c'])-\{a,a'\}\{b,b'\}\{c,c'\}.
\end{equation}
Note that this squared operator includes both commutators and anticommutators. For Pauli matrices $\lbrace\sigma_{i},\sigma_{j}\rbrace=2\delta_{ij}$, so a maximal value for the commutator implies a minimum value for the anticommutator, and vice versa. The commutators vanish while estimating the classical value and $\langle S_3\rangle_{\mathrm{LR}}=4$. For the quantum value the optimal case occurs when the commutators take the maximum amplitude $\pm2$ and the anticommutators vanish, i.e. $\langle S_{3}\rangle_{\mathrm{QM}}=4\sqrt{2}$. The ratios for the Bell operators of Eqs. \eqref{eq:M3} and \eqref{eq:S3} are given by
\begin{equation}
R(M_3)=2\,\mbox{ and }\, R(S_3)=\sqrt{2} \ .
\end{equation}
It is known that Mermin inequality generated by the Bell operator \eqref{eq:M3} can be violated by biseparable states, whereas Svetlichny inequality defined by the operator \eqref{eq:S3} can not. Bell inequalities generated by operators like $S_3$ are called \emph{multipartite Bell inequalities}. This topic is analysed in detail by Collins \emph{et. al.} \cite{Collins02bis}. 

These inequalities are already well tested experimentally. Violation of $M_3$ inequality has been reported in Ref. \cite{Pan00,Erven14}. Violation of $S_3$ inequality has been reported in Ref. \cite{Lavoie09}.

\subsection{Multipartite inequalities: Mermin polynomials}

There exists an entire family of $n$-qubit inequalities first discovered by Mermin \cite{Mermin90,Werner01}. Let us change the notation of observables $\{a,b,c...\} \equiv \{a_1,a_2,a_3...\}$, which is more convenient to treat the multipartite case. 

\begin{definition}[Mermin Polynomials]
Defining $M_1 \equiv a_1$, the Mermin polynomials are obtained recursively as
\begin{equation}
M_n=\frac{1}{2} M_{n-1}(a_n+a'_n) + \frac{1}{2}M'_{n-1}(a_n-a'_n),
\label{eq:Mermin_n}
\end{equation}
where $M'_k$ is obtained from $M_k$ by interchanging $a_n$ and $a'_n$ observables.
\label{def:Mermin}
\end{definition}

In particular, we have
\begin{align}
M_2 &=\frac{1}{2} \left(a_1 a_2 + a'_1 a_2 + a_1 a'_2 - a'_1 a'_2\right),\\
M_3 &=\frac{1}{2} \left(a_1 a_2 a'_3 + a_1 a'_2 a_3 + a'_1 a_2 a_3 - a'_1 a'_2 a'_3 \right),
\end{align}
which are actually the Bell polynomial of CHSH inequality (Eq. \eqref{eq:CHSH}) and the three-Mermin polynomial introduced in Eq. \eqref{eq:M3} up to a constant factor.

It was proven in Ref. \cite{Cereceda01} that all Mermin operators have a square form composed by the identity and commutators. Let us now proceed with our version of the proof. The square of Mermin operators can be written in terms of commutators and anticommutators as
\begin{align}
M^2_n & = \frac{1}{4} \left(M^2_{n-1}(2+\{a_n,a'_n\} )+M'^2_{n-1}(2-\{a_n,a'_n\})- [M_{n-1},M'_{n-1}][a_n,a'_n]\right), \label{eq:Mn2} \\ 
M'^2_n & = \frac{1}{4} \left(M'^2_{n-1}(2+\{a_n,a'_n\} )+M^2_{n-1}(2-\{a_n,a'_n\})- [M_{n-1},M'_{n-1}][a_n,a'_n]\right).
\end{align}
Furthermore, as $M^2_{1}=M_{1}^{'2}=1$ and assuming it is true for $M_{n}^2=M_{n}^{'2}$, for $n+1$:
\begin{align}
M^2_{n+1} & = 
\frac{1}{4}\left(4 M^2_{n}-[M_{n},M'_{n}][a_{n+1},a'_{n+1}]\right), \nonumber\\
M'^2_{n+1} & =  
\frac{1}{4}\left(4 M'^2_{n} - [M_{n},M'_{n}][a_{n+1},a'_{n+1}]\right), \nonumber\\
\Rightarrow M^2_{n+1} = M'^2_{n+1} \ .
\end{align}
So we have proved by induction that $M_{n}^2=M_{n}^{'2}$ for every $n$. Therefore, Eq. \eqref{eq:Mn2} can be simplified to
\begin{equation}
M^2_n = M^2_{n-1} - \frac{1}{4} [M_{n-1},M'_{n-1}] [a_n,a'_n],
\end{equation}
and from the definition \ref{def:Mermin},
\begin{equation}
[M_{n-1},M'_{n-1}] = [M_{n-2},M'_{n-2}] + M^2_{n-2} [a_{n-1},a'_{n-1}] \ .
\end{equation}

Given that $[M_1,M'_1]=[a_1,a'_1]$ every operator $M^2_n$ can be expressed as a sum of products of an even number of commutators. Thus the operator $M^2_n$ reads
\begin{equation}
M^2_n = 1 + \sum_{s=1}^{\left[\frac{n}{2}\right]} \frac{(-1)^s}{2^{2s}} \sum_{i_j \in D} \prod_{j=1}^{2s} [a_{i_j},a'_{i_j}],
\label{eq:Mermin2}
\end{equation}
where $D$ is the set of $n$ operators taken in groups of $2s$ elements. This result is implicitly presented in Ref. \cite{Werner01}. The classical and quantum values arise immediately. On one hand, $\langle M_n\rangle_{\mathrm{LR}}=1$, as the second term in Eq.  \eqref{eq:Mermin2} is always zero due to the presence of commutators. On the other hand, for the quantum value every commutator takes $\pm 2$, conveniently chosen to maximize the quantum value. Thus,
\begin{equation} 
\langle M^2_n\rangle_{\mathrm{QM}} = 1 + \binom{n}{2} + \binom{n}{4}+ \cdots = 2^{n-1}.
\label{msquaredquant}
\end{equation}
The quantum value for $M_n$ is, therefore, $\langle M_n\rangle_{\mathrm{QM}}=\sqrt{\langle M^2_n\rangle_{\mathrm{QM}}}=2^{\frac{n-1}{2}}$, which matches the rate computed by Werner and Wolf \cite{Werner01}. Let us note that when computing this last step it is assumed that the maximum eigenvalue of a sum of matrices is equal to the sum of the maximum eigenvalues, a fact that is not true in general but is true in this case.

The optimal states for the Mermin inequalities are the GHZ-type states \cite{Mermin90,Werner01}. For $n=2,3$ these states can be considered as maximally entangled. However, for $n\ge 4$ it is not the case \cite{Higuchi00,Scott04} if one considers the mean entropy of a reduced density matrix, averaged over all possible choices of $[n/2]$ subsystems, which define the reduced state ($[\cdot]$ denotes the integer part of a number). Therefore, Mermin inequalities 
provide an example for which the maximal violation does not correspond to maximally entangled states. Let us mention that the experimental violation of Mermin inequalities has been verified up to 14 qubits with ion traps \cite{Lanyon14} and the $M_3, M_4$ and $M_5$ cases have been implemented on a 5 superconducting qubits quantum computer designed by IBM \cite{Alsina16,GS18}.

\section{Bell inequalities for three outcomes \label{sec:BIqutrits}}

In this section we present Bell inequalities for three outcomes and two settings, i.e. $\mathcal{B}_{n23}$. These inequalities are applied to \textit{qutrit} states, i.e. quantum states of local dimension 3. We can study the quantum mechanical violation of Bell inequalities by Hermitian operators, as we did for two outcomes, or by simply unitary operators. By considering the last, we assume complex outcomes associated to the third root of unity. In this way, settings turn from Hermitian to unitary operators with eigenvalues $\{1,\omega,\omega^2\}$, where $\omega=\exp(2\pi i/3)$. Note that for qubits the Pauli matrices are both Hermitian and unitary, while for qutrits a choice between one of these properties has to be made. This particular choice could seem odd since it implies that we are measuring complex values. However, any operator that can be expressed as a linear combination (with real or complex coefficients) of rank one projectors forming a POVM allows for a physical interpretation. 

\subsection{Two parties with Hermitian operators}

Let's start with a Bell inequality written in terms of Hermitian operators as a natural generalization of the two outcomes Bell inequality. 

Collins \emph{et al.} proposed a Bell inequality for two parties, two settings and $d$ outcomes, i.e. $\mathcal{B}_{22d}$, known as CGLMP inequality \cite{CGLMP}. The violation of these inequalities have been verified experimentally \cite{Vaziri02}. In the case of three outcomes, the inequality is given by
\begin{multline}
p(a=b)+p(b=a'+1)+p(a'=b')+p(b'=a) \\
-\big(p(a=b-1)+p(b=a')+p(a'=b'-1)+p(b'=a-1)\big)\leq 2 \ ,
\label{eq:CGLMP}
\end{multline}
where $a,a',b,b'=0,1,2$ and the sum inside probabilities is modulo $d=3$. Unlike the Bell inequalities introduced in the previous section, CGLMP is given in terms of probabilities instead of expected values of operators. However, it is straightforward to write the above inequality with operators. First, it is convenient to write it again but for different outputs choice, in particular $a,a',b,b'=-1,0,1$:
\begin{multline}
p(a+b=-1) + p(a+b'=0) + p(a'+b=0) + p(a'+b'=-1) \\
-\big(p(a+b=1) + p(a+b'=1) + p(a'+b=1) + p(a'+b'=1)\big) \leq 2 \ .
\end{multline}
Then, we apply the definition of an expected value and take into account that probabilities should be normalized,
\medmuskip=0mu
\begin{align}
a&\equiv\langle a\rangle= (+1) p(a=1) + (0)p(a=0)+(-1)p(a=-1) = p(a=1)-p(a=-1), \nonumber\\
a^2&\equiv\langle a^2\rangle = (+1)^2 p(a=1) + (0)^2 p(a=0)+(-1)^2 p(a=-1) = p(a=1)+p(a=-1),\nonumber\\
1&\equiv \langle\mathbb{I}\rangle= p(a=1) + p(a=0)+ p(a=-1).
\end{align}
\medmuskip=4mu plus 2mu minus 4mu 
Finally, as stated by Bell's hypothesis, the measurements by two observers are independent, e.g. $p(a=b-1)=p(a=0)p(b=1)+p(a=1)p(b=-1)+p(a=-1)p(b=0)$. So CGLMP inequality for three outcomes can be represented by the Bell operator
\begin{multline}
\mathcal{B}_{223}=2-3(a^2+b'^2)+\frac{9}{4}\left(a^2b^2-a'^2b^2+a^2b'^2+a'^2b'^2\right) \\
+\frac{3}{4}\left(ab+a^2b-a'b-a'^2b-ab^2+a'b^2+ab'-a^2b'+a'b'+a'^2b'+ab'^2-a'b'^2\right).
\label{eq:B223}
\end{multline}
Notice that this Bell operator includes the square of settings operators. As the local dimension is 3, any operator $\mathcal{O}$ fulfills $\mathcal{O}^2\neq 1$ and $\mathcal{O}^3=\mathbb{I}$, so the complete basis needed is $\{\mathbb{I},\mathcal{O},\mathcal{O}^2\}$.

The optimal settings are computed using the method described in Ref. \cite{CGLMP,Acin02} which consist on applying a phase matrix which components are $U(\theta)_{ii}=e^{i\theta_{i}}$ and $U(\theta)_{ij}=0$ followed by a Fourier transform and maximize numerically $\langle\mathcal{B}\rangle_{\mathrm{QM}}$ to obtain the optimal $\theta_{i}$. This method is summarized in App. \ref{app:OddsEnds} and the values found for this inequality are $\vec{\theta}_{a}=\vec{0}$, $\vec{\theta}_{a'}=(0,\pi/3,2\pi/3)$, $\vec{\theta}_{b}=(0,-\pi/6,-\pi/3)$ and $\vec{\theta}=(0,\pi/6,\pi/3)_{b'}$. The corresponding maximal violation found is $\langle\mathcal{B}\rangle_{\mathrm{QM}}=2(5-\gamma^2)/3\simeq 2.92$ for the optimal state $|\psi\rangle=(|00\rangle+\gamma|11\rangle+|22\rangle)/\sqrt{2+\gamma^2}$ where $\gamma=(\sqrt{11}-\sqrt{3})/2\simeq 0.79$. The violation rate for this quasi-Bell state reads $R_{223}=(5-\gamma^2)/3\simeq 1.46$. In Ref. \cite{Acin02} the ratios for CGLMP inequalities are found up to $d=8$ levels.

The optimal settings can be conveniently expressed in terms of the eight Gell-Mann matrices $\lambda_{i}$, the traceless generators of SU(3) \cite{Gell-Mann62} that are defined in App. \ref{app:Feynman}. The optimal settings for the Bell inequality $\mathcal{B}_{223}$ are
\begin{align}
&A=B=\lambda_3 \ , \nonumber\\
&A'=B'=\frac{2}{3}\left(\lambda_{1}+\lambda_{6}\right)+\frac{1}{6}\left(\lambda_{3}+\sqrt{3}\lambda_{8}\right),
\end{align}
where we have used capital letters to remark that these are the settings that lead to the maximal violation of the inequality.

\subsection{Two parties with unitary operators}

The Bell operator of Eq. \eqref{eq:B223} has a rather long and unenlightening form. In this section, we consider a different form of this inequality by using unitary non-Hermitian operators. Any operator $\mathcal{O}$ can be decomposed into its \emph{Hermitian} and \emph{anti-Hermitian} part, i.e. $\mathcal{O}=\mathcal{O}_{H}+i\mathcal{O}_{A}$ where
$\mathcal{O}_{H}=\frac{1}{2}\left(\mathcal{O}+\mathcal{O}^\dagger\right)$ and  $\mathcal{O}_{A}=\frac{1}{2i}\left(\mathcal{O}-\mathcal{O}^\dagger\right)$
are Hermitian operators and, therefore, have real eigenvalues. Similarly, we can apply the same definition to Bell operators.
\begin{definition}[Hermitian and anti-Hermitian Bell operators]
Given a Bell operator $\mathcal{B}$, we can decompose it into its Hermitian and anti-Hermitian parts,
\begin{equation}
\mathcal{B}_{H}\equiv\frac{1}{2}\left(\mathcal{B}+\mathcal{B}^\dagger\right),\quad \mathcal{B}_{A}\equiv\frac{1}{2i}\left(\mathcal{B}-\mathcal{B}^\dagger\right),
\end{equation}
each one having real eigenvalues.
\end{definition}

It turns out that $\mathcal{B}_{223}$ operator of Eq. \eqref{eq:B223} can be written in a more elegant form by using the anti-Hermitian part of a non-Hermitian operator,
\begin{equation}
\mathcal{B'}_{223}=\left[a(\omega b-b')+a'(\omega b'-b)\right]_{A},
\label{eq:B223A}
\end{equation}
where $\omega=e^{\frac{2\pi i}{3}}$. This form appears to be a direct generalization of the CHSH operator \eqref{eq:CHSH} with different signs and relative phases added. One can check that if one of the terms reaches the maximum value $\sqrt{3}$, the other one is forced to be zero. The classical and quantum values ar $\langle\mathcal{B'}_{223}\rangle_{\mathrm{LR}}=\sqrt{3}\simeq 1.73$ and $\langle\mathcal{B'}_{223}\rangle_{\mathrm{QM}}=(\sqrt{3}+\sqrt{11})/2\simeq 1.45$, and the ratio is given by $R(\mathcal{B'}_{223})=(5-\gamma^2)/3\simeq 1.46$. The violation rate is therefore the same as for CGLMP inequality as expected, because it is the same inequality albeit written in a different language.

Let us now find the optimal settings for the operator of Eq. \eqref{eq:B223A}. The convenient representation for unitary operators are the generalized Pauli matrices which form the Weyl-Heisenberg group. For $d=3$ these matrices are
\begin{equation}
X=\left(\begin{matrix}
0&0&1\\1&0&0\\0&1&0
\end{matrix}\right), \qquad Z=\left(\begin{matrix}
1&0&0\\0&\omega&0\\0&0&\omega^2
\end{matrix}\right).
\end{equation}
An orthonormal basis is given by the nine elements
\begin{equation}
X^{j}Z^{k}=\sum_{i=0}^{2}|i+j\rangle\omega^{ik}\langle k| \ .
\end{equation}
By numerical optimization we found that the optimal settings are
\begin{align}
A&=B=X \ ,\nonumber\\
A'&=B'=\frac{1}{3}\left(-X+2\omega XZ+2\omega^2 XZ^2\right).
\end{align}
In matrix notation, $A'$ has a simple structure
\begin{equation}
A'=\left(\begin{matrix}
0&0&1\\-1&0&0\\0&-1&0
\end{matrix}\right).
\end{equation}
The optimal settings for all the complex CGLMP inequalities, when take the form $\{A=B=X, A'=B'\}$, are called \emph{multiplets of optimal settings} (MOS) and are briefly described in App. \ref{app:OddsEnds}. The $d$ dimensional CGLMP inequalities are discussed in more detail in Sec. \ref{sec:CGLMPd}.

Let us investigate the square of the operator $\mathcal{B}_{223}$ introduced in \eqref{eq:B223}. Making use of the identity for the Hermitian and anti-Hermitian parts of an operator $\mathcal{O}$,
\begin{equation}
 (\mathcal{O}_A)^2 = \frac{1}{4}(\mathcal{O}\mathcal{O}^\dagger + \mathcal{O}^\dagger\mathcal{O}) - \frac{1}{2} (\mathcal{O}^2)_H,
\label{cglmpim2}
\end{equation}
it is easy to show that $\mathcal{B}_{223}\mathcal{B}_{223}^\dagger$ has an interesting structure:
\begin{equation}
\mathcal{B}_{223}\mathcal{B}_{223}^\dagger = 3 + (1+\{\{a,a'\}\})(1+\{\{b,b'\}\}).
\label{eq:B223sq}
\end{equation}
Here we call $\{\{a,a'\}\}$ \emph{complex anticommutator}: 
\begin{equation}
\{\{a,a'\}\}= aa'^\dagger + a'a^\dagger,
\end{equation}
and attains its maximum value 2 both for MOS and \emph{Mutually Unbiased Bases} (for details, see App. \ref{app:OddsEnds}). However, its classical value can also be equal to 2 by using $a=a'=1$. Thus the form \eqref{eq:B223sq} does not allow us to distinguish between classical and quantum values.

\subsection{Three parties}

A three-party Bell inequality was proposed by Ac\'in \textit{et al.} in Ref. \cite{Acin04}. 
In the probability formalism it reads 
\begin{multline}
p(a+b+c=0) + p(a+b'+c'=1) +p(a'+b+c'=1)+p(a'+b'+c=1) \\
-2p(a'+b'+c'=0) - p(a'+b+c=2) - p(a+b'+c=2)- p(a+b+c'=2) \leq 3 \ .
\label{eq:BI_3qutrits}
\end{multline}

The analysis here is very similar to the CGLMP case: the maximal violation is given by a quasi maximally entangled state $|\psi\rangle= (|000\rangle+\delta |111\rangle+|222\rangle)/\sqrt{2+\delta^2}$, where now $\delta\simeq 1.186$. The quantum value is approximately $4.37$ and the violation rate is $R=(5-\gamma^2)/3\simeq 1.46$, the same as for 2 qutrits. The corresponding Hermitian Bell operator has a rather long form, so we will not reproduce it here. 

The optimal settings can be expressed in terms of the Gell-Mann matrices as
\begin{align}
A&=B=C=\lambda_3,\nonumber \\
A'&=B'=C'=\frac{1}{\sqrt{3}} (\lambda_2+\lambda_4+\lambda_6) .
\end{align}

Let us now consider the case of unitary settings having complex eigenvalues. The Bell operator associated to inequality \eqref{eq:BI_3qutrits} can be expressed as the Hermitian part of an operator,
\begin{align}
\mathcal{B}_{333}&=1+\frac{2}{3}  \left[abc +2a'b'c' +\omega(a'b'c+a'b c'+ab'c')-\omega^2(a'b c+a b'c +a b c')\right]_H.
\end{align}
One can also drop the additive and multiplicative terms and study the simplified operator
\begin{align}
\mathcal{B'}_{333}&=\left[abc +2a'b'c' +\omega(a'b'c+a'b c'+ab'c')-\omega^2(a'b c+a b'c +a b c')\right]_H.
\label{eq:3qutrits}
\end{align}
Here, the classical value is $\langle\mathcal{B'}_{333}\rangle_{\mathrm{LR}}=3$ and the quantum value is $\langle\mathcal{B'}_{333}\rangle_{\mathrm{QM}}=(3/4)(1+\sqrt{33})\simeq 5.058$, which yields to the ratio $R(\mathcal{B'}_{333}) = (1/4)(1+\sqrt{33})\simeq 1.686$. The optimal settings are given by
\begin{align}
A&=B=C=X,& \nonumber \\
A'&=B'=C'=Z.
\end{align}
Note that the settings are mutually unbiased (see App. \ref{app:OddsEnds}). Now the violation rate is greater because the additive constant term has been eliminated. This appears somewhat arbitrary but it is more convenient to compare  inequalities  for two and three qutrits without additive terms. In this way, it is expected that the rate of violation increases with the number of particles, as it happens for qubits. 

Intriguingly, the 3-qutrit operator \eqref{eq:3qutrits} can be derived from the 2-qutrit CGLMP operator \eqref{eq:B223A} and adding a third party such that the resulting 3-qutrit operator is symmetric. Starting from Eq. \eqref{eq:B223A},
\begin{align}
\left[\omega (ab) - (a'b+ab') + \omega (a'b')\right]_A & \leq \sqrt{3}, \nonumber\\ 
\left[-i (\omega (ab) - (a'b+ab') + \omega (a'b'))\right]_H &\leq \sqrt{3}, \nonumber\\
\left[\frac{\omega^2-\omega}{\sqrt{3}} (\omega (ab) - (a'b+ab') + \omega (a'b'))\right]_H &\leq \sqrt{3}, \nonumber\\
\left[(1-\omega^2)(ab)+(\omega-\omega^2)(a'b+ab')+ (1-\omega^2)(a'b'))\right]_H &\leq 3, \nonumber\\
\left[(ab)-\omega^2(ab+a'b+ab') + \omega(a'b+ab')+ (\omega+2)(a'b')\right]_H &\leq 3, \nonumber\\ 
\left[(ab)-\omega^2(ab+a'b+ab') + \omega(a'b+ab'+a'b')+ 2(a'b')\right]_H &\leq 3. 
\end{align}
This form of the 2-qutrit CGLMP inequality suggests an 8-term symmetric inequality for three qutrits, where all terms with the same number of primes should have the same coefficients. By inserting $c$ and $c'$ according to this last requirement we have
\begin{equation}
\left[(abc)-\omega^2(abc'+a'bc+ab'c)+ \omega(a'bc'+ab'c'+a'b'c)+2(a'b'c')\right]_H \leq 3,
\end{equation}
which is actually the symmetric 3-qutrit inequality of Eq. \eqref{eq:3qutrits}.

\subsection{Larger number of parties}

In the case of four parties, two settings and three outcomes we have found the following symmetric Bell operator
\begin{align}
\mathcal{B}_{423}= \Big[&2(abcd) + (a'bcd+ab'cd+abc'd+abcd') \nonumber\\ &+\omega(a'b'cd+a'bc'd+a'bcd'+ab'c'd+ab'cd'+abc'd') \nonumber\\
&+(a'b'c'd+a'bc'd'+a'b'cd'+ab'c'd')+2 (a'b'c'd')\Big]_A,
\label{eq:B423}
\end{align}
which produces $\langle \mathcal{B}_{423}\rangle_{\mathrm{LR}}=3\sqrt{3}\simeq 5.19$, $\langle \mathcal{B}_{423}\rangle_{\mathrm{QM}}\simeq 9.77$ and $R(\mathcal{B}_{423})\simeq 1.879$ for the optimal settings
\begin{align}
A&=B=C=D=X, \nonumber \\
A'&=B'=C'=D'=Z,
\end{align}
which are again mutually unbiased settings. The optimal state is the GHZ of four parties and dimension 3 $|\Psi_{4}^{+}\rangle=(|0000\rangle+|1111\rangle+|2222\rangle)\sqrt{3}$. The possibility to construct Bell inequalities maximally violated by maximally entangled states is discussed in Sec. \ref{sec:mapping}.

For 6 parties we have also found a symmetric Bell operator. To simplify the notation, the polynomials having terms with the same number of primes are denoted by its number of primes in parenthesis, for example
\begin{equation}
(1') \equiv a'bcdef+ab'cdef+abc'def+abcd'ef+abcde'f+abcdef',
\label{eq:primenotation}
\end{equation}
$(2')$ will be composed by all combinations containing to primed setting, etc.
In this notation, the 6 parties operator reads
\begin{equation}
\mathcal{B}_{623}=-\omega(0')+(1')-(2')+\omega(3')-(4')+(5')-\omega(6').
\label{eq:B623}
\end{equation}
For this inequality, $\langle \mathcal{B}_{623}\rangle_{\mathrm{LR}}=9\sqrt{3}\simeq 15.59$, $\langle \mathcal{B}_{623}\rangle_{\mathrm{QM}}\simeq 32.82$ and $R(\mathcal{B}_{623})\simeq 2.11$, with MOS optimal settings. The maximal violation is given by a \emph{quasi} GHZ state, as for the case of 2 and 3 qutrits. 

Let us summarize the results for the symmetric Bell operators for $n$-qutrit systems studied in this section. Unfortunately,  we could not find a 5-qutrit inequality that follows similar patterns. The inequalities considered are those determined by the coefficients of Tab. \ref{tab:BIqutrits_coef}, and the results are summarized in Tab. \ref{tab:BI_qutrits}.

\begin{table}[t!]
\centering
\begin{tabular}{c  c  c  c  c  c}
\toprule
\backslashbox{\textbf{Terms}}{\textbf{Parties}} & \textbf{2} & \textbf{3} & \textbf{4} & \textbf{5}  & \textbf{6} \\
\midrule
(0') & $\omega$ & $1$    & $2$   & $\omega^2$  & $-\omega$ \\
(1') & $1$ & $-\omega^2$ & $1$   & $-\omega^2$ & $1$ \\
(2') & $\omega$ & $\omega$    & $\omega$   & $-\omega^2$ & $-1$ \\
(3') &     & $2$    & $1$   & $-\omega^2$ & $\omega$ \\ 
(4') &     &        & $2$   & $\omega^2$  & $-1$ \\  
(5') &     &        &       & $\omega^2$  & $1$ \\
(6') &     &        &       &        & $-\omega$ \\
\bottomrule
\end{tabular}
\caption{Coefficients for symmetric Bell inequalities of the form $\mathcal{B}_{n23}$ with $n=2,\cdots,6$, where $\omega=e^{2\pi i/3}$. The primed notation $(k')$ identifies all terms having $k$ primed settings, as the example given in Eq. \eqref{eq:primenotation}.}
\label{tab:BIqutrits_coef}
\end{table}

The main patterns that can be seen from Tab. \ref{tab:BI_qutrits} are
\begin{itemize}
\item[\emph{(i)}] 
For an even number of qutrits the classical values $\langle B\rangle_{\mathrm{LR}}$ arise from the anti-Hermitian part of an operator while for odd number of qutrits one takes its Hermitian part. The following relation between the minimal and the maximal classical values holds
\begin{equation}
\langle\mathcal{B}\rangle^{min}_{\mathrm{LR}}=-2\langle\mathcal{B}\rangle_{\mathrm{LR}},
\end{equation}
where $\langle\mathcal{B}\rangle^{min}_{\mathrm{LR}}$ corresponds with the minimum value found and $\langle\mathcal{B}\rangle_{\mathrm{LR}}$ is the maximum value.

\item[\emph{(ii)}] 
There is a factor of $\sqrt{3}$ between the maximal value of the Hermitian and anti-Hermitian parts. There is also a factor of $\sqrt{3}$ between the maximal value of two consecutive numbers of qutrits. The maximal value of the Hermitian parts are the same for $n$ and $n+1$ qutrits if $n$ is even, result that is reproduced by the anti-Hermitian parts if $n$ is odd.

\item[\emph{(iii)}] 
The quantum value $\langle B\rangle_{\mathrm{QM}}$ of a non-Hermitian operator $\mathcal{B}$ is computed as the maximum over quantum values of the Hermitian and anti-Hermitian parts, i.e.
\begin{equation}
\langle\mathcal{B}\rangle_{\mathrm{QM}}=\mathrm{max}\{\langle\left[\mathcal{B}\right]_H\rangle_{\mathrm{QM}},\langle\left[\mathcal{B}\right]_A\rangle_{\mathrm{QM}}\}.
\end{equation}
The rate of violation increases with the number of qutrits except for the 5-qutrit case, which do not follow the same patterns as the other inequalities studied.

\item[\emph{(iv)}] 
The optimal settings are either MUB or MOS, with the exception of the 5-qutrit inequality.

\item[\emph{(v)}] 
The optimal states have entanglement properties close to a GHZ state or exactly those of a GHZ state. In Tab. \ref{tab:BI_qutrits}, the closeness to the GHZ state is measured by the purity of the reduced matrix $\rho$ over $\lfloor n/2 \rfloor$ particles. The GHZ state of $n$ qutrits has reductions to two parties with purity \footnote{The purity of subsystems from a composite state is an entanglement measure, being maximal, $P=1$, if the state is separable and minimal, $P=1/d$ if all possible bipartitions are maximally entangled. It is defined in next Chapter \ref{Ch:HDet} in definition \ref{def:purity}. Here, its label $\gamma$ has been changed to $P$ to not get confused with $\gamma$ factor found in inequality \eqref{eq:B223}} $P={\rm Tr} \rho^2=1/3$, whereas an absolutely maximally entangled state, i.e. an state with all its bipartitions maximally entangled, has $P=1/3^{[n/2]}$.
\end{itemize}

\begin{table}[t!]
\centering
\begin{tabular}{c  c  c  c  c  c}
\toprule
\textbf{Qutrits} & \textbf{2} & \textbf{3} & \textbf{4} & \textbf{5} & \textbf{6} \\
 \midrule
$\langle B_A\rangle_{\mathrm{LR}}$ & \boldmath{$\sqrt{3}$} & $3\sqrt{3}$ & \boldmath{$3\sqrt{3}$} & $9\sqrt{3}$&  \boldmath{$9\sqrt{3}$}\\
$\langle B_A\rangle_{\mathrm{LR}}^{min}$ & $-2\sqrt{3}$ & $-3\sqrt{3}$ & $-6\sqrt{3}$ & $-9\sqrt{3}$ & $-18\sqrt{3}$\\
$\langle B_H\rangle_{\mathrm{LR}}$ & $3$ & \boldmath{$3$} & $9$ & \boldmath{$9$} & $27$\\
$\langle B_H\rangle_{\mathrm{LR}}^{min}$ & $-3$ & $-6$ & $-9$ & $-18$ & $-27$ \\
$\langle B\rangle_{\mathrm{QM}}$ & $ 2.524 $ & $5.058$ & $9.766$ & $15.575$ & $32.817$ \\
$\mathrm{R}$ & $1.457$ & $1.686$ & $1.879$ & $1.731$ & $2.105$ \\
$\mathit{Settings}$ & MOS & MUB & MUB & \it{Num.} & MOS \\
P & $0.347$ & $0.342$ & $1/3$ & $0.351$ & $0.334$\\ 
\bottomrule
\end{tabular}
\caption{Main results for inequalities from 2 to 6 qutrits.
Here, $\langle\mathcal{B}\rangle_{\mathrm{LR}}$ and $\langle\mathcal{B}\rangle_{\mathrm{LR}}^{min}$ denote, respectively, the maximum and minimum classical values for optimizations of anti-Hermitian and Hermitian parts of an operator. The quantity that we take as the extremal classical bound is marked in bold.
$\langle\mathcal{B}\rangle_{\mathrm{QM}}$ stands for the quantum value, $R(\mathcal{B})$ is the rate between LR and QM maximum values and \emph{Settings} denotes the optimal settings. Moreover, we compute the purity $P$ of the single party reductions of the optimal state. The 5-qutrit inequality do not follow the same patters of the others, which is remarked with the fact that the optimal settings found are a numerical approximate solution (\emph{Num}).
}
\label{tab:BI_qutrits}
\end{table}

\section{Bell inequalities for two parties and arbitrary dimension \label{sec:CGLMPd}}

In this section, we extend the results found in Sec. \ref{sec:BIqutrits} for two parties and $d$ outcomes. These are actually the set of CGLMP inequalities proposed in Ref. \cite{CGLMP}. In probability language can be written as

\begin{align}
\mathcal{B}_{22d} =\sum_{k=0}^{[d/2]-1} \left( 1-\frac{2k}{d-1} \right)&\Bigg(p(a=b+k)+p(b=a'+k+1)+ p(a'=b'+k) \nonumber \\
& +p(b'=a+k) -\Big(p(a=b-k-1) + p(b=a'-k) \nonumber\\
&+p(a'=b'-k-1) + p(b'=a-k-1)\Big)\Bigg)\leq 2 \ .
\end{align}
 
Let us write these inequalities in term of operators. In order to do this let us start from a different form for CGLMP inequality for $d=3$ \eqref{eq:B223A} presented for example in Ref. \cite{Chen02},
\begin{equation}
\mathcal{B''}_{223} = \left[ab+ab'+a'b-a'b'\right]_H +\frac{1}{\sqrt{3}} \left[-ab+ab'+a'b-a'b'\right]_A \leq 2 \ .
\label{eq:B223bis}
\end{equation}     
In order to transform from probabilities to operators we have to establish a match between the number of variables and the number of equations. The variables here are the joint probabilities $p (a=b+k)$, with $k$ running from 0 to $d-1$, so there are $d$ unknowns. We need therefore $d$ equations. One equation is given by the normalization condition, i.e., the sum of probabilities is 1. For $d=2$, a second equation is enough, and that is the definition of expectation value of the product. For $d=3$ there are 3 equations. Apart from the normalization of probabilities, two extra equations are needed, and those can be the Hermitian and anti-Hermitian parts of the expected value of the product, as in Eq. \eqref{eq:B223bis}. It appears to be an accident that the CGLMP for $d=3$ can be expressed solely with the anti-Hermitian part by inserting powers of $\omega$ as in Eq. \eqref{eq:B223A}.

For $d=4$ we add the Hermitian part of the expected values of the squares of products, 
\begin{multline}
\mathcal{B'}_{224}=\frac{1}{3}\Big( 2  \left[ab+ab'+a'b-a'b'\right]_H + 2 \left[-ab+ab'+a'b-a'b'\right]_A  \\
+  \left[(ab)^2+(ab')^2+(a'b)^2-(a'b')^2\right]_H \Big),
\label{eq:B224}
\end{multline}
and for $d=5$ we add their anti-Hermitian part,
\begin{multline}
\mathcal{B'}_{225}=\frac{1}{2}\Big(\left[ab+ab'+a'b-a'b'\right]_H + \left[(ab)^2+(ab')^2+(a'b)^2-(a'b')^2 \right]_H\Big) \\
\hspace{4cm}+(-s_1+3s_2)\left[-(ab)^2+(ab')^2+(a'b)^2-(a'b')^2\right]_A \\
+ \frac{2}{5} \Big( (3s_1+s_2)\left[-ab+ab'+a'b-a'b'\right]_A \Big),
\label{eq:B225}
\end{multline} 
where $s_1$ and $s_2$ are the imaginary parts of $e^{2\pi i/5}$ and $e^{4\pi i/5}$ respectively. The classical values for these operators are $\langle\mathcal{B}_{224}\rangle_{\mathrm{LR}}=2$ and $\langle \mathcal{B}_{225}\rangle_{\mathrm{LR}}=2$.

In general, for any number of outcomes $d$, the Bell operator can be written as
\begin{equation}
\mathcal{B'}_{22d} = N \left( \sum_{k=1}^{[d/2]} r_{k,d}\mathrm{H}_{(ab)^k} + \sum_{k=1}^{[(d-1)/2]} i_{k,d} \mathrm{A}_{(ab)^k} \right) \le 2,
\end{equation}
where $r_{k,d}$ and $i_{k,d}$ are constants related to real and imaginary parts of $\omega$ (in general related to both of them), $N$ is a normalization constant such that the maximal classical value of $\mathcal{B'}_{22d}$ is 2, and
\begin{align}
\textrm{H}_{(ab)^k}&\equiv \left[(ab)^k+(ab')^k+(a'b)^k-(a'b')^k \right]_H,  \\
\textrm{A}_{(ab)^k}&\equiv  \left[-(ab)^k+(ab')^k+(a'b)^k-(a'b')^k \right]_A.
\end{align}

All these inequalities are maximally violated by $d$-dimensional MOS as defined in App. \ref{app:OddsEnds}. The numerical violation ratios increase with $d$, and can be found for example in Ref. \cite{Acin02}.

\section{Bell inequalities from maximally entangled states\label{sec:mapping}}

So far, we have seen how Bell inequalities are maximally violated by maximally entangled states or almost maximally entangled states. For instance, $\mathcal{B}_{CHSH}$ upper quantum bound is obtained with a singlet state, $\mathcal{B}_{423}$ with a GHZ state an all other exposed inequalities by a small deformation of a GHZ state. The connection between maximal entanglement and non-locality is manifested although they are not equivalent phenomena \cite{Acin1,Acin2}. However, let us now present an idea to generate Bell inequalities based on a mapping from maximally entangled states to Bell operators.

Let us start with a simple example. The two-qubit state
\begin{equation}
| \psi \rangle = \frac{1}{2} \left( |0_A0_B\rangle + |0_A1_B\rangle + |1_A0_B\rangle - |1_A1_B\rangle \right)
\label{eq:mes}
\end{equation}
belongs to the set of maximally entangled Bell states. The CHSH Bell operator can be obtained from this state by identifying first and second particle with observables for Alice and Bob, respectively. We identify symbol $0$ with non-primed settings and symbol $1$ with primed settings, i.e.
\begin{equation}
\begin{array}{ c c  c c c c}
   | 0_A \rangle & \rightarrow & a \ , & | 0_B \rangle & \rightarrow & b \ ,\\   
   | 1_A \rangle & \rightarrow & a' \ , &
   | 1_B \rangle & \rightarrow & b' \ .
\end{array}
\label{eq:legend}
\end{equation}

By removing the normalization term, the CHSH operator arises
\begin{equation}
\mathcal{B}_{CHSH} = ab + ab' + a'b - a'b' .
\end{equation}
Furthermore, the maximally entangled state of Eq. \eqref{eq:mes} is the optimal state for a suitable choice of the measurement settings. This fact motivates us to study new multipartite Bell inequalities generated from multipartite quantum states.

The general strategy is to construct Bell inequalities associated to some distinguished maximally entangled states. Starting from the Bell state for two qutrits,
\begin{equation}
|\Psi_3^+\rangle=\frac{1}{\sqrt{3}}(|00\rangle+|11\rangle+|22\rangle),
\end{equation}
and applying the Fourier transform \eqref{eq:FT2} to the second party we obtain 
\begin{align}
|\Phi_{3}\rangle&=\left(\mathbb{I}\otimes F_{3}\right)|\Psi_3^+\rangle \nonumber\\
&= \frac{1}{\sqrt{3}}\left(|00\rangle+|01\rangle+|02\rangle + |10\rangle + \omega|11\rangle+\omega^2|12\rangle+|20\rangle+\omega^2|21\rangle+\omega^4|22\rangle\right),
\label{eq:IF_GHZ}
\end{align}
where $\omega=e^{2\pi i/3}$ and, therefore, $\omega^4=\omega$. From this state, a new Bell operator for 2 qutrits and 3 settings arises,
\begin{align}
\mathcal{B}_{233} &=\left[ab+ab'+ab''+a'b+a''b+\omega\left(a'b'+a''b''\right)+ \omega^2\left(a'b''+a''b'\right)\right]_{H} \nonumber\\
&= \left[\,\vec{a}\cdot (F_3\vec{b})\,\right]_H,
\end{align}
where $\vec{a}=(a,a',a'')$ and $\vec{b}=(b,b',b'')$. This operator has a classical value $\langle \mathcal{B}_{233}\rangle_{\mathrm{LR}}=9/2$ and it is maximally violated by a GHZ state with the violation ratio $R(\mathcal{B}_{233})=2/\sqrt{3} \cos (\pi/18)\simeq 1.14$ for the MUB optimal settings
\begin{align}
A&=B=X \ , \nonumber \\
A'&=B'=Z \ , \nonumber \\
A''&=B''=X^2 Z^2 \ .
\end{align}

We can apply the same strategy for four qutrits starting with the GHZ state $|\Psi_{4}^{+}\rangle=(|0000\rangle + |1111\rangle +|2222\rangle)/\sqrt{3}$. Acting with Fourier transform $F_3$ on three parties we obtain a locally equivalent state
\begin{equation}
|\Phi_{4}\rangle=\left(\mathbb{I}\otimes F_3\otimes F_3\otimes F_3\right)|\Psi_{4}^{+}\rangle,
\label{eq:IFFF_GHZ}
\end{equation}
which leads to the Bell operator of four parties,three settings and three outputs
\begin{equation}
\mathcal{B}_{433} = \left[\,\vec{a}\cdot (F_3\vec{b})\cdot (F_3\vec{c})\cdot (F_3\vec{d})\,\right]_H,
\end{equation}
where $\vec{a}=(a,a',a'')$, $\vec{b}=(b,b',b'')$, etc., and the generalized inner product of four vectors is defined as $x_{1}\cdot x_{2}\cdot x_{3}\cdot x_{4}=\sum_{j=0}^{2} x_{1}^j x_{2}^j x_{3}^j x_{4}^j$. The optimal state is precisely $|\Psi_{4}^{+}\rangle$ with a larger violation ratio than for the operator \eqref{eq:B423}.

\begin{table}[t!]
\centering
\begin{tabular}{c  c  c  c }
\toprule
\textbf{Qutrits} & \textbf{2} & \textbf{4(}$\mathbf{\mathit{|\Psi_{4}^{+}\rangle}}$\textbf{)} & \textbf{4(}$\mathbf{\mathit{|\Omega_{4,3}\rangle}}$\textbf{)} \\
\midrule
$\langle \mathcal{B}_A\rangle_{\mathrm{LR}}$ & $3\sqrt{3}$ & $9\sqrt{3}$ & $9\sqrt{3}$ \\
$\langle \mathcal{B}_A\rangle_{\mathrm{LR}}^{min}$ & $-3\sqrt{3}$ & $-9\sqrt{3}$ & $-9\sqrt{3}$\\
$\langle \mathcal{B}_H\rangle_{\mathrm{LR}}$ & \boldmath{$4.5$} & \boldmath{$13.5$} & \boldmath{$13.5$} \\
$\langle \mathcal{B}_H\rangle_{\mathrm{LR}}^{min}$ & $-4.5$ & $-27$ & $-27$ \\
$\langle \mathcal{B}\rangle_{\mathrm{QM}}$ & $ 5.117 $ & $26.025$ & $25.372$ \\
R & $1.137$ & $1.928$ & $1.879$\\
\textit{Settings} & MUB & \it{Num.} & MUB and \it{Num.} \\
$P$ & 1/3 & 1/3 & 1/3 \\
\bottomrule
\end{tabular}
\caption{Characterization of Bell inequalities for 2 and 4 parties, 3 settings and 3 outcomes. For all the cases the optimal states are generalized Bell states (two parties) and GHZ
states (four parties). Abbreviations and symbols are considered as in Tab. \ref{tab:BI_qutrits}.}
\label{tab:settings}
\end{table}

\subsection{Bell inequalities from AME states}

An absolutely maximally entangled state (AME) of $n$ particles is a state with every reduction, up to $\lfloor n/2 \rfloor$ particles, maximally mixed  \cite{Helwig12,Goyeneche15,Gaeta15,Helwig13}. For more details about AME states and their properties, see Chapter \ref{Ch:AME}. Let us now try the strategy described above for the AME of 4 qutrits
\begin{equation}
|\Omega_{4,3}\rangle=\frac{1}{9}\sum_{i,j,k,l=0}^2 \omega^{j(i-k)+l(i+k)}|ijkl\rangle,
\label{eq:AME43_BI}
\end{equation}

The recipe to construct the Bell operator consists in taking representation \eqref{eq:AME43_BI} which contains $3^4=81$ terms with coefficients of the form $\{1,\omega,\omega^2\}$. In the next step one uses the same legend as the one introduced in Eq. \eqref{eq:legend} with the terms $|2\rangle_{M} \rightarrow m''$, as it was done above for GHZ inequalities. This procedure leads us to a Bell operator for four parties, three settings and three outcomes, which can be written in a compact way as
\begin{equation}
\mathcal{B'}_{433} = \sum_{i,j,k,l=0}^2 \omega^{j(i-k)+l(i+k)}a_{i}b_{j}c_{k}d_{l} \ .
\label{eq:Bp433}
\end{equation}
where $a_0 = a, a_1 = a', a_2 = a''$, and the same for the rest of the observables.

After transformations $d' \rightarrow  \omega d'$ and $d' \leftrightarrow d''$, 
numerical optimization produces the following optimal settings 
\begin{align}
A &=B=C=D=X \ ,        \nonumber\\
A' &=C'=D'=X^2Z^2 \ ,   \qquad   B'=X \ ,  \nonumber\\
A'' & =C''=D''=Z \ ,    \qquad\quad  B''=\mathcal{N} \ ,
\end{align}
where $\mathcal{N}$ is certain  matrix obtained numerically. The optimal settings are not symmetric because the AME state is not symmetric under interchange of particles.

Numerical optimization suggests that the optimal state is not AME. Surprisingly, it has almost the same entanglement properties as the GHZ state, namely its purity is $P=1/3$ for the density matrices of reductions to 2 parties, and $P=1/3$ for three of the possible reductions to one party, while the fourth one (party B) has $P=1$, indicating that party B is in a product state with the other three. The same violation ratio as for four qutrits inequality of Eq.  \eqref{eq:B423} is obtained. This result, and the fact that the optimal settings include $B=B'$ suggests that the third setting is not adding anything new and that this inequality is essentially the same as in the case of two settings.

Finally,  Tab. \ref{tab:settings} summarizes the results for the 3-settings qutrit inequalities arising from entangled states.

\section{Conclusions \label{sec:conclusionsBI}}

We have used the formalism of unitary matrices with complex roots of unity as eigenvalues to express known Bell inequalities in a different way. We have also used this formalism to construct novel Bell inequalities of multipartite systems, three settings and three outcomes. We have shown that the two-party and three-party inequality from Ref. \cite{CGLMP} and Ref. \cite{Acin04} are closely related. Furthermore, we have extended these cases to 4 and 6 parties and, less convincingly, to 5 parties. We obtained regular patterns for this set of inequalities, as shown in Tab. \ref{tab:BI_qutrits}. Two of the most striking patterns are a) the structure of the classical bounds and a simple arithmetic progression of the number of particles, and b) the fact that the inequalities tend to have a maximal quantum bound for settings that are either MUBs or multiplets of optimal settings (MOS).

We have also introduced a mapping from entangled states to Bell operators that enable us to build some new Bell inequalities for qutrits. In particular, we have constructed inequalities for two and four parties with three settings which are maximally violated by states with the same entanglement properties as the GHZ state. We have also shown that a Bell inequality generated by a given quantum state is not necessarily maximally violated by the same state. We gave as an example of this fact the inequality \eqref{eq:Bp433} generated by an absolutely maximally entangled state that is actually maximally violated by a GHZ-like state. This method has the potential to generate a wide range of Bell inequalities for an arbitrarily large number of parties, settings and outcomes.

Let us also mention here some important questions that remain open. Concerning the approach to Bell inequalities from squares of operators represented by commutators and anticommutators, it would be interesting to find a procedure to determine whether a given Bell operator allows such a form.  By analyzing the mapping between states and Bell operators, one can raise the question of whether a maximally entangled state is necessary to produce a tight Bell inequality in the case of two outcomes. In addition, the mathematical characterization of the entire set of MOS is a pending task. Finally, it would be interesting to have a generating polynomial for Bell inequalities with three outcomes in the same way that we have the Mermin polynomials for two outcomes.


\chapterimage{Hdet} 
\chapter{Hyperdeterminant in Spin Chains \label{Ch:HDet}}


\vspace{-1.5cm}
\begin{flushright}
\begin{minipage}{0.6\textwidth}
\textit{The miracle of the appropriateness of the language of mathematics for the formulation of the laws of physics is a wonderful gift which we neither understand nor deserve.}
\begin{flushright}
--Eugene P. Wigner, \\
``The unreasonable effectiveness of mathematics in the natural sciences'', 1960.
\end{flushright}
\end{minipage}
\end{flushright}
\vspace{1cm}

Entanglement has been extensively studied in the context of condensed matter quantum systems \cite{EntCM15}. It has proven useful to provide a deeper understanding of quantum phase transitions, as well as to validate the faithfulness of numerical approximations such as tensor networks \cite{Cirac09}. 

Most of the studies of entanglement are related to correlations among bipartitions of a system. As a relevant example, we may consider the quantum correlations between two separate parts of a quantum system on a lattice using entanglement entropy as a figure of merit. It has been found that most systems of interest obey the so-called area law for the scaling of the entanglement entropy as the size of the part increases \cite{Bombelli86,Srednicki93,Amico08,Eisert10,Hastings07}. In contrast, other studies analyse the multipartite entanglement in spin chains using as a figure of merit, the tangle, finding what they called an \textit{avalanche of entanglement} at the phase transition point.

In this chapter, we focus on the study of entanglement in spin-$\frac{1}{2}$ chains. These one-dimensional systems present quantum phase transitions. The characterization of such critical behaviour is determined by conformal symmetry. Indeed, at quantum phase transitions the system displays conformal invariance, and its analytic structure provides very powerful instruments to characterize correlations. 

In contrast to the bipartite entanglement studies, we are interested in multipartite figures of merit. Since we restrict our study to spin chains of four sites, we choose the hyperdeterminant and two polynomial invariants, $S$ and $T$, to quantify the multipartite entanglement. The hyperdeterminant is a mathematical construction introduced by Cayley in the XIX century that serves the purpose of describing multipartite entanglement \cite{Cayley45}. The complexity to compute hyperdeterminants is remarkable and makes it difficult to apply it systematically to the study of quantum systems. Here, we shall introduce the basic properties of hyperdeterminants, its analysis for some special states and its behaviour at a phase transition \cite{Osterloh02,Latorre04,Latorre09,Alsina17,Miyake02}.

Hyperdeterminant is also used in other fields. There is a connection between the hyperdeterminant, the $S$ invariant and the theory of elliptic curves \cite{Gibbs10}. There is also a known connection between hyperdeterminants and string theory: see for instance \cite{Duff07,Borsten10}.

The content of the chapter is organized as follows. In Sec. \ref{sec:FigEnt} we define some figures of merit to quantify bipartite and multipartite entanglement, including the hyperdeterminant and its generalization for mixed states. In Sec. \ref{sec:examples} we present some examples of four-partite entanglement, in particular, pure states that have interesting properties, random states and ground states of random Hamiltonians. In sections \ref{sec:Ising}, \ref{sec:XXZ} and \ref{sec:HS} we analyze the multipartite entanglement in some spin chains (some details of the computation are shown in App. \ref{app:OddsEnds}). Finally, the conclusions are exposed in Sec. \ref{sec:conclusionsHDet}. All results can be found in Ref. \cite{HDet}.

\section{Figures of merit for multipartite entanglement \label{sec:FigEnt}}

To define figures of merit to quantify the entanglement of a system, we should first find a proper description of a quantum state. Appendix \ref{app:quantum_gates} introduce the Bloch sphere representation but it is limited to one qubit. In general, one should use a mathematical formulation that can describe any state, \emph{pure} or \emph{mixed}, of $n$ particles of any dimension $d$. This is accomplished with the density matrix formulation:
\begin{definition}[Density Matrix]
Given a mixed quantum state composed of $M$ states $|\psi_{i}\rangle$ with probability $p_{i}$, its density matrix is defined as
\begin{equation}
\rho \equiv \sum_{i=1}^{M}p_{i}|\psi_{i}\rangle\langle\psi_{i}|,
\end{equation}
where $\sum_{i=1}^{M}p_{i}=1$. If the state is pure, i.e. $M=1$, $\rho=|\psi\rangle\langle\psi|$.
\label{def:density_matrix}
\end{definition}

The density matrix operator is Hermitian and normalized, i.e. $\rho^{\dagger}=\rho$ and $\mathrm{Tr}\rho=1$. It can be proven that $\rho^2\leq \rho$, saturating the inequality for pure states. We can use this property to define a figure of merit to quantify how much pure is a state:
\begin{definition}[Purity]
Given a density matrix $\rho$,
\begin{equation}
\gamma \equiv \mathrm{Tr}\rho^2.
\end{equation}
Its bounds are $\frac{1}{d}\leq \gamma \leq 1$ for a totally mixed state and a pure state respectively.
\label{def:purity}
\end{definition}

Once the concept of density matrix has been introduced, we proceed to discuss some figures of merit to quantify quantum entanglement.

\subsection{Bipartite entropies}

Entropy is used in classical information theory to quantify the average information content of a system, e.g. a message. It was introduced by Claude Shannon in 1948 \cite{Shannon48} and, for that reason, classical entropy is often known as Shannon entropy.

The quantum mechanical extension of Shannon entropy is the Von Neumann entropy, named in honor to John Von Neumann who developed the the quantum theory of measurement and, together with Lev Landau, introduced the density matrix formalism \cite{Landau27,vonNeumann18,Bengtsson06}. 
\begin{definition}[Von Neumann entropy]
\begin{equation}
S(\rho)\equiv-\mathrm{Tr}\rho\log\rho,
\end{equation} 
By taking the logarithm in $d$ basis the entropy is normalized, i.e. $0\leq S\leq 1$, and its bounds correspond to product state and maximally entangled state respectively.
\end{definition}

Given a bipartite system in a Hilbert space $\mathscr{H}_{AB}=\mathscr{H}_{A}\otimes\mathscr{H}_{B}$, Von Neumann entropy counts the amount of entanglement between $A$ and $B$ or, in other words, the amount of quantum correlations between these two subsystems. This entropy is computed using the \emph{reduce density matrix} of one of these subsystems. For a pure state involving the two subsystems $\rho_{AB}=|\psi\rangle_{AB}\langle\psi|$,
\begin{equation}
\rho_{A}=\mathrm{Tr}_{B}\rho_{AB}=\sum_{i=0}^{d}{}_{B}\langle e_{i}|\psi\rangle_{AB}\langle\psi|e_{i}\rangle_{B},
\end{equation}
where $|e_{i}\rangle=|0\rangle,|1\rangle,\cdots|d\rangle$ are the computational basis states. This operation is called \emph{partial trace}, as system $B$ is traced out from the composite system $AB$. In addition, we can diagonalize $\rho_{A}$ and compute the entropy as
\begin{equation}
S_{A}=-\mathrm{Tr}(\rho_{A}\log\rho_{A})=-\sum_{i=1}^{d^m}\lambda_{i}\log\lambda_{i},
\end{equation}
where $m$ is the number of qudits in $A$ subsystem and $\lambda_{i}$ are the eigenvalues of $\rho_{A}$, i.e. $\rho_{A}=\sum_{i=1}^{d^m}\lambda_{i}|\lambda_{i}\rangle\langle\lambda_{i}|$.

This entanglement measure has deep connections with the conformal symmetry recovered in some quantum phase transitions. Let's consider the entropy in a spin chain corresponding to the reduced density matrix of a block of size $L$ out of $N$, $S(\rho_{L})=-\mathrm{Tr}(\rho_{L}\log\rho_{L})$, where $\rho_{L}=\mathrm{Tr}_{N-L}|\Psi_{g}\rangle\langle\Psi_{g}|$ with $|\Psi_{g}\rangle$ being the ground state of the system. It turns out that entropy scales at a quantum phase transition as \cite{Callan94,Holzhey94,VLRK03,Calabrese04}
\begin{equation}
S(\rho_{L})\sim\frac{c}{3}\log L,
\end{equation} 
where $c$ is the central charge that defines the universality class of the model. Away from criticality, the entropy saturates to a constant that depends on the correlation length present in the system.

\begin{theorem}[Schmidt decomposition]
Given a composite system $\mathscr{H}_{AB}=\mathscr{H}_{A}\otimes\mathscr{H}_{B}$ with $\mathscr{H}_{A}=\mathbb{C}^{\otimes d_{A}}$ and $\mathscr{H}_{B}=\mathbb{C}^{\otimes d_{B}}$, a pure state of this system, $|\psi\rangle_{AB}$, can be written in terms of orthonormal states $|u_{i}\rangle_{A}$ and $|v_{i}\rangle_{B}$ such that
\begin{equation}
|\psi\rangle_{AB}=\sum_{i=1}^{\chi}\alpha_{i}|u_{i}\rangle_{A}|v_{i}\rangle_{B},
\end{equation}
where $\alpha_{i}$ are positive real numbers satisfying $\sum_{i=1}^{\chi}\alpha_{i}^{2}=1$ known as Schmidt coefficients and $\chi$ is the \emph{Schmidt rank}.
\end{theorem}

The proof of this theorem is based on the singular value decomposition of a matrix. Let's write $|\psi\rangle_{AB}$ state in terms of two orthonormal basis $\{|i\rangle_{A}\}$ and $\{|j\rangle_{B}\}$ in their respective subspaces:
\begin{equation}
|\psi\rangle_{AB}=\sum_{i=0}^{d_{A}-1}\sum_{j=0}^{d_{B}-1}a_{ij}|i\rangle_{A}|j\rangle_{B}.
\end{equation} 
Let's define a matrix $A$ which entries, $A_{ij}$, are the coefficients $a_{ij}$. After computing the singular value decomposition of $A$, that is $A = U D V^{\dagger}$, we obtain the matrices $U$, of dimension $\mathrm{dim}(U)=d_{A}\times d_{A}$, the matrix $V$, of dimension $\mathrm{dim}(V)=d_{B}\times d_{B}$ and the diagonal matrix $D$, of dimension $d=d_{A}\times d_{B}$ with non-negative real elements entries $\alpha_{k}$. Then,
\begin{align}
|\psi\rangle_{AB}&=\sum_{i=0}^{d_{A}-1}\sum_{j=0}^{d_{B}-1}\left(\sum_{k=1}^{d} u_{ik}\alpha_{k}v_{kj}\right)|i\rangle_{A}|j\rangle_{B}= \nonumber \\
&= \sum_{k=1}^{d}\alpha_{k}\left(\sum_{i=0}^{d_{A}-1}u_{ik}|i\rangle_{A}\right)\otimes \left(\sum_{j=0}^{d_{B}-1}v_{kj}|j\rangle_{B}\right) \equiv   \sum_{i=1}^{d}\alpha_{i}|u_{i}\rangle_{A}|v_{i}\rangle_{B}, 
\end{align}
where $u_{ij}$ and $v_{ij}$ are the coefficients of $U$ and $V$ matrices respectively. $\blacksquare$

As stated above, entropy is used to elucidate if a composite system is separable and, in case it is not, how much entangled it is. Thus, we can apply Schmidt theorem to quantify how much separable is a quantum state. The Schmidt rank is just the number of non-zero elements of $D$, i.e. $\chi\leq \mathrm{min}\{d_{A},d_{B}\}$. 

The Schmidt decomposition is very useful to compute the reduce density matrices of a bipartite system:
\begin{equation}
\rho_{AB}=|\psi\rangle_{AB}\langle\psi| = \sum_{i=1}^{d}|\alpha_{i}|^2|u_{i}\rangle_{A}\langle u_{i}|\otimes |v_{i}\rangle_{B}\langle v_{i}|,
\end{equation}
from which we can obtain the reduced density matrices, $\rho_{A}=\sum_{i=1}^{d_{A}}|\alpha_{i}|^{2}|u_{i}\rangle_{A}\langle u_{i}|$ and $\rho_{B}=\sum_{i=1}^{d_{B}}|\alpha_{i}|^{2}|v_{i}\rangle_{B}\langle v_{i}|$. We can then exact a corollary of this theorem that relates Schmidt decomposition with Von Neumann entropy:
\begin{corollary}
Schmidt decomposition shows that the reduced density matrix $\rho$ on either subsystem $A$ and $B$ have the same spectrum. The Von Neumann entropy computed in the diagonal basis can be obtained directly from the Schmidt coefficients, i.e. $S = -\sum_{i=1}^{d}|\alpha_{i}|^2\log|\alpha_{i}|^2$.
\end{corollary}
This corollary establishes that the Von Neumann entropy is a well-defined measure of entanglement. For a separable state $\chi=1$, so $\alpha_{i}=1$ and $S=0$; for an entangled state $\chi>1$, so $S>0$; and for a maximally entangled state $\chi = \mathrm{min}\{d_{A},d_{B}\}$, with equal $\alpha_{i}$ and, consequently, $S=1$.

In addition, we can make a connection between Schmidt rank $\chi$ and purity $\gamma$. If the state $|\psi\rangle_{AB}$ is separable, then $\chi=1$ which means that only one eigenvalue is different from zero, so $\alpha_{1}=1$. Then, if we compute the purity of $\rho_{A}$ and $\rho_{B}$ we will obtain $\gamma=1$; both subsystems are pure. On the contrary, if $|\psi\rangle_{AB}$ is entangled, then $\chi>1$ and more than one eigenvalue $\alpha_{i}$ is different from zero, so $\gamma<1$ and $\rho_{A}$ and $\rho_{B}$ are mixed states. 

A generalization of Von Neumann entropy are R\'enyi and Tsallis entropies \cite{Renyi61,Tsallis88,Gour10}:
\begin{definition}[R\'enyi entropy]
\begin{equation}
S^{R}_{\alpha}(\rho_{A})\equiv\frac{1}{1-\alpha}\log\mathrm{Tr}\left(\rho_{A}^{\alpha}\right), \ \alpha\neq 1, \ \alpha>0 \ .
\label{eq:Renyi}
\end{equation}
\end{definition}
\begin{definition}[Tsallis entropy]
\begin{equation}
S^{T}_{\alpha}(\rho_{A})\equiv\frac{1}{1-\alpha}\left(\mathrm{Tr}\left(\rho_{A}^{\alpha}\right)-1\right), \ \alpha\neq 1, \ \alpha>0 \ .
\label{eq:Tsallis}
\end{equation}
\end{definition}
Both are entanglement monotones for $\alpha<1$, i.e. nonincreasing on average under LOCC transformations. They also obey scaling properties and have been used to study entanglement in spin chains \cite{Franchini07}.

There are many other figures of merit to quantify bipartite entanglement. Nevertheless, some of them do not show scaling properties or fail to grab the subtleties of phase transitions.

It is reasonable to look for a complete characterization of quantum correlations beyond the one provided by entanglement entropies. It is often argued that there is a need for new measures of {\sl genuine multipartite entanglement}. There is some ambiguity in the literature about this term. It is often referred as multipartite entanglement the study of correlations between two parties of a large system of particles \cite{Facchi06,Facchi08,Facchi10,Bayat17}. On the other hand, genuine multipartite entanglement can be referred as anything which analyses correlations beyond two parties. There is a second more stringent definition that states that measures of genuine multipartite entanglement should not involve any partial trace of the system. This definition makes it very hard if not impossible to conduct studies in large systems. An example of a measure of strict multipartite entanglement could be the study of  Bell inequalities involving every party in a system, for instance those discussed in the previous Chapter \ref{Ch:Bell_Ineq}. In the following subsections it is discussed two examples of figures of merit to quantify entanglement beyond two parties.

\subsection{3-tangle}

The \textit{tangle} is a measure of multipartite entanglement for  systems with an even number of qubits. 
\begin{definition}[Tangle]
\begin{equation}
\tau_{N}\equiv|\langle\tilde{\psi}|\psi\rangle|^2, \quad |\tilde{\psi}\rangle\equiv\sigma_{y}^{\otimes n}|\psi\rangle,
\end{equation}
where $|\psi\rangle$ is a multiqubit state written in terms of the computational basis. 
\end{definition}
For $n=2$, the tangle is the square of the \emph{concurrence}, another figure of merit for bipartite entanglement that will be discussed in detail in Chapter \ref{Ch:MaxEnt}. For $n=3$, it was proposed an extension using the definition of \emph{residual tangle} \cite{Coffman00,Gregg07}
\begin{definition}[3-tangle]
The generalization of the tangle for three parties $A$, $B$ and $C$ can be expressed as the residual tangle
\begin{equation}
\tau_{ABC}\equiv 4\det\rho_{A}-\tau_{AB}-\tau_{AC},
\end{equation}
where $\tau_{AB}$ and $\tau_{AC}$ are the tangles of subsystems $AB$ and $AC$.
\end{definition}
Providing that a three-qubit quantum state can be written as
\begin{equation}
|\psi\rangle = \sum_{i,j,k=0,1}b_{ijk}|i\rangle|j\rangle|k\rangle,
\end{equation}
where $b_{ijk}\in\mathbb{C}$, the above 3-tangle can be written as
\begin{equation}
\tau\equiv\tau_{ABC}=2\vert b_{i_1 j_1 k_1}b_{i_2 j_2 k_2}b_{i_3 j_3 k_3}b_{i_4 j_4 k_4}\epsilon^{i_1 i_2}\epsilon^{j_1 j_2}\epsilon^{i_3 i_4}\epsilon^{j_3 j_4}\epsilon^{k_1 k_3} \epsilon^{k_2 k_4} \vert,
\end{equation}
where a $\epsilon^{00}=\epsilon^{11}=0$ and $\epsilon^{01}=\epsilon^{10}=1$. Note that this contraction introduces minus signs, as opposed to pure contractions of subsystems which only involve the always positive Kronecker delta. The tangle is invariant under local unitary transformations on any party. It is a figure of genuine multipartite entanglement that involves no partition of the system. There are other works that study the multipartite entanglement in spin chains for an arbitrary, but finite, number of particles using the Meyer-Wallach measure of global entanglement \cite{Radgohar18}.

The introduction of 3-tangle in this discussion is motivated, apart from its applications in the study of entanglement in spin chains \cite{Bayat17}, for its connections with the next figure of merit, the hyperdeterminant.

\subsection{Hyperdeterminant}

The hyperdeterminant is the generalization of a determinant for matrices of higher dimensions. It was first introduced by Cayley \cite{Cayley45} in 1845 to characterize the conditions for a system of linear equations to have a non-trivial solution. In particular, Cayley provided an analytic expression to compute the hyperdeterminant of a $2\times 2\times 2$ matrix. Later on, Schl\"afli made the extension to the $2\times 2\times 2\times 2$ matrices \cite{Schlafli52}. Since then, many mathematicians have studied this function and its connections with different mathematical branches. In fact, hyperdeterminants can be defined in different ways \cite{Gelfand94}.

For the purpose of this chapter, we are interested in obtaining an analytical expression for the hyperdeterminant of $2\times 2\times 2\times 2$ matrix. The $n$-hyperdeterminant, i.e. the hyperdeterminant of a $2^{\times n}$ matrix, will be denoted as $\hdet_{n}$. Then, it is possible to define $\hdet_{4}$ recursively from $\hdet_{3}$ and $\hdet_{2}$, using the connection with the discriminants of a polynomial.
\begin{definition}[Polynomial discriminant]
Given a polynomial of degree $n$
\begin{equation}
P_{n}(x)\equiv a_{0} + a_{1}x + \cdots a_{n-1}x^{n-1} + a_{n} x^{n},
\end{equation}
its discriminant can be defined as
\begin{equation}
\Delta(P_{n}(x))\equiv a_{n}^{2n-2}\prod_{i<j}^{n}\left(r_{i}-r_{j}\right)^2,
\end{equation}
where $r_{i}$ are the polynomial roots and $a_{n}^{2n-2}$ is a normalization factor.
\end{definition}

A discriminant could be complex or real, depending on the coefficients of the polynomial. If the coefficients are real numbers, then the discriminant is always real. In that case, it is zero if at least two roots are equal; it is positive if there exist $2k$ pairs of conjugate roots for $0\leq k\leq n/2$ where $n$ is the degree of the polynomial; and it is negative if there exist $2k+1$ pairs of conjugate roots for $0\leq k\leq (n-2)/4$ \cite{Gelfand94}. For that reason, it is introduced the absolute value in the below definitions of $\hdet_{n}$, as it is done in previous works with the tangle \cite{Coffman00}.

Let's start with a generic $2\times 2$ matrix $C$ which entries are $c_{ij}$. Then the $\hdet_{2}$ is
\begin{equation}
\hdet_{2}=|\det(C)|=|c_{00}c_{11}-c_{01}c_{10}|.
\end{equation}
If we identify the $c_{ij}$ coefficients with a two qubits wave function, i.e. $|\psi\rangle=\sum_{ij}c_{ij}|ij\rangle$, then $\hdet_{2}$ corresponds to the concurrence for two qubits. Next, let's replace each $c_{ij}$ coefficient in $\hdet_{2}$ expression with $b_{ij0}+b_{ij1}x$. The discriminant of the resulting polynomial $P_{3}(x)$ is actually $\hdet_{3}$:
\begin{align}
P_{3}(x)&= \hdet_{2} /. c_{ij}\rightarrow b_{ij0}+b_{ij1}x, \label{eq:P3} \\
\hdet_{3}&=|\Delta\left(P_{3}(x)\right)|, 
\end{align}
where $/.$ stands for ``replace'' and $B$ is a matrix which entries are $b_{ijk}$. It turns out that if we identify these $b_{ijk}$ as the coefficients of a three qubits wave function, i.e. $|\psi\rangle=\sum_{ijk}b_{ijk}|ijk\rangle$, then $\hdet_{3}=\tau$. The next iteration gives the expression for $\hdet_{4}$.

\begin{definition}[Hyperdeterminant of $n=4$]
Let's construct a polynomial of degree 4 by replacing each $b_{ijk}$ coefficient in $\hdet_{3}$ of Eq. \eqref{eq:P3} by $t_{ijk0}+t_{ijk1}x$. Its discriminant gives an expression for $\hdet_{4}$:
\begin{align}
P_{4}(x)&\equiv \hdet_{3} /. b_{ijk}\rightarrow t_{ijk0}+t_{ijk1}x, \\
\hdet_{4}&\equiv\frac{1}{256}|\Delta\left(P_{4}(x)\right)|,
\end{align}
where we can identify $t_{ijkl}$ elements with the coefficients of a four qubits state $|\psi\rangle=\sum_{ijkl}t_{ijkl}|ijkl\rangle$.
\end{definition}

The hyperdeterminant is a mathematical figure that can be used to quantify multipartite entanglement if we construct it with wave function coefficients. For that reason, from now on, we will label each $\hdet_{4}$ with the corresponding quantum state that has been used to construct it, that is $\hdet_{4}(|\psi\rangle)$ is the 4-hyperdeterminant of the state $|\psi\rangle$.

For $n=4$, hyperdeterminant can be also defined in terms of fundamental invariants \cite{Luque03}. These polynomials are invariant under the SLOCC group $SL(\mathbb{C},2)^{4}$ and can be used to classify multipartite entanglement as well. Most of the 18 invariants are related to bi-partitions of the system but, in particular, two of them can measure global correlations involving every spin in the system. These two polynomial invariants, called $S$ and $T$, are also related with $\hdet_{4}$ and can be obtained from the coefficients of the polynomial $P_{4}(x)$ defined above.
\begin{definition}[$S$ and $T$ invariants]
From the polynomial $P_{4}(x)=a_{0}x^4+4a_{1}x^3+6a_{2}x^2+4a_{3}x+a_{4}$, which coefficients $a_{i}$ are obtained from the three qubits wave function coefficients $b_{ijk}$, the invariants $S$ and $T$ take the form
\begin{align}
S &\equiv 3a_{2}^2-4a_{1}a_{3}+a_{0}a_{4}, \\
T &\equiv -a_{2}^3+2a_{1}a_{2}a_{3}-a_{0}a_{3}^2-a_{1}^2a_{4}+a_{0}a_{2}a_{4}.
\end{align}
Then, the $\hdet_{4}$ can be obtained from
\begin{equation}
\hdet_{4}(|\psi\rangle)\equiv S^3-27 T^2.
\label{eq:HDetST}
\end{equation}
\end{definition}

Notice that the relation \eqref{eq:HDetST} reveals a possible cancellation between these two invariants that leads to an $\hdet_{4}=0$. We can observe multipartite entanglement of the form that $S$ and $T$ invariants can capture it but, however, $\hdet_{4}$ could be blind to it. 


The hyperdeterminant is invariant under local changes of basis. That is, given a state $|\varphi\rangle$ and a state $|\tilde{\varphi}\rangle=U_1\otimes\cdots\otimes U_n|\varphi\rangle$, where $U_i$ are independent unitary changes of each local basis,
\begin{equation}
\hdet_{n}(|\varphi\rangle)=\hdet_{n}(|\tilde{\varphi}\rangle).
\end{equation}
This immediately shows that the hyperdeterminant provides a possible figure or merit to quantify  multipartite entanglement. 

It is worth remarking that the $\hdet_{4}$ vanishes for  quantum states that can be written as the product states on any bipartition. That is, for a state made out of four parties,
\begin{equation}
\begin{array}{lll}
|\psi\rangle =|\varphi\rangle_{1}|\phi\rangle_{234} & \Rightarrow &  \mathrm{HDet}_{4}(|\psi\rangle)=0,\\
|\psi\rangle =|\varphi\rangle_{12}|\phi\rangle_{34} & \Rightarrow & \mathrm{HDet}_{4}(|\psi\rangle)=0,
\end{array}
\label{eq:12_34}
\end{equation}
with the same result for any permutation of indices. In the first case, when the state is a product state of 1-qubit and a generic state of the rest, the invariants $S$ and $T$ are zero, so is the hyperdeterminant. This brings the idea that the hyperdeterminant is only sensitive to genuine 4-party entanglement. In the second case, where the state can be separable in two halves, some more basic polynomial invariants are proportional to the concurrence, but it remains true that  $S$ and $T$  are zero, as well as the hyperdeterminant. 

\subsubsection{Definition of hyperdeterminant for mixed states}

The above definition of hyperdeterminant is only valid for pure states. We propose to extend it for mixed states following a similar definition as the one used for Entanglement of Formation \cite{Wootters98}.

\begin{definition}[Hyperdeterminant for mixed states]
For all $\xi$ possible decompositions of a density matrix $\rho=\sum_{i}p^{\xi}_{i}|\psi^{\xi}_{i}\rangle\langle\psi^{\xi}_{i}|$,
\begin{equation}
\hdet_{n}(\rho)\equiv \min_{\{\xi\}}\sum_{i}p^{\xi}_{i}\hdet_{n}(|\psi^{\xi}_{i}\rangle),
\label{eq:HDetth2}
\end{equation}
\end{definition}
Similarly, we can extend the above definition to the invariants $S$ and $T$.

The construction of hyperdeterminants for density matrices brings the possibility of defining the hyperdeterminant for  thermal states. Let us consider the density matrix of a system of $n$ spins  in equilibrium with a thermal reservoir
\begin{equation}
\rho_{\beta}=\frac{e^{-\beta \mathcal{H}}}{\mathcal{Z}}=\frac{1}{\mathcal{Z}}\sum_{i=0}^{2^n-1}e^{-\beta E_{i}}|E_{i}\rangle
\langle E_i| ,
\label{eq:ther}
\end{equation}
where $\mathcal{Z}=\mathrm{Tr}\left(e^{-\beta \mathcal{H}}\right)$ is the partition function and $|E_{i}\rangle$ is the state with energy $E_{i}$. We shall define the hyperdeterminant of the above thermal state as 
\begin{equation}
\hdet_n(\rho_{\beta})\equiv \frac{1}{\mathcal{Z}}\sum_{i=0}^{2^n-1}e^{-\beta E_{i}}\hdet_n(|E_{i}\rangle) 
\label{eq:Th}
\end{equation}
where $\hdet_{n}(|E_{i}\rangle)$ is the hyperdeterminant of the state $|E_{i}\rangle$.

In case of degeneracy, a linear superposition of states with same energy is also an eigenstate of the system. Then, the most general state can be written as
\begin{equation}
|\psi\rangle_{th}=\frac{1}{\mathrm{Tr}\left(e^{-\beta \mathcal{H}}\right)}\sum_{i}e^{-\beta E_{i}}\left(\sum_{j}a^{i}_{j}|E^{i}_{j}\rangle\right),
\end{equation}
where the first summation is over all different values of $E_{i}$ and the second corresponds to the linear superposition of eigenstates with the same eigenvalue $E_{i}$, with $\sum_{j}|a^{i}_{j}|^2=1$. Then, taking the definition for $\hdet_{n}$ for mixed states from Eq.  \eqref{eq:HDetth2},
\begin{equation}
\hdet_{n}(\rho_{\beta})= \min_{\{a^{i}_{j}\}}\hdet_{n}(|\psi\rangle_{th}).
\label{eq:Th2}
\end{equation}
A similar definitions hold for thermal values of $S$ and $T$ invariants. 

\section{Examples of four-partite entanglement \label{sec:examples}}

\subsection{Special states}


There are states for which $\hdet_{4}$ vanishes as a consequence of a cancellation between $S$ and $T$ invariants following the relation \eqref{eq:HDetST}. The most relevant example is the GHZ state \cite{GHZ89},
\begin{equation}
|GHZ\rangle=\frac{1}{\sqrt{2}}\left(|0000\rangle + |1111\rangle\right),
\label{ghz} 
\end{equation}
which has $S=1/(2^6 3)$, $T=-1/(2^9 3^3)$ and zero $\hdet_{4}$. This result shows that $\hdet_{4}$ captures a different type of entanglement that the one associated to superposition of fully orthogonal states. 

There are other special states that have the same values as above for $S$ and $T$ invariants. One example are the cluster states $|C_{1}\rangle$, $|C_{2}\rangle$ and $|C_{3}\rangle$ \cite{Briegel01,Gour10},
\begin{align}
|C_{1}\rangle &=\frac{1}{2}\left(|0000\rangle +|0011\rangle + |1100\rangle - |1111\rangle\right), 
\\
|C_{2}\rangle &=\frac{1}{2}\left(|0000\rangle +|0110\rangle + |1001\rangle - |1111\rangle\right), \\
|C_{3}\rangle &=\frac{1}{2}\left(|0000\rangle +|0101\rangle + |1010\rangle - |1111\rangle\right), 
\label{cluster} 
\end{align}
which maximizes the Von Neumann entropy of two of their three bipartition. Other example is the $|YC\rangle$ state \cite{Yeo06},
\begin{equation}
|YC\rangle=\frac{1}{\sqrt{8}}\left(|0000\rangle -|0011\rangle -|0101\rangle + |0110\rangle +|1001\rangle + |1010\rangle + |1100\rangle +|1111\rangle \right),
\label{yc} 
\end{equation}
which can perform a faithful teleportation of an arbitrary two-qubit entangled state. These states bring the idea that invariants $S$ and $T$ measure some kind of entanglement, but the hyperdeterminant makes a further selection. 

The 4-qubit $W$ state \cite{Dur00},
\begin{equation}
|W\rangle= \frac{1}{2}\left(|0001\rangle + |0010\rangle + |0100\rangle + |1000\rangle\right),
\label{eq:W}
\end{equation}
has $S=T=0$. Again, W-ness is a different kind of entanglement as the one capture by $\hdet_4=0$.

States that maximize $\hdet_{4}$ have been studied previously. Numerical analysis shows that a state with maximum $\hdet_{4}$ is \cite{Osterloh06,Goyeneche15}
\begin{equation}
|HD\rangle=\frac{1}{\sqrt{6}}\left(|1000\rangle+|0100\rangle +|0010\rangle +|0001\rangle+\sqrt{2}|1111\rangle\right),
\label{hd} 
\end{equation}
with $\hdet_4=1/(2^8 3^9)\simeq 1.98 \ 10^{-7}$, $S=0$ and $T=-1/(2^4 3^6)$. Another state with the same values for $\hdet_{4}$, $S$ and $T$ corresponds to the state $|L\rangle$ \cite{Gour10}
\begin{multline}
 |L\rangle =  \frac{1}{\sqrt{12}}\Big(\left(1+w\right)\left(|0000\rangle+|1111\rangle\right) + \left(1-w\right)\left(|0011\rangle+|1100\rangle\right)  \\   
 +  w^2\left(|0101\rangle+|0110\rangle+|1001\rangle+|1010\rangle\right)\Big),
\end{multline}
where $w=e^{\frac{2\pi i}{3}}$. This state also maximizes the average Tsallis entropy for $0<\alpha<2$ and $\alpha>2$. 

Other relevant states are the nine families of quadripartite entangled states defined by Verstraete \emph{et al.} in \cite{Verstraete02}. There is only one family of states with $\mathrm{HDet}_{4}$ different from zero:
\begin{multline}
G_{abcd}=\frac{a+d}{2}\left(|0000\rangle+|1111\rangle\right)+\frac{a-d}{2}\left(|0011\rangle+|1100\rangle\right)  \\ 
+ \frac{b+c}{2}\left(|0101\rangle+|1010\rangle\right)+\frac{b-c}{2}\left(|0110\rangle+|1001\rangle\right),
\label{eq:Gabcd}
\end{multline}
whose  values for $S$, $T$ and $\mathrm{HDet}_{4}$ are given by
\begin{align}
&
\begin{aligned}
S=\frac{1}{12}\left((b^2 - c^2)^2 (a^2 - d^2)^2 + (a^2 - b^2) (b^2 - c^2) (a^2 -d^2) (c^2 - d^2) \right. \\
+ \left.(a^2 - b^2)^2 (c^2 - d^2)^2\right), 
\end{aligned} \label{eq:Gabcd_S}\\
& 
T=\frac{1}{1728}\left((a c + b d)^2 + (a b + c d)^2-2 (b c + a d)^2\right) \nonumber\\
&\Big( \left((a c + b d)^2 + (a b + c d)^2-2 (b c + a d)^2 \right)^2
-9 (b - c)^2 (b + c)^2 (a - d)^2 (a + d)^2\Big), \label{eq:Gabcd_T}
\\
&\mathrm{HDet}_{4}=\frac{1}{256}(a^2 - b^2)^2 (a^2 - c^2)^2 (b^2 - c^2)^2 (a^2 - d^2)^2 (b^2 - d^2)^2 (c^2 - d^2)^2.
\label{eq:Gabcd_hdet}
\end{align}

Notice that if two parameters are equal, $\mathrm{HDet}_{4}$ become zero. We will see that the ground state of $XXZ$ model is of this type.

There are three families of states with $S$ and $T$ non zero in general. These are the state
\begin{equation}
L_{abc_{2}}=\frac{a+b}{2}\left(|0000\rangle+|1111\rangle\right)+\frac{a-b}{2}\left(|0011\rangle+|1100\rangle\right)+c\left(|0101\rangle+|1010\rangle\right)+|0110\rangle,
\end{equation}
with
\begin{equation}
S=\frac{1}{12}(a^2 - c^2)^2 (c^2 - b^2)^2, \qquad T=\frac{1}{216}(a^2 - c^2)^3 (c^2 - b^2)^3, 
\end{equation}
the state
\begin{equation}
L_{a_{2}b_{2}}=a\left(|0000\rangle+|1111\rangle\right)+b\left(|0101\rangle+|1010\rangle\right)+|0110\rangle+|0011\rangle,
\end{equation}
with
\begin{equation}
S=\frac{1}{12}(a - b)^4 b^4, \qquad   T = -\frac{1}{216}(a - b)^6 b^6 \ ,
\end{equation}
and the state
\begin{equation}
L_{a_{2}0_{3\oplus\bar{1}}}=a\left(|0000\rangle+|1111\rangle\right)+|0011\rangle+|0101\rangle+|0110\rangle,
\end{equation}
with
\begin{equation}
S= \frac{1}{12}a^8 \ ,   \qquad T= -\frac{1}{216}a^{12} \ .  
\end{equation}
For all of them, $\mathrm{HDet}_{4}$ is zero due to an exact cancellation between $S$ and $T$ invariants that arise from Eq. \eqref{eq:HDetST}.

Finally, the  families
\begin{align}
&\begin{aligned}
L_{ab_{3}} = a\left(|0000\rangle+|1111\rangle\right)+\frac{a+b}{2}\left(|0101\rangle+|1010\rangle\right)+ \frac{a-b}{2}\left(|0110\rangle+|1001\rangle\right) \ , \\ 
+\frac{i}{\sqrt{2}}\left(|0001\rangle+|0010\rangle+|0111\rangle+|1011\rangle\right) \ ,\\
\end{aligned} \\
&L_{a_{4}}=a\left(|0000\rangle+|0101\rangle+|1010\rangle+|1111\rangle\right)+i\left(|0001\rangle-|1011\rangle\right)+|0110\rangle \ ,  \\
&L_{0_{5\oplus\bar{3}}}=|0000\rangle+|0101\rangle+|1000\rangle+|1110\rangle \ ,    \\ 
&L_{0_{7\oplus\bar{1}}}=|0000\rangle+|1011\rangle+|1101\rangle+|1110\rangle \ ,  \\
&L_{0_{3\oplus\bar{1}}0_{3\oplus\bar{1}}}=|0000\rangle+|0111\rangle \ , 
\end{align}
have $S$ and $T$ equal to zero.

\subsection{Random states}

In order to obtain a better picture of what are the typical values for $\hdet_{4}$, $S$ and $T$ invariants, we compute them for random pure states. The very definition of a random state depends on the prior which is accepted. Here, we take as a prior two distributions of coefficients in the computational basis: a flat distribution and a Haar distribution. 

\begin{definition}[Flat and Haar distributed states]
Given an $n$ qubits state written in terms of the computational basis states $|\varphi_{i}\rangle$ and the coefficients $z_{i}\in\mathbb{C}$, the state
\begin{equation}
|\psi\rangle\equiv\sum_{i=0}^{2^n-1}z_{i}|\varphi_{i}\rangle,
\end{equation}
is flat distributed if $\mathrm{Re}(z_{i})$ and $\mathrm{Im}(z_{i})$ are independent and identically distributed (i.i.d.) values following a uniformly distribution on $[0,1]$ and Haar distributed if $z_{i}$ are i.i.d. complex Gaussian variables with zero median and unit variance, i.e. $N(0,1)$.
\end{definition}

We have generated $10^5$ random 4-qubit states with a flat and Haar prior on the coefficients and plotted $\hdet_{4}$ in Fig. \ref{fig:hdet-random} in comparison with ground state of random matrix  Hamiltonians that satisfy the GOE, GUE and GSE distributions (see next subsection \ref{sec:GSrandom}). The mean value of $\hdet_{4}$ is around $\sim 1.2 \ 10^{-9}$, two orders of magnitude lower than the maximum possible value ($1.98 \ 10^{-7}$ for $|HD\rangle$ state). Moreover, only $2\%$ of the states have $\hdet_{4}$ greater than $10^{-8}$. Similar results were obtained in \cite{Alsina17}. This result is to be compared with the entanglement entropy of such states for a random bipartition, where maximal volume entropy is found \cite{Alsina17}. 

The $\hdet_{4}$ distribution obtained is not the same for flat and Haar distributed random states: the second have lower values of $\hdet_{4}$. Therefore, the hyperdeterminant is a more subtle figure of merit that is not maximal for most states, except for a small subset of random states, and can distinguish between two random priors.

A way to understand the scarce abundance of high hyperdeterminant states is based on the comparison between the multipartite and the bipartite entanglements. The latter is measured mainly by the Von Neumann entropy, where one does not encounter cancellations coming from the different terms of the reduced density matrix. On the contrary, to obtain high hyperdeterminant values, requires a fine tuning, as illustrates the $\hdet_{4}$ expression for $G_{abcd}$ state of Eq. \eqref{eq:Gabcd_hdet}. Random states do not propitiate this fine tuning which leads to low values for the hyperdeterminant. 

\subsection{Ground state of random Gaussian Hamiltonians }
\label{sec:GSrandom}

\begin{figure}[t!]
\centering
\includegraphics[width=0.6\textwidth]{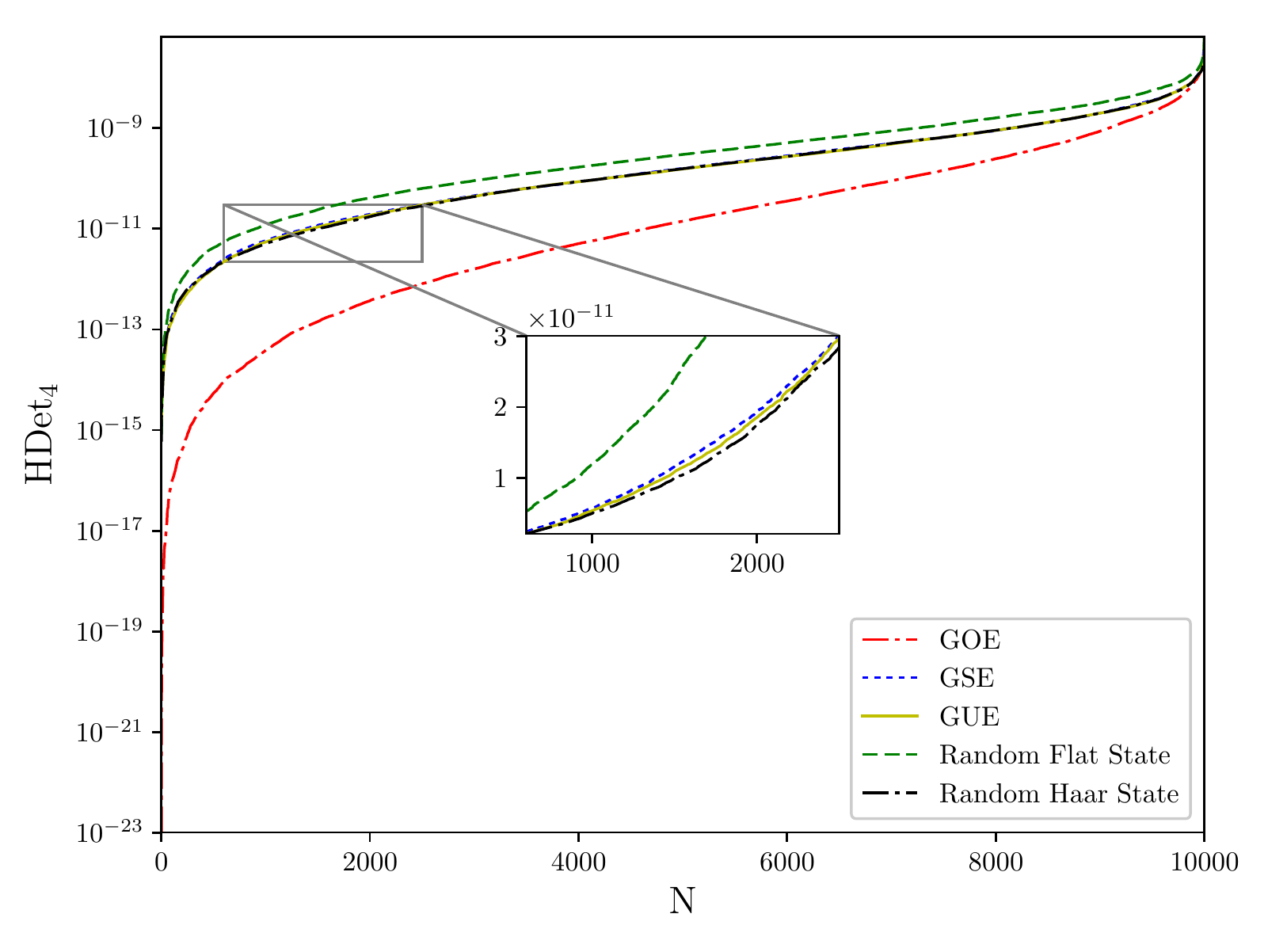}
\caption{$\hdet_{4}$ for $10^5$ random Hamiltonians distributed following random distributions corresponding to GOE, GUE and GSE introduced in Def. \ref{def:GOE}, Def. \ref{def:GUE} and Def. \ref{def:GSE}. These distributions are compared with $\hdet_{4}$ of flat and Haar distributed random states.}
\label{fig:hdet-random}
\end{figure}

Random matrices are closely related with several physical fields \cite{Random06}. For that reason, we also analyse the ground state of random Hamiltonians constructed artificially with random matrices. In particular, we construct random matrices of dimension $2^4\times 2^4$ whose entries are random numbers distributed following three types of Gaussian ensembles.

\begin{definition}[Gaussian Orthogonal Ensamble (GOE) \label{def:GOE}]
Symmetric $N\times N$ matrix which diagonal entries are i.i.d. with distribution $N(0,1)$ and the off-diagonal entries are i.i.d. (subject to the symmetry) with distribution $N(0,\frac{1}{2})$
\end{definition}
\begin{definition}[Gaussian Unitary Ensamble (GUE) \label{def:GUE}]
Hermitian $N\times N$ matrix which diagonal entries are i.i.d. with distribution $N(0,1)$ and the off-diagonal entries are i.i.d. (subject to being Hermitian) with distribution $N_{2}(0,\frac{1}{2})$, i.e. its corresponding real and imaginary parts are distributed following a $N(0,\frac{1}{2})$ distribution.
\end{definition}
\begin{definition}[Gaussian Symplectic Ensamble (GSE) \label{def:GSE}]
Self-dual $N\times N$ matrix which diagonal entries are i.i.d. with distribution $N(0,1)$ and the off-diagonal entries are i.i.d. (subject to being self-dual) with distribution $N_{4}(0,\frac{1}{2})$, i.e. its corresponding quaternion units are distributed following a $N(0,\frac{1}{2})$ distribution.
\end{definition}

Figure \ref{fig:hdet-random} shows the values of $\hdet_{4}$ for the ground state of $10^5$ random Hamiltonians for the three Gaussian distributions. For GUE and GSE, the mean value for $\hdet_{4}$ is slightly lower than for a random state and have the same value as Haar distributed random states, whereas for GOE is much smaller. This result is independent of the number of distributions considered, which suggests the existence of a probability density related to  $\hdet_{4}$.

We have introduce $\hdet_{4}$ basic properties and typical values for random distributions. The next part of this chapter is to study four-partite entanglement in some well-known spin models, using as a figure of merit $\hdet_{4}$, $S$ and $T$ invariants. 

\section{The transverse Ising model \label{sec:Ising}}

One of the most studied  one-dimensional quantum spin models is the transverse Ising model \cite{Dutta15}. This model is described by the Hamiltonian
\begin{definition}[Transverse Ising Model]
\begin{equation}
\mathcal{H}_{\mathrm{Ising}}\equiv-J\sum_{i=1}^n  \sigma_{i}^{x}\sigma_{i+1}^{x} - \lambda\sum_{i=1}^n \sigma_{i}^{z}.
\label{eq:HIsing}
\end{equation}
where $J$ is the coupling constant and $\lambda$ is the transverse field strength.
\end{definition} 
In this chapter, we study the ferromagnetic interaction, i.e. $J>0$, and without lost of generality it can be set $J=1$ and $\lambda\geq 0$. We also consider periodic boundary conditions, i.e. $\sx_{n}\sx_{1}$.

The non-commuting transverse field term introduces quantum fluctuations in the model causing a quantum phase transition from an ordered phase (magnetization different from zero) to a disordered paramagnetic phase (magnetization is zero), at critical value of $\lambda=\lambda_{c}$. For infinite chains, $\lambda_{c}=1$ is the critical point where conformal invariance is restored. At $\lambda=0$ there are two degenerate ground states with ferromagnetic ordering, $|\rightarrow\rightarrow\cdots\rightarrow\rangle$ and $|\leftarrow\leftarrow\cdots\leftarrow\rangle$ -- where $|\rightarrow\rangle$ and $|\leftarrow\rangle$ are the spin states in the $\sigma^{x}$ basis -- and at $\lambda>\lambda_{c}$ the external field strength wins over the neighbouring interaction $J$ and the system lies in the paramagnetic phase. For finite chains in the ferromagnetic phase, a non vanishing value of $\lambda$ breaks the degeneracy of the ground state and produces an exponentially small energy gap between the two lowest energy states. On the other hand, the critical value $\lambda_c$ moves away from its value in the following sense. The entropy of the Ising spin chain peaks around the quantum phase transition. As long as the length of the chain increases, the critical point approaches to $1$. The entanglement entropy near $\lambda=1$ scales logarithmically following the conformal scaling law  with central charge $c=\frac{1}{2}$ till the correlation length bounds the entropy.

\subsection{Eigenstates}

The analytic expressions of $\mathrm{HDet}_{4}$, $S$ and $T$ invariants for all the  eigenstates are summarized in Tab. \ref{Tab:IsingTh} (see App. \ref{app:OddsEnds} for details). One can distinguish three types of behaviours: 
{\sl i}) $\mathrm{HDet}_{4}$ is different from zero, {\sl ii}) $\mathrm{HDet}_{4}$ zero, due to a cancellation of non-vanishing $S$ and $T$ invariants, and {\sl iii}) $\mathrm{HDet}_{4}$, $S$ and $T$ are all zero.

\begin{table}[t!]
\centering
\begin{tabular}{@{}lccc}
\toprule
\textbf{State} & $\mathbf{HDet_{4}}$ & $\mathbf{\mathit{S}}$ & $\mathbf{\mathit{T}}$ \\
\midrule
$|\Psi_{0}\rangle,|\Psi_{15}\rangle$ & $H(\alpha_+,\beta_+,\gamma_+)$ & $S(\alpha_+,\beta_+,\gamma_+)$ & $T(\alpha_+,\beta_+,\gamma_+)$ \\
$|\Psi_{1}\rangle,|\Psi_{5}\rangle,|\Psi_{10}\rangle,|\Psi_{14}\rangle$, & 0 & $\frac{1}{(1+\lambda^2)^{2}}\frac{1}{2^6 3}$ & $\frac{1}{(1+\lambda^2)^{3}}\frac{1}{2^9 3^3}$ \\
$|\Psi_{2}\rangle,|\Psi_{13}\rangle$ & $H(\alpha_-,\beta_-,\gamma_-)$ & $S(\alpha_-,\beta_-,\gamma_-)$ & $T(\alpha_-,\beta_-,\gamma_-)$ \\
$|\Psi_{3}\rangle,|\Psi_{4}\rangle,|\Psi_{7}\rangle,|\Psi_{8}\rangle,|\Psi_{11}\rangle,|\Psi_{12}\rangle$ & 0 & 0 & 0\\
$|\Psi_{6}\rangle,|\Psi_{9}\rangle$ & 0 & $\frac{1}{2^6 3}$ & $-\frac{1}{2^9 3^3}$\\
\bottomrule
\end{tabular}
\caption{Summary of the values of $\mathrm{HDet}_{4}$, $S$ and $T$ invariants for the 15 transverse Ising model eigenstates $|\Psi_{k}\rangle$ with $0\leq k \leq 15$ as a function of $\lambda$ for $0\leq\lambda\leq 2/\sqrt{3}$. Functions $H(\alpha_\pm,\beta_\pm,\gamma_\pm)$, $S(\alpha_\pm,\beta_\pm,\gamma_\pm)$ and $T(\alpha_\pm,\beta_\pm,\gamma_\pm)$ are written in Eq. \eqref{eq:STHDet_Ising} and Eq. \eqref{eq:abc+-_Ising} and the explicit eigenstates $|\Psi_{k}\rangle$ in App. \ref{app:OddsEnds}.}
\label{Tab:IsingTh}
\end{table}

To illustrate this result, let us write explicitly an  eigenstate for each type of  behaviour.
Let's start with eigenstates with zero $\mathrm{HDet}_{4}$, $S$ and $T$. An example is given by eigenstate $|\Psi_{3}\rangle$:
\begin{equation}
|\Psi_{3}\rangle=|\Psi^{-}\rangle_{13}|00\rangle_{24},
\end{equation}
where $|\Psi^-\rangle=(|01\rangle-|10\rangle)/\sqrt{2}$. The subscripts 13 and 24 stand for the spins represented by the corresponding state. Both invariants and $\mathrm{HDet}_{4}$ are zero when the state can be factorized in two bipartitions, which is this case.

For the first excited state, $S$ and $T$ are non zero but $\mathrm{HDet}_{4}=0$:
\begin{equation}
|\Psi_{1}\rangle =\frac{1}{2\sqrt{(\lambda+\sqrt{\lambda'})^2+1}}\left(\lambda+\sqrt{\lambda'}\right)\left(|00\rangle|\Psi^{+}\rangle+|\Psi^{+}\rangle|00\rangle\right) +|11\rangle|\Psi^{+}\rangle+|\Psi^{+}\rangle|11\rangle,
\end{equation}
where $|\Psi^{+}\rangle=(|01\rangle+|10\rangle)/\sqrt{2}$ and $\lambda'=1+\lambda^2$. Notice that this state is a combination of two $|W\rangle$-type states. For that reason, it keeps the same properties as $W$ states explained in the previous section.

There are other kind of states where $S \neq 0$ and $T  \neq 0$, but $\mathrm{HDet}_{4}=0$, in particular
\begin{align}
|\Psi_{6}\rangle&=\frac{1}{\sqrt{2}}\left(|0011\rangle-|1100\rangle\right),\\
|\Psi_{9}\rangle&=\frac{1}{\sqrt{2}}\left(|0101\rangle-|1010\rangle\right).
\label{eq:MaxEnt2}
\end{align}
These states have the same values of $S$ and $T$ as the GHZ state and are not separable in two bipartitions but they entangle half of the system with the other half. In fact, they represent the two ways of maximally entangle two spins in one direction with the other two in the opposite direction. If we define the states $|\rightrightarrows\rangle\equiv|00\rangle$ and $|\leftleftarrows\rangle\equiv|11\rangle$, then $|\Psi_{6}\rangle=\frac{1}{\sqrt{2}}\left(|\rightrightarrows\rangle|\leftleftarrows\rangle-|\leftleftarrows\rangle|\rightrightarrows\rangle\right)$ and $|\Psi_{9}\rangle=\frac{1}{\sqrt{2}}\left(|\rightrightarrows\rangle_{13}|\leftleftarrows\rangle_{24}-|\leftleftarrows\rangle_{13}|\rightrightarrows\rangle_{24}\right)$, which are $|\Psi^-\rangle$ states.

There are four states with non-zero $\mathrm{HDet}_{4}$: ground state and second, thirteenth and fifteenth excited states. The corresponding functions of $S(\alpha,\beta,\gamma)$, $T(\alpha,\beta,\gamma)$ and $H(\alpha,\beta,\gamma)$ shown in Tab. \ref{Tab:IsingTh} are
\begin{align}
S(\alpha,\beta,\gamma)&=\frac{\Gamma(\alpha,\beta,\gamma)}{12\mathcal{N}(\alpha,\beta,\gamma)^2},\nonumber\\
T(\alpha,\beta,\gamma)&=\frac{\left(4 \beta^2 (\alpha + \gamma^2)-(\alpha- \gamma^2)^2\right)\left(\Gamma (\alpha,\beta,\gamma)-768 \alpha \beta^4\gamma^2\right)}{216\mathcal{N}(\alpha,\beta,\gamma)^3},\nonumber\\
H(\alpha,\beta,\gamma)&= S(\alpha,\beta,\gamma)^3-27T(\alpha,\beta,\gamma)^2,
\label{eq:STHDet_Ising}
\end{align}
where $\Gamma(\alpha,\beta,\gamma)= \alpha^2 (\alpha-4 \beta^2)^2 -4 \alpha(\alpha^2 -2\alpha \beta^2 -56 \beta^4) \gamma^2 +2 (3 \alpha^2 + 4 \alpha \beta^2 + 8\beta^4)\gamma^4 -4 (\alpha + 2 \beta^2) \gamma^6 + \gamma^8$ and $\mathcal{N}(\alpha,\beta,\gamma)=(1 + \alpha^2 + 4\beta^2 + 2\gamma^2)^2$, which is actually the fourth power of the norm of these states as a function of $\alpha$, $\beta$ and $\gamma$ parameters. These parameters are also functions of $\lambda$ and for those states with non-zero $\mathrm{HDet}_{4}$ are
\begin{align}
\alpha_{\pm}&= \frac{1}{\lambda}\left(2 \lambda^3 +\sqrt{2}\lambda^2\sqrt{\lambda' \pm\sqrt{\lambda''}}-\sqrt{2}\sqrt{\lambda' \pm\sqrt{\lambda''}} \left(1 \mp \sqrt{\lambda''}\right) - \lambda\left(1 \mp2\sqrt{\lambda''}\right)\right),\nonumber\\
\beta_{\pm}&= \lambda+\frac{1}{\sqrt{2}}\sqrt{\lambda' \pm\sqrt{\lambda''}}, \nonumber\\
\gamma_{\pm}&= 1 + \frac{\sqrt{2} \lambda}{\sqrt{\lambda' \pm\sqrt{\lambda''}}},
\label{eq:abc+-_Ising}
\end{align}
where $\lambda''=1+\lambda^4$.

The ground state $|\Psi_{0}\rangle$ and the second excited state $|\Psi_{2}\rangle$ can be written in terms of the above parameters:
\begin{equation} 
|\varphi_{\pm}\rangle \propto \alpha_{\pm}|0000\rangle+ 2\beta_{\pm}|\Psi^{+}\rangle_{13}|\Psi^{+}\rangle_{24}+\gamma_{\pm}(|0101\rangle+|1010\rangle)+|1111\rangle,
\label{eq:36}
\end{equation}
where $+$ and $-$ stand for $|\Psi_{0}\rangle$ and $|\Psi_{2}\rangle$ respectively.  This equation shows  how rich is the quadripartite entanglement in these states. They contain all entanglement forms seen previously: part of the state is separable into two subsystems, other part of the state entangles maximally two spins in $|0\rangle$ state with two spins in $|1\rangle$ state and also contain the states with all spins aligned.

\begin{figure}[t!]
\centering
\includegraphics[width=0.6\textwidth]{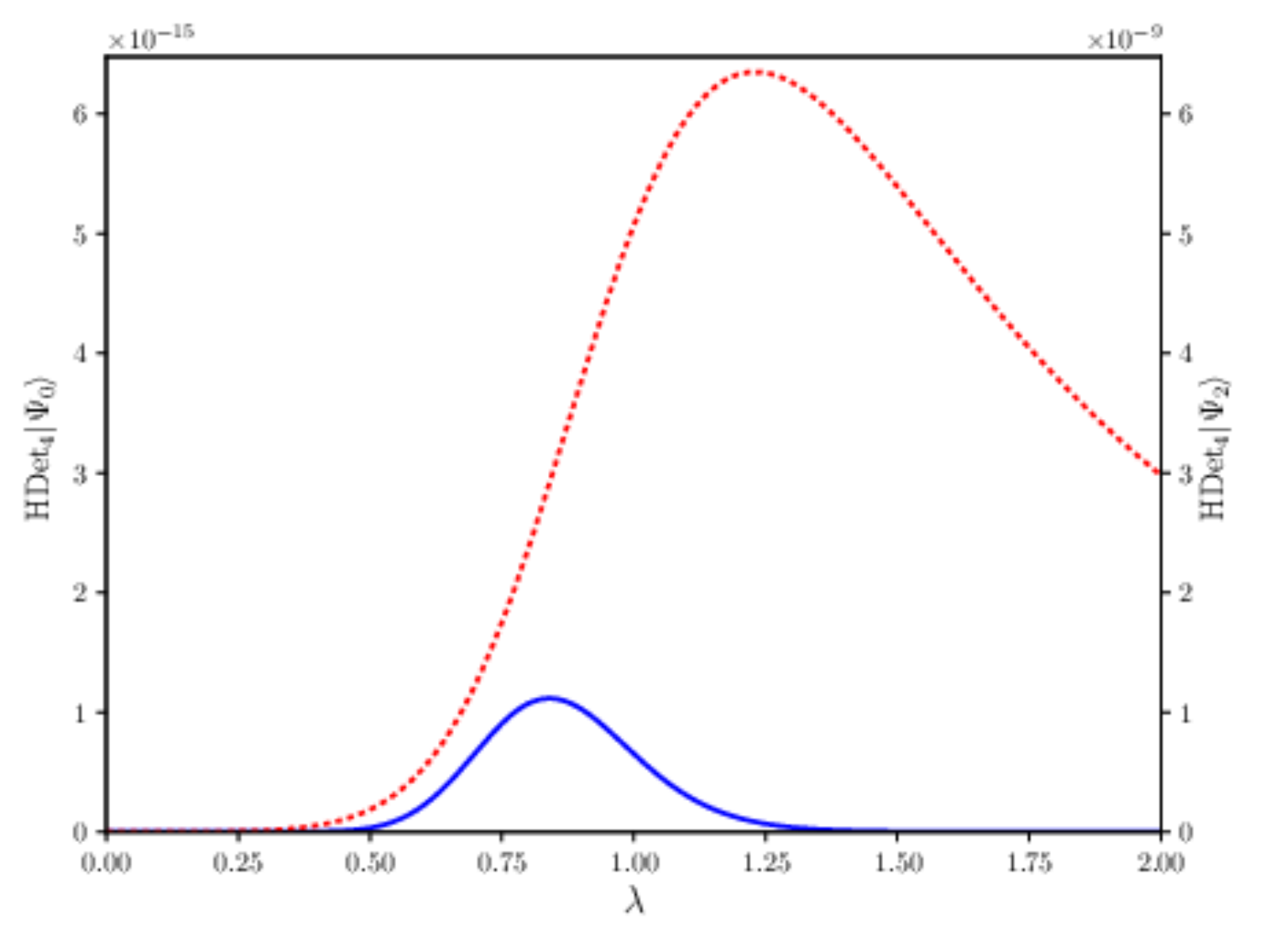}
\caption{$\mathrm{HDet}_{4}$ for the ground state $|\Psi_{0}\rangle$ (left axis, blue solid curve) and second excited state $|\Psi_{2}\rangle$ (right axis, dotted red curve) of the Ising model of $n=4$ spins as a function of the $\lambda$ field parameter. The $\mathrm{HDet}_{4}$ of the second excited state is seven orders of magnitude greater than the ground state's.}
\label{Fig:IsThHdet}
\end{figure}

Figure \ref{Fig:IsThHdet} shows $\mathrm{HDet}_{4}$ for the ground state and the second excited state. Both curves have peaks at  different values of $\lambda$: the ground state $\mathrm{HDet}_{4}$ peaks at $\lambda\sim 0.8$, close to the critical point, which for a chain of $n=4$ sites is $\lambda\simeq 0.7$, while the $\mathrm{HDet}_{4}$ of the second excited state peaks at $\lambda\sim 1.2$, where it is not the second excited state anymore, as $|\Psi_{2}\rangle$ intersects with $|\Psi_{3}\rangle$ at $\lambda=2/\sqrt{3}\sim 1.15$. The order of magnitude of the peaks are also different: when the ground state has $\mathrm{HDet}_{4} \propto 10^{-16}$, the second excited state has $\mathrm{HDet}_{4} \propto 10^{-9}$, similar to the mean value of $\mathrm{HDet}_{4}$ for a random state. Moreover, the excited state peak is broader than the ground state peak. Then, even both states have the same analytic structure, the differences in the coefficients of the wave function lead to a difference of seven orders of magnitude between their corresponding $\mathrm{HDet}_{4}$.

\section{The Heisenberg \texorpdfstring{$XXZ$}{} model \label{sec:XXZ}}

Together with transverse Ising model, Heisenberg model is another well known spin chain model. The $XXZ$ chain is a generalization of this model which Hamiltonian can be written as
\begin{definition}[$XXZ$ model]
\begin{equation}
\mathcal{H}_{XXZ}\equiv J\sum_{i=1}^n \left( \sx_{i}\sx_{i+1}+\sy_{i}\sy_{i+1}+ \Delta \sz_{i}\sz_{i+1} \right), 
\label{eq:XXZ}
\end{equation}
where $\Delta$ is the anisotropy parameter and $J$ is the coupling constant.
\end{definition}

We set $J=1$, which entails a ferromagnetic ground state for $\Delta<-1$ and an anti-ferromagnetic ground state for $\Delta>1$, also called N\'eel phase.

This model is critical in the region $\Delta\in(-1,1]$, known as the $XY$ phase \cite{Baxter85}. Its entropy scales following a conformal scaling law with a central charge $c=1$, so it belongs to a different universality class than the Ising model. Then, this model present two quantum phase transitions, at $\Delta=1$ and at $\Delta=-1$. The first one is a Kosterlitz-Thouless where the gap scales as $e^{-\pi^2/2\sqrt{2(\Delta-1)}}$ for $\Delta$ slightly larger than one \cite{Cabra04}. The second transition at $\Delta = -1$ belongs to the Dzhaparidze-Nersesyan-Pokrovsky-Talapov 
universality class \cite{DN78,Pokrovsky79}, where the entropy scales as $S \simeq \frac{1}{2} \log L$ at $\Delta \rightarrow -1^{+}$ \cite{Chen13}.  

\subsection{Eigenstates}

We diagonalize the $XXZ$ Hamiltonian with $n=4$ spins and periodic boundary conditions. All energy spectrum is shown in Tab. \ref{Tab:XXZstates} in App. \ref{app:OddsEnds}. Analogously with Ising spectrum, the level order will depend on the value of $\Delta$. For $\Delta<-1$, the ground state is degenerate and corresponds to the states with all spins aligned (ferromagnetic phase). For $\Delta>-1$ the ground state is unique and has energy $-2(\Delta+\sqrt{8+\Delta^2})$. At the isotropic point $\Delta =1$, it describes a 
resonating valence bound state, which will be explained at the end of this subsection. 

The expressions of  $S$,  $T$ and  $\hdet_{4}$ for the states obtained after the diagonalization are summarized in Tab. \ref{Tab:XXZ_hdet}. All states exhibit $\hdet_{4}=0$ either because $S$ and $T$ vanish, or because they cancel each other.

There are three types of states that lead to null $S$ and $T$ invariants. Similarly with  Ising model, some states are separable into two subsystems. For example, one of the states with zero energy can be written as
\begin{equation}
|\Psi(E=0)\rangle=\frac{1}{\sqrt{2}}\left(|0111\rangle-|1101\rangle\right)=|\Psi^-\rangle_{13}|11\rangle_{24},
\end{equation}
where $|0\rangle\equiv|\uparrow\rangle$ and $|1\rangle\equiv|\downarrow\rangle$ are the eigenstates of $\sz$.

There are two eigenstates that are product states in the $XXZ$ spectrum: $|0000\rangle$ and $|1111\rangle$, all spins are aligned, both with energy $4\Delta$. They correspond to the ground states for $\Delta<-1$ and the most excited states for $\Delta>1$ respectively.

Finally, the third type of states with $S=T=0$ are $W$-like. For instance, one of the states with energy 4 is
\begin{equation}
|\Psi(E=4)\rangle=\frac{1}{2}\left(|0111\rangle +|1011\rangle +|1101\rangle +|1110\rangle\right).
\end{equation}

\begin{table}[t!]
\centering
\begin{tabular}{@{}lccc}
\toprule
\textbf{Energy} & $\mathbf{\mathit{S}}$ & $\mathbf{\mathit{T}}$ & $\mathbf{\hdet_{4}}$ \\
\midrule
$-4(2)$, 4(2), 0(6), 4$\Delta$(2) & 0 & 0 & 0 \\
0, $-4\Delta$ &  $\frac{1}{2^63}$ & $-\frac{1}{2^9 3^3}$ & 0 \\
$-2\left(\Delta-\sqrt{8+\Delta^2}\right)$ & $S_{+}$ & $T_{+}$ & 0\\
$-2\left(\Delta+\sqrt{8+\Delta^2}\right)$ & $S_{-}$ & $T_{-}$ & 0\\
\bottomrule
\end{tabular}
\caption{$S$, $T$ and $\hdet_{4}$ of $XXZ$ model for states obtained after the Hamiltonian diagonalization. All states lead to zero $\hdet_{4}$ and only four states have $S$ and $T$ invariants different from zero. The values in parenthesis represent the degeneracy and $S_{\pm}$ and $T_{\pm}$ expressions  correspond with Eq. \ref{eq:ST_XXZ}.
}
\label{Tab:XXZ_hdet}
\end{table}

Only four energies have $S$ and $T$ different from zero. Two of them, with energies 0 and $-4\Delta$, correspond with the two ways of maximally entangle two sets of spins in opposite directions. These are the same states as the Ising model but in $\sz$ basis, i.e. $|\upuparrows\rangle\equiv|00\rangle$ and $|\downdownarrows\rangle\equiv|11\rangle$. Then, these states become $\frac{1}{\sqrt{2}}\left(-|\upuparrows\rangle_{12}|\downdownarrows\rangle_{34}+|\downdownarrows\rangle_{12}|\upuparrows\rangle_{34}\right)$ and $\frac{1}{\sqrt{2}}\left(-|\upuparrows\rangle_{13}|\downdownarrows\rangle_{24}+|\downdownarrows\rangle_{13}|\upuparrows\rangle_{24}\right)$. Both states have $S$ and $T$ constant and with the same value as in the Ising model case, i.e. $S=1/(2^6 3)$ and $T=-1/(2^9 3^3)$.

There are two states with $S$ and $T$ that depend on $\Delta$. The one that has an energy $-2\left(\Delta+\sqrt{8+\Delta^2}\right)$, corresponds to the ground state for $\Delta> -1$:
\begin{equation}
|\phi_{1}\rangle=\frac{1}{\mathcal{N}}\left(|0011\rangle +|0110\rangle +|1100\rangle +|1001\rangle -\frac{1}{2}\left(\Delta+\sqrt{8+\Delta^2}\right)\left(|0101\rangle+|1010\rangle\right)\right),
\label{eq:XXZ_gs}
\end{equation}
where $\mathcal{N}=8 + \Delta(\Delta +\sqrt{8 + \Delta^2})$. Invariants $S$ and $T$ are non zero as long as $\Delta\neq 1$. When $\Delta=1$ it becomes a resonating valence bound state, as it is shown at the end of this subsection. The other state has energy $-2\left(\Delta-\sqrt{8+\Delta^2}\right)$ and corresponds to the state with higher energy for $\Delta<1$:
\begin{equation}
|\phi_{2}\rangle=\frac{1}{\mathcal{N'}}\left(|0011\rangle +|0110\rangle +|1100\rangle +|1001\rangle -\frac{1}{2}\left(\Delta-\sqrt{8+\Delta^2}\right)\left(|0101\rangle+|1010\rangle\right)\right).
\label{eq:XXZ_e15}
\end{equation}
where $\mathcal{N'}=8 + \Delta(\Delta -\sqrt{8 + \Delta^2})$. This state has $S$ and $T$ different from zero as long as $\Delta\neq -1$. 

\begin{figure}
\centering
\includegraphics[width=0.6\textwidth]{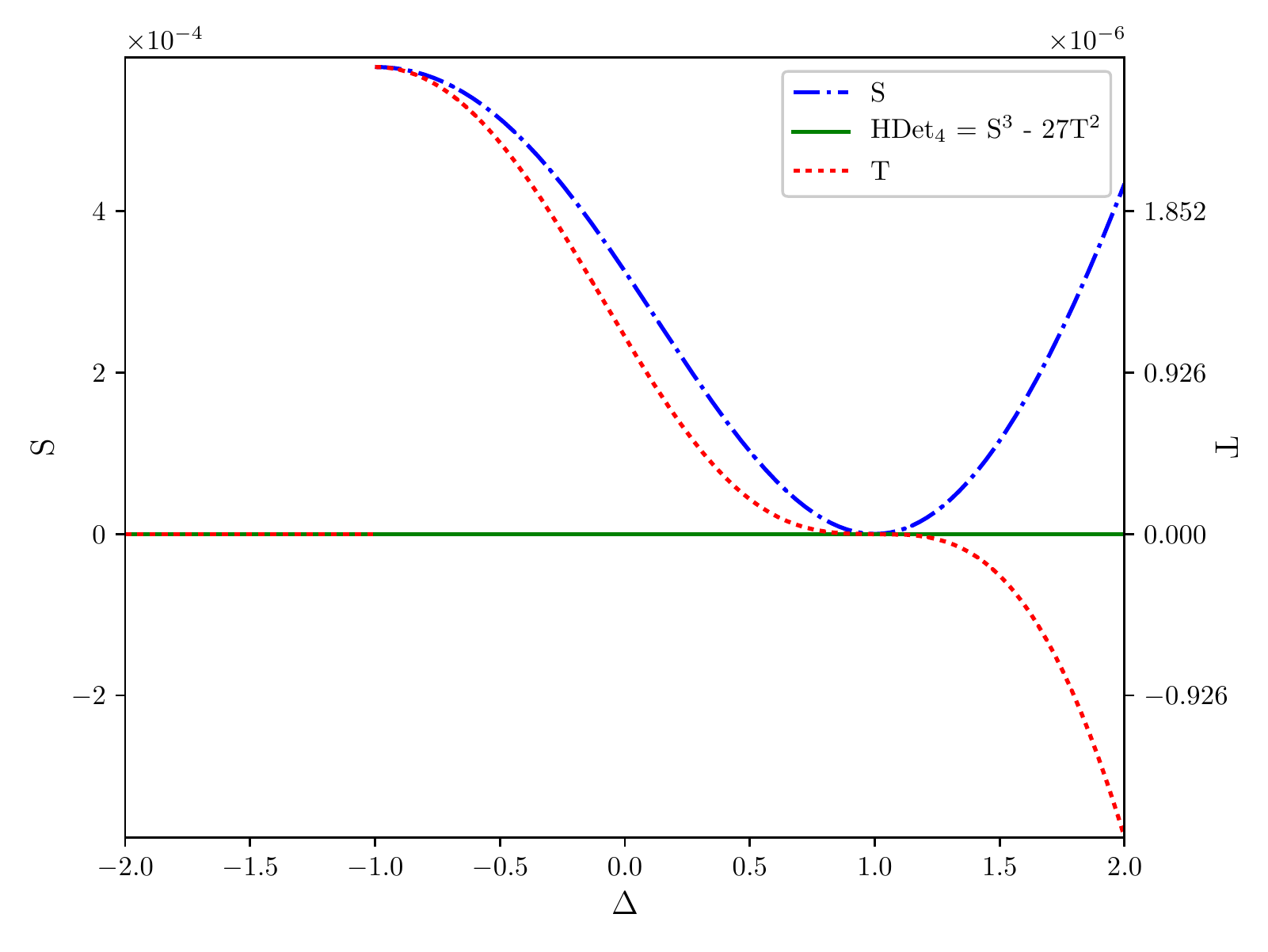}
\caption{$S$ and $T$ invariants of the ground state of $n=4$ $XXZ$ spin chain. $\hdet_{4}$ is always zero but the $S$ and $T$ invariants are able to detect the transition points at $\Delta=-1$ and $\Delta=1$.}
\label{Fig:XXZ}
\end{figure}

The expressions for the invariants of these two states are
\begin{align}
S_{\pm}&=\frac{1}{2^8\cdot 3}\frac{\left(\Delta\pm\sqrt{8+\Delta^2}\right)^4\left(4-\Delta\left(\Delta\mp\sqrt{8+\Delta^2}\right)\right)^2}{\left(8+\Delta\left(\Delta\pm\sqrt{8+\Delta^2}\right)\right)^4}, \\
T_{\pm}&=\frac{1}{2^{12}\cdot 3^3}\frac{\left(\Delta\pm\sqrt{8+\Delta^2}\right)^6\left(4-\Delta\left(\Delta\mp\sqrt{8+\Delta^2}\right)\right)^3}{\left(8+\Delta\left(\Delta\pm\sqrt{8+\Delta^2}\right)\right)^6},
\label{eq:ST_XXZ}
\end{align}
and are shown in Fig. \ref{Fig:XXZ}. Invariants for these two states seem to be sensible to the transition points $\Delta=1$ and $\Delta=-1$. 

As a final remark, notice that the above states correspond to the $G_{abcd}$ state of Eq. \eqref{eq:Gabcd} with $a=-d$, which makes $S$ and $T$ proportional to $(a^2 - b^2)(a^2 - c^2)$.

The $XXZ$ model for $\Delta=1$ corresponds with the isotropic Heisenberg model, also known as the $XXX$ or simply Heisenberg model. Its Hamiltonian is invariant under the rotation group, which allows for an easy derivation of energy spectrum. For $n=4$ spins, the Hamiltonian can be written in terms of spin operators $\vec{S}=\frac{1}{2}(\sx,\sy,\sz)$ as 
\begin{align}
\mathcal{H}_{XXX}&=
4\left(\vec{S}_{1}\cdot\vec{S}_{2} +\vec{S}_{2}\cdot\vec{S}_{3} +\vec{S}_{3}\cdot\vec{S}_{4} +\vec{S}_{4}\cdot\vec{S}_{1}\right)  \nonumber\\
&= 2\left(\left(\vec{S}_{1}+\vec{S}_{2}+\vec{S}_{3}+\vec{S}_{4}\right)^2 -\left(\vec{S}_{1}+\vec{S}_{3}\right)^2 -\left(\vec{S}_{2}+\vec{S}_{4}\right)^2\right) \nonumber\\
 &= 2\left( s(s+1) - s_{13}(s_{13}+1)-s_{24}(s_{24}+1)\right),
\end{align}
where $s$ is the total spin and $s_{13}$ and $s_{24}$ are the total spin for particles 1 and 3, and 2 and 4 respectively. 

Tab. \ref{Tab:XXZ_Spin} shows the different values of $s_{13}$, $s_{24}$ and $s$ and the corresponding energy of $\mathcal{H}_{XXX}$. When the total spin is zero, the state is called a Resonating Valence Bound \cite{Anderson73}. There are two of them in Heisenberg spin chain:
\begin{align}
|\phi_{1}\rangle &=
\frac{1}{2\sqrt{2}}\left(|0011\rangle +|0110\rangle +|1100\rangle +|1001\rangle -2\left(|0101\rangle+|1010\rangle\right)\right), \label{eq:phi1}\\
|\phi_{2}\rangle &=\frac{1}{2}\left(|0011\rangle -|0110\rangle -|1001\rangle +|1100\rangle\right). \label{eq:phi2}
\end{align}
The first one corresponds to the ground state whereas the second is a lineal combination of states with zero energy. Both have the property $S=T=0$. To check if this is a general property of resonating valence bound states, we have checked that the state
\begin{equation}
|\phi\rangle=\cos\theta|\phi_{1}\rangle + e^{i\varphi}\sin\theta|\phi_{2}\rangle
\end{equation}
also have $S$ and $T$ zero $\forall$ $\theta,\varphi$.

\begin{table}[t!]
\centering
\begin{tabular}{@{}cccc}
\toprule
\textbf{Energy} & $\mathbf{\mathit{s_{13}}}$ & $\mathbf{\mathit{s_{24}}}$ & $\mathbf{\mathit{s}}$ \\
\midrule
$-8$ & 1 & 1 & 0 \\
$-4$ & 1 & 1 & 1 \\
$0$ & 0 & 1 & 1 \\
$0$ & 1 & 0 & 1 \\
$0$ & 0 & 0 & 0 \\
4 & 1 & 1 & 2 \\
\bottomrule
\end{tabular}
\caption{Energies for the $n=4$ Heisenberg model ($XXZ$ model with $\Delta=1$) according to the total spin of their particles. When the total spin is zero, it is called a Resonating Valence Bound state.}
\label{Tab:XXZ_Spin}
\end{table}

\subsection{Degeneracy}

We can also check what is the effect of degeneracy on $\hdet_{4}$. Although all states of $XXZ$ Hamiltonian have $\hdet_{4}=0$, linear combinations of states with the same energy, which is also an eigenstate, could have $\hdet_{4}\neq 0$ or modify the values of $S$ and $T$ invariants. 

As example, let us analyse the case of Heisenberg model. As it is shown in Tab. \ref{Tab:XXZ_Spin}, there are four different energies in this particular case. The ground state is not degenerate, so the values of $\hdet_{4}$, $S$ and $T$ invariants remain the same as computed above; $\hdet_{4}=0$, $S=S_{-}$ and $T=T_{-}$.

The state with energy $E=-4$ has degeneracy 3. Any state with the form
\begin{multline}
|\Psi(E=-4)\rangle = \frac{1}{\mathcal{N}}\left(a(|0111\rangle -|1011\rangle +|1101\rangle -|1110\rangle) + b(|0101\rangle -|1010\rangle) \right. \\
\left. + c(|0001\rangle -|0010\rangle +|0100\rangle -|1000\rangle)\right)
\end{multline}
is also an eigenstate. This state has $\hdet_{4}=0$ due to an exact cancellation between the two invariants:
\begin{eqnarray}
S(E=-4)&=&\frac{(b^2 - 4 ac)^4}{192 (2 a^2 + b^2 + 2 c^2)^4}, \nonumber\\
T(E=-4)&=&-\frac{(b^2 - 4 ac)^6}{13824 (2 a^2 + b^2 + 2 c^2)^6}.
\end{eqnarray}

The state with energy $E=4$ has degeneracy 5. Then, any state with the form
\begin{multline}
|\Psi(E=4)\rangle = \frac{1}{\mathcal{N}}\left(a(|0111\rangle +|1011\rangle +|1101\rangle +|1110\rangle) + b|0000\rangle+ \right. \\
\hspace{3cm} \left. c(|0001\rangle +|0010\rangle +|0100\rangle +|1000\rangle) + d|1111\rangle  + \right. \\
\left. e(|0011\rangle +|0110\rangle +|1100\rangle +|1001\rangle +|0101\rangle+|1010\rangle)\right)
\end{multline}
is also an eigenstate. In this case, $\hdet_{4}$ could be different from zero. We do not include the expressions of the invariants as they are cumbersome and not very illustrative.

Finally, the state with energy $E=0$ has degeneracy 7. In this case, $\hdet_{4}=0$ again for the cancellation between $S$ and $T$ invariants.

\subsection{Thermal state}

\begin{figure}[t!]
\centering
\includegraphics[width=0.6\textwidth]{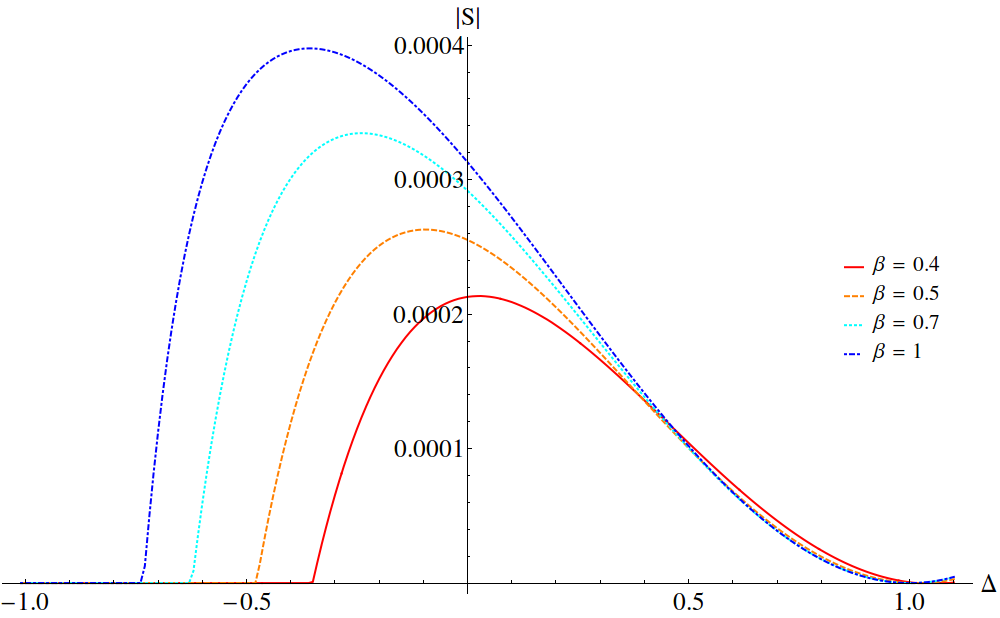}
\caption{$S$ invariant for the $XXZ$ spin chain model as a function of $\Delta$ for different values of $\beta=1/T$. The amount of entanglement quantified by the $S$ invariant tends to zero as the temperature increases, as expected.
}
\label{Fig:XXZth}
\end{figure}

The $S$ invariant for a thermal states of the $XXZ$ spin chain with $n=4$ sites is computed using definition of Eq. \eqref{eq:Th2} and plotted in Fig. \ref{Fig:XXZth}. As $\beta$ decreases, the amount of entanglement quantified by this invariant decreases until zero. As expected, multipartite entanglement is lost at high temperatures.

Furthermore, discontinuity at $\Delta=-1$ softens and moves to higher $\Delta$ as temperature increases. On the contrary, the vanishing $S$ at $\Delta=1$ remains independently of the $\beta$ values. 

\section{The generalized Haldane-Shastry wave functions \label{sec:HS}}

The Haldane-Shastry model (HS) \cite{Haldane88,Shastry88} describes a chain of equally spaced spin-$\frac{1}{2}$ particles in a circle with pairwise interactions inversely proportional to the square of the distance between the spins. 
\begin{definition}[Haldane-Shastry model]
The Haldane-Shastry Hamiltonian represents a $n$ spin-$1/2$ chain with interaction
\begin{equation}
\mathcal{H}_{HS} \equiv  \frac{\pi^2}{n^2}  \sum_{i > j  }^n  \frac{ \vec{S}_i \cdot \vec{S}_j} { \sin^2 \left(  \frac{ \pi (i-j)}{n} \right)} \ ,
\label{hs1}
\end{equation}
where $\vec{S}_i = \frac{1}{2} \vec{\sigma}_i$.
\end{definition}

The ground state of HS Hamiltonian can be written as \cite{Cirac10}
\begin{equation}
\psi(s_{1},\cdots ,s_{n})\propto \delta_{s}e^{i\frac{\pi}{2} \sum_{i:\mathrm{odd}}s_{i}}\prod_{i>j}^{n}\left\vert\sin\left(\frac{\pi(i-j)}{n}\right)\right\vert^{s_{i}s_{j}/2}.
\label{eq:HS_sin}
\end{equation}
where the spin at the site $i=1, \dots, n$ is given by $s_i/2$ with $s_{i}=\pm 1$, $\delta_{s}=1$ if $\sum_{i=1}^{n}s_{i}=0$ and $\delta_{s}=0$ otherwise. The latter condition implies  that the total third component of the spin vanishes, that is $\langle \sum_i S^z_i \rangle =0$, but the HS state is also a singlet of the rotation group, $\langle (\sum_i \vec{S}_i )^2 \rangle =0$.  

The HS wave function has a huge overlap with the ground state of the isotropic Heisenberg model. In fact,  for $n=4$ sites these two wave functions are the same. The HS Hamiltonian belongs to the same universality class as the isotropic Heisenberg model, which is described by the Wess-Zumino-Witten model $SU(2)_1$  that has a central charge $c=1$.  

The wave function (\ref{eq:HS_sin}) was generalized in Ref. \cite{Cirac10} to the following one 
\begin{equation}
\psi(s_{1},\cdots ,s_{n})\propto \delta_{s}e^{i\frac{\pi}{2} \sum_{i:\mathrm{odd}}s_{i}}\prod_{i>j}^{n}\left\vert\sin\frac{\pi(i-j)}{n}\right\vert^{ \alpha s_{i}s_{j}}, 
\label{eq:HS_sin2}
\end{equation}
and was used as a variational ansatz for the ground state of the $XXZ$ model in the critical regime. The relation between the anisotropy parameter $\Delta$ and the parameter $\alpha$ was found to be $\Delta=-\cos(2\pi\alpha)$, with  $0 < \alpha \leq \frac{1}{2}$, corresponding to the critical region $-1 < \Delta \leq 1$. The cases $\alpha = 0, \frac{1}{4}$ provide the exact solution of the $XXZ$ model for $\Delta =-1$ and $\Delta=0$, while  $\alpha = \frac{1}{2}$, is the HS wave function (\ref{eq:HS_sin}). 

\begin{figure}[t!]
\centering
\includegraphics[width=0.55\textwidth]{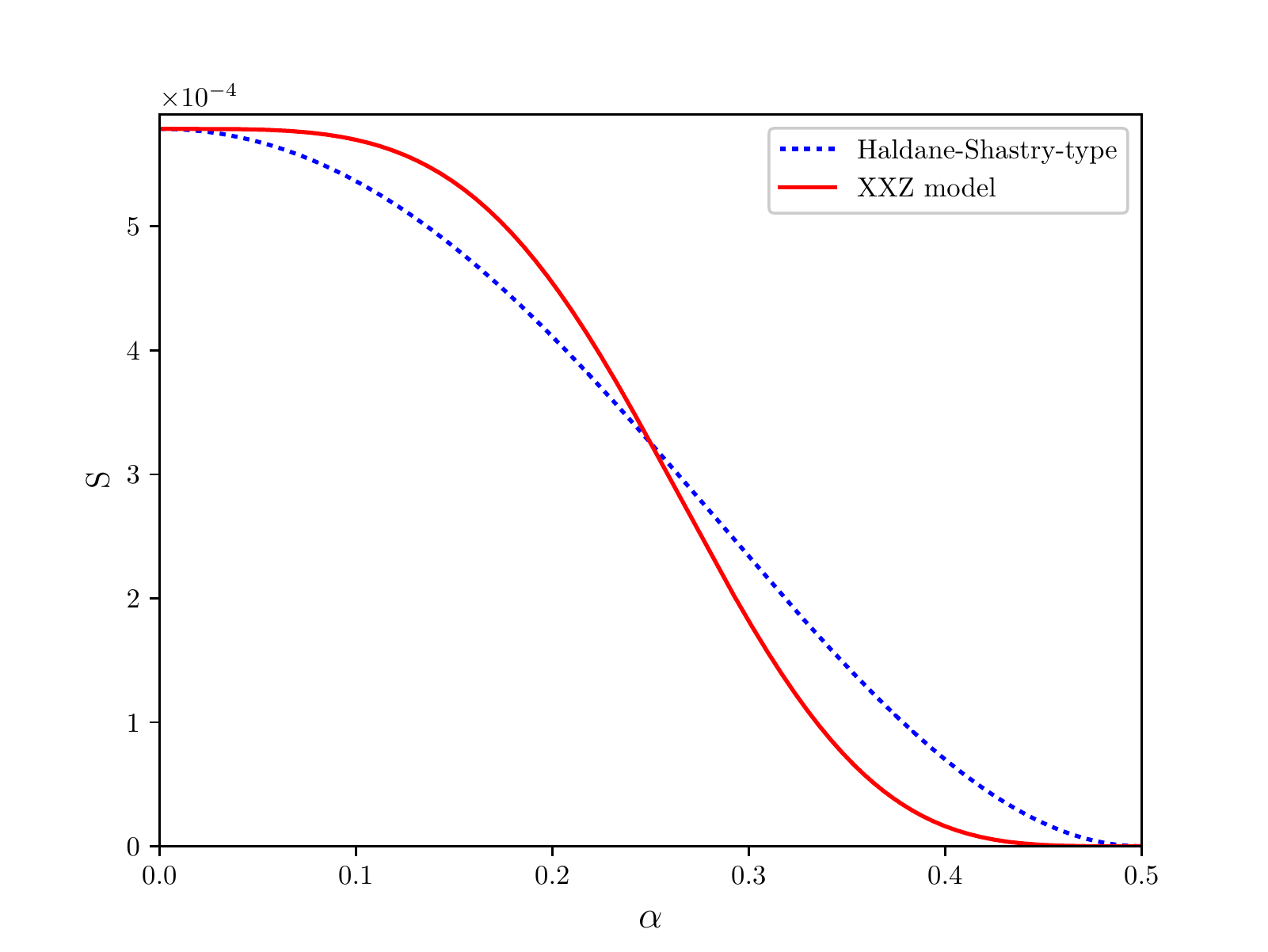}
\caption{Comparison of the $S$ invariant of the ground state of the $XXZ$ model and the wave function (\ref{eq:HS_sin2}) 
for $n=4$ spins. Both wave functions coincide for $\Delta =-1, 0, 1$ which correspond to $\alpha =0, \frac{1}{4}, \frac{1}{2}$. } 
\label{Fig:S_alpha}
\end{figure}

\subsection{Ground state and \texorpdfstring{$S$}{} and \texorpdfstring{$T$}{} invariants}

In the ground state of the HS  model is
\begin{equation}
|\Psi\rangle_{HS}=\frac{1}{\mathcal{N}}\left(4^{-\alpha}(|0011\rangle+|0110\rangle+|1001\rangle+|1100\rangle)-(|0101\rangle+|1010\rangle)\right),
\label{eq:HScomp}
\end{equation}
where $\mathcal{N}=\sqrt{1+3(4^{-2\alpha})+4^{-\alpha}}$ and we have used the computational basis states $|0\rangle$ and $|1\rangle)$ to describe the spins $s_{i}=\pm 1$. This type of wave function have $\hdet_{4}=0$ as a consequence of the cancellation between $S$ and $T$ invariants
\begin{align}
S_{HS}&=\frac{4^{4\alpha-3}\left(16^\alpha-4\right)^2}{3(2+16^\alpha)^4},\nonumber\\
T_{HS}&=-\frac{8^{4\alpha-3}\left(16^\alpha-4\right)^3}{27(2+16^\alpha)^6}.
\label{eq:ST_HS}
\end{align}
Thus, as in the  $XXZ$ model, we shall study the $S$ and $T$ behaviours instead of $\hdet_{4}$ which vanishes identically.

Figure \ref{Fig:S_alpha} shows the $S$ invariant as a function of $\alpha$ parameter. As expected, it matches with the $XXZ$ $S$ invariant at $\alpha=0,\frac{1}{4},\frac{1}{2}$. Also, $\alpha=\frac{1}{4}$ is the inflexion point: for $\alpha<\frac{1}{4}$ we get $S_{XXZ}>S_{HS}$ whereas for $\alpha>\frac{1}{4}$ the results is $S_{XXZ}<S_{HS}$. Similar results are found for $T$ invariant.

\subsection{Dimerized wave function}

We can modify the interaction strength between the spins introducing a new parameter $\delta$,
in the  wave function:
\begin{definition}[Dimerized HS wave function]
\begin{equation}
\psi_{\delta}(s_{1},\cdots ,s_{n})\propto \delta_{s}e^{i \frac{\pi}{2} \sum_{i:\mathrm{odd}}s_{i}}\prod_{i>j}^{n}\left\vert 2\sin\left(\theta_{i}-\theta_{j}\right)\right\vert^{\alpha s_{i}s_{j}},
\label{eq:HSdelta}
\end{equation}
with $\theta_{j}=\pi/n\left(j+\delta(-1)^j\right)$ for $j=1,\cdots ,n$.
\end{definition}

In terms of the computational basis states, the wave function become
\begin{equation}
|\Psi_{\delta}\rangle \propto a_{1}\left(|0011\rangle+|1100\rangle\right) +a_{2}\left(|0101\rangle+|1010\rangle\right) +a_{3}\left(|0110\rangle+|1001\rangle\right),
\end{equation}
where
\begin{align}
a_{1}&=-2^{-\alpha}\left|\frac{\cos\left(\pi(3+2\delta)/4\right)}{\cos(\pi\delta/2)-\sin(\pi\delta/2)}\right|^{2\alpha},\nonumber\\
a_{2}&=|\cos(\pi\delta)|^{-2\alpha},\nonumber\\
a_{3}&=-4^{-\alpha}\left|1-\frac{2}{1+\tan(\pi\delta/2)}\right|^{2\alpha}.
\end{align}
The invariants $S$ and $T$ become
\begin{align}
S&=\frac{\left(a_{1}^4+\left(a_{2}^2-a_{3}^2\right)^2-2a_{1}^2\left(a_{2}^2+a_{3}^2\right)\right)^2}{192\left(|a_{1}|^2+|a_{2}|^2+|a_{3}|^2\right)^4} \ ,\\
T&=-\frac{\left(a_{1}^4+\left(a_{2}^2-a_{3}^2\right)^2-2a_{1}^2\left(a_{2}^2+a_{3}^2\right)\right)^3}{13824\left(|a_{1}|^2+|a_{2}|^2+|a_{3}|^2\right)^6} \ .
\end{align}

\begin{figure}[t!]
\centering
\begin{minipage}{0.55\textwidth}
\centering
\includegraphics[width=\textwidth]{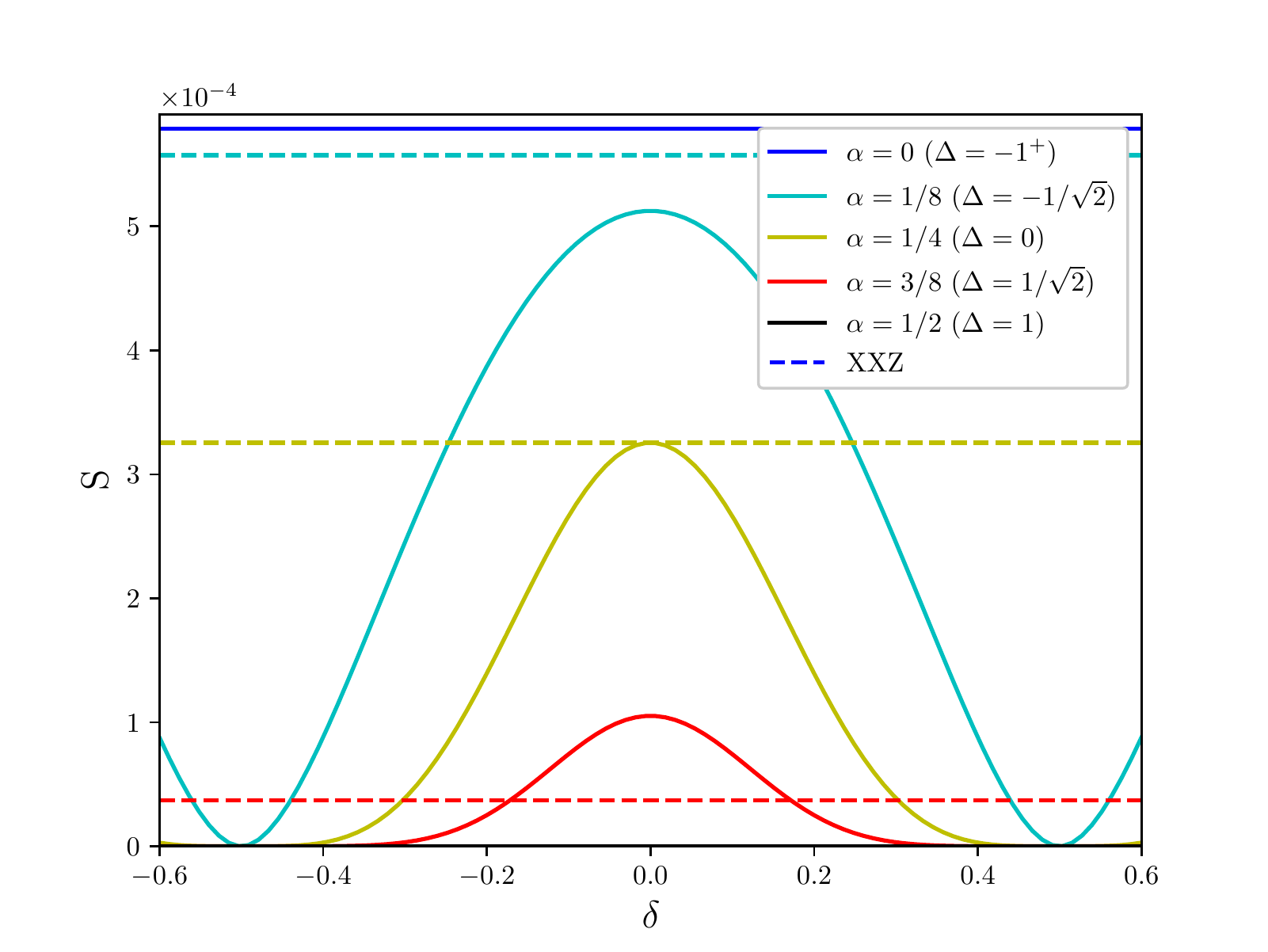}
\end{minipage}%
\begin{minipage}{0.35\textwidth}
\centering
\includegraphics[width=\textwidth]{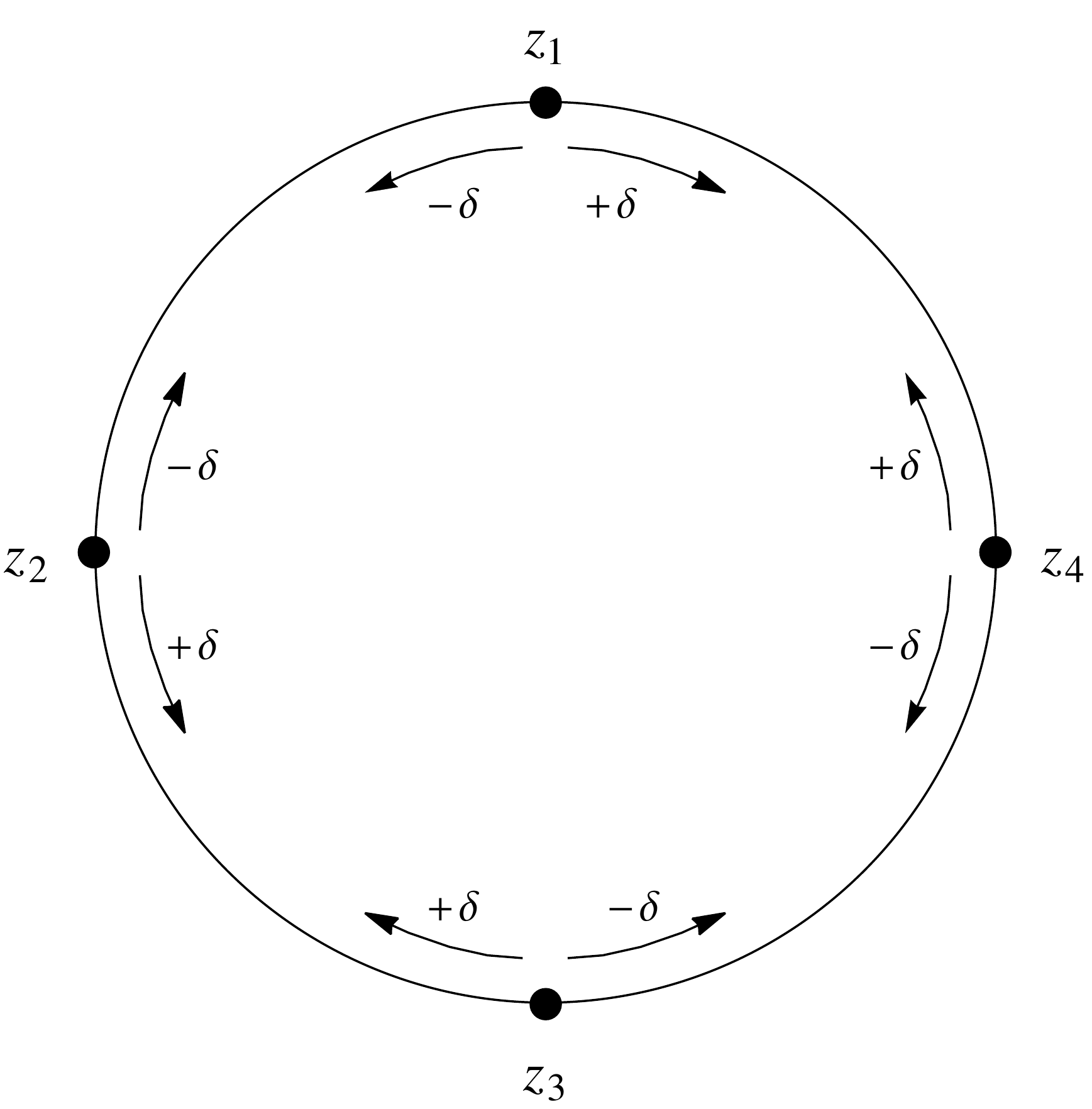}
\end{minipage}
\caption{\textit{Left}: $S$ invariant as a function of $\delta$ parameter for different values of $\alpha$. \textit{Right}: Diagrammatic representation of the $n=4$ Haldane-Shastry spin chain with the dimerization parameter $\delta$. For $\delta>0$ spins 1 and 4 and 2 and 3 are attracted each other, while for $\delta<0$ the attraction is between spins 1 and 2 and 3 and 4. For $|\delta|=\frac{1}{2}$, two consecutive spins are at the same position and the ground state is divided into two singlet states (dimer). As a consequence, $S$ and $T$ invariants become zero.}
\label{Fig:S_delta}
\end{figure}

Figure \ref{Fig:S_delta} left shows the $S$ invariant as a function of $\delta$ parameter for different $\alpha$ values. It matches with $XXZ$ model at $\alpha=0,\frac{1}{2}$ and shows a periodicity $S(\alpha,\delta)=S(\alpha,\delta\pm 1)$. Its maximum are located at $\delta=\pm m$ and its minimum at $\delta=\pm \frac{m}{2}$ for integer $m$. Moreover, maximum for $\alpha=\frac{1}{4}$ matches with $S$ invariant for the $XXZ$ model at $\Delta=0$, as expected. In fact, it is enough to consider $\delta\in[-\frac{1}{2},\frac{1}{2}]$. 

We can write the wave function of Eq. \eqref{eq:HSdelta} using the complex numbers $z_{j}=e^{2 i \theta_{j}}$. Then, $z_{j}$ correspond with the position of local spins, so at $\delta=\frac{1}{2}(-\frac{1}{2})$, spins 1 and 4 (1 and 2) and 2 and 3 (3 and 4) are at the same position and the state is a product of two singlets, i.e. a dimer, as it is shown diagrammatically in Fig. \ref{Fig:S_delta} right. Then, the state of four spins is separable into two subsystems and $S$ and $T$ become zero.

\section{Conclusions \label{sec:conclusionsHDet}}

\begin{figure}[t!]
\centering
\includegraphics[width=0.8\textwidth]{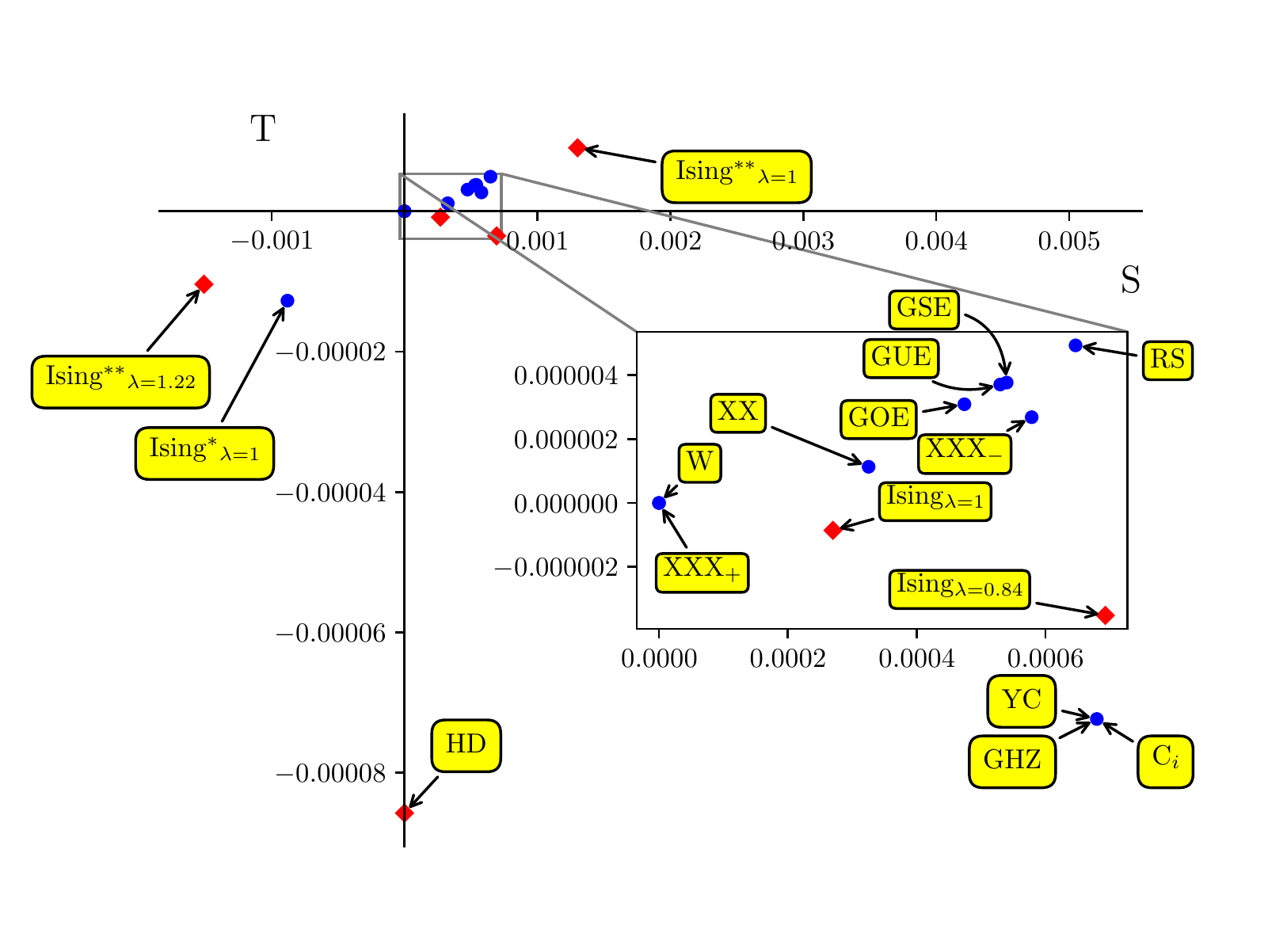}
\caption{\emph{Entanglement landscape.} This plot shows the amount of entanglement of several wave functions analysed in this chapter and quantified using $S$ and $T$ invariants. For the  Ising model we plot ground state, $1^{st}$ and $2^{nd}$ excited states -- denoted respectively with $*$ and $**$. For the $XXZ$ model, we plot $\Delta=0$, that is, Heisenberg model, and $\Delta=\pm 1$, labelled with $XXX_\pm$. RS stands for the mean value of a random state and GOE, GUE and GSE for the mean values of the ground state of random matrix Hamiltonians. Due to relation \eqref{eq:HDetST}, some states have zero $\hdet_{4}$. Then we indicate with red diamond points the states with $\hdet_{4}\neq 0$.}
\label{Fig:landscape}
\end{figure}

In this chapter we have studied the quadripartite entanglement of several quantum states of four spin-$\frac{1}{2}$ models, in particular transverse Ising model, $XXZ$ model and generalized Haldane-Shastry model. We have also studied random pure states and ground states of Gaussian Hamiltonians. 

As a figure of merit to quantify multipartite entanglement, we have used the Schl\"afli hyperdeterminant $\hdet_{4}$ \cite{Schlafli52}, which is an extension of the $2\times 2\times 2$ dimensional Cayley's hyperdeterminant \cite{Cayley45}. The hyperdeterminant can also be constructed from the two polynomial invariants $S$ and $T$ as $\hdet_{4} = S^3 - 27 T^2$. The latter quantities provide a more refined  characterization of the quadripartite entanglement, particularly in those cases where $\hdet_{4}$ vanishes. 

An overview of the results is shown in the $S-T$ diagram plotted in Fig. \ref{Fig:landscape}.  We found that $\hdet_{4}$ is sensible to different priors on such random states. Flat and Haar distributed coefficients in these states exhibit different mean values of $\hdet_{4}$. It can also be observed a difference between GOE and the other two Gaussian distributions, GUE and GSE, in terms of their mean values of $\hdet_{4}$.
 
For the Ising model, we found that ground state $\hdet_{4}$ shows a pronounced peak at $\lambda\simeq 0.84$, which lies near the critical point of the model for $n=4$ spins, located at $\lambda\simeq 0.7$. The $XXZ$ model exhibit vanishing values of $\hdet_{4}$ for all non-degenerate states. This fact is due to an exact cancellation between the $S$ and $T$ invariants as a consequence of the relation $\hdet_{4} = S^3 - 27 T^2$. In the whole critical regime $- 1 < \Delta \leq 1$, one has $S \geq 0$, with a discontinuity at the point $\Delta =-1$ and achieving $S=0$ for $\Delta=1$. In the anti-ferromagnetic regime $\Delta >1$, one has that $S < 0$. All together suggests that $\hdet_{4}$ and $S$ invariant are able to catch a quantum phase transition.

The results obtained for the generalized Haldane-Shastry type model are similar to  those of the  $XXZ$ model in the critical regime. We also introduce a dimerization factor $\delta$ and study the multipartite entanglement as a function of this coefficient. The result shows that $S$ and $T$ invariants are maximum when $\delta=0$ and zero when $\delta=\frac{1}{2}$, which corresponds to two consecutive spins at the same physical position: the state becomes a product state of two singlets (dimer). Again, $S$ and $T$ invariants seem to be sensible to phase changes.

In summary, we have shown that $\hdet_{4}$ is a useful tool to characterize multipartite entanglement in several wave functions. For random distributed states, it is sensible to the prior used. This analysis can be extended to other priors than the ones used in this work. In the analysis of ground state of 4 spin chains, it is able to detect phase transitions even for  such a small number of particles. A direct extension to higher values of the spin or  more sites seems at the moment out of reach, but it suggests new tool to characterize multipartite entanglement along this direction. 

\chapterimage{XY_image} 
\chapter{Quantum Phase Transition in a Quantum Computer \label{Ch:Ising}}


\vspace{-1.5cm}
\begin{flushright}
\begin{minipage}{0.6\textwidth}
\textit{I therefore believe it's true that with a suitable class of quantum machines you could imitate any quantum system, including the physical world.}
\begin{flushright}
--Richard P. Feynman, \\
``Simulating physics with computers'', 1982.
\end{flushright}
\end{minipage}
\end{flushright}
\vspace{1cm}

In recent years quantum computing has dived fully into the experimental realm. Since Richard Feynman made the observation that opens this chapter, there have been many improvements in the control of quantum devices. Now, quantum computers have become a reality, although we are still far from a universal quantum processor. Many prototypes are already available, but they are too noisy to be used beyond proofs of concept.

The explosion of the quantum computing field in this first decade of the XXI century is entailing the creation of dozens of new companies, in general start-ups with an academic origin in the universities where this field was born. Well-known technological companies have also joined to this race for the universal quantum processor: some from the beginning, others as the progress in the experimental part have become more relevant. It turns out to be interesting that two representatives of these company models were the first ones to offer cloud-based quantum computation platforms: the multinational company IBM and the start-up Rigetti Computing. Both are betting for superconducting qubits, although their devices characterization is different. As more quantum computer prototypes are coming out, it will be important, from research and economic point of view, to find methods to compare them and test their quality.

Furthermore, the scientific community is still working on Feynman's original aim for the quest of a quantum computer \cite{Feynman82}: the simulation of quantum systems. Many classical techniques have been developed in that direction. For instance, quantum Monte Carlo methods \cite{Kalos62,Hammond94,Blunt14} or tensor networks algorithms \cite{Orus14}. However, the firsts are concerned from the well-known sign problem, the seconds are only efficient for slightly entangled systems \cite{Vidal03} whereas strongly correlated quantum systems, such as those displaying frustration, will need a quantum computer to be efficiently simulated \cite{Ortiz01}. There are some works that propose quantum algorithms to construct arbitrary Slater determinants, both in one and two dimensions, to simulate the dynamics of the ground state of fermionic Hamiltonians, in particular, the Hubbard model \cite{Wecker15,Jiang18}. Other proposals introduce the concept of \emph{compressed quantum computation}, i.e. simulation of $n$-spin chain using $\log n$ qubits \cite{Kraus11,Hebenstreit17}. 

In this chapter, we describe the implementation of a four-qubit experiment that could be interesting both as a proposal for testing and comparing device quality and for its implications in condensed matter physics. The main result is the performance of an \emph{exact} simulation of a one-dimensional spin chain with an $XY$-type interaction. The design of a quantum circuit that diagonalizes the Hamiltonian follows the method of Ref. \cite{Verstraete09} which implements the same steps as the analytical solution of the model. Therefore, the same idea can be extended to other integrable models like the Kitaev-honeycomb model, which a circuit has already been proposed \cite{Schmoll17}. Because this circuit solves the model, it provides access to the whole spectrum and not only to the ground state: time evolution and thermal states can be simulated exactly as well. This introduces a new approach in quantum simulation if an exact circuit is found for those nontrivial models, such as the Heisenberg model, which has an ansatz to be solved. In particular, for one-dimensional spin chains, the Bethe ansatz \cite{Bethe31} is the most successful method and several proposals exist to simulate and extend it to two-dimensions using tensor network techniques \cite{Murg12}. As the one-dimensional $XY$ model has analytic solutions for an arbitrary number of spins and the circuit proposed can be efficiently generalized to a larger number of qubits, the methods outlined in this chapter can be used to benchmark a quantum computer by seeing how this compares against known solutions.

This chapter is structured as follows. The first section describes briefly the characteristics of the $XY$ model and solves it analytically. Next, in Sec. \ref{sec:circuit}, we introduce the method proposed in Ref. \cite{Verstraete09} to construct an efficient circuit that diagonalizes this Hamiltonian: the number of gates scales as $n^2$ and the circuit depth as $n\log n$. Section \ref{sec:time} gives a specific example of how to simulate time evolution using the circuit derived in the previous section and in Sec. \ref{sec:thermal}, two methods to simulate the expected value of an operator for finite temperature. The description of the experimental setups and their results, published in Ref. \cite{Cervera18}, are explained in sections \ref{sec:devices} and \ref{sec:results} respectively. Finally, the conclusions are exposed in Sec. \ref{sec:conclusion}.

\section{The \texorpdfstring{$XY$}{} model}

This model is one of the most used toy models in condensed matter physics because it contains quantum phase transitions. It is the generalization of other famous models such as transverse Ising model of Eq. \eqref{eq:HIsing}, introduced in the previous chapter, or $XX$ model. The $XY$ Hamiltonian describes a spin chain with nearest-neighbour interaction plus a transverse field. For one-dimensional systems, this Hamiltonian can be written as
\begin{definition}[$XY$ model Hamiltonian]
\begin{equation}
\mathcal{H}_{XY}\equiv J\sum_{i=1}^{n}\left(\frac{1+\gamma}{2}\sx_{i}\sx_{i+1} + \frac{1-\gamma}{2}\sy_{i}\sy_{i+1}\right) + \lambda\sum_{i=1}^n\sz_{i} \ ,
\label{eq:HXY_original}
\end{equation}
where $J$ will determine the behaviour of the ordered phase, ferromagnetic for $J<0$ or anti-ferromagnetic for $J>0$, $\gamma$ is the anisotropic parameter and $\lambda$ the transverse field strength. 
\end{definition}
The spin chain described by this Hamiltonian can be open or can have periodic boundary conditions, i.e. last spin interacts with first spin by adding the term $\sigma_n\sigma_1$. With respect to this chapter, it is considered periodic boundary conditions. The Ising Hamiltonian corresponds with $\gamma=1$ and the $XX$ Hamiltonian with $\gamma=0$.

At $J=\lambda$ there is a quantum phase transition between ferromagnetic (or anti-ferromagnetic) and paramagnetic phases. This transition belongs to the same universality class as the Ising model quantum phase transition. At $\gamma=0$ there is an anisotropic transition, between ordered phases in $x$ and $y$ directions. More details about the phases of this model can be found in Ref. \cite{Sachdev09,Dutta15}.

The analytical solution of this model without transverse field was first introduced by Lieb, Schultz and Mattis in 1961 \cite{Lieb61} and later, in 1962, Katsura solved it with the external field \cite{Katsura62}.

\subsection{Analytical solution}

It is convenient to write the Hamiltonian of Eq. \eqref{eq:HXY_original} in terms of $\sigma^{\pm}=\sx\pm i\sy$ operators. 
Then, the Hamiltonian reads
\begin{equation}
\mathcal{H}_{XY}=\frac{J}{2}\sum_{i=1}^{n}\left(\sM_{i}\sm_{i+1} +\sm_{i}\sM_{i+1}+ \gamma\left(\sM_{i}\sM_{i+1}+\sm_{i}\sm_{i+1}\right)\right) + \lambda\sum_{i=1}^n\sz_{i} \ .
\label{eq:HXY_sMsm}
\end{equation}

The first step to solve this model consists on applying the Jordan-Wigner transformation \cite{Jordan28}, which maps the spin operators $\mathbf{\sigma}$ into fermionic modes $c$:

\begin{definition}[Jordan-Wigner transformation]
\begin{equation}
c_{j}\equiv\left(\prod_{l=1}^{j-1}(-2\sz_{l})\right)\sm_{j}, \qquad c_{j}^{\dagger}\equiv\sM_{j}\left(\prod_{l=1}^{j-1}(-2\sz_{l})\right).
\label{eq:JW}
\end{equation}
\end{definition}
These new operators $c_{j}$ and $c_{j}^{\dagger}$ are the fermionic annihilation and creation operators respectively acting on the vacuum $|\Omega_{c}\rangle$, $c_{i}|\Omega_{c}\rangle=0$, and following the anticommutation rules $\{c_{i},c_{j}\}=\{c_{i}^{\dagger},c_{j}^{\dagger}\}=0$ and $\{c_{i},c_{j}^{\dagger}\}=\delta_{ij}$. After this transformation, the Hamiltonian becomes
\begin{align}
\mathcal{H}_{XY}&=\frac{J}{2}\sum_{i=1}^{n-1}\left(c_{i}^{\dagger}c_{i+1}+c_{i+1}^{\dagger}c_{i}+ \gamma\left(c_{i}^{\dagger}c_{i+1}^{\dagger}+c_{i+1}c_{i}\right)\right)+\lambda\sum_{i=1}^{n-1}\left(c_{i}^{\dagger}c_{i+1}-1/2\right) - \nonumber\\
&-\frac{J}{2}\left(c_{1}^{\dagger}c_{n}+ c_{n}^{\dagger}c_{1}+ \gamma\left(c_{n}^{\dagger}c_{1}^{\dagger}+c_{1}c_{n}\right)\right) + \frac{J}{2}Q_{n}\left(c_{n}c_{1}^{\dagger}+c_{n}^{\dagger}c_{1} + \gamma\left(c_{n}^{\dagger}c_{1}^{\dagger}+c_{1}c_{n}\right)\right),
\label{eq:HXY_JW}
\end{align}
where $Q_n=\prod_{l=1}^{n-1}\left(1-2c_{j}^{\dagger}c_{j}\right)$. Notice that the second line terms do not contain the sum over all sites, so they will be negligible in the thermodynamic limit, i.e. for $n\rightarrow\infty$.

\subsubsection{Fourier Transform}

The next step to diagonalize the Hamiltonian is to apply the translational invariance using the well known Fourier transform. To simplify the discussion, we will solve this Hamiltonian in the thermodynamic limit, i.e. neglecting the second line terms of Eq. \eqref{eq:HXY_JW}:
\begin{equation}
\mathcal{H}_{XY}=\frac{J}{2}\sum_{i=1}^{n}\left(c_{i}^{\dagger}c_{i+1}+c_{i+1}^{\dagger}c_{i}+ \gamma\left(c_{i}^{\dagger}c_{i+1}^{\dagger}+c_{i+1}c_{i}\right)\right)+ \lambda\sum_{i=1}^{n}\left(c_{i}^{\dagger}c_{i+1}-1/2\right) \ ,
\label{eq:HXY_JW_th}
\end{equation}
where we have consider $n-1\simeq n$ since we are in the thermodynamic limit.
The exact solution with periodic and anti-periodic boundary conditions can be found in Ref.  \cite{Katsura62,Eriksson08,DPF09}.

\begin{definition}[Fourier transform]
\begin{equation}
b_{k}\equiv\frac{1}{\sqrt{n}}\sum_{j=1}^{n}\exp\left(i\frac{2\pi j}{n}k\right)c_{j}, \ 
b_{k}^{\dagger}\equiv\frac{1}{\sqrt{n}}\sum_{j=1}^{n}\exp\left(i\frac{2\pi j}{n}k\right)c_{j}^{\dagger}, \quad k=-\frac{n}{2}+1,\cdots,\frac{n}{2} \ .
\label{eq:FT}
\end{equation}
\end{definition}

After the Fourier transform, the Hamiltonian becomes
\begin{equation}
\mathcal{H}_{XY}=\sum_{k=-n/2+1}^{n/2}\left(\left(J\cos\left(\frac{2\pi k}{n}\right)+\lambda\right)b_{k}^{\dagger}b_{k}+ J\gamma\frac{e^{i\frac{2\pi k}{n}}}{2}\left(b_{k}^{\dagger}b_{-k}^{\dagger}+b_{k}b_{-k}\right)\right)-\lambda \frac{n}{2}.
\label{eq:HXY_fourier}
\end{equation}
This Hamiltonian is already diagonal for $\gamma=0$, which corresponds to the $XX$ model. For other cases, that include Ising model ($\gamma=1$), it is necessary a last transformation.

\subsubsection{Bogoliubov transformation}

Hamiltonian of Eq. \eqref{eq:HXY_fourier} mixes $k$ and $-k$ modes, so it is necessary to find a transformation such that
\begin{align}
a_{k}&=A_{k}b_{k}+B_{k}b_{-k}^{\dagger},  \nonumber\\
a_{-k}&=C_{k}b_{k}^{\dagger}+D_{k}b_{-k},
\end{align}
where $a_{k}$ and $a_{-k}$ are new operators that also obey fermionic anticommutation relations. This implies that $|A_{k}|^2+|B_{k}|^2=1$, $|C_{k}|^2+|D_{k}|^2=1$ and $A_{k}C_{k}+B_{k}D_{k}=0$.

Before applying such a transformation, it is useful to write the above Hamiltonian in terms of only positive $k$ modes, i.e. sum up half of the modes:
\begin{multline}
\mathcal{H}_{XY}=\sum_{k=0}^{n/2}\Bigg(\left(J\cos\left(\frac{2\pi k}{n}\right)+\lambda\right)\left(b_{k}^{\dagger}b_{k}+ b_{-k}^{\dagger}b_{-k}\right) \\ 
+iJ\gamma\sin\left(\frac{2\pi k}{n}\right)\left(b_{k}^{\dagger}b_{-k}^{\dagger}+b_{k}b_{-k}\right)\Bigg) -\lambda \frac{n}{2} \ .
\label{eq:HXY_fourier2}
\end{multline}
Thus, we can write the Hamiltonian in a matrix form
\begin{align}
\mathcal{H}_{XY}=&\sum_{k=0}^{n/2}\left(\begin{matrix}b_k^\dagger & b_{-k}\end{matrix}\right) \left(\begin{matrix}
J\cos\left(\frac{2\pi k}{n}\right)+\lambda & iJ\gamma\sin\left(\frac{2\pi k}{n}\right) \\ -iJ\gamma\sin\left(\frac{2\pi k}{n}\right) & -J\cos\left(\frac{2\pi k}{n}\right)-\lambda\end{matrix}\right) \left(\begin{matrix}b_k \\ b_{-k}^\dagger\end{matrix}\right) \nonumber\\
\equiv& \sum_{k=0}^{n/2}\Psi_{k}^\dagger \mathbf{H}_{k}\Psi_{k} \ ,
\label{eq:HXY_matrix}
\end{align}
where we have used the fermionic anticommutation relations and $\sum_{k=0}^{n/2}\cos(2\pi k/n)=0$ to include the constant term. We have also introduced a two-component fermion field $\Psi_{k}=\left(\begin{matrix} b_k & b_{-k}^\dagger\end{matrix}\right)^{T}$. 

To diagonalize this Hamiltonian we have to find the eigenvalues of the $\mathbf{H}_{k}$ matrix and the transformation that leads to this diagonalization, i.e.
\begin{equation}
\mathcal{H}_{XY}=\sum_{k=0}^{n/2}\Psi_{k}^\dagger\mathbf{H}_{k}\Psi_{k}=\sum_{k=0}^{n/2}\Psi_{k}\mathbf{U}_{k}^\dagger\left(\begin{matrix} \omega_k & 0 \\ 0 & -\omega_k \end{matrix}\right)\mathbf{U}_{k}\Psi_{k},
\end{equation}
where 
\begin{equation}
\omega_k=\sqrt{\left(J\cos\left(\frac{2\pi k}{n}\right)+\lambda\right)^2 +J^2\gamma^2\sin^2\left(\frac{2\pi k}{n}\right)}
\end{equation}
are the eigenvalues of $\mathbf{H}_{k}$. We can now define the transformed two-component fermion field
\begin{equation}
\Phi_{k}\equiv\left(\begin{matrix}a_{k} \\ a_{-k}^{\dagger}\end{matrix}\right) = \mathbf{U}_{k}\Psi_{k} 
\label{eq:Bog_modes}
\end{equation}
with
\begin{equation}
\mathbf{U}_{k}=\left(\begin{matrix}A_{k} & B_{k} \\ C^{*}_{k} & D^{*}_{k}\end{matrix}\right).
\end{equation}
Using the constraints for the coefficients of $\mathbf{U}_{k}$ and matching Eq. \eqref{eq:Bog_modes} with \eqref{eq:HXY_matrix} we can find the coefficients of the Bogoliubov transformation \cite{Valatin58,Bogoliubov58}:

\begin{definition}[Bogoliubov transformation]
\renewcommand\arraystretch{1.5}
\begin{equation}
\begin{matrix}
a_{k}\equiv\cos(\theta_{k}/2)b_{k}+i\sin(\theta_{k}/2)b_{-k}^{\dagger}\\
a_{-k}\equiv-i\sin(\theta_{k}/2)b_{k}^{\dagger}+\cos(\theta_{k}/2)b_{-k}
\end{matrix} \quad \mathrm{with} \ \theta_{k}=2\arctan\left(\frac{J\gamma\sin\left(\frac{2\pi k}{n}\right)}{J\cos\left(\frac{2\pi k}{n}\right) + \lambda}\right).
\label{eq:Bolgoliubov}
\end{equation}
\end{definition}

After this transformation we have finally diagonalized the $XY$ Hamiltonian,
\begin{equation}
\mathcal{H}_{XY}=\sum_{k=-n/2+1}^{n}\omega_{k}\left(a_{k}^{\dagger}a_{k}-\frac{1}{2}\right) \ .
\label{eq:HXY_diag}
\end{equation}
The fact that it is possible to arrive to this diagonal Hamiltonian means that the $XY$ model is integrable.

\section{Quantum circuit to diagonalize the XY Hamiltonian\label{sec:circuit}}

Once we know how to diagonalize the $XY$ Hamiltonian, we can proceed to design and construct a quantum circuit that implements this diagonalization process. Although the circuit presented is designed to solve this Hamiltonian, the key idea is general and the process can be generalized to other models, specially those that are exactly solvable.

Let's first consider the existence of a quantum circuit that \textit{disentangles} a given Hamiltonian and transforms its entangled eigenstates into product states. This circuit will be represented by an unitary operation $U_{dis}$ that transforms the Hamiltonian $\mathcal{H}$ into a non-interacting one, i.e. $\widetilde{\mathcal{H}}=\sum_{i}\epsilon_{i}\sigma_{i}^{z}$. 

\begin{definition}[Disentangling operation]
\begin{equation}
\widetilde{\mathcal{H}}\equiv U_{dis}^{\dagger}\mathcal{H}U_{dis} \ .
\end{equation}
\end{definition}
This diagonal Hamiltonian contains the energy spectrum $\epsilon_{i}$ of the original one and its eigenstates correspond to the computational basis states. Then, we will have access to the whole spectrum of the model by just preparing a product state and applying $U_{dis}$.

In general, to find these disentangling unitaries will be a hard task, probably as hard as finding a method to diagonalize analytically the Hamiltonian. However, for models that can be solved analytically we can try to map the corresponding steps of the diagonalization process into quantum gates that perform the same operations on qubits. For the case of $XY$ Hamiltonian, we have already reviewed these steps: {\sl i)} Implement the Jordan-Wigner transformation to map the spins into fermionic modes. {\sl ii)} Perform the Fourier transform to get fermions to momentum space. {\sl iii)} Perform a Bogoliubov transformation to decouple the modes with opposite momentum. Thus, the construction of the disentangling gate can be done by pieces:
%
\begin{equation}
U_{dis}=U_{JW}U_{FT}U_{Bog}.
\end{equation}

In the following subsections, the quantum gates needed to implement the above transformation are derived.

\subsection{Jordan-Wigner transformation}

The steps followed to diagonalize the $XY$ Hamiltonian have been applied to an infinite spin chain. Current quantum devices  are finite and, in particular, the explicit circuit that will be used as example is composed of up to 4 qubits, small number to consider thermodynamic limit. For that reason, we can add some modifications to the original Hamiltonian of Eq. \eqref{eq:HXY_original} in order to cancel the periodic boundary terms and solve the system as it was infinite. This will introduce some finite-size effects that will become negligible for higher qubit circuits. Then, technically we will diagonalize a modified $XY$ Hamiltonian that becomes indeed the $XY$ Hamiltonian for large $n$.

\begin{definition}[Modified $XY$ Hamiltonian]
\begin{align}
\mathcal{H}_{XY}\equiv J\sum_{i=1}^{n}&\left( \frac{1+\gamma}{2}\sx_{i}\sx_{i+1} + \frac{1-\gamma}{2}\sy_{i}\sy_{i+1}\right) + \lambda\sum_{i=1}^n\sz_{i} \nonumber \\
&+J\left(\frac{1+\gamma}{2}\sy_{1}\sz_{2}\cdots\sz_{n-1}\sy_{n} + \frac{1-\gamma}{2}\sx_{1}\sz_{2}\cdots\sz_{n-1}\sx_{n}\right) \ .
\end{align}
\end{definition}

Now, we proceed with the Jordan-Wigner transformation of Eq. \eqref{eq:JW}. Notice that the second line added in the above Hamiltonian cancels the periodic boundary conditions, $\sigma_{n}^{x}\sigma_{1}^{x}$ and $\sigma_{n}^{y}\sigma_{1}^{y}$, after this transformation, leading directly the Hamiltonian in the thermodynamic limit of Eq. \eqref{eq:HXY_JW_th}. Thus we can continue applying the explained steps to diagonalize the Hamiltonian without worrying about the boundary terms.

Let's analyse what is the effect of this transformation in the wave function:
\begin{equation}
|\psi\rangle=\sum_{i_{1},\cdots,i_{n}=0,1}\psi_{i_{1}\cdots i_{n}}|i_{1}\cdots i_{n}\rangle = \sum_{i_{1},\cdots,i_{n}=0,1}\psi_{i_{1}\cdots i_{n}} (c_{1}^{\dagger})^{i_{1}}\cdots (c_{n}^{\dagger})^{i_{n}}|\Omega_{c}\rangle \ .
\end{equation}
Notice that the coefficients $\psi_{i_{1}\cdots i_{n}}$ do not change. Then it will not be necessary to implement any gates on the quantum register to perform this transformation. However, for now on we should take into account we are dealing with fermionic modes, so any swap between two occupied modes will carry a minus sign. In terms of quantum gates, this is translated into the use of {\sl fermionic} SWAP gate (fSWAP) each time we exchange two modes:

\begin{definition}[Fermionic-SWAP gate]
\begin{equation}
\mathrm{fSWAP}\equiv\left(\begin{array}{cccc}
1&0&0&0\\0&0&1&0\\0&1&0&0\\0&0&0&-1
\end{array}\right) \ .
\label{fSWAP}
\end{equation}
\end{definition}

This gate corresponds with the usual SWAP gate followed or preceded by a CZ gate.

\subsection{Fourier transform}

Once we have the fermionic modes, we get them to momentum space using the Fourier transform of Eq. \eqref{eq:FT}. For $n=2^m$ for some integer $m$, this transformation can be implemented with a log-depth circuit and using at most two-body quantum gates. This method is called {\sl fast Fourier transform} and consists in two parallel Fourier transformations over $n/2$ sites, the even and the odd sites \cite{Ferris14}:
\begin{equation}
\sum_{j=0}^{n-1}e^{\frac{2\pi i k}{n}j}c_{j}^{\dagger}
=\sum_{j'=0}^{\frac{n}{2}-1}e^{\frac{2\pi i k}{n/2}j'}c_{2j'}^{\dagger}+e^{\frac{2\pi i k}{n}}
e^{\frac{2\pi i k}{n/2}j'}c_{2j'+1}^{\dagger} \ .
\end{equation}
To implement such a transformation we need a combination of a two-qubit gate, a `beam-splitter' $F_2$, and one-qubit gate, the `phase-delay' $\omega_{n}^{k}$, which applies the so-called twiddle-factor $e^{2\pi i k/n}$:
\arraycolsep=1.8pt\def\arraystretch{1.2}
\begin{equation}
F_{2}=\left(\begin{array}{cccc} 1&0&0&0\\ 0& \frac{1}{\sqrt{2}}& \frac{1}{\sqrt{2}} &0\\
0& \frac{1}{\sqrt{2}}& -\frac{1}{\sqrt{2}} &0\\ 0&0&0& -1\end{array}\right), \qquad  \omega_{n}^{k}=\left(\begin{array}{cc}1&0\\0& e^{\frac{2\pi ik}{n}} \end{array}\right),
\end{equation}
where the fermionic anticommutation relation has been taken into account in the $-1$ element of the $F_{2}$ matrix. 

All together, the Fourier transform gate becomes

\begin{definition}[Fourier transform gate]
\arraycolsep=1.8pt\def\arraystretch{1.2}
\begin{equation}
F^{n}_{k}\equiv\left(\begin{array}{cccc} 1&0&0&0\\ 0& \frac{1}{\sqrt{2}}& \frac{e^{\frac{2\pi ik}{n}}}{\sqrt{2}} &0\\
0& \frac{1}{\sqrt{2}}& -\frac{e^{\frac{2\pi ik}{n}}}{\sqrt{2}} &0\\ 0&0&0& -e^{\frac{2\pi ik}{n}}\end{array}\right).
\label{F_gate}
\end{equation}
\end{definition}
The explicit decomposition of this gate in terms of common quantum gates is shown in Fig.\ref{Fig:Qgates} left.

\subsection{Bogoliubov transformation}

Bogoliubov transformation described in Eq. \eqref{eq:Bolgoliubov} is actually a rotation that involve Fourier operators $b$ with opposite momenta. In particular, is a rotation round $x$ axis that can be carried with the two-qubit rotational gate

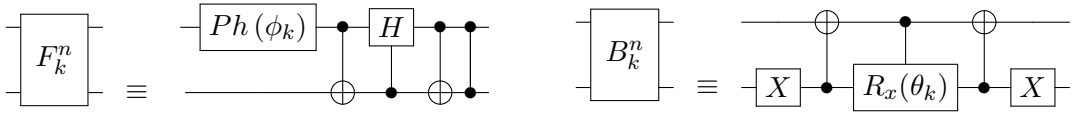
\begin{figure}[t!]
\begin{minipage}{0.5\textwidth}
\centering
\[
\Qcircuit @C=0.5em @R=1em 
{
&\multigate{1}{F_{k}^{n}}& \qw &\push{\rule{0em}{0em} \quad\quad \rule{0em}{0em}}& \gate{Ph\left(\phi_{k}\right)} & \ctrl{1} & \gate{H} & \ctrl{1} & \ctrl{1} & \qw \\
&\ghost{F_{k}^{n}}& \qw& \push{\rule{0em}{0em}\equiv\quad\rule{0em}{0em}} & \qw &\targ & \ctrl{-1} & \targ & \ctrl{-1} & \qw 
}
\]
\label{Fig:Fk}
\end{minipage}%
\begin{minipage}{0.5\textwidth}
\centering
\[
\Qcircuit @C=0.5em @R=1em 
{
&\multigate{1}{B_{k}^{n}} & \qw & \push{\rule{0em}{0em}\quad \ \rule{0em}{0em}}& & \qw & \targ & \ctrl{1} & \targ & \qw & \qw \\
&\ghost{B_{k}^{n}} & \qw & \push{\rule{0em}{0em}\equiv\rule{0em}{0em}} & & \gate{X} &  \ctrl{-1} & \gate{R_{x}(\theta_{k})} & \ctrl{-1} & \gate{X} & \qw
}
\]
\label{Fig:Bk}
\end{minipage}
\caption{Decomposition of the building block of Fourier transform gate (\textit{left}), where $\phi_{k}=2\pi k/n$, and Bogoliubov gate (\textit{right}). Details about the decomposition of controlled-Hadamard and controlled-$R_x$ gates can be found in App. \ref{app:quantum_gates}.}
\label{Fig:Qgates}
\end{figure}

\begin{definition}[Bogoliubov gate]
\begin{equation}
B_{k}^{n}\equiv\left(\begin{matrix}\cos\left(\frac{\theta_{k}}{2}\right)&0&0& i\sin\left(\frac{\theta_{k}}{2}\right)\\0&1&0&0\\0&0&1&0\\i\sin\left(\frac{\theta_{k}}{2}\right)&0&0&\cos\left(\frac{\theta_{k}}{2}\right)\end{matrix}\right).
\label{B_gate}
\end{equation}
\end{definition}
Its decomposition in basic gates is shown in Fig.\ref{Fig:Qgates} right. 

Thus, the initial qubits are in a computational basis state that represents one eigenstate of the diagonal Hamiltonian of Eq. \eqref{eq:HXY_diag}. Then, the $U_{dis}$ circuit applies the inverse Bogoliubov transformation so each qubit represents one Fourier mode. Finally, the inverse Fourier transform maps the Fourier modes to the fermionic modes $c$. Undo the Jordan-Wigner transformation is just a conceptual operation because the wave function is not affected by this operation as has been explained before.

\subsection{\texorpdfstring{$n=4$}{} spin chain}

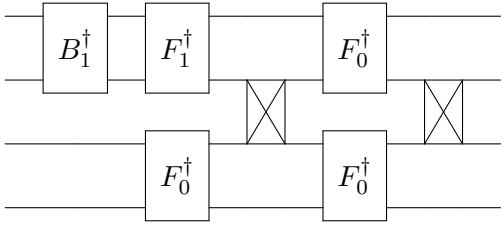
\begin{figure}[t!]
\centering
\[
\Qcircuit @C=0.5cm @R=.5cm {
& \multigate{1}{B_{1}^{\dagger}} & \multigate{1}{F_{1}^{\dagger}}  & \qw & \qw & \multigate{1}{F_{0}^{\dagger}} & \qw & \qw & \qw \\
& \ghost{B_{1}^{\dagger}} & \ghost{F_{1}^{\dagger}} & \qw & \link{1}{-1} & \ghost{F_{0}^{\dagger}} & \qw & \link{1}{-1} &  \qw \\
& \qw & \multigate{1}{F_{0}^{\dagger}}  & \qw & \link{-1}{-1} & \multigate{1}{F_{0}^{\dagger}} & \qw & \link{-1}{-1} &  \qw \\
& \qw & \ghost{F_{0}^{\dagger}}  & \qw & \qw & \ghost{F_{0}^{\dagger}} & \qw & \qw & \qw 
  \gategroup{2}{4}{3}{5}{.0em}{-}\gategroup{2}{7}{3}{8}{.0em}{-}
  }
\]
\caption{
Quantum circuit that transforms computational basis states into $XY$ Hamiltonian eigenstates. The two-qubit gates $F_{1}^\dagger$ and $F_{0}^\dagger$ apply the inverse Fourier transform and the $B_{1}^\dagger$ the inverse Bogoliubov transformation. Gates represented with crosses correspond with the fSWAP gates that take care of the fermion anticommutation relations and can be removed depending on the connectivity of the quantum chip.
\vspace{-0.3cm}}
\label{Fig:circuit}
\end{figure}

The explicit circuit for an $n=4$ chain is shown in Fig. \ref{Fig:circuit}. As an example, let's compute the ground state of the model. First, we prepare the initial state as the ground state for the diagonal Hamiltonian $\widetilde{\mathcal{H}}$:
\begin{equation}
|gs\rangle=\left\{\begin{array}{llll}|0000\rangle & \mathrm{for} & \lambda>1, & \forall J,\gamma \ ,  \\
|0001\rangle & \mathrm{for} & \lambda<1, & J>0,\forall\gamma \ \mathrm{and} \ J<0,\gamma=0 \ , \\
|0010\rangle & \mathrm{for} & \lambda<1, & J<0,\gamma\neq0 \ .
\end{array}\right. 
\end{equation}

The circuit strategy consists in undoing the steps that diagonalize the $XY$ Hamiltonian. Thus, we first undo Bogoliubov transformation by applying $(B_{k}^{n})^\dagger$ gates, followed by undoing the Fourier transform using the $(F_{k}^{n})^\dagger$ gates and finally undo the Jordan-Wigner transformation which, fortunately, does not need from any gate as has been explained in the previous section.

For $n=4$, the Bogoliubov modes are $\pm 3\pi/2$ and $\pm \pi/2$, so we need two Bogoliubov gates. We have removed the $B_{0}^\dagger$ gate from the circuit of Fig. \ref{Fig:circuit} because it corresponds with the identity gate. The circuit also contains fSWAP gates represented with crosses. These will be necessary if even and odd qubits are not physically connected and, as much, they will increase the total number of gates in $n^2$. 

\section{Time evolution \label{sec:time}}

Once we have the $U_{dis}$ circuit, we have access to the whole $XY$ spectrum by only implementing this gate over the computational basis states. This allows us to perform \emph{exactly} time evolution, where the characterization of all states is needed. 

The time evolution of a given state driven by a time-independent Hamiltonian is described using the time evolution operator:
\begin{definition}[Time evolution quantum state]
\begin{align}
U(t)&\equiv e^{-it\mathcal{H}}, \\
|\psi(t)\rangle&=U(t)|\psi_{0}\rangle=\sum_{i}e^{-it \epsilon_{i}}|E_{i}\rangle\langle E_{i}|\psi_{0}\rangle,
\label{eq:time_state}
\end{align}
where $|\psi_{0}\rangle$ is the initial state and $\epsilon_{i}$ are the energies of the Hamiltonian states $|E_{i}\rangle$. 
\end{definition}
Then, if $|\psi_{0}\rangle$ is an eigenstate of $\mathcal{H}$ there is no change in time (\textit{steady state}) and, therefore, the expected value of an observable $\mathcal{O}$ will be constant in time. On the contrary, and if $[\mathcal{H},\mathcal{O}]\neq 0$, the expected value will show an oscillation in time given by
\begin{equation}
\langle\mathcal{O}(t)\rangle=\sum_{i,j}e^{-it(\epsilon_{i}-\epsilon_{j})} \langle\psi_{0}|E_{j}\rangle\langle E_{j}|\mathcal{O}|E_{i}\rangle\langle E_{i}|\psi_{0}\rangle.
\label{eq:time_obs}
\end{equation}

We can take advantage from the fact that the eigenstates of the non-interacting Hamiltonian $\widetilde{\mathcal{H}}$ are the computational basis states and, as we have solved the model, we also know all energies $\epsilon_{i}$. Then, it is straightforward to construct the time evolution of a given state $|\psi_{0}\rangle$ by only expressing it in the computational basis and adding the corresponding factors $e^{-it\epsilon_{i}}$. After that, we only need to implement $U_{dis}$ gate over this state to obtain the time evolution driven by the $XY$ Hamiltonian.

As example, let's compute the time evolution of the expected value of transverse magnetization for the $n=4$ anti-ferromagnetic Ising Hamiltonian, that is $J=\gamma=1$. In particular, let's take all spins aligned in the positive $z$ direction as initial state, i.e. $|\uparrow\uparrow\uparrow\uparrow\rangle$, which in the computational basis is the $|0000\rangle$ state. First, we have to express this state in the $\widetilde{\mathcal{H}}$ basis, which using $U_{dis}^{\dagger}$ becomes
\begin{equation}
|\psi_{0}\rangle=U_{dis}^{\dagger}|0000\rangle=\cos(\phi/2)|0000\rangle+i\sin(\phi/2)|1100\rangle,
\end{equation}
with $\phi=\arctan(1/\lambda)$. Then, we apply the time evolution operator  to obtain $|\psi(t)\rangle$:
\begin{equation}
|\psi(t)\rangle=
\left(\cos\phi|00\rangle+ ie^{4it\sqrt{1+\lambda^2}}\sin\phi|11\rangle\right)\otimes|00\rangle.
\label{eq:time_Ising}
\end{equation}
To prepare this state, we just need to apply a $R_{Y}(\phi)$ gate on the first qubit to introduce the $\phi$ angle, followed by a phase gate to introduce the evolution phase $e^{4it\sqrt{1+\lambda^2}}$ and a CNOT gate between first and second qubits.

Analytically, 
\begin{equation}
\langle\sigma_{z}\rangle=\frac{1+2\lambda^2+\cos\left(4t\sqrt{1+\lambda^2}\right)}{2+2\lambda^2},
\end{equation}
from which we can obtain the expected value of transverse magnetization, $M_{z}=\frac{1}{2}\langle\sigma_{z}\rangle$. 

\section{Thermal simulation \label{sec:thermal}}

When a quantum system is exposed to a heat bath its density matrix at thermal equilibrium is characterized by thermally distributed populations of its quantum states following a Boltzmann distribution:
\begin{definition}[Density matrix thermal state]
\begin{equation}
\rho(\beta)\equiv \frac{e^{-\beta\mathcal{H}}}{\mathcal{Z}}=\frac{1}{\mathcal{Z}}\sum_{i}e^{-\beta\epsilon_{i}}| E_{i}\rangle\langle E_{i}| \ ,
\end{equation}
where $\beta=1/(k_{B} T)$, $\mathcal{Z}=\sum_{i}e^{-\beta\epsilon_{i}}$ is the partition function and $\epsilon_{i}$ and $|E_{i}\rangle$ are the energies and eigenstates of the Hamiltonian $\mathcal{H}$. 
\end{definition}

The expected value of some operator $\mathcal{O}$ for finite temperature is computed as
\begin{equation}
\langle\mathcal{O}(\beta)\rangle=\mathrm{Tr}[\mathcal{O}\rho(\beta)]=\frac{1}{\mathcal{Z}}\sum_{i}e^{-\beta\epsilon_{i}}\langle E_{i}|\mathcal{O}|E_{i}\rangle \ .
\end{equation}

Simulate thermal evolution according to Ising Hamiltonian is, again, straightforward once we have $U_{dis}$ gate because it consists on preparing the corresponding state in the $\widetilde{\mathcal{H}}$ basis and apply $U_{dis}$ circuit. In the case of thermal evolution, $|E_{i}\rangle$ states are the states of the computational basis, so no further gates are needed to initialize qubits apart from the corresponding combination of $X$ gates to prepare the initial product state.

At that point, we can perform an \textit{exact simulation} or \textit{sampling}. In the first case, we run the circuit to obtain the expected value of the observable taking as initial state all states in the computational basis and average them with their corresponding energies. This is done classically once we have the expected values of each state. On the other hand, we can perform a more realistic simulation by sampling all states according to Boltzmann distribution. First, we need to prepare classically a random generator that returns one of the computational states following the distribution $e^{-\beta\epsilon_{i}}$. Then, we run the circuit many times and compute the expected value of the operator by preparing as initial state the one returned by the generator each time.

The first method demands more runs of the experiment, to be precise $N\times 2^n$, needed for the computation of each expected value. As the averaging part is done classically, no statistical errors arise from it. For the second method, with only $N$ runs we will obtain a value for the observable with a statistical error of $1/\sqrt{N}$.

\section{Experimental implementation \label{sec:devices}}

\subsection{IBM Quantum Experience}

\begin{figure}
\centering  
\subfigure[$ \ $\emph{Tenerife}]
{\includegraphics[width=0.2\columnwidth]{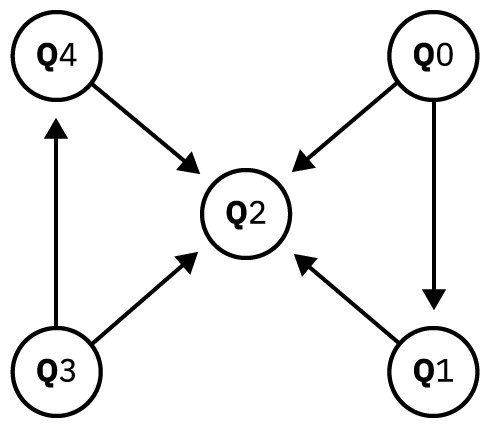}}\hspace{2cm}
\subfigure[$ \ $\emph{Yorktown}]
{\includegraphics[width=0.2\linewidth]{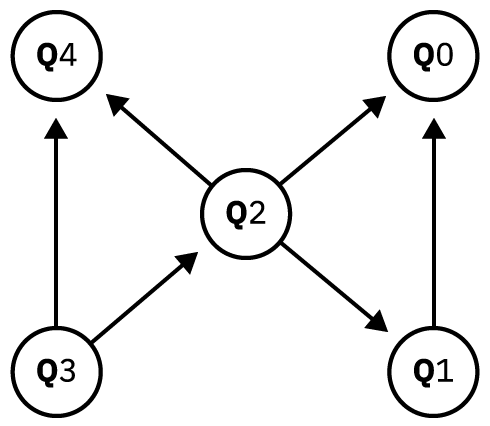}}\\
\subfigure[$ \ $\emph{Rueschlikon}]
{\includegraphics[width=0.6\linewidth]{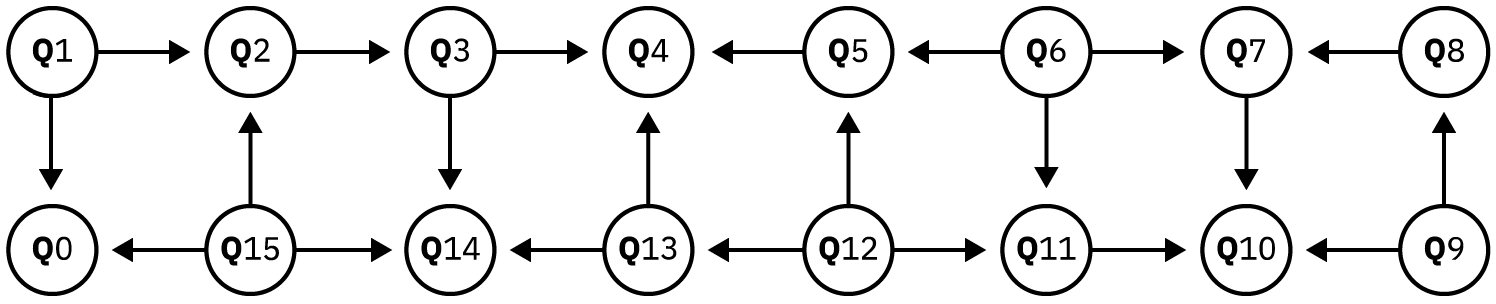}}\\
\subfigure[$ \ $\emph{Melbourne}]
{\includegraphics[width=0.6\linewidth]{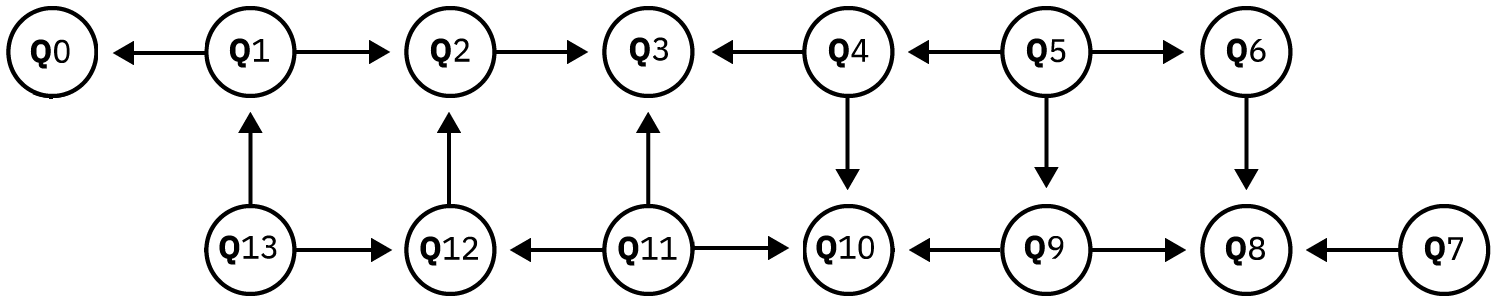}}
\caption{IBM quantum chips. Arrows between qubits indicate the directionality of CNOT gates: control $\rightarrow$ target. (Image source: \emph{QISKit} Github \url{https://github.com/Qiskit/} with modifications in \emph{Melbourne} caption).}
\label{Fig:IBM_backends}
\end{figure}

Since 2016, IBM company is providing universal quantum computer prototypes based on superconducting transmon qubits which are accessible on the cloud, both interactively in their web page, the \textit{Quantum Composer}, and using a software development kit called \textit{QISKit}. 

Currently, there are four quantum devices available: two 5-qubit chips, \emph{Tenerife} \cite{Tenerife} and \emph{Yorktown} \cite{Yorktown}, a 16-qubit chip, \textit{Rueschlikon} \cite{Rueschlikon} and a 14-qubit chip, \emph{Melbourne} \cite{Melbourne}. These devices are in their first version now, in 2019, but, except \emph{Melbourne}, they are actually a second generation of the first prototypes: \emph{ibmqx2}, \emph{ibmqx4} and \emph{ibmqx3}/\emph{ibmqx5} respectively.

All backends work with a universal gate set composed by one-qubit unitary gate $U_{3}$ and a two-qubit gate, the CNOT gate. More information about quantum gates can be found in the App. \ref{app:quantum_gates}. Other basic gates are also configured in their low level quantum language, \emph{QISKit Terra}, such as SWAP, $S$ or $H$ gates. However, it is important to keep in mind which is the basic gate set, as all other quantum gates will be decomposed in terms of the basic set automatically when the circuit is run, increasing the expected circuit depth. 

The differences between devices, apart from the number of qubits, come from the qubits connectivity and the role that each qubit plays when a CNOT gate is applied: control or target. Figure \ref{Fig:IBM_backends} shows the connectivity of the available devices. Each qubit in the 5-qubit devices is connected with another two except the central one which is connected with the other four. Qubits in the 16-qubit and 14-qubit devices are connected with three neighbours in a ladder-type geometry. The one-directionality of the CNOT gate and the qubits connectivity are crucial for the quantum circuit implementation. If the circuit demands interaction between qubits that are not physically connected, we should implement SWAP gates which will increase our circuit depth and the probability of errors in our final result. Moreover, each time we need to implement a CNOT gate using as a control qubit a physical qubit which is actually a target, we have to invert the CNOT direction using Hadamard gates which, again, will increase the circuit depth and the error probability.

For our propose, \emph{Rueschnikon} and \emph{Melbourne} are the best choices for the implementation of the $n=4$ circuit. We can use any of the squares and identify upper qubits as 0 and 2 and lower qubits as 1 and 3, according to the circuit of Fig. \ref{Fig:circuit}.

\subsection{Rigetti Computing: Forest}

At the end of 2017, Rigetti Computing launched a 19-qubit processor, \emph{Acorn} \cite{Acorn}, that can be used in the cloud through a development environment called \emph{Forest} \cite{Rigetti}. \textit{Forest} includes a python toolkit, \emph{pyQuil}, that allows the users to program, simulate and run quantum algorithms similar to IBM's \emph{QISKit}. The chip is made of 20 superconducting transmon qubits but, for some technical reasons, qubit 3 is off-line and cannot interact with its neighbors, so it is actually a 19-qubit device. In June 2018, they launched a new chip, \emph{Agave} \cite{Agave}, made up of 8 qubits and recently, in November 2018, another chip of 16 qubits, \emph{Aspen-1}.

\begin{figure}[t!]
\centering  
\subfigure[$ \ $\emph{Acorn}]
{\includegraphics[width=0.55\columnwidth]{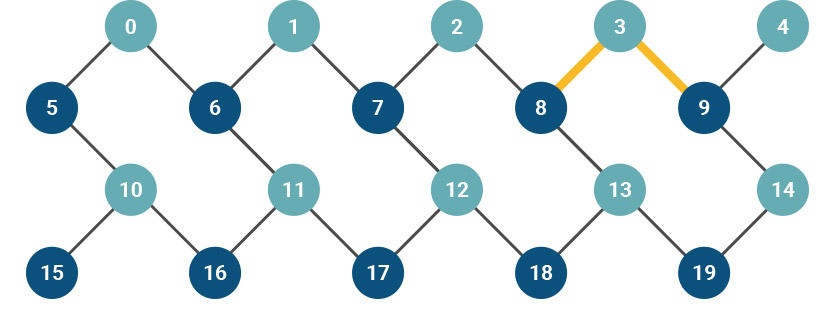}}\hspace{2cm}\\
\subfigure[$ \ $\emph{Agave}]
{\includegraphics[width=0.25\linewidth]{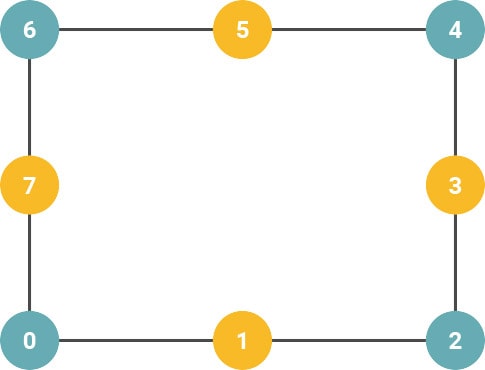}}
\subfigure[$ \ $\emph{Aspen-1}]
{\includegraphics[width=0.45\linewidth]{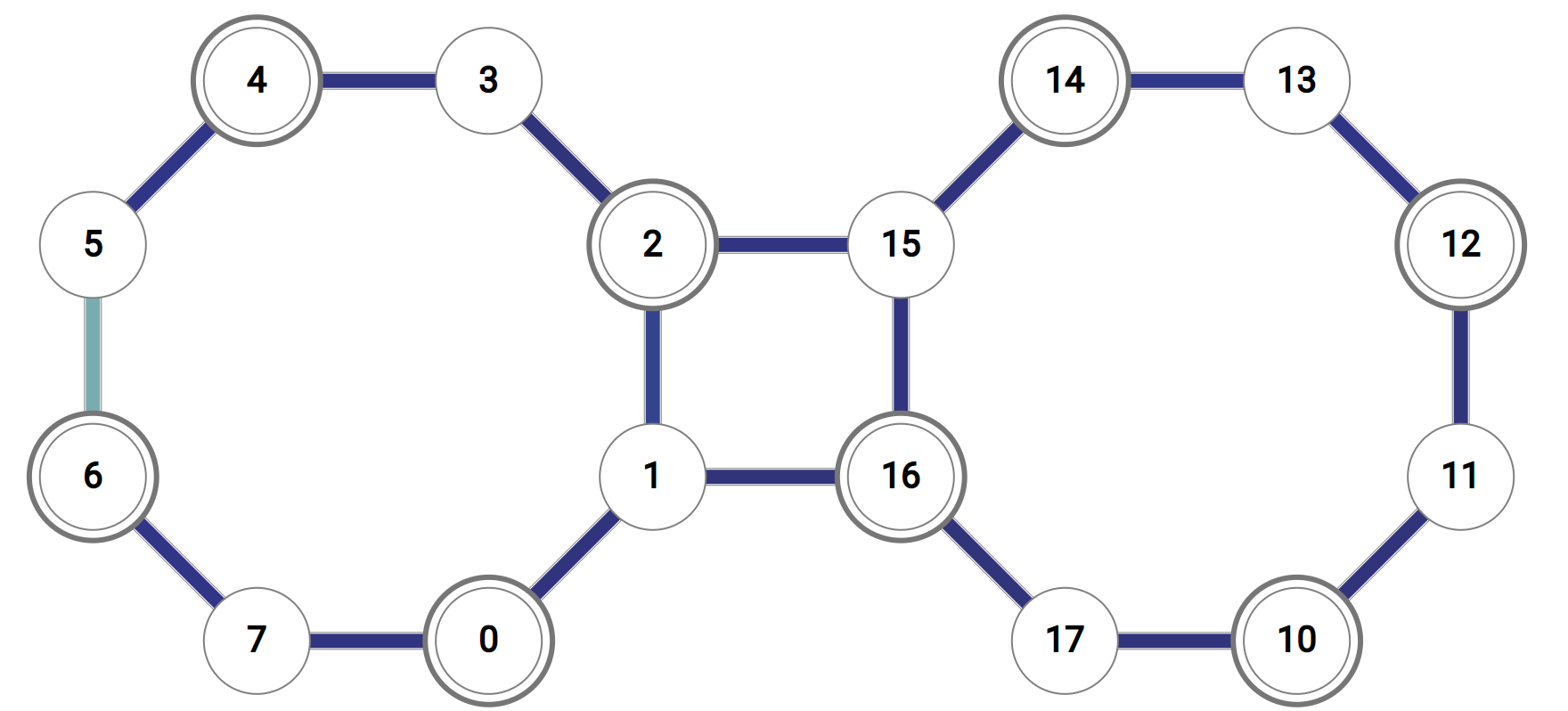}}
\caption{Rigetti quantum chips architecture (Image source: Rigetti Computing web page \url{https://www.rigetti.com/}).}
\label{Fig:Rigetti_backends}
\end{figure}

Rigetti's basic gate set is formed by three one-qubit rotational gates, $R_{z}(-\phi)$ and $R_{x}(\pm\pi/2)$ and a two-qubit gate, CZ. The minus sign added in the angle of $R_{z}$ gate is because Rigetti defines rotational gates as $e^{-i\frac{\theta}{2}\sigma_{i}}$ in contrast with the definition used in this thesis, $e^{i\frac{\theta}{2}\sigma_{i}}$. The use of the CZ gate instead of CNOT has the advantage of bi-directionality, as the result is the same independently of which is the control qubit. For that reason, the connectivity of the devices shown in Fig. \ref{Fig:Rigetti_backends} does not specify the direction of the two-qubit gate. 

The qubit topology is very different from IBM's devices. In \emph{Acorn} chip, qubits are connected following a zigzag-type geometry, in \emph{Agave}, qubits form a rectangle and in \emph{Aspen-1} they are located in two rings of 8 qubits each that are joined with two connections. Then, for the circuit of $n=4$ spins, we can not do without the fSWAP gates, which means that the circuit depth will be greater than the 16-qubits and 14-qubits IBM devices. On the other hand, it will be comparable with the 5-qubits devices, which also needs from these gates.

\section{Results\label{sec:results}}

The experimental results presented below were taken in a period between March and May 2018 in \emph{ibmqx4} (now \emph{Yorktown}), \emph{ibmqx5} (now \emph{Rueschlikon}) and \emph{Acorn} devices. They were published in Ref. \cite{Cervera18} and the program used for IBM devices was awarded and now is used as a tutorial \cite{IBMtutorial}. Some properties and, specially, post-processing tasks offered by these two companies have changed recently, so the results if the circuits are run at the present moment could be different from the ones obtained.

\begin{figure}[t!]
\centering
\includegraphics[width=0.6\textwidth]{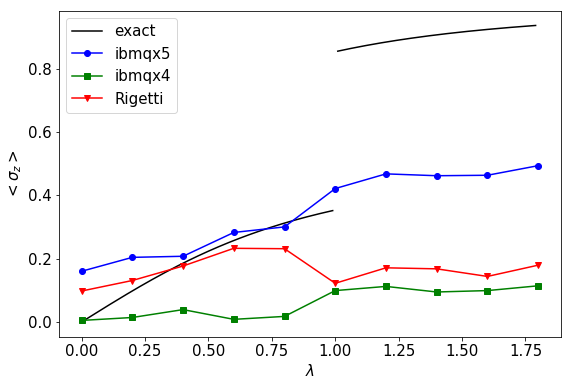}
\caption{Expected value of $\langle\sigma_z\rangle$ of the ground state of an $n=4$ Ising spin chain as a function of transverse field strength $\lambda$. Solid line represents the exact result in comparison with the experimental simulations represented by scatter points. The best simulation comes from \textit{ibmqx5} device, which is an expected result since the number of gates used is lesser than with the other devices because of qubits connectivity.
}
\label{Fig:mag}
\end{figure}

Let's set a particular case of the $XY$ model, the anti-ferromagnetic Ising spin chain, to do the experiments and to compare the performance of the three devices. Figure \ref{Fig:mag} shows the results of the exact simulation of ground state transverse magnetization. All points contain a statistical error of $1/\sqrt{N}$ with $N=1024$ which comes from the average over all runs to compute the expected value. The other error sources are discussed qualitatively in the following paragraphs.

The best performance comes from the \emph{ibmqx5} device. This is an expected result as we do not need from fSWAP gates because the qubits connectivity. On the other hand, Rigetti's device, \emph{Acorn}, perform better than the \emph{ibmqx4}, even though the number of gates is very similar. Again, it is important to point out that these results could change if we run the experiment at present. In fact, the results obtained after running a quantum circuit could differ depending on the time of the day that they were taken. Each quantum device is calibrated every few hour so the results are expected to be better immediately after this calibration rather than hours later.

The simulation approaches better to the prediction for low $\lambda$. The explanation could come from how affect the experimental error sources to the magnetization. Assuming that two-qubit gates implementation take several hundreds of ns and single qubit gates around one hundred of ns, errors coming from decoherence are expected to be low, as these times are around 50 $\mu$s. On the other hand, errors coming from the gate implementation are cumulative and probably the most important error source. It is not negligible neither errors coming from qubits readout, which can induce a bit flip.

The analysis of the results become more clear if we look at the exact ground state wave function:
\begin{align}
|gs\rangle=\frac{1}{\mathcal{N}}
\left\{\begin{array}{ll}
\sqrt{2}\alpha\left(|00\rangle|\Psi^{-}\rangle + |\Psi^{-}\rangle|00\rangle
\right)+ |11\rangle|\Psi^{-}\rangle + |\Psi^{-}\rangle|11\rangle
  & \mathrm{for} \  \lambda<1,\\
\alpha\left(|0011\rangle-|0110\rangle+|1001\rangle+|1100\rangle\right) +2|1111\rangle  & \mathrm{for} \ \lambda>1, 
\end{array}\right.
\label{eq:gs}
\end{align}
where $\alpha=\lambda-\sqrt{1+\lambda^2}$, $\mathcal{N}=2\sqrt{2}\sqrt{1+\lambda\alpha}$ and $|\Psi^{-}\rangle=(|01\rangle-|10\rangle)/\sqrt{2}$. As $\lambda$ increases, the amplitude for the states proportional to $\alpha$ goes to zero. That means that any error occurring for $\lambda>1$ is dramatic as it will affect the state with higher probability amplitude, the $|1111\rangle$. Then, any error in that regime will inevitably cause a decrease in magnetization. On the other hand, errors in some states for $\lambda<1$ can be compensated in average for the other elements with the same probability amplitude.

\begin{figure}[t!]
\centering
\includegraphics[width=0.8\textwidth]{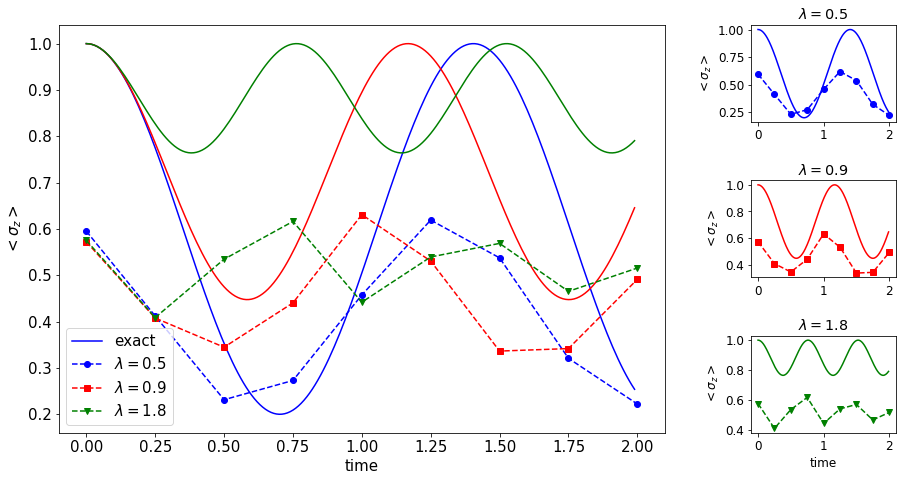}
\caption{Time evolution simulation of transverse magnetization, $\langle\sigma_z\rangle$, for the state $|\uparrow\uparrow\uparrow\uparrow\rangle$ of an $n=4$ Ising spin chain. Left plot compares the exact result with the experimental run in the \textit{ibmqx5} chip for different values of $\lambda$. Right plots detailed the results for each $\lambda$ to compare them with the theoretical values. Although the magnetization is lesser than expected, the oscillations follow the same theoretical pattern.}
\label{Fig:mag_time}
\end{figure}

Similar results are obtained for the time evolution simulation. Figure \ref{Fig:mag_time} shows the results for the simulation of the $|\uparrow\uparrow\uparrow\uparrow\rangle$ state transverse magnetization as it was explained in Sec. \ref{sec:time}. As for the preparation of the initial state it is necessary to implement more gates, only the results for the \emph{ibmqx5} device are shown, which is the one that could afford this extra circuit depth.

As expected from the previous result, points that represent higher magnetization carry more errors respect to the predicted theoretical values. However, it is remarkable that the relations among the different points along the values of transverse magnetic field are proportionally correct. The oscillations take place for lower values of $\langle\sigma_{z}\rangle$, have lower amplitudes and are a little bit shifted to the left. Even though, they cross each other at the corresponding points and increase and decrease proportionally to the exact result. That is a clear indicator that the error sources in the quantum device are systematic, as the result does not depend on the state preparation.

As a final remark, notice that we compute the transverse magnetization instead of the staggered magnetization, i.e. $M_{x}=\sum_{i}(-1)^{i}\sigma^{x}_{i}$, which is the order parameter for the anti-ferromagnetic Ising model. For the purpose of these experiments, it is more natural to compute $\langle\sigma_{z}\rangle$, since the states obtained with these quantum devices are expressed in the $\sigma_z$ basis. However, it will be straightforward to compute $M_{x}$ as the only change needed appears in the classical post-processing part.

\section{Conclusions}\label{sec:conclusion}

In this chapter, it has been implemented the exact simulation of a one-dimensional Ising spin chain with a transverse field in some quantum computer prototypes: two from IBM and one from Rigetti computing. The method to construct a quantum operation that diagonalize exactly the $XY$ Hamiltonian has been reviewed, providing the explicit circuit for the simulation of an $n=4$ spin chain. It has been also introduced novel approaches to simulate time and thermal evolution using the circuit obtained, in particular, to compute the ground state transverse magnetization and the time evolution of the state of all spins aligned.

The circuit presented allows computing all eigenstates of the $XY$ Hamiltonian by just initializing the qubits in one of the states of the computational basis. It is then an implementation of a Slater determinant with a quantum computer. Because of the one-dimensional $XY$ model is an exactly solvable model, which means that we can compute analytically all the states and energies for any number of spins, and the circuit is efficient, the number of gates scales as $n^2$ and the circuit depth as $n\log n$, it can represent a method to test quantum computing devices of any size. As has been shown, it is also a hard test because the simulation of the phase transition surrounding and time evolution require a high qubits control.

The best performance has been obtained with the \emph{ibmqx5} chip, although the error respect to the theoretical prediction is large in the paramagnetic phase of the model. A possible reason why this chip shows better results than the others comes from the number of gates used in the quantum circuit, as the qubits connectivity in that device allows us to save all the fSWAP gates. On the other hand, Rigetti's chip performs better than the \emph{ibmqx4} chip, even though both implemented circuits have the same gate depth. However, the results of these few qubits experiments could change totally if we run the circuits again. The results shown were obtained a few months before this thesis was written and, from then on, the quantum devices have changed their properties. In conclusion, this work represents just a proof of concept of how quantum computers can be tested and compared.

The paramagnetic phase is difficult to simulate due to the fact that any error that can induce a qubit bit flip will produce a decrease in magnetization, as can be traced out from the ground state wave function of Eq. \eqref{eq:gs}. However, and taking into account this fact, the time evolution simulation is reasonably good, as the expected oscillations for different transverse magnetic field strengths are shifted to the left and have lower amplitude and magnetization, but are also proportional to each other as are the theoretical values.

As a final remark, this circuit is also interesting from a point of view of condensed matter physics as specific methods to simulate exactly time and thermal evolution are provided. This can open the possibility of simulating other interesting models: integrable, like Kitaev Honeycomb model \cite{Schmoll17}, or with an ansatz, like the Heisenberg model \cite{Bethe31}.

\hyphenpenalty=10000
\chapterimage{Entangled} 

\chapter{Absolute Maximal Entanglement in Quantum Computation \label{Ch:AME}}
\hyphenpenalty=50

\vspace{-1.5cm}
\begin{flushright}
\begin{minipage}{0.6\textwidth}
\textit{La mode est architecture: c'est une question de proportions.}
\begin{flushright}
--Coco Chanel
\end{flushright}
\end{minipage}
\end{flushright}
\vspace{1cm}

The proliferation of quantum computing devices has caused the necessity of benchmark methods to test them. Current quantum computers are typically characterized by its number of qubits and its connectivity and its performance is measured with gate fidelities, coherence and relaxation times. However, the results obtained are far below the expected accuracy if errors of gates were to be taken at face value and considered independent, as has been already shown in the results of the previous chapter.

Several proposals exist to benchmark quantum computers. As an example, corporations like IBM have defined a figure of merit called \emph{quantum volume} to quantify the quality of their devices \cite{Volume}. This method follows the ideas of randomize benchmarking \cite{Knill08}, another method used to extract qubit gate fidelities. In the previous chapter, we have introduced the simulation of exactly solvable models in a quantum computer, which can be also used as a benchmark method since we can compare the result obtained with the correct solution computed analytically. However, all these methods do not take into account the probably principal resource of quantum computation: entanglement. Although one expects to develop some amount of entanglement in the protocols presented above, are quantum computers able to generate as much entanglement as we will require? The only way to test it is by forcing quantum devices to generate highly entangled states and observe if they are capable to support them.

We know that entanglement is at the core of quantum advantage. Or, in other words, quantum advantage is a consequence of high entanglement generation. This is not surprising since Bell inequalities are violated by high entangled states: quantum physics cannot be described classically because of the existence of entanglement. 

In this chapter, we present some quantum circuits that generate \textit{Absolutely Maximally Entangled (AME) states}, i.e. states that maximally entangle all their bipartitions. The circuits introduced are composed by few CZ gates and one-qubit gates that can be performed in parallel. This proposal is distinctly different from bosonic sampling \cite{Aaronson13}, where large entanglement is developed along the circuit to make it impossible to be faithfully reproduced by classical simulation. In a sense, maximally entangled states are a test for a useful quantum computer, not for quantum advantage.

The existence of this kind of states is limited: for qubits, they only exist for $n=2,3,5$ and 6 parties. For that reason, we also propose to simulate AME states of higher dimensions using qubits. The results will maximally entangle some parties although not of them. We derive some interesting properties of these circuits, for example, that the entropy is majorized after each entangling gate is applied. This characteristic could be a consequence of circuit optimality.

Beside benchmarking interest, AME states define an interesting mathematical problem itself and attractive practical applications. These include quantum secret sharing \cite{Helwig12,HC13}, open destination quantum teleportation \cite{HC13} and quantum error correcting codes \cite{Scott04}. The last one is a fundamental ingredient for building a quantum computer. In addition, there is a natural link between AME states and holography through error correcting codes \cite{Latorre15,Pastawski15}.

The structure of this chapter is organized as follows. First, we introduce a short review of AME states that includes its definition and the most fundamental properties. Second, in Sec. \ref{sec:graph}, we present the graph state formulation that we will use to construct the quantum circuits for AME states. These circuits are shown in Sec. \ref{sec:AMEgraph} and the simulation of AME states of $d>2$ with qubits in Sec. \ref{sec:AMEqubits}. For its interest in error correcting codes, we present an example of an AME state of minimal support in Sec. \ref{sec:AMEminimal}. Finally, we introduce the entropy majorization analysis in Sec. \ref{sec:maj} and close with the conclusions in Sec. \ref{sec:AMEcon}.

\section{Absolutely Maximally Entangled states}

The formal definition of an Absolutely Maximally Entangled (AME) state is the following:
\begin{definition}[Absolutely Maximally Entangled states]
An AME($n,d$) state is a $n$ qudit state with local dimension $d$ whose all possible bipartitions to $\lfloor n/2\rfloor$ parties are maximally entangled, i.e. all reduced density matrices are proportional to the identity.
\end{definition}
Such states are maximally entangled when considering the entropy of reductions as a measure of multipartite entanglement. Thus, all bipartitions of an AME state have entropy
\begin{equation}
S=\left\lfloor\frac{n}{2}\right\rfloor \ ,
\end{equation}
taking the $\log$ in $d$ basis.

Bell states and GHZ state are AME states for the bipartite and tripartite cases respectively and for any dimension $d$. However, the GHZ states for $n\geq 4$ are not AME states. The existence of AME($n,d$) states $\forall \ n$ and $d$ is a hard open problem. Only for qubits, $d=2$, the problem is fully solved: an AME($n$,2) exist only for $n=2,3,5,6$ \cite{Helwig13,Huber16}.

AME states connect to different mathematical ideas. One example is the family of AME states of minimal support which are one-to-one related to a special class of maximum distance separable codes \cite{Huffman03,MDS}, index unity orthogonal arrays \cite{Goyeneche14,Seveso18}, permutation multi-unitary matrices when $n$ is even \cite{Goyeneche15} and to a set of $m=n-\lfloor n/2\rfloor$ mutually orthogonal Latin hypercubes of size $d$ defined in dimension $\lfloor n/2\rfloor$ \cite{Goyeneche18}.
\begin{definition}[AME states of minimal support]
An AME($n,d$) state has minimal support if it can be written as a linear combination of $d^{\lfloor n/2\rfloor}$ fully separable orthogonal pure states, e.g. computational basis states. 
\end{definition}
All coefficients of every AME state having minimal support can be chosen to be positive and identically equal to $d^{-\lfloor n/2\rfloor/2}$. By contrast, AME states of non minimal support will need non-trivial phases to have all reduced density matrices proportional to the identity. In other words, non minimal support AME require destructive interference. 

The study of AME states was initially motivated by the identification of \emph{$k$-uniform} states \cite{Gisin98,Higuchi00,Facchi08,Facchi10,Arnaud13,Brown05}. These states are maximally entangled in $k$ arbitrary subsystems. Then, an AME state is an extremal case of a $k$-uniform state with $k=\lfloor n/2\rfloor$. For that reason, AME states were previously known as \emph{Maximally Multipartite Entangled} states \cite{Facchi08} or \emph{perfect maximally multipartite entangled states} \cite{Facchi09}. The reason why, for a fixed dimension, AME states are more difficult to obtain as $n$ increases come to the fact that more states need to be included in the Maximally Entangled set \cite{Vicente13}. Any state outside this set can be obtained via LOCC from one of the states within the set. This is not an equivalent definition of an AME state, as states in a Maximally Entangled set are not necessarily AME states, but all AME states are part of one of these sets.

\section{Graph states \label{sec:graph}}

Graph states are a type of pure quantum states that can be constructed from a graph following the below recipe.
\begin{definition}[Graph state]
Given a graph of $\mathcal{V}=\{v_{i}\}$ vertices connected by $\mathcal{E}=\{e_{ij}=\{v_{i},v_{j}\}\}$ edges, its corresponding graph state is constructed as
\begin{equation}
|G\rangle\equiv\prod_{i<j}^{n}\mathrm{CZ}_{ij}^{A_{ij}}(F_{d}|\bar{0}\rangle)^{\otimes{n}},
\end{equation} 
where $F_{d}$ is the Fourier gate, CZ is the generalized Control-Z gate and $|\bar{0}\rangle$ is the zeroth qudit state. The components of the adjancency matrix $\mathcal{A}$, $A_{ij}$, provide the weights of each edge $e_{ij}$, where weight zero means no edge and $A_{ii}=0$.
\end{definition}

The definition of the generalized CZ gate and Fourier gate $F_{d}$ for qudits are
\begin{equation}
\mathrm{CZ}_{ij}=\sum_{k=0}^{d-1}\omega^{kl}|\bar{k}\rangle\langle \bar{k}|_{i}\otimes|\bar{l}\rangle\langle \bar{l}|_{j} \ ,
\label{eq:CZgate}
\end{equation}
with $\omega=e^{2\pi i/d}$, and
\begin{equation}
F_{d}=\frac{1}{\sqrt{d}}\sum_{k=0}^{d-1}\omega^{kl}|\bar{k}\rangle\langle \bar{l}|
\label{eq:Fgate}
\end{equation}
respectively. The state $|\bar{\psi}\rangle$ is a $|\psi\rangle$ state for qudits. For now on, to distinguish between qubits and qudits states, we will write a bar on qudit states and keep the usual notation, with no bar, for qubits.

Following the above definition, the explicit construction of a graph state from its corresponding graph is straightforward. First, each vertex corresponds with the qudit state $|\bar{\psi}_{0}\rangle = F_{d}|\bar{0}\rangle$, and second, each edge corresponds with a CZ gate applied between two vertices.

Notice that after applying the Fourier gates $F_d$ we obtain a state with all basis elements and, since CZ gates only introduce relative phases between these elements, the final state of a graph contains a superposition of $d^n$ computational states. 

Graph states can also be described using stabilizer states \cite{stab}. They have many applications, especially in quantum error correcting codes \cite{Looi08} and one-way quantum computing \cite{RB01}. A graphical interpretation of entanglement in graph states is provided in Ref. \cite{Helwig13} and multipartite entanglement properties in qubit graph states as well as its optimal preparation have been studied in Ref. \cite{Hein04,Cabello09,Cabello11}.

\section{AME states from graphs \label{sec:AMEgraph}}

We can obtain an AME state from its corresponding graph. It will be a particular form of an AME state of maximal support since, by construction, we have the superposition of all qudit basis elements. 

\begin{figure}[t!]
\begin{minipage}{0.6\textwidth}
\centering
\[ \qquad 
\Qcircuit @C=1.5em @R=.7em @!R {
\lstick{\ket{0}} & \gate{H} & \ctrl{1} & \qw & \qw & \qw & \ctrl{4} & \qw \\
\lstick{\ket{0}} & \gate{H} & \ctrl{-1} & \ctrl{1} & \qw & \qw & \qw & \qw \\
\lstick{\ket{0}} & \gate{H} & \qw & \ctrl{-1} & \ctrl{1} & \qw & \qw & \qw \\
\lstick{\ket{0}} & \gate{H} & \qw & \qw & \ctrl{-1} & \ctrl{1} & \qw & \qw \\
\lstick{\ket{0}} & \gate{H} & \qw & \qw & \qw & \ctrl{-1} & \ctrl{-4}  & \qw
}
\]
\end{minipage}
\begin{minipage}{.4\textwidth}
\centering
\includegraphics[width=0.7\textwidth]{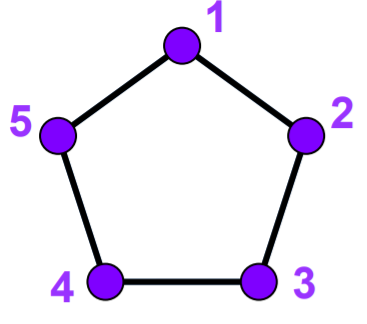}
\end{minipage}
\caption{Quantum circuit to generate AME(5,2) and its corresponding graph.}
\label{Fig:AME52}
\end{figure}

As an example, consider the graph of Fig. \ref{Fig:AME52} right. It is a graph of five vertices and five edges so we need five one-qudit gates and five CZ gates to construct the corresponding graph state. If we are dealing with qubits, each $F_{2}$ gate is actually a Hadamard gate so to obtain the quantum circuit we have just to apply Hadamard gates on all qubits and CZ gates according to the edges of the graph, as shows Fig. \ref{Fig:AME52} left. This circuit generates an AME(5,2) state and its graph can be used to construct any AME(5,$d$) by using $F_{d}$ and generalized CZ instead of Hadamards and qubit CZ gates.

We will be interested on finding the optimal AME graph states, i.e. those graphs with minimum number of edges and coloring index \cite{Cabello09}. Less number of edges is translated into less operations to generate these AME states and colouring index is related with the number of operations that can be performed in parallel, so it is proportional to the circuit depth.

Finding AME graph states is in general a hard task as we increase the local dimension $d$ and the number of parties $n$. Fortunately there are some methods and properties to find these graphs for specific values of $d$ and $n$ \cite{Cabello11}. 

The first interesting property is that there are some graph states that work for any dimension $d$. In particular, the graph states shown in Fig. \ref{Fig:AME5d} and Fig. \ref{Fig:AME6d} work for any prime dimension $d$. The graph state of Fig. \ref{Fig:AME4d} also fulfils this property but for prime dimension $d\geq 3$.

For a non-prime local dimension there exist some methods to find AME graph states. One of those consists on taking the prime factorization $d=d_{1}d_{2}\cdots d_{m}$ and look for the AME($n,d_{i}$) states independently. If they exist, the AME($n,d$) is just the tensor product of the AME($n,d_{i}$) states. In case prime factorization of $d$ includes a power of some factor, we can construct an AME state by defining artificially each party using qudits of lower dimension and by performing CZ gates between these qudits. These method has been used to find an the AME(4,4) state using qubits instead of ququarts (qudits of $d=4$) as it is illustrated in Fig. \ref{Fig:AME44_circuit}. The local dimension of each party, $d=4$, is achieved with the state of two qubits.

\section{Simulation of AME states with qubits \label{sec:AMEqubits}}

The construction of a quantum circuit that generates a qubit AME state starting from its graph is straightforward. As explained in the previous section, we just have to perform Hadamard gates on all qubits and CZ gates according to graph edges. Both gate operations are common in current quantum devices. However, if we are interested in generating an AME state of $d>2$ we may need a qudit quantum computer that can perform quantum operations beyond binary quantum computation. Currently, available quantum computers only work with two-level systems and, for that reason, we propose to simulate these AME states of greater dimension using qubits instead of qudits. To do so, we will translate the local dimensions $d$ into multiqubit states, namely
\begin{align}
|\bar{0}\rangle &\equiv |00\rangle, \nonumber\\
|\bar{1}\rangle &\equiv |01\rangle, \nonumber\\
|\bar{2}\rangle &\equiv |10\rangle, \nonumber\\
|\bar{3}\rangle &\equiv |11\rangle.
\label{eq:ABC}
\end{align}
For $d>4$, we will need to increase the number of qubits accordingly, i.e. we will need $m=\lceil\log_{2} d\rceil$ qubits to describe each qudit. 

Since we have the graphs for these states, the challenge will be to simulate the effect of the generalized CZ gate of Eq. \eqref{eq:CZgate} and the Fourier gate of Eq. \eqref{eq:Fgate} with qubit gates. To be precise, we are not interested in the exact Fourier gate but on generating the state $|\bar{\psi}_{0}\rangle = F_{d}|\bar{0}\rangle$. For that propose, we will look for an initialization gate $U_{d}^{in}$ that acts on qubits in the state $|0\rangle$ and obtains the $|\psi_{0}\rangle$ state, i.e. the $|\bar{\psi}_{0}\rangle$ state written in terms of qubits according to the mapping of Eq. \eqref{eq:ABC}.

If $d$ is a power of 2, the $|\bar{\psi}_{0}\rangle$ state can be generated easily using only Hadamard gates. In particular, for $d=4$,
\begin{multline}
|\bar{\psi}_{0}\rangle=F_{4}|\bar{0}\rangle=\frac{1}{2}\left(|\bar{0}\rangle+|\bar{1}\rangle+|\bar{2}\rangle+|\bar{3}\rangle\right) \rightarrow \\
\rightarrow |\psi_{0}\rangle= U_{4}^{in}|00\rangle = (H\otimes H)|00\rangle=\frac{1}{2}\left(|00\rangle+|01\rangle+|10\rangle+|11\rangle\right).
\end{multline}
Notice that $F_{4}\neq U_{4}^{in}=(H\otimes H)$ but for our propose it does not matter, since we just want to obtain the $|\bar{\psi}_{0}\rangle$ state with qubits.

\begin{figure}[t!]
\[ \qquad 
\Qcircuit @C=1.5em @R=.7em @!R {
 & & & & & \lstick{\ket{0}} & \multigate{1}{U^{in}_{3}} & \qw & & & \lstick{\ket{0}} & \gate{R_y(\theta)} &  \ctrl{1} & \targ & \qw \\
\lstick{\ket{\bar{0}}} & \gate{F_{3}} & \qw & \equiv & & \lstick{\ket{0}} & \ghost{U^{in}_{3}} & \qw & = &
 & \lstick{\ket{0}} & \qw & \gate{H} & \ctrl{-1} & \qw
}
\]
\caption{Quantum circuit to obtain $|\bar{\psi}_{0}\rangle$ qutrit state using two qubits, i.e. to generate $|\psi_{0}\rangle=\left(|00\rangle+|01\rangle+|10\rangle\right)/\sqrt{3}$ state. The angle of the rotational gate is $\theta=-2\arccos(1/\sqrt{3})$.}
\label{Fig:F3}
\end{figure}
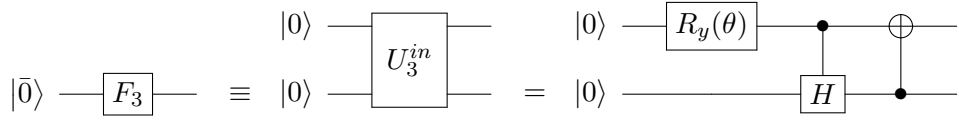

For $d=3$ the $|\bar{\psi}_{0}\rangle$ can be obtained from the $U_{3}^{in}$ gate defined in Fig. \ref{Fig:F3},
\begin{equation}
|\bar{\psi}_{0}\rangle = F_{3}|\bar{0}\rangle=\frac{1}{\sqrt{3}}\left(|\bar{0}\rangle+|\bar{1}\rangle+|\bar{2}\rangle\right) \rightarrow |\psi_{0}\rangle =U_{3}^{in}|00\rangle =  \frac{1}{\sqrt{3}}\left(|00\rangle+|01\rangle+|10\rangle\right).
\end{equation}

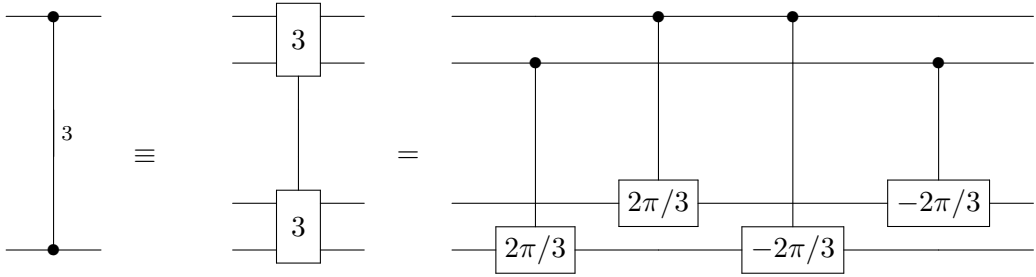
\begin{figure}[t!]
\[
\Qcircuit @C=1.5em @R=.01em @!R 
{
 & \ctrl{3} & \qw  & &  &
 & \multigate{1}{3} & \qw & & & \qw  & \ctrl{4} & \ctrl{5} & \qw  & \qw  \\
 &  &  &  &  &  
 &  \ghost{3} & \qw &  &  & \ctrl{4} & \qw & \qw &  \ctrl{3} & \qw  \\
 &  & &  & & & \qwxo{\hspace{1pt}} & & & & & & & & \\
  & \qwxo{3} & & \equiv & & & \qwxo{\hspace{1pt}} & & = & & & & & & \\
 &  &  &  &  &
   & \multigate{1}{3} \qwxo{\hspace{1pt}} & \qw &  &  & 
   \qw & \gate{2\pi/3} & \qw & \gate{-2\pi/3} & \qw \\
 & \ctrl{-3} & \qw & & &  &
  \ghost{3} & \qw & & & \gate{2\pi/3} & \qw & \gate{-2\pi/3} & \qw & \qw 
}
\]
\caption{Generalized CZ gate for qutrits, $d=3$, performed with four qubits. First two CPh gates and last two CPh gates can be implemented in parallel, so the circuit depth is just 2 CPh gates.}
\label{Fig:CZ3}
\end{figure}

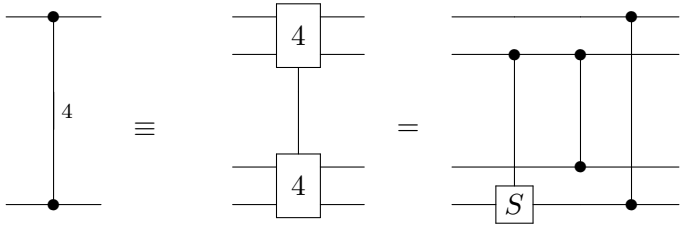
\begin{figure}[t!]
\[ 
\Qcircuit @C=1.5em @R=.01em @!R 
{
 & \ctrl{3} & \qw & & &
 & \multigate{1}{4} & \qw & & & \qw & \qw & \ctrl{5} & \qw   \\
 &  &  &  &  &  
 &  \ghost{4} & \qw &  &  & \ctrl{4} & \ctrl{3}& \qw & \qw   \\
 &  & &  & & & \qwxo{\hspace{1pt}} & & & & & & &  \\
  & \qwxo{4} & & \equiv & & & \qwxo{\hspace{1pt}} & & = & & & & & \\
 &  &  &  &  &
   & \multigate{1}{4} \qwxo{\hspace{1pt}} & \qw &  &  & 
   \qw & \control \qw & \qw & \qw \\
 & \ctrl{-3} & \qw & & &  &
  \ghost{4} & \qw & & & \gate{S} & \qw & \control \qw & \qw 
}
\]
\caption{Generalized CZ gate for ququarts, $d=4$, performed with four qubits. First gate is a controlled-S gate, which is actually a CPh gate with $\theta=\pi/2$. Last two CZ gates can be implemented in parallel, so the circuit depth is just 2 gates.}
\label{Fig:CZ4}
\end{figure}

In general, to obtain the circuit to produce the $|\psi_{0}\rangle$ states will be hard except if $d$ is a power of 2, as explained above. On the contrary, finding a circuit that implements a generalized CZ gate is more intuitive since this gate only introduces a phase in some qudit states. We can reproduce this effect by using controlled-Phase gates CPh($\theta$).

Figure \ref{Fig:CZ3} shows the circuit to implement generalized CZ gate for qutrits with qubits. We will need four qubits and four CPh gates to achieve the expected result of this gate. The quantum circuit to implement the generalized CZ gate for ququarts is shown in Fig. \ref{Fig:CZ4}. Only three gates are needed: two-qubit CZ gates and a controlled-$S$ gate, which is a CPh with $\theta=\pi/2$.

\begin{figure}[t!]
\centering
\begin{minipage}{.6\textwidth}
\centering
\[ \qquad 
\Qcircuit @C=1em @R=.1em 
{
 \lstick{\ket{0}^{\otimes m}} & \gated{{\color{violet}U_d^{in}}}& \gated{d}  & \cw & \cw & \cw & \gated{d} & \cw \\
 & & \qwxo{\hspace{1pt}} & & & &\qwxo{\hspace{1pt}} & \\
 \lstick{\ket{0}^{\otimes m}} & \gated{{\color{violet}U_d^{in}}}&\gated{d} \qwxo{\hspace{1pt}} & \gated{d} & \cw & \cw & \qwxo{\hspace{1pt}} \cw & \cw \\
  &  & & \qwxo{\hspace{1pt}} & & & \qwxo{\hspace{1pt}} & \\
  \lstick{\ket{0}^{\otimes m}} & \gated{{\color{violet}U_d^{in}}}& \cw & \gated{d} \qwxo{\hspace{1pt}} & \gated{d} & \cw & \cw \qwxo{\hspace{1pt}} & \cw\\
 &  & & &  \qwxo{\hspace{1pt}} & & \qwxo{\hspace{1pt}} & \\
 \lstick{\ket{0}^{\otimes m}} & \gated{{\color{violet}U_d^{in}}}& \cw  & \cw & \gated{d} \qwxo{\hspace{1pt}} & \gated{d} & \cw \qwxo{\hspace{1pt}} & \cw\\
  &  & &  & & \qwxo{\hspace{1pt}} & \qwxo{\hspace{1pt}} & \\
  \lstick{\ket{0}^{\otimes m}} & \gated{{\color{violet}U_d^{in}}}& \cw  & \cw & \cw & \gated{d} \qwxo{\hspace{1pt}} & \gated{d} \qwxo{\hspace{1pt}}& \cw 
}\]
\end{minipage}\hspace{-2cm}
\begin{minipage}{.4\textwidth}
\centering
\includegraphics[width=0.7\textwidth]{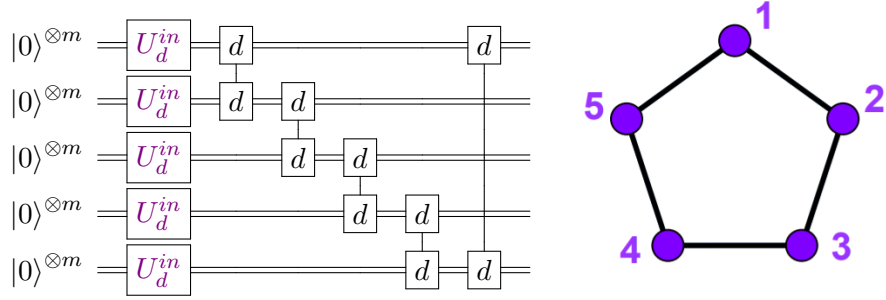}
\end{minipage}
\caption{Graph state (\textit{right}) that generates an AME(5,d) state and its corresponding circuit (\textit{left}) using qubits instead of qudits. The number of qubits needed to represent each qudit is $m=\lceil\log_{2}d\rceil$. First, qubits are prepared in the basis superposition state using $U_{d}^{in}$, which corresponds to $H$ for qubits, $U_{3}^{in}$ of Fig. \ref{Fig:F3} for qutrits and $U_{4}^{in}=H\otimes H$ for ququarts. Then, CZ gates are performed between qudits, which for $d=3$ and $d=4$ can be implemented with the circuit of Fig. \ref{Fig:CZ3} and \ref{Fig:CZ4} respectively.}
\label{Fig:AME5d}
\end{figure}

\begin{figure}[t!]
\centering
\begin{minipage}{0.8\textwidth}
\[
\Qcircuit @C=0.8em @R=.1em 
{
 \lstick{\ket{0}^{\otimes m}} & \gated{{\color{violet}U_d^{in}}}& \gated{d}  & \cw & \cw & \cw & \cw & \gated{d} & \gated{d} & \cw & \cw & \cw \\
 & & \qwxo{\hspace{1pt}} & & & & & \qwxo{\hspace{1pt}} & \qwxo{\hspace{1pt}} & & & & \\
 \lstick{\ket{0}^{\otimes m}} & \gated{{\color{violet}U_d^{in}}}& \gated{d} \qwxo{\hspace{1pt}} & \gated{d} & \cw & \cw & \cw & \qwxo{\hspace{1pt}} \cw & \cw \qwxo{\hspace{1pt}} & \gated{d} & \cw & \cw \\
  &  & & \qwxo{\hspace{1pt}} & & & & \qwxo{\hspace{1pt}} & \qwxo{\hspace{1pt}} & \qwxo{\hspace{1pt}} & \\
  \lstick{\ket{0}^{\otimes m}} & \gated{{\color{violet}U_d^{in}}}& \cw & \gated{d} \qwxo{\hspace{1pt}} & \gated{d} & \cw & \cw & \cw \qwxo{\hspace{1pt}} & \cw \qwxo{\hspace{1pt}} & \cw \qwxo{\hspace{1pt}} & \gated{d} & \cw \\
 &  & & &  \qwxo{\hspace{1pt}} & & & \qwxo{\hspace{1pt}} & \qwxo{\hspace{1pt}} & \qwxo{\hspace{1pt}} & \qwxo{\hspace{1pt}} & \\
 \lstick{\ket{0}^{\otimes m}} & \gated{{\color{violet}U_d^{in}}}& \cw  & \cw & \gated{d} \qwxo{\hspace{1pt}} & \gated{d} & \cw & \cw \qwxo{\hspace{1pt}} & \cw \qwxo{\hspace{1pt}} & \gated{d} \qwxo{\hspace{1pt}} & \cw \qwxo{\hspace{1pt}} & \cw\\
  &  & &  & & \qwxo{\hspace{1pt}} & & \qwxo{\hspace{1pt}} & \qwxo{\hspace{1pt}} & & \qwxo{\hspace{1pt}} & & \\
  \lstick{\ket{0}^{\otimes m}} & \gated{{\color{violet}U_d^{in}}}& \cw  & \cw & \cw & \gated{d} \qwxo{\hspace{1pt}} & \gated{d} & \cw \qwxo{\hspace{1pt}}& \gated{d} \qwxo{\hspace{1pt}} & \cw & \cw \qwxo{\hspace{1pt}} & \cw \\
   &  & &  & &  & \qwxo{\hspace{1pt}} & \qwxo{\hspace{1pt}} & & & \qwxo{\hspace{1pt}}\\
  \lstick{\ket{0}^{\otimes m}} & \gated{{\color{violet}U_d^{in}}}& \cw  & \cw & \cw & \cw & \gated{d} \qwxo{\hspace{1pt}} & \gated{d} \qwxo{\hspace{1pt}} & \cw & \cw & \gated{d} \qwxo{\hspace{1pt}} & \cw
}
\]
\end{minipage}\hspace{-2cm}
\begin{minipage}{0.2\textwidth}
\centering
\includegraphics[width=1.3\textwidth]{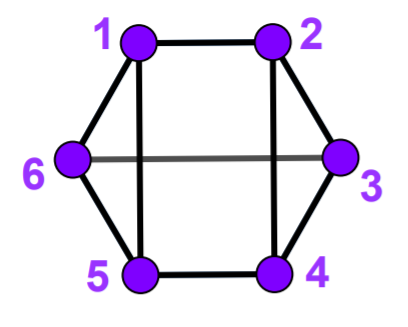}
\end{minipage}
\caption{Graph state (\textit{right}) that generates an AME(6,$d$) state and its corresponding circuit (\textit{left}) using qubits instead of qudits. The number of qubits needed for represent each qudit is $m=\lceil\log_{2}d\rceil$. Qubits are prepared using $U_{d}^{in}$ gates and CZ gates of dimension $d$ are simulated using the circuits of Fig. \ref{Fig:CZ3} and Fig. \ref{Fig:CZ4}.}
\label{Fig:AME6d}
\end{figure}

\begin{figure}[t!]
\centering
\begin{minipage}{0.6\textwidth}
\centering
 \[
\Qcircuit @C=1em @R=.1em 
{
 \lstick{\ket{0}^{\otimes m}} & \gated{{\color{violet}U_d^{in}}}& \gated{d} & \cw & \cw & \cw & \gated{d} & \cw    \\
 & & \qwxo{\hspace{1pt}} & & & & \qwxo{\hspace{1pt}} &  \\
 \lstick{\ket{0}^{\otimes m}} & \gated{{\color{violet}U_d^{in}}}&\gated{d} \qwxo{\hspace{1pt}} & \gated{d} & \gated{d} & \cw & \cw \qwxo{\hspace{1pt}} & \cw  \\
  &  &  & \qwxo{\hspace{1pt}} & \qwxo{\hspace{1pt}} & & \qwxo{\hspace{1pt}} & \\
  \lstick{\ket{0}^{\otimes m}} & \gated{{\color{violet}U_d^{in}}}& \cw & \cw \qwxo{\hspace{1pt}} &   \cw\qwxo{\hspace{1pt}} & \gated{d} & \gated{d} \qwxo{\hspace{1pt}} & \cw \\
 &  &  & \qwxo{\hspace{1pt}} & \qwxo{\hspace{1pt}} & \qwxo{\hspace{1pt}} & & \\
 \lstick{\ket{0}^{\otimes m}} & \gated{{\color{violet}U_d^{in}}}& \cw  & \gated{d}\qwxo{\hspace{1pt}} & \gated{d}\qwxo{\hspace{1pt}} & \gated{d}\qwxo{\hspace{1pt}} & \cw & \cw
}
\]
\end{minipage}\hspace{-2cm}
\begin{minipage}{0.4\textwidth}
\centering
\includegraphics[width=0.6\textwidth]{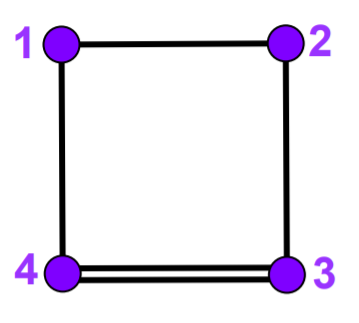}
\end{minipage}
\caption{Graph state (\textit{right}) that generates an AME(4,$d$) state for any prime dimension $d\geq 3$ and its corresponding circuit (\textit{left}) using qubits instead of qudits.}
\label{Fig:AME4d}
\end{figure}

\begin{figure}[t!]
\begin{minipage}{0.6\textwidth}
\centering
\[\qquad
\Qcircuit @C=1em @R=.1em 
{
& &\lstick{\ket{0}} & \gate{{\color{violet}H}} & \qw & \qw & \qw & \qw & \ctrl{3} & \ctrl{4} & \qw & \qw & \qw  \\
& & \lstick{\ket{0}} & \gate{{\color{violet}H}} & \ctrl{1} & \qw & \qw & \qw & \qw & \qw & \qw & \ctrl{5} & \qw  
\inputgroupv{1}{2}{0.8em}{0.8em}{{\color{red}A}}\\
& & \lstick{\ket{0}} & \gate{{\color{violet}H}} & \ctrl{-1} & \ctrl{3} & \qw & \qw & \qw & \qw & \qw & \qw & \qw  \\
& & \lstick{\ket{0}} & \gate{{\color{violet}H}} & \qw & \qw & \qw & \ctrl{4} & \ctrl{-3} & \qw & \qw & \qw & \qw  
\inputgroupv{3}{4}{0.8em}{0.8em}{{\color{red}B}}\\
& & \lstick{\ket{0}} & \gate{{\color{violet}H}} & \qw & \qw & \qw & \qw & \qw & \ctrl{-4} & \ctrl{2} & \qw & \qw  \\
& & \lstick{\ket{0}} & \gate{{\color{violet}H}} & \qw & \ctrl{-3} & \ctrl{2} & \qw & \qw & \qw & \qw & \qw & \qw 
\inputgroupv{5}{6}{0.8em}{0.8em}{{\color{red}C}}\\
& & \lstick{\ket{0}} & \gate{{\color{violet}H}} & \qw & \qw & \qw & \qw & \qw & \qw & \ctrl{-2} & \ctrl{-5} & \qw  \\
& & \lstick{\ket{0}} & \gate{{\color{violet}H}} & \qw & \qw & \ctrl{-2} & \ctrl{-4} & \qw & \qw & \qw & \qw & \qw 
\inputgroupv{7}{8}{0.8em}{0.8em}{{\color{red}D}}\\
}
\]
\end{minipage}
\begin{minipage}{0.4\textwidth}
\centering
\includegraphics[width=0.75\textwidth]{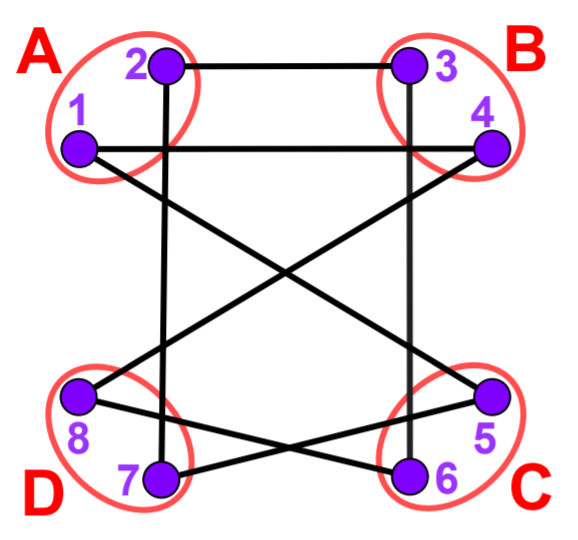}
\end{minipage}
\caption{Quantum circuit to produce the AME(4,4) state with qubits (\textit{left}) and its corresponding graph (\textit{right}). Parties $A$, $B$, $C$ and $D$ are maximally entangled between them. Notice that this circuit do not correspond with an AME(8,2) state, since this AME state do not exist.}
\label{Fig:AME44_circuit}
\end{figure}

At this point, all ingredients to construct the AME states for qubits and to simulate AME states of $d>2$ has been introduced. Circuits of Fig. \ref{Fig:AME5d} and Fig. \ref{Fig:AME6d} can be used to simulate any AME(5,$d$) and AME(6,$d$) state with qubits providing $U_d^{in}$ and CZ gates. Similarly, circuit of Fig. \ref{Fig:AME4d} can be used to simulate any AME(4,$d$) state for prime $d\geq3$. Finally, Fig. \ref{Fig:AME44_circuit} shows explicitly the circuit and the graph to obtain the AME(4,4) state.

\section{AME states circuits of minimal support \label{sec:AMEminimal}}

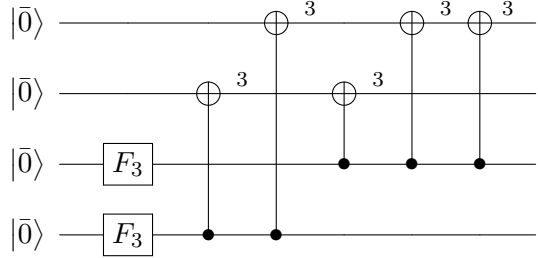
\begin{figure}[t!]
\centering
{\[
\qquad \Qcircuit @C=1.5em @R=1em @!R {
\lstick{\ket{\bar{0}}} & \qw & \qw & \targ & \qw_{3} & \targ & \targ_{3} & \qw_{3}\\
\lstick{\ket{\bar{0}}} & \qw & \targ & \qw_{3} & \targ & \qw_{3} & \qw & \qw \\
\lstick{\ket{\bar{0}}} & \gate{F_{3}} & \qw & \qw & \ctrl{-1} & \ctrl{-2} & \ctrl{-2} & \qw\\
\lstick{\ket{\bar{0}}} & \gate{F_{3}} & \ctrl{-2} & \ctrl{-3} & \qw & \qw & \qw & \qw
}
\]
}
\caption{Quantum circuit required to generate the state
$|\Omega_{4,3}\rangle$ (4 qutrits) based on the Fourier gate $F_{3}$ and C$_{3}$--adder gate for qutrits. 
}
\label{Fig:AME43}
\end{figure}

For qutrits, it is known an AME state of minimal support that can be written as \cite{Goyeneche15}
\begin{equation}
|\Omega_{4,3}\rangle=\frac{1}{3}\sum_{i,j=0,1,2}|\overbar{i}\rangle|\overbar{j}\rangle|\overbar{i+j}\rangle+|\overbar{i+2j}\rangle,
\label{eq:AME43}
\end{equation}
where all operations are computed mod(3). The quantum circuit that generates this state is shown in Fig. \ref{Fig:AME43}. The quantum gates required to construct this circuit are the Fourier transform gate for qutrits $F_{3}$ and the C$_3$--adder gate
\begin{equation}
\overbar{\mathrm{C}}_{3}|\overbar{i}\rangle|\overbar{j}\rangle=|\overbar{i}\rangle|\overbar{i+j}\rangle,
\end{equation}
which is the generalization of CNOT gate for qutrits and it is represented with the CNOT symbol with the superscript 3.

The simulation of the $F_{3}|\bar{0}\rangle$ state using qubits has been already explained in the previous section.  
The construction of the C$_{3}$--adder gate is more cumbersome and we leave the details to the App. \ref{app:OddsEnds}. The strategy that we use consists on using controlled gates that allow us to perform the sums separately for each control state. If the control qutrit is in the state $|\bar{0}\rangle$, we should apply the identity, so no gates are needed in that case. If the control qutrit is in the state $|\bar{1}\rangle$, i.e. $|01\rangle$, then we should implement CNOT and Toffoli gates (CCNOT) that take as a control qubit the second qubit, i.e. they will not affect the state of the second pair of qubits in case the first two are in a different state. Similarly, if the qutrit state is $|\bar{2}\rangle$, i.e. $|10\rangle$, we should search for a sequence of CNOT and CCNOT gates that implement the corresponding sums using as a control qubit the first qubit.

\begin{figure}[t!]
\centering
\[
\Qcircuit @C=0.7em @R=.7em @!R {
& \ctrl{3} & \qw & & & &  \qw & \qw & \qw & \ctrl{3} & \ctrl{2} & \ctrl{2} & \qw & \ctrl{3} & \qw & & &  \qw & \qw & \qw & \ctrl{3} & \ctrl{2} & \ctrl{2} & \qw  \\
 & & & &  & &  \ctrl{1} & \ctrl{1} & \ctrl{1} & \qw & \qw & \qw & \ctrl{1}& \qw & \qw &  & &  \ctrl{1} & \ctrl{1} & \ctrl{1} & \qw & \qw & \qw  & \qw \\
 & & & \push{\rule{.3em}{0em}=\rule{.3em}{0em}}  & & & \targ & \ctrl{1} & \targ & \qw_{a} & \targ & \ctrl{1}_{a} & \ctrl{-1} & \qw & \qw & \push{\rule{.3em}{0em}=\rule{.3em}{0em}} & & \targ & \ctrl{1} & \targ & \qw_{b} & \targ & \ctrl{1}_{b} & \qw \\
& \targ & \qw_{3} & & & & \qw & \targ & \ctrl{-1}_{a} & \targ & \ctrl{-1} & \targ & \qw_{a} & \ctrl{-3} & \qw & & & \qw & \targ & \ctrl{-1}_{b} & \targ & \ctrl{-1} & \targ & \qw_{b}
}
\]
\caption{C$_{3}$--adder implemented with approximate Toffoli gates of Fig. \ref{Fig:Toffaprox}. The C$_{3}$--adder that uses the CCNOT$_a$ gates needs extra controlled-Z gates to cancel out the minus signs introduced by the Toffoli gate approximation.}
\label{Fig:C3adder_approx}
\end{figure}
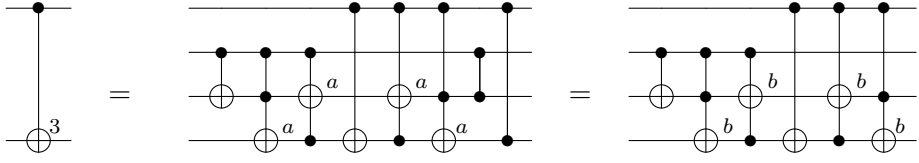

The resulting circuit is shown in Fig. \ref{Fig:C3adder_approx}, where we have used approximate CCNOT gates CCNOT$_{a}$ and CCNOT$_{b}$ described in Fig. \ref{Fig:Toffaprox} of App. \ref{app:quantum_gates}, instead of usual CCNOT gates in order to reduce significantly the circuit depth \cite{Barenco95}. This circuit is divided in two sectors, each one performs the C$_{3}$--adder gate if the controlled qubit is $|\bar{1}\rangle$, the first 3 gates, or $|\bar{2}\rangle$, the last 3 gates. Any of those gates affect the qubit state if the control qutrit is in the $|\bar{0}\rangle$ state. 

Clearly, the C$_{3}$ gate is the responsible for the growth of circuit depth. However, we can implement the first two adders using two CNOT gates each one taking advantage that the target qutrit is in the state $|\bar{0}\rangle$, i.e. qubits are in the state $|00\rangle$.

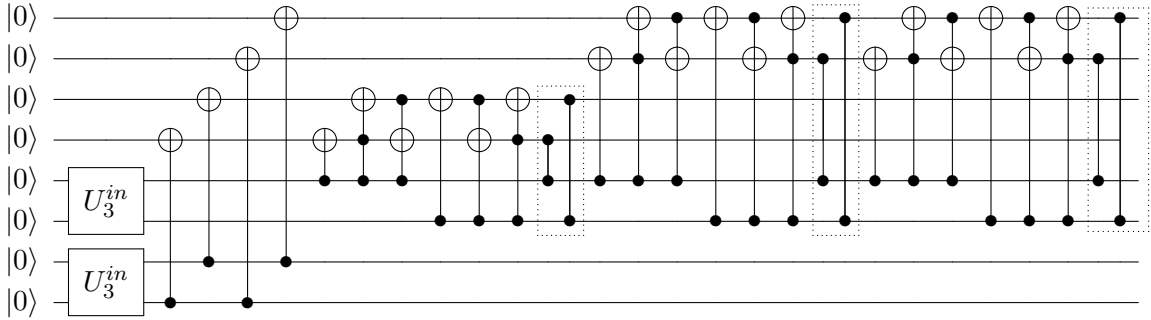
\begin{figure}[t!]
\centering
\[
\Qcircuit @C=0.5em @R=0.5em @!R {
\lstick{\ket{0}} & \qw & \qw & \qw & \qw & \targ & \qw & \qw & \qw & \qw & \qw & \qw & \qw & \qw & \qw & \targ & \ctrl{1} & \targ & \ctrl{1} & \targ & \qw & \ctrl{5} & \qw & \targ &\ctrl{1} & \targ & \ctrl{1} & \targ & \qw & \ctrl{5}& \qw \\
\lstick{\ket{0}} & \qw & \qw & \qw & \targ & \qw & \qw & \qw & \qw & \qw & \qw & \qw & \qw & \qw & \targ & \ctrl{-1} & \targ & \qw & \targ & \ctrl{-1} & \ctrl{2} & \qw & \targ & \ctrl{-1} & \targ & \qw & \targ & \ctrl{-1} & \ctrl{2} & \qw & \qw \\
\lstick{\ket{0}} & \qw & \qw & \targ & \qw & \qw & \qw & \targ & \ctrl{1} & \targ & \ctrl{1} & \targ & \qw & \ctrl{3} & \qw & \qw & \qw & \qw & \qw & \qw & \qw & \qw & \qw & \qw & \qw & \qw & \qw & \qw & \qw & \qw & \qw \\
\lstick{\ket{0}} & \qw & \targ & \qw & \qw & \qw & \targ & \ctrl{-1} & \targ & \qw & \targ & \ctrl{-1} & \ctrl{1} & \qw & \qw & \qw & \qw & \qw & \qw & \qw & \qw & \qw & \qw & \qw & \qw & \qw & \qw & \qw & \qw & \qw \\
\lstick{\ket{0}} & \multigate{1}{U_{3}^{in}} & \qw & \qw & \qw & \qw & \ctrl{-1} & \ctrl{-1} & \ctrl{-1} &  \qw & \qw & \qw & \ctrl{-1} & \qw & \ctrl{-3} & \ctrl{-3} & \ctrl{-3} & \qw & \qw & \qw & \ctrl{-2} & \qw & \ctrl{-3} & \ctrl{-3} & \ctrl{-3} & \qw & \qw & \qw & \ctrl{-2} & \qw & \qw  \\
\lstick{\ket{0}}  & \ghost{U_{3}^{in}} & \qw & \qw & \qw & \qw  & \qw & \qw & \qw & \ctrl{-3} & \ctrl{-2} & \ctrl{-2} & \qw & \ctrl{-3} & \qw & \qw & \qw & \ctrl{-5} & \ctrl{-4} & \ctrl{-4} & \qw & \ctrl{-5} & \qw & \qw & \qw & \ctrl{-5} & \ctrl{-4} & \ctrl{-4} & \qw & \ctrl{-5} & \qw & \\
\lstick{\ket{0}} & \multigate{1}{U_{3}^{in}} & \qw & \ctrl{-4} & \qw & \ctrl{-6} & \qw & \qw & \qw & \qw & \qw & \qw & \qw & \qw & \qw & \qw & \qw & \qw & \qw & \qw & \qw & \qw & \qw & \qw & \qw & \qw & \qw & \qw & \qw & \qw & \qw \\
\lstick{\ket{0}}  & \ghost{U_{3}^{in}} & \ctrl{-4} & \qw & \ctrl{-6} & \qw & \qw & \qw & \qw & \qw & \qw & \qw & \qw & \qw & \qw & \qw & \qw & \qw & \qw & \qw & \qw & \qw & \qw & \qw & \qw & \qw & \qw & \qw & \qw & \qw & \qw
\gategroup{3}{13}{6}{14}{0.7em}{.}
\gategroup{1}{21}{6}{22}{0.7em}{.}
\gategroup{1}{29}{6}{31}{0.7em}{.}
}
\]
\caption{Circuit for the construction of the AME(4,3) state using two qubits to represent one qutrit. The controlled-Z gates (framed with dots), are only necessary if we are using the approximation of Toffoli gate CCNOT$_a$.}
\label{Fig:AMEqubits}
\end{figure}

All together, the final circuit for the construction of the $|\Omega_{4,3}\rangle$ state using two qubits to represent each qutrit is shown in Fig. \ref{Fig:AMEqubits}, where CZ gates are framed because they are only necessary if we are implementing the CCNOT$_a$. 

\section{Entanglement majorization in AME states circuits \label{sec:maj}}

In this section, we analyse how entanglement is created along the circuits that generate AME states. Following the majorization arrow idea \cite{Latorre02}, we check if after each entangling gate applied the entanglement grows or remains equal in all system bipartitions, that is, if at each step of the circuit the eigenstates of the reduced density matrices majorize the eigenstates of the previous step.

\begin{definition}[Majorization]
Given two vectors $\mathbf{a},\mathbf{b}\in \mathbb{R}^{d}$ with their components ordered in decreasing order, namely $\mathbf{a}^{\downarrow}$ and $\mathbf{b}^{\downarrow}$ with $a_{i+1}^{\downarrow}\geq a_{i}^{\downarrow}$ and similarly for $\mathbf{b}^{\downarrow}$, it is said that $\mathbf{a}$ \emph{majorizes} $\mathbf{b}$, i.e. $\mathbf{a}\succ\mathbf{b}$, iff
\begin{equation}
\sum_{i=1}^{k}a_{i}^{\downarrow}\geq\sum_{i=1}^{k}b_{i}^{\downarrow} \quad \mathrm{for} \ k=1,\cdots, d \ ,
\label{eq:majorization}
\end{equation} 
and $\sum_{i=1}^{d}a_{i}=\sum_{i=1}^{d}b_{i}$.
\end{definition} 

At some step $s$ during the computation, the circuit has generated a quantum state with density matrix $\rho_{s}$. We then compute the reduce density matrix of every of its bipartitions in two subsystems, $A$ and $B$, i.e. $\rho_{A}^{s}=\mathrm{Tr}_{B}\rho_{s}$, and diagonalize this matrix to obtain its eigenvalues $\mathbf{\lambda^s}=\{\lambda_{i}^{s}\}$. We will establish that this circuit obeys majorization iff $\mathbf{\lambda^s}\succ\mathbf{\lambda^{s+1}}$, i.e.
\begin{align}
\sum_{i=1}^{k}\left(\lambda_{i}^{\downarrow}\right)^{s} &\geq  \sum_{i=1}^{k}\left(\lambda_{i}^{\downarrow}\right)^{s+1} \quad \mathrm{for} \ k=1,\cdots ,d^{m}-1 \quad \forall A, s \ ,
\end{align}
where $m=n-\floor{n/2}$ is the number of qudits in $A$ bipartition. We do not consider last summation $k=d^m$ because the eigenvalues of a density matrix are normalized to the unity. Since there are $\left(\begin{array}{c}n\\ \floor{n/2} \end{array}\right)$ bipartitions, this analysis leads to a total number of $\left(\begin{array}{c}n\\ \floor{n/2} \end{array}\right)(d^{m}-1)$ inequalities that must be fulfilled.

We can apply less strict tests if we just look at the majorization of other figures of merit to quantify bipartite entanglement, for instance Von Neumann entropy or purity, which in terms of $\lambda_{i}$ are defined as $S=-\sum_{i}\lambda_{i}\log_{d}\lambda_{i}$ and $\gamma=\sum_{i}\lambda_{i}^2$ respectively. Both functions in terms of $\lambda_{i}$ are convex, so we can apply the Karamata's inequality \cite{Kadelburg05} to prove that
\begin{align}
\mathbf{\lambda^s}\succ\mathbf{\lambda^{s+1}} &\Rightarrow  S^{s}\leq S^{s+1} \ , \label{eq:majS}\\
\mathbf{\lambda^s}\succ\mathbf{\lambda^{s+1}} &\Rightarrow  \gamma^{s}\geq \gamma^{s+1} \ . \label{eq:majP}
\end{align}
For details, see App. \ref{app:OddsEnds}. Thus, we can first do one of these less restrictively tests and if the above inequalities are not fulfilled in all steps, then we can conclude that there is no majorization in terms of eigenvalues.

\begin{figure}
\centering     
\includegraphics[width=7cm,height=5cm,keepaspectratio]{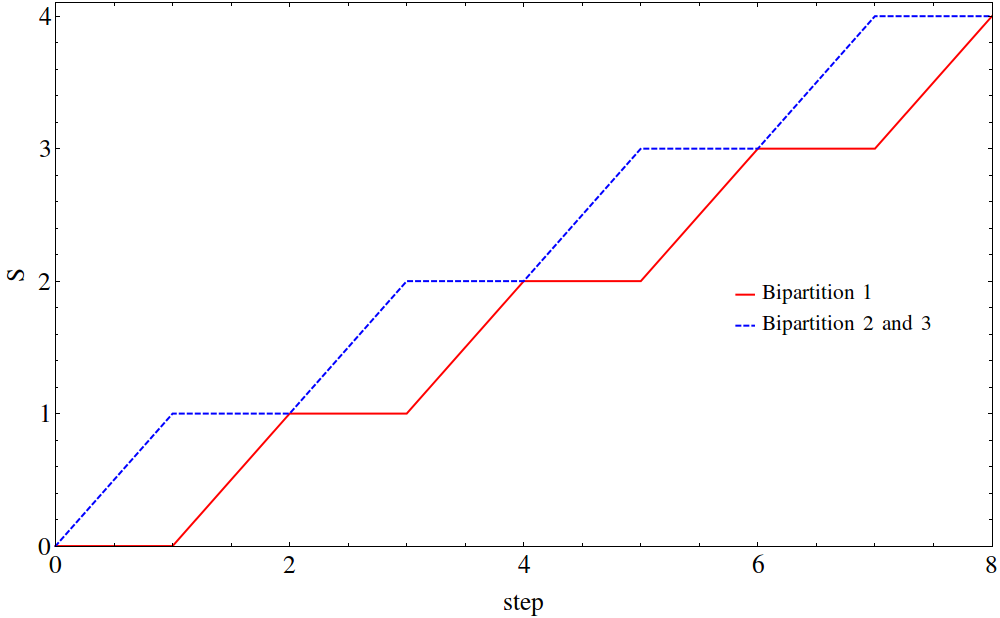}
\includegraphics[width=7cm,height=5cm,keepaspectratio]{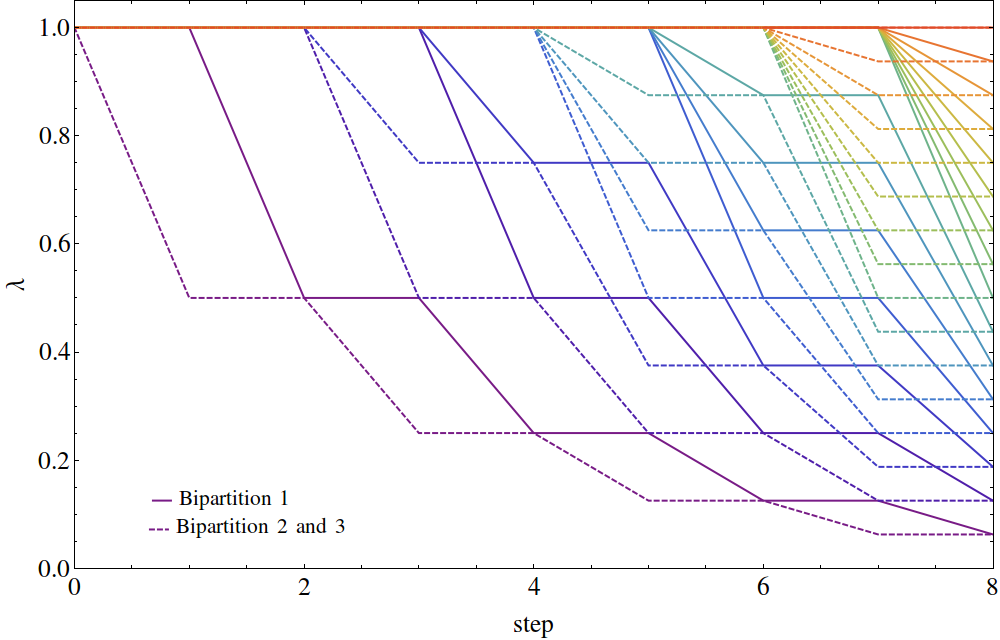}
\caption{Majorization in AME(4,4) state circuit of Fig. \ref{Fig:AME44_circuit}. \textit{Left}: entropy increases at each step $s$ in the three bipartitions until it reach the maximum value $S=2\log_{2}4=4$. \textit{Right}: majorization in terms of eigenvalues $\lambda$ of the reduce density matrices. At the end of the computation, all eigenvalues are the same, which leads to reduce density matrices proportional to the identity. Notice that we only compute three bipartitions of the systems, the ones corresponding to $AB-CD$, $AC-BD$ and $AD-BC$. The other possible bipartitions are not maximally entangled due to the AME(8,2) state do not exist. }
\label{Fig:major}
\end{figure}

As an example, Fig. \ref{Fig:major} shows the majorization of AME(4,4) state of Fig. \ref{Fig:AME44_circuit} in terms of entropy and eigenvalues of the reduce density matrix for each bipartition. The circuit majorizes since entropy never decrease and eigenvalues never increase at each step. At the end of the computation, all bipartition have reached the maximum value of $S=2\log_{2}4=4$ and all eigenvalues are the same, which means that the reduce density matrices are proportional to the identity, as expected for an AME state.

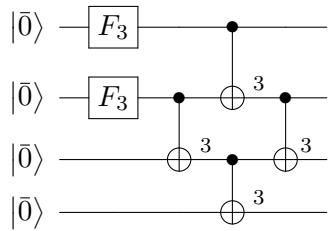
\begin{figure}[t!]
\centering
 \[ \qquad
\Qcircuit @C=1em @R=1em 
{
 \lstick{\ket{\overbar{0}}} & \gate{F_3} & \qw & \ctrl{1} & \qw & \qw \\
 \lstick{\ket{\overbar{0}}} & \gate{F_3} & \ctrl{1} & \targ & \ctrl{1}_{3} & \qw \\
 \lstick{\ket{\overbar{0}}} & \qw & \targ & \ctrl{1}_{3} & \targ & \qw_3 \\
 \lstick{\ket{\overbar{0}}} & \qw & \qw & \targ & \qw_{3} & \qw 
}\]
\caption{Quantum circuit to obtain an AME(4,3) of minimal support.
}
\label{Fig:AME43_optim}
\end{figure}

After analysing the circuit to construct the state $|\Omega_{4,3}\rangle$ written in Fig. \ref{Fig:AME43}, we found that it do not majorize, i.e. when the four C$_3$--adder is applied, the entropy of one of the bipartitions decrease before reach the maximum value after the application of the last C$_3$--adder gate. For that reason, we conclude that this circuit is not optimal and it is possible to obtain an AME(4,3) state of minimal support with less gates. In particular, we found many equivalent circuits that can generate this kind of state with only four C$_3$--adder gates. An example is shown in Fig. \ref{Fig:AME43_optim}. 

Table \ref{Tab:major} summarizes the results on majorization test for all AME circuits written in this work up to $n=6$ and $d=4$. All AME circuits majorize except AME(6,2) and AME(6,4). To check if this is a property of AME states of even dimension, one should compute greater AME states of greater $d$.

\begin{table}
\centering
\begin{tabular}{c|c|c|c}
\toprule
\diagbox[]{$\mathbf{\mathit{n}}$}{$\mathbf{\mathit{d}}$} & \textbf{2} & \textbf{3} & \textbf{4} \\
\midrule
2 & \checkmark & \checkmark & \checkmark  \\
3 & \checkmark & \checkmark & \checkmark   \\
4 & $\cancel{\exists}$ & \checkmark & \checkmark  \\
5 & \checkmark & \checkmark & \checkmark   \\
6 & $\times$ & \checkmark & $\times$   \\
\bottomrule
\end{tabular}
\caption{Majorization of AME states of $n$ parties and dimension $d$. Checkmarks \checkmark  indicate that the corresponding circuits majorize and crosses $\times$ that they do not.}
\label{Tab:major}
\end{table}

\section{Conclusions \label{sec:AMEcon}}

In this chapter, we have presented the explicit circuits to generate AME states with a quantum computer. These circuits has been obtained from graphs since there is a one-to-one correspondence between graph edges and vertices and quantum gates. 

AME states do not exist for any number of parties and any local dimension. For that reason, we have also proposed to simulate AME states of $d>2$ using qubits. To do so, we have mapped each qudit with a multiqubit state, e.g. $|\bar{0}\rangle\rightarrow|00\rangle$, $|\bar{1}\rangle\rightarrow|01\rangle$, $|\bar{2}\rangle\rightarrow|10\rangle$, ... Then, we have deduce the qubit gates that simulate the effect of $F_{d}$ and generalized CZ gates.

The generation of highly entangled states is probably one of the hardest test that can perform a quantum computer. The circuits proposed in this chapter are composed by few one and two-qubit gates, which allows to discard as much as possible errors coming from decoherence times and gate fidelities. In addition, we have shown that almost all circuits majorize the entropy in all their bipartitions, that is, after each entangling gate is applied, the entropy of all bipartitions increases or remains equal. This fact can be understood as the circuit is entangling in the most possible optimal way, using the minimal number of gates to achieve the maximal entropy.

As a final remark, we should mention that we have tested the circuit of Fig. \ref{Fig:AME52} in current quantum computers\footnote{The implementation was done in May 2018 in IBM devices and in December 2018 in Rigetti devices.} (IBM's and Rigetti's quantum devices described in the previous chapter). Actually, we implemented an optimal version of that circuit that adapts to the devices architecture and gate set and reduces the number of Hadamard gates in a way that no extra gates are needed for the experimental implementation. The results were not distinguishable from noise. This fact emphasizes the conclusion that these circuits are indeed a hard test, as even one of the most simple version -- consisting only in five basic entangling gates and three one-qubit gates -- cannot be implemented successfully.


\chapterimage{MaxEnt_image} 

\chapter{Maximal Entanglement in Particle Physics \label{Ch:MaxEnt}}

\vspace{-1.5cm}
\begin{flushright}
\begin{minipage}{0.6\textwidth}
\textit{All things physical are information-theoretic in origin.}
\begin{flushright}
--John A. Wheeler, \\
``Information, Physics, Quantum: the search for links'', 1989.
\end{flushright}
\end{minipage}
\end{flushright}
\vspace{1cm}

In the previous chapters, we have seen different properties and applications that are related to entanglement. Besides the examples given in the introduction chapter, in chapter \ref{Ch:Bell_Ineq}, we have shown that entanglement is necessary to discriminate classical from quantum physics using Bell inequalities. Entanglement also plays a crucial role in condensed matter field, as has been pointed out in chapters \ref{Ch:HDet} and \ref{Ch:Ising}.
 
It is clear that entanglement is at the core of understanding and exploiting quantum physics. It is therefore natural to analyse the generation of entanglement at its most fundamental origin, namely the theories of fundamental interactions in particle physics. If the quantum theory of electromagnetism, QED, would never generate entanglement among electrons, Nature would never display a violation of a Bell inequality. This implies that
entanglement must be generated by quantum unitary evolution at the fundamental level.

A deeper question emerges in the context of high energy physics. Is maximal entanglement (MaxEnt) possible at all? In other words, are the laws of Nature such that MaxEnt
can always be realized? One can imagine a QED-like theory where entanglement could be generated, but in a way that would be insufficient to violate Bell inequalities. Then, it would be formally possible to think of the existence of an underlying theory of hidden variables. On the other hand, if MaxEnt is realized in QED, it is then possible to design experiments that will discard classical physics right at the level of the scattering of elementary particles. Taking a step further, one can ask what are the consequences of imposing that the laws of Nature are able to realize maximally entangled states. Can this requirement be promoted to a principle, and to which extent is this principle consistent with fundamental symmetries such as gauge invariance?

The quest for simple postulates to describe  the fundamental interactions observed in Nature resulted in the common acceptance of the gauge principle, that is, the invariance of the physical laws over internal local rotations for specific symmetry groups. Leaving aside quantum gravity, the Standard Model describes electroweak and strong interactions by means of a Lagrangian which is largely dictated by gauge symmetry requirements. It is natural to pursue further the search for yet an even simpler principle. A possible candidate to formulate a basic underlying axiom for local symmetries is provided by Information Theory. We may recall the words of J. A. Wheeler, {\it ``all things physical are information-theoretic in origin''} that substantiate his philosophy of {\sl ``it from bit''} \cite{Wheeler}. Indeed, it is conceivable that our equations are just a set of operations to implement basic information ideas and protocols.

The exploration of a concrete example of the {\sl ``it from bit''} philosophy based on a maximal entanglement principle is the content of this chapter. We shall show first than in QED only two mechanisms can generate MaxEnt in the high energy scattering of fermions prepared in an initial helicity product state. These are {\it i)} $s$-channel processes where the virtual photon carries equal overlaps of the helicities of the final state particles, and {\it ii)} processes that display interference between $t$ and $u$ channels. We will then illustrate the deep connection between maximal entanglement and the structure of the electron-photon interaction vertex in QED. Indeed, maximal entanglement in most channels is related to the exact form of the QED vertex. As a consequence, imposing that the laws of nature are able to deliver maximal entanglement is tantamount to imposing the QED vertex. We shall finally analyse the consequences of imposing MaxEnt on the weak interaction and discover some surprising constraints on the parameters of the Standard Model. The formalism used is only valid for asymptotically free particles, i.e. leptons and bosons. However, we also explore if color elements factorize from helicity states in QCD processes.

In this chapter, we just focus only on processes with two particles, for which maximal entanglement and maximal entropy are equivalent things. So we shall use MaxEnt to refer indistinctly to both concepts. For systems with more than two particles, the classification of entanglement becomes richer and does not necessarily correspond to the entropy of the subsystem, as has been shown before. The results of this chapters have been published in Ref. \cite{MaxEnt}.

Some previous works have studied the role of entanglement in particle physics. In Ref. \cite{ALP01} it was shown that orthopositronium can decay into 3-photon states that can be used to perform Bell-like experiments that discard classical physics faster than the standard 2-particle Bell inequality. Bell inequalities have also been discussed in kaon physics \cite{Benatti97,Bertlmann01,Bertlmann01bis,Bertlmann06,Bramon07} and its relation with the characterization of $T$-symmetry violation \cite{Bernabeu11} as well as in neutrino oscillations \cite{Banerjee15}. How entanglement varies in an elastic scattering process has been studied using the $S$-matrix formalism in \cite{Peschanski16} and in Deep Inelastic Scattering in \cite{Kharzeev17}. Also, a discussion of quantum correlations in the CMB radiation has been brought to the domain of Bell inequalities \cite{Maldacena15} and the role of entanglement suppression in strong interactions has been recently analyzed in \cite{Martin}.

The outline of this chapter is as follows. We first introduce the figure of merit that we will use to quantify entanglement in scattering processes. Then, in Sec. \ref{sec:QED}, we study how entanglement is generated in QED scattering processes. In Sec. \ref{sec:MaxEnt} we conjecture that MaxEnt could be at the core of fundamental interactions and, in particular, in Sec. \ref{sec:uQED} we investigate to which extent MaxEnt can be used as a constraining principle on the structure of the QED interactions. In Sec. \ref{sec:Weak} we assess some of the implications of MaxEnt for the weak interactions and in Sec. \ref{sec:QCD} we study gluon-gluon scattering process. Finally, the conclusions are exposed in Sec. \ref{sec:summaryMaxEnt}. In addition, App. \ref{app:Feynman} provides the formalism and conventions used and a summary of Feynman rules necessary to compute the scattering process amplitudes. In App. \ref{app:QED} we show a detailed analysis of entanglement generation in all tree-level QED processes.

\vfill

\section{Quantifying two-particle entanglement}

In Chapter \ref{Ch:HDet}, we have introduced several figures of merit to quantify multipartite entanglement. In particular, for two-party entanglement, we defined bipartite entropies, such Von Neumann or R\'enyi entropies, purity and Schmidt rank. We also mentioned some relations between other multipartite figures of merit and \emph{concurrence}. For instance, the $n=2$ tangle corresponds with the square of the concurrence and $\hdet_{2}$ is equivalent to the concurrence for two qubits.

To study the generation of MaxEnt we should choose a figure of merit to quantify bipartite entanglement and in this chapter we decided to use the concurrence \cite{Hill97}. It is equivalent to use Von Neumann entropy but it is easier to compute analytical expressions using this other entanglement monotone.

Starting with a general two qubit state
\begin{equation}
|\psi\rangle=\alpha|00\rangle+\beta|01\rangle+\gamma|10\rangle+\delta|11\rangle,
\end{equation}
with $\alpha,\beta,\gamma,\delta\in\mathbb{C}$ and $|\alpha|^2+|\beta|^2+|\gamma|^2+|\delta|^2=1$, we compute the reduce density matrix $\rho_{A}=\mathrm{Tr}_{B}\rho_{AB}$,
\begin{equation}
\rho_{A}=\left(\begin{array}{cc}
|\alpha|^2+|\beta|^2 & \alpha\gamma^{*}+\beta\delta^{*} \\
\alpha^{*}\gamma+\beta^{*}\delta & |\gamma|^2+|\delta|^2
\end{array}\right).
\end{equation}
A MaxEnt state is the one whose reduced density matrix is proportional to the identity. Thus, computing the eigenvalues of $\rho_{A}$,
\begin{equation}
\lambda_{\pm}=\frac{1}{2}\pm\frac{1}{2}\sqrt{1-4|\alpha\delta-\beta\gamma|^2},
\end{equation}
we can arise which are the constraints on $\alpha,\beta,\gamma$ and $\delta$ coefficients to obtain a MaxEnt state. If the eigenvalues are equal, i.e. $|\alpha\delta-\beta\gamma|=1/2$, then the state $|\psi\rangle$ is maximally entangled. If one of the eigenvalues is zero, i.e. $|\alpha\delta-\beta\gamma|=0$, then the state produced $|\psi\rangle$ is a product state. As one can notice, all information about the entanglement is stored in $|\alpha\delta-\beta\gamma|$ coefficient, which is called concurrence.
\begin{definition}[Concurrence]
Given a two-party pure state $|\psi\rangle=\alpha|00\rangle+\beta|01\rangle+\gamma|10\rangle+\delta|11\rangle$ with $\alpha,\beta,\gamma,\delta\in\mathbb{C}$ and $|\alpha|^2+|\beta|^2+|\gamma|^2+|\delta|^2=1$, the concurrence is defined as
\begin{equation}
\Delta\equiv 2|\alpha\delta-\beta\gamma|.
\end{equation}
Its upper bound $\Delta=1$ corresponds with a MaxEnt state whereas its lower bound $\Delta=0$ with a product state.
\end{definition}

Here we shall study scattering process where the incoming particles are in a product state of their helicities, that is, the incoming particles are not entangled ($\Delta=0$). In general, the outgoing state will be a superposition of all possible helicity combinations, and thus the scattering amplitude will include each possible combination of outgoing helicities. Using the $\mathcal{S}$ matrix, for an initial particles with helicities right ($R$) and left ($L$) respectively, the final state become
\medmuskip=2mu
\begin{align}
|\psi\rangle_{RL} &\equiv \mathcal{S}|\psi\rangle_{i} = \mathcal{S}|RL\rangle \nonumber  \\
& \sim \mathcal{M}_{|RL\rangle\rightarrow|RR\rangle} |RR\rangle + \mathcal{M}_{|RL\rangle\rightarrow|RL\rangle}|RL\rangle+ \mathcal{M}_{|RL\rangle\rightarrow|LR\rangle}|LR\rangle+ \mathcal{M}_{|RL\rangle\rightarrow|LL\rangle}|LL\rangle \nonumber \\
\medmuskip=4mu plus 2mu minus 4mu 
& \equiv \alpha_{RL}|RR\rangle+\beta_{RL} |RL\rangle+\gamma_{RL} |LR\rangle+\delta_{RL} |LL\rangle \ ,
\label{eq:Mmatrix}
\end{align}
\medmuskip=4mu plus 2mu minus 4mu 
since $\mathcal{M}_{|i\rangle\rightarrow|f\rangle}\sim \langle f|\mathcal{S}|i\rangle$ where $|f\rangle$ are all possible final states.
In App. \ref{app:Feynman} we detail how to compute these helicity amplitudes both for fermions and bosons.

The above formalism is only valid if the final states are asymptotically free. Otherwise, we can not apply the $\mathcal{S}$ matrix formulation described in App. \ref{app:Feynman}. This fact limits our analysis to leptons, photons and weak bosons and excludes quarks and gluons. 

We note that in high energy scattering a generic outgoing state will involve all possible outcomes of the process being analysed. The reduction to a two-level system therefore corresponds to a post-selection of results. This is the correct description that delivers the probabilities which we could insert into a Bell inequality, once the final state has been identified.

\section{MaxEnt generation in QED \label{sec:QED}}

\begin{figure}[t]
\centering
\includegraphics[width=0.4\textwidth]{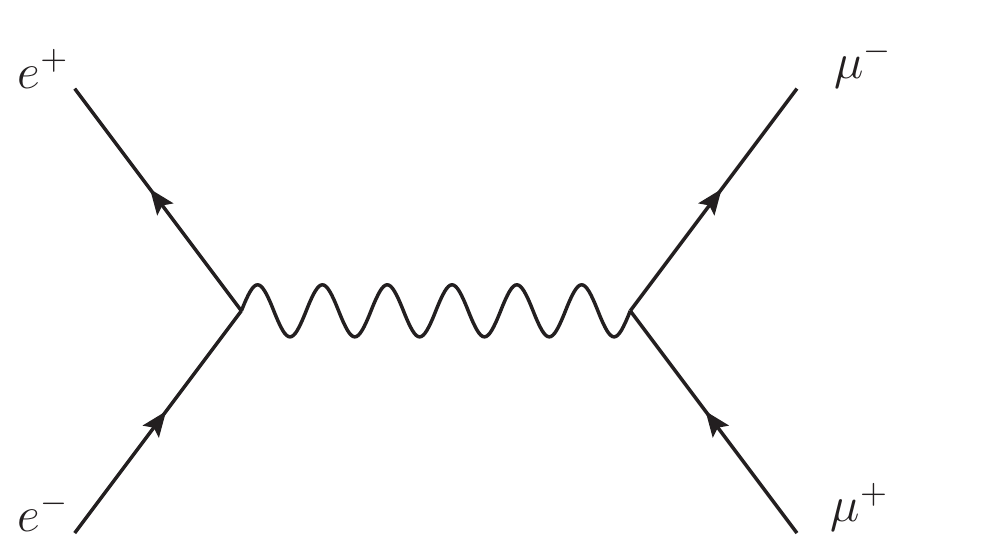}\\
\caption{Feynman diagram for electron-positron scattering into a muon-antimuon pair at tree-level.}
\label{Fig:epem_to_mupmum}
\end{figure}

Let us start our discussion with the analysis of how entanglement is generated in electron-positron annihilation into a muon-antimuon pair, $e^ - e^ + \rightarrow \mu^ - \mu^ +$, described at tree-level in QED by a single $s$-channel diagram as shows Fig. \ref{Fig:epem_to_mupmum}). As in the rest of this chapter, we will work on the center-of-mass frame. In App. \ref{app:Feynman}, we introduce some basic Feynman rules to compute the scattering amplitudes of all processes studied in this chapter.

It is convenient to first focus on the current generated at the interaction vertices. If the incoming particles propagate along the $z$-direction, the incoming current associated to two incoming particles in a $RL$ helicity product state will be 
\begin{equation}
\bar{v}_\uparrow \gamma^ \mu u_\uparrow = 2 p_0 (0, 1,i,0),
\end{equation}
where $p_0$ is the electron's energy. The outgoing particles will then be described by a current as a function of $\theta$, the scattering angle. As shown in App. \ref{app:QED}, we find that at high energies the leading contribution only appears for incoming $RL$ (and $LR$) helicities,
\begin{equation}
|\psi\rangle_{RL}\sim (1+\cos\theta) |RL\rangle +(-1+\cos\theta)|LR\rangle,
\label{eq:e+e-tomu+mu-}
\end{equation}
up to a prefactor which is not relevant here. Therefore, for a scattering angle $\theta=\pi/2$ the final state becomes maximally entangled and proportional to $|RL\rangle-|LR\rangle$, with $\Delta_{RL}=1$. This result illustrates how MaxEnt can be generated in a high energy scattering process. While scattering amplitudes in general carry a non-trivial angular dependence, it is always possible to perform the measurement in the specific direction where MaxEnt is obtained, not unlike the way maximally entangled states are obtained in quantum optics by parametric down conversion. Let us also note that the dominant terms in the $e^ - e^ + \rightarrow \mu^- \mu^+$ scattering  at high energies are easily described by chirality conservation. This is not the case at lower energies, where the emergence of entanglement is more complex. 

For incoming particles in the $RR$ helicity product state, all terms in the amplitude are suppressed by a power of $p_0$ as compared to the $RL$ case. Nevertheless, MaxEnt is found for every angle $\theta$ and incoming momenta $p^0$. An experiment that prepares $RR$ incoming states will therefore always result in MaxEnt.

It is instructive to revisit the computation of the $RL$ case focusing on the currents associated to the virtual photon. The incoming current (in the $z$-direction) corresponds to
\begin{equation}
j^{\mu\,(RL)}_{\mathrm{in}}= 2 p_0 (0,1,i,0),
\end{equation}
and at high energies the non-vanishing outgoing currents at $\theta=\pi/2$ read 
\begin{align}
j^{\mu\,(RL)}_{\mathrm{out}}&= 2 p_0 (0,0,-i,-1), \\
j^{\mu\,(LR)}_{\mathrm{out}}&= 2 p_0 (0,0,i,-1).
\end{align}
Thus the third component of $j_{\mathrm{in}}$ carries equal overlap (with different sign) of the two possible helicity combinations for the outgoing state. In a sense, the photon cannot distinguish between those two options. This is the basic element that leads to MaxEnt generation in $s$-channel processes.

\begin{figure}[t]
\centering
  {	\includegraphics[width=.25\columnwidth]{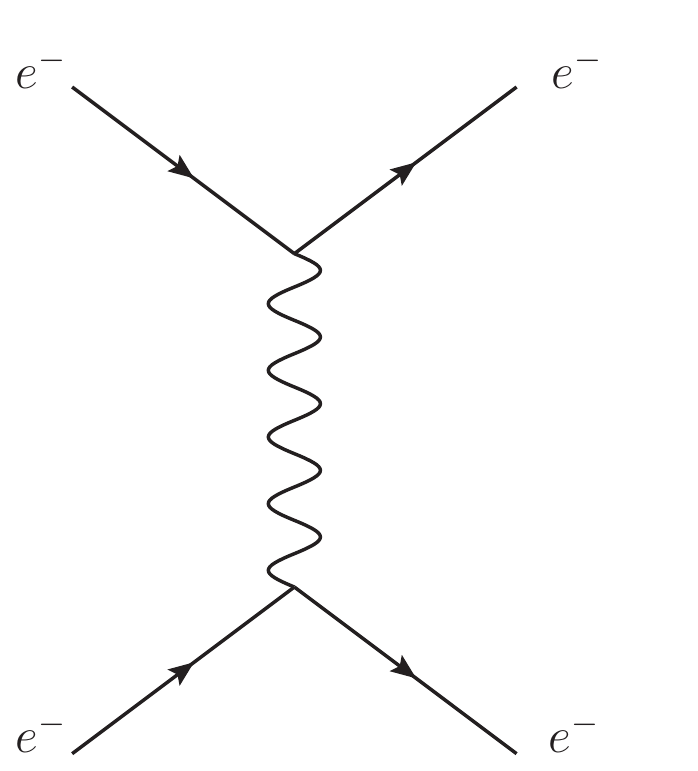}}
  {\includegraphics[width=.25\columnwidth]{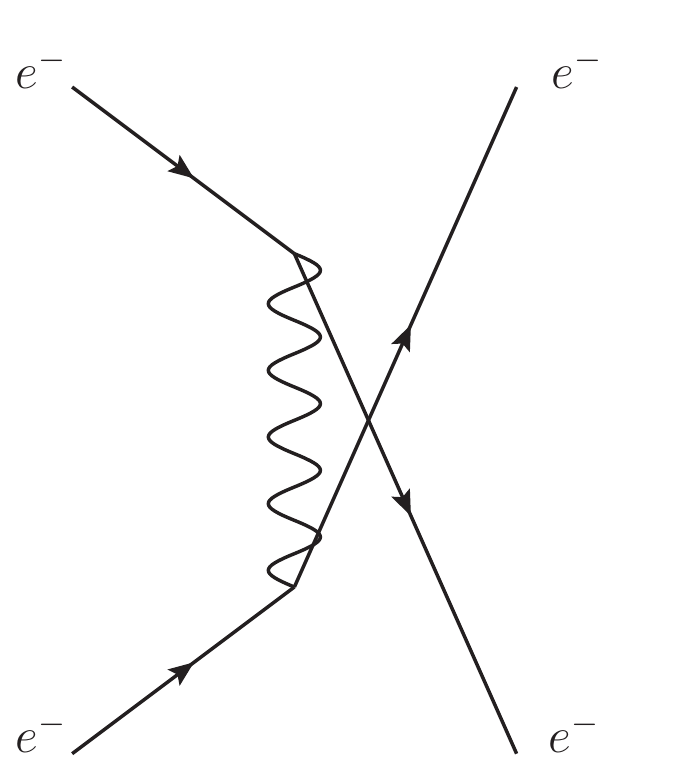}}\\
\vspace{-0.2cm}
\caption{Feynman diagrams for
  M{\o}ller scattering, $e^-e^-\to e^-e^-$,
  in the $t$ (\textit{left}) and $u$ (\textit{right}) channels.
  }
  \label{Fig:moeller}
\end{figure}

Entanglement can also be generated in QED through a completely different mechanism. Let us consider M\o ller scattering, i.e. $e^-e^-\rightarrow e^-e^-$, which receives contributions only from $t$- and $u$-channel diagrams (see Fig.\ref{Fig:moeller}). For this process, the computation of the amplitude shows that no entanglement is generated at high energies within each $t$ or $u$ channel separately, and that the only entangled state is produced by their superposition, resulting in the amplitude
\begin{equation}
|\psi\rangle_{RL} \sim \frac{t}{u}|LR\rangle - \frac{u}{t}|RL\rangle,
\end{equation}
leading to a concurrence
\begin{equation}
\Delta_{RL} \xrightarrow[]{{p_0\to \infty}}
  2 \frac{u^ 2 t^ 2}{u^ 4+t^ 4} \xrightarrow[]{t=u} 1 \ .
\label{eq:Delta_Moller}
\end{equation} 
Therefore, MaxEnt ($\Delta_{RL}=1$) is realized when $t=u$, which corresponds again to  the scattering angle $\theta=\pi/2$. The indistinguishability of $u$ and $t$ histories is now at the heart of entanglement. This also implies that entanglement will not be generated in processes such as $e^- \mu^ - \to e^- \mu^ -$ where the same $u/t$ interference cannot take place. Including electron mass $m_e$ effects, the concurrence $\Delta_{RL}$ reads
\begin{equation}
\Delta_{RL}=\Bigg|\frac{2tu\left(tu+m_{e}^{2}\frac{(t-u)^{2}}{t+u}\right)}{2m_{e}^{2}(t-u)^{2}(2m_{e}^{2}-2(t+u)+\frac{tu}{t+u})+(t^{4}+u^{4})}\Bigg| \, ,
\label{eq:exactDelta_Moller}
\end{equation} 
which shows the more powerful result that, for all energies, the scattering angle $\theta=\pi/2$ (when $t=u$) leads to MaxEnt, $\Delta_{RL}|_{\theta=\pi/2}=  1$ for all $p_0$.

In the case of incoming particles in an $RR$ product state, no entanglement is generated in the high energy limit, since the amplitude is dominated by the final state which also lives in the $RR$ sector, as required by helicity conservation. On the other hand, at very low energies the calculation of the concurrence gives
\begin{equation}
\Delta_ {RR}  \xrightarrow[]{|\vec{p}| \ll m_e, \theta = \pi/2}
 1 + O\left(|\vec p|^ 2/m_e^ 2\right) \, .
\label{eq:Moller_LE}
\end{equation}  
The combination of Eqns.\eqref{eq:Delta_Moller} and \eqref{eq:Moller_LE} illustrates the remarkable fact that two electrons can always get entangled at low energies, irrespectively of their initial helicities. It also justifies that at low energies we easily find entangled fermions and we can describe their interactions with effective models such as the Heisenberg model. Electron-electron interaction is different from all other processes due to the indistinguishability of the particles.

The way in which MaxEnt is generated in QED scattering processes can be studied more thoroughly. It is indeed possible to show that MaxEnt also arises in Bhabha scattering and in pair annihilation of electron-positron to two photons. Table \ref{Tab:Ent} shows the MaxEnt states that can be obtained in all tree-level QED processes, both at high and low energies. All processes, with the exception of electron-muon scattering and Compton scattering (photon-electron scattering) can generate maximally entangled states in some energy limit and at a given scattering angle. In two cases, MaxEnt is generated independently of the scattering angle: pair annihilation into photons, and electron-positron annihilation into muons, in both cases at low energy and for an initial state $|RR\rangle$. It is highly non-trivial that a single coupling, the QED vertex, can take care of generating entanglement in all these processes, and at the same time guarantee that if entanglement is present in the initial state, it will be preserved by the interaction.

\begin{table}[t!]
\centering
\def\arraystretch{1.0}
\begin{tabular}{ c c c c c}
\toprule
 \textbf{Process} & \multicolumn{2}{c}{\textbf{Initial state} $\mathbf{\mathit{|RR\rangle}}$} & \multicolumn{2}{c}{\textbf{Initial state} $\mathbf{\mathit{|RL\rangle}}$} \\
 & \textbf{HE} & \textbf{LE} & \textbf{HE} & \textbf{LE} \\
 \midrule
$e^{-}\mu^{-}\rightarrow e^{-}\mu^{-}$ & -- & -- & -- & -- \\
$e^{-}e^{+}\rightarrow \mu^{-}\mu^{+}$ & -- & ($\cos\theta|\Phi^{-}\rangle - \sin\theta|\Psi^{+}\rangle)_{\forall \theta} $ & $|\Psi^{-}\rangle_{\theta=\pi/2}$ \\
$e^{-}e^{-}\rightarrow e^{-}e^{-}$ & -- & $|\Phi^{-}\rangle_{\theta=\pi/2}$ & $|\Psi^{-}\rangle_{\theta=\pi/2}$ & $|\Psi^-\rangle_{\theta=\pi/2}$ \\
$e^{-}e^{+}\rightarrow e^{-}e^{+}$ & -- & -- & $|\Psi^{+}\rangle_{\theta=\pi/2}$ & -- \\
$e^{-}e^{+}\rightarrow \gamma\gamma$ & -- & $|\Phi^{-}\rangle_{\forall \theta}$ & $|\Psi^{-}\rangle_{\theta=\pi/2}$ & -- \\
\bottomrule
 &\multicolumn{2}{c}{\textbf{Initial state} $\mathbf{\mathit{|R+\rangle}}$} & \multicolumn{2}{c}{\textbf{Initial state} $\mathbf{\mathit{|R-\rangle}}$} \\
 & \textbf{HE} & \textbf{LE} & \textbf{HE} & \textbf{LE} \\
 \midrule
$e^{-}\gamma\rightarrow e^{-}\gamma$ & -- & -- & -- & -- \\
\bottomrule
\end{tabular}
\caption{Maximally entangled states ($\Delta =1$) for tree-level QED processes, both in the high and low energy limits (HE and LE respectively). The states are written in terms of the Bell basis, i.e. $|\Phi^\pm\rangle\sim|RR\rangle\pm|LL\rangle$ and $|\Psi^{\pm}\rangle\sim|RL\rangle\pm|LR\rangle$. For the processes in the upper part of the table, the initial state is expressed in terms of the helicities of the fermions, $R$ and $L$. For Compton scattering, $e^{-}\gamma\rightarrow e^{-}\gamma$, the initial state is expressed in terms of the helicity of the electron and the polarization of the photon, $|+\rangle$ or $|-\rangle$. The scattering angle where the entangled state is produced is indicated in the subscript. A dash indicates that MaxEnt cannot be reached for any value of the scattering angle $\theta$.
}
\label{Tab:Ent}
\end{table}

\section{MaxEnt as a constraining principle \label{sec:MaxEnt}}

It is tantalizing to turn the discussion upside down and attempt to promote MaxEnt to a fundamental principle that constraints particle interactions. Following Wheeler's idea of looking for an Information Theory principle underlying the laws of Nature, we propose to investigate to what extent a MaxEnt principle makes sense in particle physics. Such principle would guarantee the intrinsically quantum character of the laws of Nature, allowing Bell-type experiments to be carried out violating the bounds set by classical physics. In this formulation, MaxEnt emerges as a purely information-theoretical principle that can be applied to a variety of problems.

\begin{conjecture}[MaxEnt principle]
The laws of Nature can generate maximal entanglement in scattering processes of incoming particles which are not entangled. This should happen in as many processes as possible.
\end{conjecture}

We shall, thus, construct a global figure of merit that takes into account many processes at a time. In that direction, we have chosen the concurrence, an entanglement monotone for two particle systems which are the systems that will be analysed in the following sections.

We have already shown the consistency of this principle with tree-level QED interaction. To verify the power of such a principle, we start by leaving unconstrained the coupling in QED and analyse the constraint that the MaxEnt principle dictates on it. Later on, we will perform a similar analysis focusing on the parameters of the weak interaction. We will also start to explore to extend this principle to more sophisticated processes such gluon scattering, as an example of QCD computation.

It may be argued that most interactions generate entanglement. However, it is certainly true that only a limited class of couplings can produce MaxEnt, as it will be shown. It is a natural extremization principle which is at play, as it is the case in other principles applied to describe Nature. Furthermore, MaxEnt carries the added value that physics is forced to be non-classical as Bell inequalities are violated.

\section{Unconstrained QED \label{sec:uQED}}

\subsection{Formalism}

Let's start with the tree-level QED Lagrangian. It describes a free fermion and anti-fermion by using Dirac equation, a free photon, by including Maxwell equations, and a minimal interaction term between photon and fermion. Without imposing gauge invariance, the QED Lagrangian can be written as
\begin{equation}
\mathcal{L}=\bar{\psi}\left(i\gamma^{\mu}\partial_{\mu}-m\right)\psi-\frac{1}{4}F_{\mu\nu}F^{\mu\nu}-eA_{\mu}\bar{\psi}G^{\mu}\psi,
\label{eq:QEDLagrangian}
\end{equation}
where $G^\mu$ are four $4\times4$ matrices that are included to contract them with the photon field $A_{\mu}$. By imposing gauge invariance, one finds that $G^\mu=\gamma^\mu$, as it should be. However, we keep the above general formalism that in principle allows violations of rotation and gauge invariance. While of course this theory is not realized in Nature, our goals are to determine to which extent imposing MaxEnt constrains this interaction vertex and to verify that QED can be reproduced.

The matrices $\gamma^{\mu}$ form the Clifford algebra of the $4\times 4$ matrices; $\{\gamma^{\mu},\gamma^{\nu}\}=2g^{\mu\nu}$ if the metric is $(+,-,-,-)$. The complexification of the Clifford algebra $C\ell_{1,3}(\mathbb{R})$,  $C\ell_{1,3}(\mathbb{R})_{\mathbb{C}}$, is isomorphic to the algebra of $4\times 4$ complex matrices. Then, we can use a real linear combination of the $\gamma^{\mu}$ matrices to express any $4\times 4$ complex matrix,
\begin{equation}
G^{\mu}=c_{1}^{\mu}\mathbb{I}+c_{2}^{\mu\nu}\gamma_{\nu}+ic_{3}^{\mu}\gamma^{5}+ c_{4}^{\mu\nu}\gamma^{5}\gamma_{\nu}+c_{5}^{\mu\nu\rho}\sigma_{\nu\rho},
\label{eq:generalG}
\end{equation}
where $c_{i}\in\mathbb{R}$, $\gamma^{5}=i\gamma^{0}\gamma^{1}\gamma^{2}\gamma^{3}$ and $\sigma^{\nu\rho}=-\frac{i}{2}\left[\gamma^{\nu},\gamma^{\rho}\right]$. Each of the above combination of $\gamma^\mu$ matrices are called Dirac bilinears. 

Since we have $4\cdot 16$ free real values to parametrize these four matrices, we opt to first apply the conservation of discrete symmetries to simplify the analysis. In particular, we impose parity, charge conjugation and time reversal symmetries on the Lagrangian \eqref{eq:QEDLagrangian} (see Tab. \ref{Tab:Sym}). After that, the only term that survives in Eq. \eqref{eq:generalG} is the one proportional to $\gamma_\mu$ plus some constraints on $c_{2}^{\mu\nu}$ parameters. As we only have one combination, we relabelled $c_{2}^{\mu\nu}\equiv a^{\mu\nu}$. The final form of $G^\mu$ matrices become
\begin{align}
G^{0}&=a_{00}\gamma^0, \nonumber \\
G^{i}&=a_{ij}\gamma^{j}.
\end{align}
So, in the end, we will have $9+1$ degrees of freedom in tree-level processes ($a$ parameters and the COM angle).

This change in the interaction term is translated into a change in the Feynman rule for fermion-photon vertex: instead of $-ieQ_{f}\gamma^\mu$ we have $-ieQ_{f}G^{\mu}$. Propagators and free lines are not affected since they are deduced from Dirac and Maxwell equations. The QED vertex is recovered for $a_{00}=a_{11}=a_{22}=a_{33}=1$ and $a_{ij}=0$ for $i\neq j$. This change leads to new expressions for all processes amplitudes. Our aim is to compute again all concurrences for these processes and maximize them respect $a_{\mu\nu}$ and $\theta$ parameters, i.e.
\begin{equation}
\underset{a_{\mu\nu},\theta}{\mathrm{max}}\left(\Delta_{\scriptscriptstyle e^-\mu^-\rightarrow e^-\mu^-},\Delta_{\scriptscriptstyle e^+e^-\rightarrow\mu^+\mu^-},\Delta_{\scriptscriptstyle Bhabha},\Delta_{\scriptscriptstyle Moller},\Delta_{\scriptscriptstyle Pair \ annihilation},\Delta_{\scriptscriptstyle Compton}\right),
\end{equation}
where \emph{Bhabha} scattering is $e^+e^-$ scattering, \emph{Pair annihilation} corresponds with $e^+e^-\rightarrow\gamma\gamma$ process and \emph{Compton} scattering to $e^-\gamma$ scattering.

\begin{table}[t!]
\centering
\begin{tabular}{ccccccc}
\toprule
 & $\mathbf{\mathit{\bar{\psi}\mathbb{I}\psi}}$ & $\mathbf{\mathit{\bar{\psi}\gamma^{\mu}\psi}}$ & $\mathbf{\mathit{\bar{\psi}\gamma^{5}\psi}}$ & $\mathbf{\mathit{\bar{\psi}\gamma^{5}\gamma^{\mu}\psi}}$ & $\mathbf{\mathit{\bar{\psi}\sigma^{\mu\nu}\psi}}$ & $\mathbf{\mathit{A_{\mu}}}$\\
 \midrule
$\mathcal{P}$ & +1 & $(-1)^{\mu}$ & -1 & $-(-1)^{\mu}$ & $(-1)^{\mu}(-1)^{\nu}$ & $(-1)_{\mu}$ \\
$\mathcal{T}$ & +1 & $(-1)^{\mu}$ & -1 & $(-1)^{\mu}$ & $-(-1)^{\mu}(-1)^{\nu}$ & $(-1)_{\mu}$ \\
$\mathcal{C}$ & +1 & -1 & +1 & +1 & -1 & -1 \\
$\mathcal{CPT}$ & +1 & -1 & +1 & -1 & +1 & -1 \\
\bottomrule
\end{tabular}
\caption{Transformation properties of fermion bilinears and photon field under parity, charge conjugation and time reversal symmetries. Here, $(-1)^{\mu}=+1$ if $\mu=0$ and $(-1)^{\mu}=-1$ if $\mu=1,2,3$.}
\label{Tab:Sym}
\end{table}

\subsection{An example: \texorpdfstring{$e^- e^+\rightarrow\mu^-\mu^+$}{}}

Let's compute explicitly the amplitudes of $e^-e^+\rightarrow\mu^-\mu^+$ process using this unconstrained QED interaction. We compute these amplitudes in the high energy limit, i.e. considering $m_{e}=m_{\mu}=0$.

Let us first restrict the particles to be in the $XZ$ plane. For incoming $|RL\rangle$ particles
\begin{align}
\mathcal{M}_{|RL\rangle\rightarrow|RL\rangle}&= -a_{j2}^2 - a_{j1}^2\cos\theta + a_{j1} a_{j3}\sin\theta + i\left(a_{j1} a_{j2} (1-\cos\theta) + a_{j2} a_{j3}\sin\theta\right)\, ,\nonumber \\
\mathcal{M}_{|RL\rangle\rightarrow|LR\rangle}&= a_{j2}^2 - a_{j1}^2\cos\theta + a_{j1} a_{j3}\sin\theta - i\left(a_{j1} a_{j2} (1+\cos\theta) - a_{j2} a_{j3}\sin\theta\right)\, , 
\end{align}
while all other scattering amplitudes vanish.

The two possible final states that maximize the concurrence, that is, that realize MaxEnt
are given by $|RL\rangle\pm|LR\rangle$, and therefore
$\mathcal{M}_{|RL\rangle\rightarrow|RL\rangle}=\pm\mathcal{M}_{|RL\rangle\rightarrow|LR\rangle}$. Requiring a final state that satisfies the maximal entanglement principle
we find that for a scattering angle of $\theta=\pi/2$ the following conditions must be satisfied
\begin{equation}
\begin{array}{rcl}
a_{j2}^2 - i a_{j1}a_{j2} = 0 & \longrightarrow  \ A_{22} = A_{12} = 0 & \mathrm{if} \ \ \mathcal{M}_{|RL\rangle}=\mathcal{M}_{|LR\rangle} \quad \mathrm{or}\\
a_{j1}a_{j3} + i a_{j2}a_{j3} = 0& \longrightarrow \  A_{13}=A_{23} = 0 & \mathrm{if} \ \ \mathcal{M}_{|RL\rangle}=-\mathcal{M}_{|LR\rangle},
\end{array}
\end{equation}
where $A_{kl}\equiv a_{jk}a_{jl}$ is a positive definite matrix. It is also possible to redo the same analysis but now requiring the scattered particles to lie in the $YZ$ and $XY$ planes respectively. In the case of QED, this leads to the same amplitudes because QED vertex preserves rotational symmetry, but this is not necessarily the case with a general vertex of the form $-ieG^\mu$. If the motion of the initial particles takes place in the $Y$ axis and the outgoing scattered particles lie in the $YZ$ plane, the corresponding scattering amplitudes become
\begin{align}
\mathcal{M}_{|RL\rangle\rightarrow|RL\rangle}&= -a_{j1}^2 - a_{j2}^2\cos\theta + a_{j2} a_{j3}\sin\theta - i\left(a_{j1} a_{i2} (1-\cos\theta) + a_{j1} a_{j3}\sin\theta\right),\nonumber \\
\mathcal{M}_{|RL\rangle\rightarrow|LR\rangle}&= a_{j1}^2 - a_{j2}^2\cos\theta + a_{j2} a_{j3}\sin\theta + i\left(a_{j1} a_{j2} (1+\cos\theta) - a_{j1} a_{j3}\sin\theta\right). 
\end{align}
If now we request MaxEnt to be realized at an scattering angle of $\theta=\pi/2$, one finds that
\begin{equation}
\begin{array}{rcl}
a_{j1}^2 + i a_{j1}a_{j2} = 0 & \longrightarrow \ A_{11} = A_{12} = 0 & \mathrm{if}  \ \ \mathcal{M}_{|RL\rangle}=\mathcal{M}_{|LR\rangle} \quad \mathrm{or}\\
a_{j2}a_{j3} - i a_{j1}a_{j3} = 0& \longrightarrow \ A_{23}=A_{13} = 0 & \mathrm{if} \ \ \mathcal{M}_{|RL\rangle}=-\mathcal{M}_{|LR\rangle}.
\end{array}
\end{equation}

Similarly, for incoming particles in $X$ axis and outgoing in the $XY$ plane, the results read instead
\begin{align}
\mathcal{M}_{|RL\rangle\rightarrow|RL\rangle}&= -a_{j3}^2 - a_{j2}^2\cos\phi + a_{j1} a_{j2}\sin\phi + i\left(a_{j2} a_{i3} (1-\cos\phi) + a_{j1} a_{j3}\sin\phi\right),\nonumber \\
\mathcal{M}_{|RL\rangle\rightarrow|LR\rangle}&= -a_{j3}^2 + a_{j2}^2\cos\phi - a_{j1} a_{j2}\sin\phi + i\left(a_{j2} a_{j3} (1+\cos\phi) - a_{j1} a_{j3}\sin\phi\right). 
\end{align}
where $\phi$ is the azimuthal angle that goes from 0 to $2\pi$. Fixing $\phi=\pi/2$ we get another set of conditions
\begin{align}
\begin{array}{rcl}
a_{j1}a_{j2} + i a_{j1}a_{j3} = 0 & \longrightarrow \  A_{12} = A_{13} = 0 & \mathrm{if} \ \ \mathcal{M}_{|RL\rangle}=\mathcal{M}_{|LR\rangle} \quad \mathrm{or}\\
a_{j3}^2 - i a_{j2}a_{j3} = 0& \longrightarrow \ A_{23}=A_{33} = 0 & \mathrm{if} \ \,
\mathcal{M}_{|RL\rangle}=-\mathcal{M}_{|LR\rangle} \, .
\end{array}
\end{align}

A crucial property of  entanglement is that it should be invariant under local unitary transformations like rotations. For this reason, it is possible to obtain the $|RL\rangle + |LR\rangle$ state in one plane and $|RL\rangle -|LR\rangle$ state in another, but both for the same scattering angle because of isometry. Therefore, there are a finite number of possible solutions that satisfy the above constraints: the one which corresponds to QED is the $|RL\rangle - |LR\rangle$ state for $XZ$ and $YZ$ plane and $|RL\rangle + |LR\rangle$ state for $XY$ plane. While in this example we have imposed that MaxEnt is realized for specific choices of the scattering angles $\theta,\phi=\pi/2$, it is conceivable that additional constraints could be obtained by exploiting the information contained in other scattering angles.

From this specific example among the list of processes that we have analysed in unconstrained QED, one can also observe that it is not possible to distinguish the overall sign of the  $a_{ij}$ coefficients, as they always appear squared or multiplied in pairs. Other processes, involving for example a final state with three particles, might be necessary in order to resolve this degeneracy.

Notice also that unconstrained QED allows angular momentum violation. Let us take as example the amplitude for an initial state $|LR\rangle$,
\begin{align}
\mathcal{M}_{|LR\rangle\rightarrow|RL\rangle}&= -a_{j2}^2 + a_{j1}^2\cos\theta - a_{j1} a_{j3}\sin\theta + i\left(-a_{j1} a_{j2} (1+\cos\theta) + a_{j2} a_{j3}\sin\theta\right)\, ,\nonumber \\
\mathcal{M}_{|LR\rangle\rightarrow|LR\rangle}&= a_{j2}^2 + a_{j1}^2\cos\theta - a_{j1} a_{j3}\sin\theta + i\left(a_{j1} a_{j2} (-1+\cos\theta) - a_{j2} a_{j3}\sin\theta\right)\, . 
\end{align}
Therefore, if the initial state is $|\Psi^{-}\rangle=\frac{1}{\sqrt{2}}\left(|RL\rangle -|LR\rangle\right)$, i.e. the singlet state, then the final state $|\psi\rangle_{\rm{final}}$ becomes
\medmuskip=0mu
\begin{align}
  |\psi\rangle_{\Psi^{-}}&{\sim} \mathcal{M}_{|RL\rangle\rightarrow|RL\rangle}|RL\rangle + \mathcal{M}_{|RL\rangle\rightarrow|LR\rangle}|LR\rangle 
  - \left(\mathcal{M}_{|LR\rangle\rightarrow|RL\rangle}|RL\rangle + \mathcal{M}_{|LR\rangle\rightarrow|LR\rangle}|LR\rangle\right) \nonumber\\
  &{\sim} \Bigg(-\sum_{j}a_{j1}^2\cos\theta + \sum_{j}a_{j1}a_{j3}\sin\theta\Bigg)\left(|RL\rangle-|LR\rangle\right) 
  \quad +i\sum_{j}a_{j1}a_{j2}\left(|RL\rangle + |LR\rangle\right),
 \end{align}
 \medmuskip=4mu plus 2mu minus 4mu 
which, in general, is not a singlet state: as long as $\sum_{j}a_{j1}a_{j2}\neq 0$, angular momentum is violated in this process $\forall\theta$.

\subsection{Final solution}

The complete application of the MaxEnt principle to uQED requires the computation of all the scattering amplitudes in the new theory and then the determination of the constraints on the $a_{\mu\nu}$ coefficients from the maximization of the concurrences. Here we have  maximized the sum of the concurrences of four different processes: Bhabha and M\o ller scattering, $ee\to \gamma\gamma$ and $e^- e^+\to \mu^- \mu^+$, accounting for all initial  helicity combinations for product states. The maximization has been performed both over the  $a_{\mu\nu}$ coefficients and over the scattering angle $\theta$. Full consistency is found between the constraints provided by each of the four processes. The solution to the maximization of the concurrence is found to be
\begin{equation}
\left(G^{0},G^{1},G^{2},G^{3}\right)=\left(\pm\gamma^{0},\pm\gamma^{1},\pm\gamma^{2},\pm\gamma^{3}\right) ,
\label{eq:solG}
\end{equation}
where all combination of $\pm$ signs between gamma matrices is a solution. This result shows that QED is indeed a solution, though not the only one, of requiring MaxEnt for the above subset of scattering processes in uQED. Some of these solutions are equivalent to QED since a global sign can be absorbed in the electric charge.

The solutions of Eq. \eqref{eq:solG}) are divided into two groups, those related to QED and those that are inconsistent with QED, for instance because they violate rotation symmetry or do not conserve the current. The latter solutions cannot be ruled out since the scattering processes considered here cannot determine the overall sign of the $\gamma^{\mu}$ matrices, as they always appear in pairs. Including further scattering or decay processes which involve three outgoing particles might remove this ambiguity and eliminate the inconsistent solutions.

Thus, with this analysis we have found $2^{4}$ possible solutions that can generate MaxEnt with a QED-type interaction. We have also checked that these solutions are isolated points in the phase space of the $a_{\mu\nu}$. As an example, let's show that QED is an isolated maximum. For that propose, we will deform the coefficients $a_{\mu\nu}$ as
\begin{align}
a_{ij}&\rightarrow  a_{ij}^{QED}+\epsilon \ \delta_{ij}, \nonumber\\
a_{00}&\rightarrow  a_{00}^{QED}+\epsilon \ \delta_{00}, \nonumber\\
\theta_{MaxEnt}&\rightarrow  \theta_{MaxEnt}^{QED}+\epsilon \ \delta\theta. 
\end{align}
Then, we compute the concurrence and expand them around its maximum, i.e.
\begin{equation}
\Delta=1 +f(\delta_{ij},\delta\theta)\epsilon + g(\delta_{ij},\delta\theta)\epsilon^{2}+\cdots \ .
\end{equation}
If $\Delta$ is an isolated maximum in the phase space of the $\delta_{ij}$, then their derivatives must be zero. After imposing this, we find that the only possible solution is $\delta_{ij}=0$ for $i\neq j$ and $\delta_{00}=\delta_{11}=\delta_{22}=\delta_{33}=\delta$, which is indeed QED.

\section{Entanglement in tree-level weak interactions \label{sec:Weak}}

The mechanism underlying MaxEnt generation in weak interactions is more subtle, due to the interplay between vector and axial currents and between $Z$ and $\gamma$ channels. The coupling of the $Z$ boson to fermions reads
\begin{equation}
i\frac{g}{\cos\theta_{W}}\gamma^{\mu}\left(g^{f}_{V}-g^{f}_{A}\gamma^{5}\right) \, ,
\end{equation}
where the axial and vector couplings are $g^{f}_{A}=T^{f}_{3}/2$ and $g^{f}_{V}= T^{f}_{3}/2 - Q_{f}\sin^2\theta_{W}$, and $\theta_W$ is the \emph{Weinberg mixing angle} or \emph{weak mixing angle}. For electrons and muons, $T_3=-1/2$ and $Q_f=-1$. Beyond tree-level, the weak mixing angle runs with the energy and is scheme dependent. The PDG average \cite{PDG18} in the on-shell scheme is $\sin^{2}\theta_{W} = 0.22343 \pm 0.00007$. Therefore, the vector coupling $|g_V|$ for electrons is smaller than the axial one $|g_A|$ by about one order of magnitude.

In the following subsections, we analyse some tree-level processes involving neutral currents. In particular, we start with the most simple one, the $Z$ boson decay into leptons, followed by weak $e^-e^+\rightarrow \mu^-\mu^+$, which involves only one channel. We complicate this last process introducing the photon channel to evaluate the possible changes in term of entanglement constraints.

\subsection{\texorpdfstring{$Z$}{} decay}

We now analyse the helicity structure of $Z$ boson decay to $e^{-}e^{+}$. As $Z$ is a massive particle, it has three possible polarizations: right- and left-handed circular polarizations, and longitudinal polarization, which we will denote as $|0\rangle$ (see App. \ref{app:Feynman}). As $m_{e}\ll m_Z$ we can neglect the electron mass. The non-vanishing  helicity amplitudes for this decay process are
\begin{align}
\mathcal{M}_{|0\rangle\rightarrow|RL\rangle}&= g_{R}m_Z\sin\theta \, , \nonumber\\
\mathcal{M}_{|0\rangle\rightarrow|LR\rangle}&= g_{L}m_Z\sin\theta \, , \nonumber\\
\mathcal{M}_{|R\rangle\rightarrow|RL\rangle}&= g_{R}m_Z\sqrt{2}\sin^{2}(\theta/2) \, , \nonumber\\
\mathcal{M}_{|R\rangle\rightarrow|LR\rangle}&= -g_{L}m_Z\sqrt{2}\cos^{2}(\theta/2)\, , \\
\mathcal{M}_{|L\rangle\rightarrow|RL\rangle}&= g_{R}m_Z\sqrt{2}\cos^{2}(\theta/2) \, ,\nonumber\\
\mathcal{M}_{|L\rangle\rightarrow|LR\rangle}&= -g_{L}m_Z\sqrt{2}\sin^{2}  (\theta/2) \nonumber \, ,
\end{align}
where we have defined $g_{R}=(g_{V}-g_{A})/2$ and $g_{L}=(g_{V}+g_{A})/2$. 

If the $Z$ boson is longitudinally polarized, the concurrence of the final leptons becomes
\begin{equation}
\Delta_{0}=\frac{2|g_{L}g_{R}|}{g_{L}^{2}+g_{R}^{2}} \, ,
\end{equation}
Then one can see that the leptons pair is maximally entangled provided that $|g_{L}|=|g_{R}|$, i.e. if $g_{A}=0$ or $g_{V}=0$. As $g^{\ell}_{A}=T^{\ell}_{3}/2\neq 0$ the only possible solution is $g^{\ell}_{V}=0$ which leads to 
\begin{equation}
\sin^2\theta_{W} = \frac{T^{\ell}_{3}}{2 Q_{\ell}} = 0.25 \rightarrow \theta_{W}=\frac{\pi}{6} \ .
\end{equation}
This value is remarkably close to the experimental value.

For a $Z$ boson initially polarized with either a right- or left-handed polarization, the concurrence becomes instead
\begin{align}
\Delta_{R}&=\frac{2|g_{L}g_{R}|\sin^{2}\theta}{|2\left(g_{L}^{2}-g_{R}^{2}\right)\cos\theta +\left(g_{L}^{2}+g_{R}^{2}\right)(1+\cos^{2}\theta)|}\, , \\
\Delta_{L}&= \frac{2|g_{L}g_{R}|\sin^{2}\theta}{|2\left(g_{L}^{2}-g_{R}^{2}\right)\cos\theta -\left(g_{L}^{2}+g_{R}^{2}\right)(1+\cos^{2}\theta)|}\, .
\end{align}
As long as $g_{R}/g_{L}=\pm\cot^{2}(\theta/2)$, for an initial right-handed polarization, or $g_{R}/g_{L}=\pm\tan^{2}(\theta/2)$, for an initial left-handed polarization, MaxEnt is achieved. However, if we assume the same relation between $g_{R}/g_{L}$ independently of the initial polarization, then only one solution is possible: $g_{R}/g_{L}=\pm 1$, i.e. the same solution as for longitudinal polarization, $g_V=0$ or equivalently $\theta_{W}=\pi/6$ for leptons. 

We have obtained the same value $\theta_{W}=\pi/6$ for the three possible initial helicities of the $Z$ boson. In the following subsections, we will analyse the consistency of this result with other more complex processes.

\subsection{Weak \texorpdfstring{$e^-e^+\rightarrow \mu^-\mu^+$}{}} 

Let us consider $e^-e^+ \to \mu^- \mu^+$ scattering  mediated by a $Z$ boson in the high energy limit, where $m_Z$ is neglected. The resulting scattering amplitudes are
\begin{align}
|\psi\rangle_{LR} \sim(1+\cos\theta) g_{L}^2|LR\rangle + (-1+\cos\theta) g_L g_R |RL\rangle \, , \\
|\psi\rangle_{RL} \sim(-1+\cos\theta) g_R g_L |LR\rangle + (1+\cos\theta) g_R^2 |RL\rangle \, ,
\end{align}
Notice that the using of right and left handed couplings $g_{R}$ and $g_{L}$ simplifies the
structure of the currents since $j_{\mathrm{in}}^{RL}\sim g_R (0,1,i,0)$ and $j_{\mathrm{in}}^{LR}\sim g_L (0,1,-i,0)$.

The corresponding concurrences for $|\vec p| \gg m_Z$ read
\begin{align}
\Delta_{RL}&\xrightarrow[]{|\vec p| \gg m_Z} \frac{ \sin^2\theta |g_L g_R|}{2(\sin^4(\theta/2) g_L^2 + \cos^4(\theta/2) g_R^2)}, \\
\Delta_{LR}&\xrightarrow[]{|\vec p| \gg m_Z}  
\frac{ \sin^2\theta |g_L g_R|}{2(\cos^4(\theta/2) g_L^2 + \sin^4(\theta/2) g_R^2)}.
\label{eq:concurrenceZ}
\end{align}
Again, by applying the MaxEnt requirement to these concurrences, we can then derive
a constraint between the couplings $g_R$ and $g_L$, and the scattering angle $\theta$. In particular, we find $\cos^2(\theta/2) g_L \pm \sin^2(\theta/2) g_R=0$ for $RL$ concurrence and $\sin^2(\theta/2) g_L \pm \cos^2(\theta/2) g_R=0$ for the $LR$ concurrence. Note that in general concurrence maximization occurs for different values of $\theta$ for each initial state.

\begin{figure}[t]
\centering
\includegraphics[width=0.45\textwidth]{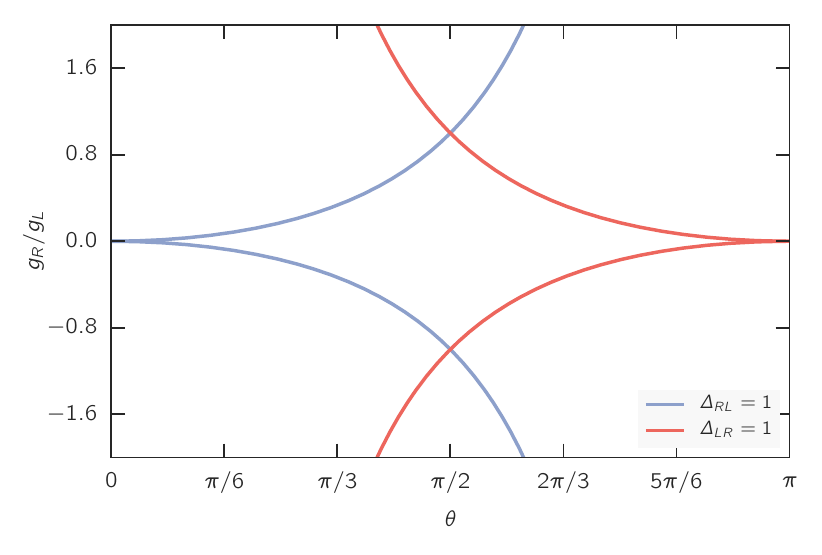}
\includegraphics[width=0.45\columnwidth]{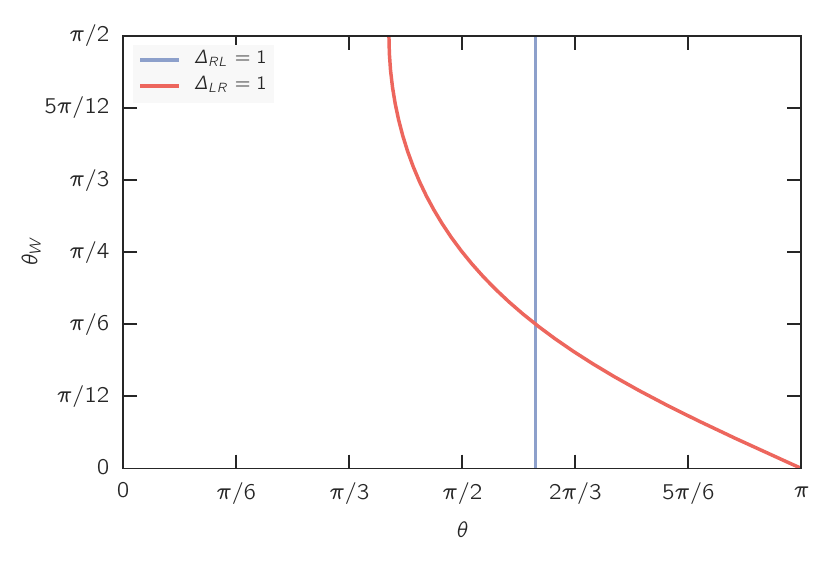}
\caption{\textit{Left}: Maximal concurrence line as a function of the scattering angle $\theta$ and the coupling ratio $g_{R}/g_{L}$ for $Z$-mediated $e^{-}e^{+}\to\mu^{-}\mu^{+}$ scattering. Blue line: electron and positron with right- and left-handed initial helicities respectively; red line: electron and positron with left- and right-handed initial helicities. Maximal entanglement is achieved at the same scattering angle $\theta$ for the two initial helicity configurations when the coupling ratio is equal to one, which leads to a weak mixing angle of $\pi/6$. \textit{Right}: Maximum concurrence line for the weak mixing angle $\theta_{W}$ as a function of scattering angle $\theta$ for the process $e^{-}e^{+}\rightarrow\mu^{-}\mu^{+}$, now including also the effects of $Z/\gamma$ interference. Imposing that MaxEnt is achieved for the same value of the scattering angle $\theta$ fixes $\theta_{W}=\pi/6$.}
\label{Fig:emu_Z}
\end{figure}

Both concurrences are simultaneously maximized for $\theta=\pi/2$, where $g_R=\pm g_L$. In fact, these are the same constraints as the ones found in $Z$ decay. To be more illustrative, we plot these two constraints, coming from initial $RL$ and $LR$ helicities, in Fig. \ref{Fig:emu_Z} left. Therefore, either the axial coupling is zero, recovering the known QED result, or the vector coupling is zero, leading to a weak mixing angle of $\sin^2\theta_W=1/4$. 

%

\subsection{Interference \texorpdfstring{$e^-e^+\rightarrow \mu^-\mu^+$}{}}

Finally, we have studied how the concurrences are modified if we include both the contribution from $\gamma$-exchange and $Z$-exchange in $e^{-}e^{+}\to\mu^{-}\mu^{+}$ scattering. Given that $m_{e},m_{\mu}\ll m_Z$, we can neglect the masses of both leptons. In this process, the amplitudes with equal initial helicities vanish, while the scattering amplitudes for opposite initial helicity configurations are given by
\begin{align}
\mathcal{M}_{|RL\rangle\rightarrow|RL\rangle}&= -\left(\frac{4\mu^{2} g_{R}^{2} }{\left(4\mu^{2}-1\right)}\sec^{2}\theta_{W} +Q^{2}\sin^{2}\theta_{W} \right)\left(1+\cos\theta\right)\, , \nonumber\\
\mathcal{M}_{|RL\rangle\rightarrow|LR\rangle}&= \left(\frac{4\mu^{2} g_{R} g_{L} }{\left(4\mu^{2}-1\right)}\sec^{2}\theta_{W}+Q^{2}\sin^{2}\theta_{W}\right)\left(1-\cos\theta\right)\, ,  
\label{eq:Memuweak}\\
\mathcal{M}_{|LR\rangle\rightarrow|RL\rangle}&= \mathcal{M}_{|RL\rangle\rightarrow|\substack{LR}\rangle} \left(g_{R}\leftrightarrow g_{L}\right)  \, ,\nonumber\\
\mathcal{M}_{|LR\rangle\rightarrow|LR\rangle}&= \mathcal{M}_{|RL\rangle\rightarrow|\substack{RL}\rangle} \left(g_{R}\leftrightarrow g_{L}\right) \,, \nonumber
\end{align}
where we have defined $\mu\equiv|\vec{p}|/m_Z$.

The purely weak scattering process $e^{-}e^{+}\rightarrow\mu^{-}\mu^{+}$, i.e. where the two currents exchange a $Z$ boson instead of a photon like in QED, can be obtained if we set $Q=0$ in the amplitudes of Eq. \eqref{eq:Memuweak}. This is a non-trivial check since the $\gamma$ contribution adds terms to both $RL$ and $LR$, which are independent of $\sin^2\theta_W$. In a way, having MaxEnt in each channel separately does not imply that we will obtain the same constraints for MaxEnt when taking into account both channels.

The introduction of the photon channel complicates the expressions for the concurrences.
They simplify if we express them in terms of $Q$ and $T_{3}$, in which case we find
\begin{align}
\Delta_{RL}&\xrightarrow[]{|\vec p| \gg m_Z}\frac{2Q\left(Q-T_{3}\right)\sin^{2}\theta}{2\left(2Q-T_{3}\right)T_{3}\cos\theta +\left(\left(Q-T_{3}\right)^{2}+Q^{2}\right)\left(1+\cos^{2}\theta\right)},\\
\Delta_{LR}&\xrightarrow[]{|\vec p| \gg m_Z}\frac{Q\left(Q-T_{3}\right)\sin^{2}\theta \ s^{2}_{W} \left(T_{3}^{2}+Q^{2}s^2_{W}-2QT_{3}s^2_{W}\right)}{2Q^{2}\left(Q-T_{3}\right)^{2}s^{4}(s^2_{W})^2+ 2\left(T_{3}^{2}+Q^{2}s^2_{W}-2QT_{3}s^2_{W}\right)^{2}c^{4}},
\end{align}
where $s^2_{W}\equiv\sin^2\theta_{W}$, $c\equiv\cos(\theta/2)$ and $s\equiv\sin(\theta/2)$.
Note that $\Delta_{RL}$ does not depend on the weak mixing angle, but $\Delta_{LR}$ does.  Taking the leptonic electric and weak isospin charges $Q=-1$ and $T_{3}=-1/2$, respectively,
we find that
\begin{align}
\Delta_{RL}^{\ell}&\xrightarrow[]{|\vec p| \gg m_Z}\frac{4\sin^2\theta}{6\cos\theta + 5 (1 +\cos^{2}\theta)}, \\
\Delta_{LR}^{\ell}&\xrightarrow[]{|\vec p| \gg m_Z}\frac{\sin^2\theta\sin^2\theta_W}{\cos^4(\theta/2) + 4\sin^4(\theta/2)\sin^4\theta_W}\, .
\end{align}
Imposing that MaxEnt should be reached for some scattering angles implies that
\begin{align}
\theta\left(\Delta_{RL}^{\ell}=1\right) &= \arccos\left(-\frac{1}{3}\right) \ \forall \ \theta_{W}, \\
\theta_{W}\left(\Delta_{LR}^{\ell}=1\right)&= \arcsin\left(\frac{1}{\sqrt{2}}\cot(\theta/2)\right).
\end{align}
The two curves are shown in the right panel of Fig.\ref{Fig:emu_Z}. If MaxEnt is realized
for the same scattering angle independently of the specific scattering initial state, then the prediction $\theta_{W}=\pi/6$ readily follows, consistently with the result that we find by requesting MaxEnt in the decays of the $Z$ boson into leptons and $e^+e^-\rightarrow\mu^+\mu^-$ scattering mediated by a $Z$ boson.

\subsection{MaxEnt generation in weak interactions}

While the application of MaxEnt to $Z$-boson mediated scattering does not fix completely the coupling structure of the weak interactions, as we mentioned its application to $Z$ decay fixes $g_V=0$ and thus $\sin^2\theta_W=1/4$. The lack of full predictability of MaxEnt in the full scattering case is due to the freedom to choose different angles for MaxEnt depending on the chirality of the initial particles. It is remarkable that we have obtained the same result for $\theta_{W}$ in different processes even in one that involves diagrams with no dependence in $\theta_{W}$. This fact emphasizes the consistency of this result.

The value of $\sin^2\theta_W=1/4$ is in agreement with the experimental value \cite{PDG18} within $\sim 10\%$. There are two possible explanations for the $\sim 10\%$ difference with respect to the experimental value of the weak mixing angle. On the one hand, this analysis has been performed at first order in perturbation theory; the full MaxEnt analysis should be performed taking into account also higher orders, which modify the amplitudes. On the other hand, it is possible that MaxEnt does not fix this parameter, but only gives us a close value, a first intuition. It is however remarkable that requesting MaxEnt simultaneously for the two initial state helicities leads either to QED or to a theory which looks surprisingly close to the weak interaction.

\section{MaxEnt in tree-level QCD: gluon scattering \label{sec:QCD}}

Let's figure out if MaxEnt is present in QCD interaction and if we can extract any information from it. As an example, we compute the amplitudes for tree-level $gg\rightarrow gg$ process. This process involve four Feynman diagrams corresponding to $s$, $t$, $u$ and 4-vertex channels. There exist an approach to compute effortlessly the polarization amplitudes for any gluon scattering process \cite{QFTSM}. This approach takes into account that all gluons are incoming in order to simplify the computations, thus outgoing gluon polarization amplitudes are flipped. However, in the following lines, we present the scattering amplitudes using the same conventions as in the other processes analysed in this chapter, so we do not use the common approach to compute these amplitudes although it is equivalent.

Gluons are not asymptotically free at finite energies. This means that we can only apply the $\mathcal{S}$ formalism if we consider infinite energy. This fact entails that it is not possible to prepare neither to observe gluon polarization states. The aim of the following analysis is to check if there is a factorization between color elements in front of polarization states and analyse if the imposition of MaxEnt leads to some constraint on these elements.

According to Feynman rules for QCD (see App. \ref{app:Feynman}), the non-zero amplitudes of $gg\rightarrow gg$ process are
\begin{equation}
\left(|RR\rangle\rightarrow|RR\rangle\right)_{s} =-ig_{s}^{2}f^{abc}f^{a'b'c}\cos\theta,
\end{equation}
for the $s$ channel,
\begin{align}
\left(|RR\rangle\rightarrow|RR\rangle\right)_{t}&= ig_{s}^{2}f^{aa'c}f^{bb'c}\frac{1}{8}\left(39-24\cos\theta+\cos(2\theta)\right)\cot^{2}(\theta/2),\\
\left(|RL\rangle\rightarrow|RL\rangle\right)_{t}&= ig_{s}^{2}f^{aa'c}f^{bb'c}\frac{1}{2}(3+\cos\theta)\cos^{2}(\theta/2)\cot^{2}(\theta/2),\\
\left(|RL\rangle\rightarrow|LR\rangle\right)_{t}&= ig_{s}^{2}f^{aa'c}f^{bb'c}\frac{1}{2}(3+\cos\theta)\sin^{2}(\theta/2)
\end{align}
for the $t$ channel,
\begin{align}
\left(|RR\rangle\rightarrow|RR\rangle\right)_{u}&= ig_{s}^{2}f^{ab'c}f^{ba'c}\frac{1}{8}\left(39+24\cos\theta+\cos(2\theta)\right)\tan^{2}(\theta/2),\\
\left(|RL\rangle\rightarrow|RL\rangle\right)_{u}&= ig_{s}^{2}f^{ab'c}f^{ba'c}\frac{1}{2}(3-\cos\theta)\cos^{2}(\theta/2),\\
\left(|RL\rangle\rightarrow|LR\rangle\right)_{u}&= -ig_{s}^{2}f^{ab'c}f^{ba'c}\frac{1}{2}(3-\cos\theta)\sin^{2}(\theta/2)\tan^{2}(\theta/2)
\end{align}
for the $u$ channel and
\begin{align}
\left(|RR\rangle\rightarrow|RR\rangle\right)_{4}&= -ig_{s}^{2}\Big(f^{abc}f^{a'b'c}\cos\theta+\nonumber\\
&+f^{aa'c}f^{bb'c}(3-\cos\theta)\cos^{2}(\theta/2)+f^{ab'c}f^{ba'c}(3+\cos\theta)\sin^{2}(\theta/2)\Big),\\
\left(|RL\rangle\rightarrow|RL\rangle\right)_{4}&= ig_{s}^{2}\cos^{4}(\theta/2)\left(f^{ab'c}f^{ba'c}+f^{aa'c}f^{bb'c}\right),\\
\left(|RL\rangle\rightarrow|LR\rangle\right)_{4}&= ig_{s}^{2}\sin^{4}(\theta/2)\left(f^{ab'c}f^{ba'c}+f^{aa'c}f^{bb'c}\right)
\end{align}
for the 4-vertex channel. Similar results are found for $|LL\rangle$ and $|LR\rangle$ initial states.


Thus, the total amplitudes become
\begin{align}
|RR\rangle\rightarrow|RR\rangle= 2ig_{s}^{2}\Big[f^{abc}f^{a'b'c}\frac{-t+u}{s}&+f^{aa'c}f^{bb'c}\left(\frac{-t+u}{s}+2\right)\frac{u}{t} \nonumber\\
&- f^{ab'c}f^{ba'c}\left(\frac{-t+u}{s}-2\right)\frac{t}{u}\Big],
\end{align}
\begin{align}
|RL\rangle\rightarrow|RL\rangle&= -2ig_{s}^{2}\left(f^{ab'c}f^{ba'c}\frac{t}{s}+f^{aa'c}f^{bb'c}\frac{u^2}{st}\right),\\
|RL\rangle\rightarrow|LR\rangle&= -2ig_{s}^{2}\left(f^{ab'c}f^{ba'c}\frac{t^2}{su}+f^{aa'c}f^{bb'c}\frac{t}{s}\right).
\label{eq:gluons_ex}
\end{align}
Then, clearly only an initial polarization state $|RL\rangle$ (or $|LR\rangle$) can generate entanglement,
\begin{equation}
|\psi\rangle_{RL}=\frac{1}{\mathcal{N}}\left[\left(f^{ab'c}f^{ba'c}\frac{t}{s}+f^{aa'c}f^{bb'c}\frac{u^2}{st}\right)|RL\rangle+ \left(f^{ab'c}f^{ba'c}\frac{t^2}{su}+f^{aa'c}f^{bb'c}\frac{t}{s}\right)|LR\rangle\right].
\end{equation}
For $t=u$, i.e. $\theta=\pi/2$, both $|RL\rangle\rightarrow|RL\rangle$ and $|RL\rangle\rightarrow|LR\rangle$ amplitudes are equal independently of structure constants. Then, MaxEnt assumption is consistent in this process although no further knowledge about structure constants can be extracted from it.

\section{Conclusions \label{sec:summaryMaxEnt}}

In this chapter, we have explored the relationship between the generation of maximally entangled states and the tree-level scattering amplitudes in QED, weak and QCD interactions. In particular, we have analysed all tree-level QED processes, $Z$ decay into leptons, weak $e^+e^-\rightarrow\mu^+\mu^-$ (including the photon interference) and $gg\rightarrow gg$ tree-level scattering. 

We found that MaxEnt is generated through two mechanisms: indistinguishability of the particles involved in the process and $s$-channel processes, where the virtual photon carries equal overlaps of the helicities of the final state particles. In addition, we found that promoting MaxEnt to a fundamental principle in the spirit of Wheeler's ``it from bit'' philosophy allows one to constrain the coupling structure describing the interactions between fermions and gauge bosons. As a matter of fact, QED couplings are found to be the solution to a MaxEnt principle once some global symmetries (C, P and T) are imposed.

Following this path, we also found that MaxEnt in weak interactions prefers a weak angle $\theta_W=\pi/6$, surprisingly close to the Standard Model value. We computed different processes to test the consistency of this result. Moreover, we checked if we can extract any information from structure constants in gluon scattering by imposing MaxEnt: the results were consistent with MaxEnt conjecture but at the same time where independent from structure constants.

In this framework, MaxEnt arises as a possible powerful information principle that can be applied to different processes, bringing in unexpected constraints on the structure of high energy interactions. To mention a few possibilities, MaxEnt may provide new insights into the all-order structure of the QED vertex, and may hint at further relations between the parameters of the Standard Model or in new physics beyond it.


\chapterimage{Book_image} 

\chapter{Conclusions and Outlook\label{Ch:Conclusions}}

\vspace{-1.5cm}
\begin{flushright}
\begin{minipage}{0.6\textwidth}
\textit{Science makes people reach selflessly for truth and objectivity; it teaches people to accept reality, with wonder and admiration, not to mention the deep joy and awe that the natural order of things brings to the true scientist.}
\begin{flushright}
-- Lise Meitner, \\
Lecture, Austrian UNESCO Commission, 1953.
\end{flushright}
\end{minipage}
\end{flushright}
\vspace{1cm}

In this thesis, we have covered several topics related to maximal entanglement: its quantification, generation and applications to quantum information and particle physics. 

In Chapter \ref{Ch:Bell_Ineq}, we have first reviewed multiparty qubit Bell inequalities from an operational perspective. We have used this approach to obtain new Bell inequalities for three outcomes. In contrast to qubit Bell inequalities, qutrit inequalities are maximally violated by a deformation of the GHZ state, which shows that maximal entanglement and non-locality are not equivalent concepts. We have extended this analysis to Bell inequalities of higher dimension and we have obtained similar results in terms of maximal violation as for the qutrit case. Moreover, we have presented a new method to obtain novel Bell inequalities from maximally entangled states.

Chapter \ref{Ch:HDet} analyses a particular figure of merit to quantify four-partite entanglement: the hyperdeterminant. For completeness, we have defined other figures of merit, among them, two polynomial invariants that are related to the hyperdeterminant. To be precise, we have used the Schl\"afli hyperdeterminant computed from the $S$ and $T$ polynomial invariants. We have studied the value of this figure in some well-known quantum states such as GHZ and W states, to conclude that hyperdeterminant only captures some types of multipartite entanglement. We have also analysed the multipartite entanglement in random states and ground states of random Hamiltonians and we have observed that hyperdeterminant is sensible to different random priors. The ground state of Ising model, $XXZ$ model and Haldane-Shastry wave function have been also analysed. Hyperdeterminant peaks pronouncedly around the phase transition point in the Ising model but, in contrast, is always zero in the $XXZ$ and Haldane-Shastry models. For that reason, we have used the other invariants, $S$ and $T$, to study these other models. We found that at transition points, $S$ and $T$ invariants become zero whereas in the critical region they have a non-vanishing value.

In Chapters \ref{Ch:Ising} and \ref{Ch:AME} we have moved the discussion to the field of quantum computation. First, in Chapter \ref{Ch:Ising}, we have performed an exact simulation of the Ising model by proposing an explicit circuit that is able to diagonalize the $XY$ Hamiltonian. The implementation of this kind of circuit allows us to have access to all eigenstates of the model by just preparing product states. This fact makes possible to implement time simulation or thermal states. We have tested the circuit for the particular case of a $n=4$ Ising spin chain in some current quantum computers: two from IBM company and one from Rigetti Computing company. The results have shown that, for low external magnetization, the simulation is close to the exact result whereas for higher magnetizations the error respect to the theoretical value is larger. The explanation lies in the structure of the ground state, where bit-flip errors provoke a large decrease of the magnetization when the external field is high. Despite the fact that decoherence and relaxation times are technically large enough to perform the circuit, we have concluded that the big error source came probably from gate fidelities and other error sources such as qubit crosstalk. However, with the simulation of time evolution, we have observed that these error sources should be systematic.

We continued the study of maximal entanglement in quantum computation in Chapter \ref{Ch:AME}. There, we proposed a hard test to be accomplished by a quantum computer: the generation of Absolutely Maximally Entangled (AME) states. Since entanglement plays a key role in quantum computing and the simulation of slightly entangled states can be done with classical methods, a quantum device must be able to generate and hold highly entangled states in order to show some advantage. We have introduced explicit circuits to generate these states that only require Hadamard and CZ gates. These circuits have been obtained from graph states. In addition, we have proposed to simulate AME states of higher dimension using qubits. We have observed that AME states circuits obey in general a majorization arrow, i.e. after each entangling gate is applied, the entropy of all bipartitions increase or remains equal respect to the previous step. We have used this property to obtain equivalent circuits that contained a smaller number of entangling gates.

Finally, in Chapter \ref{Ch:MaxEnt}, we have analysed how maximal entanglement is generated at the most fundamental level, in particular in QED, weak neutral interactions and a tree-level process in QCD (gluon scattering). In the case of QED, we have shown that maximally entangled states can be generated at tree-level and that the imposition of a maximal entanglement conjecture leads to the correct structure of the tree-level vertex without imposing gauge invariance. Next, we have also imposed maximal entanglement in the final state helicities of particles after a weak neutral process, e.g. $Z$ boson decay to leptons. The result entails that the weak mixing angle should have a value of $\pi/6$, very close to the experimental value. Finally, we have tried to extend this analysis to a QCD process, in particular, gluon scattering. However, we found that gluons can be maximally entangled in terms of helicities independently of the values of the structure constants, so no further knowledge can be obtained from maximal entanglement imposition.

Many research lines can be extracted from the work done in this thesis. Let us mention some examples. Bell inequalities are a necessary experiment to elucidate if we are in the quantum mechanics realm or not: the search for these inequalities for any number of parties and dimensions is still an open problem and it is possible to use one of the methods presented in this work to obtain new Bell inequalities. As has been already mentioned, the complete quantification of multipartite entanglement is also an open problem in quantum information. Different figures of merit quantify different kinds of entanglement, so it is necessary to study all of them and in which fields are useful, for example, to capture quantum phase transitions. In this thesis we have analyzed one figure and three models, many others can be tested. From the quantum computing side, it is mandatory to propose methods to benchmark the quantum devices that are being constructed nowadays. We have proposed two methods that can be used to understand better what are the error sources.  In closing, we have proposed a maximal entanglement conjecture. To raise it to have the status of a principle, many other tests are necessary, starting with the computation of higher order corrections and the extension to other particle processes.

With all these conclusions exposed, it is even clearer that entanglement is at the central core of Nature interactions and have important implications in quantum information. Some works presented here will continue their development after this thesis project.


\begin{appendix}

\part*{Appendices}

\renewcommand\chaptername{Appendix}

\chapterimage{QCircuit} 

\chapter{Quantum Gates\label{app:quantum_gates}}

\vspace{-1.5cm}
\begin{flushright}
\begin{minipage}{0.6\textwidth}
\textit{A classical computation is like a solo voice -- one line of pure tones succeeding each other. A quantum computation is like a symphony -- many lines of tones interfering with one another.}
\begin{flushright}
--Seth Lloyd, \\
``Programming the Universe: A Quantum Computer Scientist Takes on the Cosmos'', 2006.
\end{flushright}
\end{minipage}
\end{flushright}
\vspace{1cm}

\emph{Quantum operations}, \emph{quantum logic gates} or simply \emph{quantum gates} are actions applied on specific quantum states that modify the total quantum system. They are physically implemented by Hamiltonians that depend on the quantum device, e.g. superconducting circuits, trapped ions, photons, etc. For pure quantum states, they are commonly represented by unitary matrices that act on a small number of \emph{qudits}, the minimal units of quantum information, i.e. $d$-level quantum states. Then, quantum gates are $d^n\times d^n$ unitary matrices acting on $n$ qudits of \emph{local dimension} $d$.

To simplify the notation, each quantum gate is represented by a specific symbol, usually a box, that gird the qudits over which it acts. At the same time, each qudit is represented by a straight line. All together, they constitute a \emph{quantum circuit} which representatives are the qudits and its building blocks the quantum gates. 

It is important to point out that in this work quantum circuit diagrams are read from left to right, i.e. unitary operations are applied on qudits as long as they appear in the diagram. This contrasts with the matrix notation, which is read from right to left following the matrix multiplication rules.

\section{Basis convention}

Present quantum computation is leading by two-level quantum systems. The $d=2$ quantum information units are called \emph{qubits} and quantum gates are $2^n\times 2^n$ unitary matrices. A (pure) qubit state is usually represented by 
\begin{equation}
|\psi\rangle=\alpha|0\rangle + \beta |1\rangle,
\end{equation}
where $\alpha,\beta\in \mathbb{C}$, $|\alpha|^2+|\beta|^2=1$ and $|0\rangle$ and $|1\rangle$ are the labels of the two quantum levels. In total, a qubit has 2 degrees of freedom (in general, $n$ qubits have $2\cdot 2^n$ parameters and two constraints, $|\langle\psi|\psi\rangle|^2=1$ and a global phase). For that reason, it can be represented as a point on a sphere of unit radius called \emph{Bloch sphere}, i.e. 
\begin{equation}
|\psi\rangle=\cos(\theta/2)|0\rangle + e^{i\phi}\sin(\theta/2)|1\rangle.
\end{equation}
Orthogonal points in the Bloch sphere correspond to eigenstates of $\sx$, $\sy$ and $\sz$ matrices. In particular, the convention establishes that 
\begin{definition}
\begin{equation}
|0\rangle \equiv (0,0,1)_{\mathrm{Bloch}} \equiv
|\psi_{z}^{+}\rangle = \left(\begin{matrix}1 \\0 \end{matrix}\right), \quad
|1\rangle \equiv (0,0,-1)_{\mathrm{Bloch}} \equiv |\psi_{z}^{-}\rangle = \left(\begin{matrix}0 \\1 \end{matrix}\right).
\end{equation}
\end{definition}
Then, the other orthogonal points in terms of $\sz$ basis states are
\begin{align}
(\pm 1,0,0)_{\mathrm{Bloch}} &=
|\psi_{x}^{\pm}\rangle = \frac{1}{\sqrt{2}}\left(|0\rangle\pm|1\rangle\right) = \frac{1}{\sqrt{2}}\left(\begin{matrix}1 \\ \pm 1 \end{matrix}\right), \\
(0,\pm 1,0)_{\mathrm{Bloch}} & = |\psi_{y}^{\pm}\rangle = \frac{1}{\sqrt{2}}\left(|0\rangle\pm i|1\rangle\right) =\frac{1}{\sqrt{2}}\left(\begin{matrix}1 \\ \pm i \end{matrix}\right). 
\end{align}

Experimentally, it is usually chosen the $|0\rangle$ and $|1\rangle$ states as the natural two quantum levels of the physical systems, e.g. electron spin up and down, left and right photon polarization, etc. If, for some reason, we are interested in measuring in other directions, we should implement the corresponding rotation on the final state, which is nothing more than a rotation on the Bloch sphere of each qubit. In fact, any one-qubit unitary gate is actually a rotation in the Bloch sphere.

For other local dimensions, it is used a similar convention: quantum levels are labelled and written with the vectors $|0\rangle = \left(1,0,\cdots,0\right)$, $|1\rangle=\left(0,1,\cdots,0\right)$, $\ldots$, $|d\rangle=\left(0,0,\cdots,1\right)$.

\section{One-qubit gates}

The most used one-qubit gates are shown in Tab. \ref{Tab:1qgates}. This is not an independent gate set, since these gates can be obtained from the others. For instance, the following relations hold
\begin{align}
\mathbb{I}&=X^2=Y^2=Z^2, \\
H&=\frac{1}{\sqrt{2}}\left(X+Z\right), \\
X&=HZH,\\
S&=T^2, \\
Y&= R_z(\pi/2)R_y(2\pi)R_z(\pi/2)X, \\
Z&= R_y(\pi)X, \\
R_i(\theta_1+\theta_2)&=R_i(\theta_1)R_i(\theta_2) \ \mathrm{for} \ i=x,y,z, \label{eq:R1}\\
R_i(-\theta)&=XR_{i}(\theta)X, \ \mathrm{for} \ i=y,z, \label{eq:R2}\\
R_x(-\theta)&=YR_{x}(\theta)Y.
\end{align}

Any unitary $2\times 2$ matrix $U$ with $\det(U)=e^{i\delta}$ can be parametrized as
\begin{equation}
U=\left(\begin{matrix}
e^{i\left(\delta+\frac{\alpha+\beta}{2}\right)}\cos(\theta/2) & e^{i\left(\delta+\frac{\alpha-\beta}{2}\right)}\sin(\theta/2)\\
-e^{i\left(\delta-\frac{\alpha-\beta}{2}\right)}\sin(\theta/2) & e^{i\left(\delta-\frac{\alpha+\beta}{2}\right)}\cos(\theta/2)
\end{matrix}\right)= e^{i\delta}R_{z}(\alpha)R_{y}(\theta)R_{z}(\beta).
\label{eq:U}
\end{equation}
We can then construct any unitary gate using only rotational gates and a phase, for example
\begin{align}
X&=e^{i\frac{\pi}{2}}R_{y}(\pi)R_{z}(\pi), \\
Y&=e^{i\frac{\pi}{2}}R_{y}(-\pi), \\
Z&=e^{i\frac{\pi}{2}}R_{y}(-\pi),\\
H&=e^{i\frac{\pi}{2}}R_{y}(-\pi)R_{z}(-\pi/2),\\
S&=e^{i\frac{\pi}{4}}R_{y}(-\pi/2),\\
R_x(\theta)&=R_z(\pi/2)R_y(\theta)R_z(-\pi/2).
\end{align}

Another common way to parametrize one-qubit unitary gates is
\begin{equation}
U_{3}(\theta,\phi,\lambda)=\left(\begin{matrix}\cos(\theta/2) & -e^{i\lambda}\sin(\theta/2) \\ e^{i\phi}\sin(\theta/2) & e^{i(\lambda+\phi)}\cos(\theta/2)\end{matrix}\right)= Ph(\phi+\lambda)R_z(\lambda+\pi)R_y(\theta)R_z(-\lambda-\pi),
\end{equation}
and similarly as the other parametrization, it can generate any unitary operation, e.g. $H=U_{3}(\pi/2,0,\pi)$ or $S=U_3(0,0,\pi/2)$.

Before moving to two-qubit unitary gates, the following lemma is also useful \cite{Barenco95}:
\begin{lemma}
Any unitary matrix $U$ can be written as 
\begin{equation}
U=e^{i\delta}AXBXC,
\label{lem:unitary}
\end{equation} 
where $X$ is the Pauli matrix $\sx$ and $ABC=\mathbb{I}$.
\end{lemma}

Since $U=e^{i\delta}R_{z}(\alpha)R_{y}(\theta)R_{z}(\beta)$ and using the relations \eqref{eq:R1} and \eqref{eq:R2}:
\begin{align}
R_{z}(\alpha)R_{y}(\theta)R_{z}(\beta)&= R_{z}(\alpha)R_{y}\left(\frac{\theta}{2}\right)XXR_{y}\left(\frac{\theta}{2}\right)XXR_{z}\left(\frac{\alpha+\beta}{2}\right)XXR_{z}\left(\frac{\beta-\alpha}{2}\right) \nonumber\\
&=R_{z}(\alpha)R_{y}\left(\frac{\theta}{2}\right)XR_{y}\left(-\frac{\theta}{2}\right)R_{z}\left(-\frac{\alpha+\beta}{2}\right)XR_{z}\left(\frac{\beta-\alpha}{2}\right) \nonumber\\
&=AXBXC,
\end{align}
where $A=R_{z}(\alpha)R_{y}\left(\frac{\theta}{2}\right)$, $B=R_{y}\left(-\frac{\theta}{2}\right)R_{z}\left(-\frac{\alpha+\beta}{2}\right)$ and $C=R_{z}\left(\frac{\beta-\alpha}{2}\right)$.

\begin{table}[t!]
\centering
\setlength{\defaultaddspace}{10pt}
\addtolength{\tabcolsep}{5pt} 
\begin{tabular}{ccc}
\toprule
\textbf{Name} & \textbf{Symbol} & \textbf{Matrix} \\
\midrule
 \addlinespace
 & \Qcircuit @C=1em @R=1em @!R {&\gate{X}&\qw} & $\left(\begin{matrix}0&1\\1&0\end{matrix}\right)$\\
 \addlinespace
Pauli & \Qcircuit @C=1em @R=1em @!R {&\gate{Y}&\qw} & $\left(\begin{matrix}0&-i\\i&0\end{matrix}\right)$ \\ \addlinespace
 & \Qcircuit @C=1em @R=1em @!R {&\gate{Z}&\qw} & $\left(\begin{matrix}1&0\\0&-1\end{matrix}\right)$ \\
 \addlinespace
Hadamard & \Qcircuit @C=1em @R=1em @!R {&\gate{H}&\qw} & $\frac{1}{\sqrt{2}}\left(\begin{matrix}1&1\\1&-1\end{matrix}\right)$\\ \addlinespace
 & \Qcircuit @C=1em @R=1em @!R {&\gate{R_{x}(\theta)}&\qw} & $e^{i\frac{\theta}{2}X}=\left(\begin{matrix}\cos(\theta/2)&i\sin(\theta/2)\\i\sin(\theta/2)&\cos(\theta/2)\end{matrix}\right)$ \\ \addlinespace
Rotational & \Qcircuit @C=1em @R=1em @!R {&\gate{R_{y}(\theta)}&\qw} & $e^{i\frac{\theta}{2}Y}=\left(\begin{matrix}\cos(\theta/2)&\sin(\theta/2)\\-\sin(\theta/2)&\cos(\theta/2)\end{matrix}\right)$ \\
\addlinespace
 & \Qcircuit @C=1em @R=1em @!R {&\gate{R_{z}(\theta)}&\qw} & $e^{i\frac{\theta}{2}Z}=\left(\begin{matrix}e^{i\theta/2}&0\\0&e^{-i\theta/2}\end{matrix}\right)$ \\
 \addlinespace
  & \Qcircuit @C=1em @R=1em @!R {&\gate{Ph(\phi)}&\qw} & $\left(\begin{matrix}1&0\\0&e^{i\phi}\end{matrix}\right)$ \\ \addlinespace
  Phase & \Qcircuit @C=1em @R=1em @!R {&\gate{S}&\qw} & $Ph(\pi/2)=\left(\begin{matrix}1&0\\0&i\end{matrix}\right)$ \\ \addlinespace
   & \Qcircuit @C=1em @R=1em @!R {&\gate{T}&\qw} & $Ph(\pi/4)=\left(\begin{matrix}1&0\\0&\frac{1+i}{\sqrt{2}  }\end{matrix}\right)$ \\ \addlinespace
\bottomrule
\end{tabular}
\caption{One-qubit basic gates.}
\label{Tab:1qgates}
\end{table}

\section{Two-qubit gates}

Any $2^2\times 2^2$ unitary matrix is a two-qubit gate. However, the common gates used implement particular operations, in particular, controlled and swap operations.

Swap gates exchange the state of two-qubits following some rule. The most basic one is the SWAP gate, which exchanges the qubits amplitude if they are in a different state:
\begin{equation}
\mathrm{SWAP}(a|00\rangle+b|01\rangle+c|10\rangle+d|11\rangle)= a|00\rangle+c|01\rangle+b|10\rangle+d|11\rangle.
\end{equation}

Controlled gates implement an operation on target qubit if the control qubit is in the state $|1\rangle$. One of the most representative of this kind of gates is the controlled-$X$ gate, usually called CNOT,
\begin{equation}
\mathrm{CNOT}(a|00\rangle+b|01\rangle+c|10\rangle+d|11\rangle)= a|00\rangle+b|01\rangle+c|11\rangle+d|10\rangle,
\end{equation}
which flips, i.e. applies $X$ gate, the target qubit (in the example, the second qubit) if the control qubit is in the $|1\rangle$ state.

Table \ref{Tab:2qgates} shows a summary of the most used two-qubit gates, including their symbol, matrix and common decomposition in terms of other gates. This last thing is particularly useful for the experimental implementation as, depending on the physical platform used to run quantum circuits, some gates are more easy to implement than others. 

\begin{table}[t!]
\centering
\arraycolsep=2pt\def\arraystretch{1.1}
\begin{tabular*}{\textwidth}{c @{\extracolsep{\fill}} cccc}
\toprule
\multirow{2}{*}{\textbf{Name}} & \multirow{2}{*}{\textbf{Abbreviation}} & \multirow{2}{*}{\textbf{Circuit}} & \textbf{Common} & \multirow{2}{*}{\textbf{Matrix}} \\
 & & & \textbf{decomposition} & \\
\midrule
\addlinespace
Controlled-NOT & CNOT & \raisebox{0.5\totalheight}{\Qcircuit @C=1em @R=2.5em 
{& \ctrl{1}&\qw \\ & \targ & \qw}}& \raisebox{0.5\totalheight}{\Qcircuit @C=0.6em @R=2.2em 
{& \qw & \ctrl{1}& \qw & \qw \\ & \gate{H} & \ctrl{-1} & \gate{H}& \qw}} & 
$\left(\begin{array}{cccc}1 & 0  & 0  & 0  \\0&1&0&0\\ 0&0&0&1\\ 0&0&1&0 \end{array}\right)$ \\
\addlinespace
Controlled-Z & CZ & \raisebox{0.55\totalheight}{\Qcircuit @C=1em @R=2.7em  
{& \ctrl{1}&\qw \\ & \ctrl{-1} & \qw }} & \raisebox{0.5\totalheight}{\Qcircuit @C=0.6em @R=2.2em 
{& \qw & \ctrl{1}& \qw & \qw \\ & \gate{H} & \targ & \gate{H}& \qw}} &$\left(\begin{matrix}
1&0&0&0\\0&1&0&0\\0&0&1&0\\0&0&0&-1\end{matrix}\right)$ \\
\addlinespace
Controlled-phase & CPh($\phi$) & \raisebox{0.5\totalheight}{\Qcircuit @C=1em @R=2em  
{& \ctrl{1}&\qw \\ & \gate{Ph(\phi)} & \qw}} & \raisebox{0.5\totalheight}{\Qcircuit @C=0.6em @R=2em  
{& \ctrl{1}&\qw &\qw \\ & \ctrl{-1} & \gate{Ph(\phi+\pi)} & \qw}} &$\left(\begin{matrix}
1&0&0&0\\0&1&0&0\\0&0&1&0\\0&0&0&e^{i\phi}\end{matrix}\right)$ \\
\addlinespace
 & & & & \\
 \addlinespace
\multirow{2}{*}{Controlled-unitary} & \multirow{2}{*}{CU} & \multirow{2}{*}{\raisebox{0.5\totalheight}{\Qcircuit @C=1em @R=2.2em 
{& \ctrl{1}&\qw \\ & \gate{U} & \qw}}} & See lemma \ref{lem:CU} & \multirow{2}{*}[3ex]{$\left(\begin{matrix}
1&0&0&0\\0&1&0&0\\0&0&u_{00}&u_{01}\\0&0&u_{10}&u_{11}\end{matrix}\right)$} \\
 & & & and Fig.\ref{Fig:CU} & \hphantom{0} \\
\addlinespace
\addlinespace
 & & & & \\
Swap & SWAP & \raisebox{0.65\totalheight}{\Qcircuit @C=1em @R=3em 
{& \qswap & \qw \\ & \qswap \qwx & \qw }} & \raisebox{0.5\totalheight}{\Qcircuit @C=0.3cm @R=2.2em 
{& \ctrl{1}&\targ & \ctrl{1} & \qw \\ & \targ & \ctrl{-1} & \targ & \qw}} &$\left(\begin{matrix}
1&0&0&0\\0&0&1&0\\0&1&0&0\\0&0&0&1\end{matrix}\right)$ \\
\addlinespace
\bottomrule
\end{tabular*}
\caption{Two-qubit basic gates.}
\label{Tab:2qgates}
\end{table}

\subsection{Controlled-unitary gates decomposition}

Let's start with a lemma that follows the previous one and it is related with qubit gate decomposition \cite{Barenco95}:
\begin{lemma}
Any controlled-unitary gate can be performed with CNOT and rotational gates with the circuit: 
%
\[\Qcircuit @C=1em @R=2em {
& \ctrl{1} & \qw & \push{\rule{0em}{0em} \ \qquad \rule{0em}{0em}} & \qw & \ctrl{1} & \qw & \ctrl{1} & \gate{Ph(\delta)} & \qw \\
& \gate{U} & \qw & \push{\rule{0em}{0em} = \quad \rule{0em}{0em}}& \gate{C}& \targ & \gate{B} & \targ & \gate{A} & \qw
}\]
where $A$, $B$ and $C$ gates are the ones obtained in Lemma \ref{lem:unitary}.
\label{lem:CU}
\end{lemma}

If the first qubit is in the $|0\rangle$ state, the second qubit remains in the same state and, as $ABC=\mathbb{I}$, no operation is implemented in this qubit. But if the first qubit is in the $|1\rangle$ state, a $X$ gate is applied in the second leading to the operation $AXBXC$ on this qubit. Finally, if a $Ph(\delta)$ is applied on the first qubit, it adds a phase on states $|10\rangle$ and $|11\rangle$ which, together with the previous action, leads to the implementation of $U=e^{i\delta}AXBXC$.

\subsubsection{Special cases}

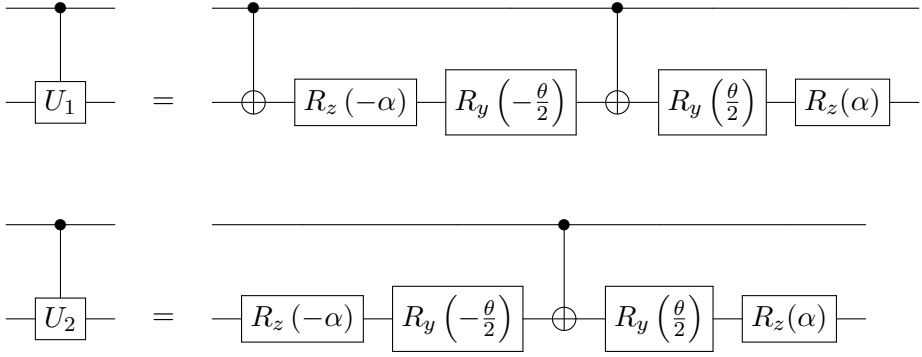
\begin{figure}[t!]
\centering
\[\Qcircuit @C=1em @R=2em {
& \ctrl{1} & \qw & \push{\rule{0em}{0em} \ \qquad \rule{0em}{0em}} & \ctrl{1} & \qw & \qw & \ctrl{1} & \qw & \qw & \qw \\
& \gate{U_1} & \qw & \push{\rule{0em}{0em} = \quad \rule{0em}{0em}}& \targ & \gate{R_z\left(-\alpha\right)} & \gate{R_y\left(-\frac{\theta}{2}\right)} & \targ & \gate{R_y\left(\frac{\theta}{2}\right)} & \gate{R_z(\alpha)} & \qw
}\]\\
\[\Qcircuit @C=1em @R=2em {
& \ctrl{1} & \qw & \push{\rule{0em}{0em} \ \qquad \rule{0em}{0em}} & \qw & \qw & \ctrl{1} & \qw & \qw & \qw \\
& \gate{U_2} & \qw & \push{\rule{0em}{0em} = \quad \rule{0em}{0em}} & \gate{R_z\left(-\alpha\right)} & \gate{R_y\left(-\frac{\theta}{2}\right)} & \targ & \gate{R_y\left(\frac{\theta}{2}\right)} & \gate{R_z(\alpha)} & \qw
}\]
\caption{Basic gates decomposition of a controlled-unitary operations of the form of Eq. \eqref{eq:U_v1} and Eq. \eqref{eq:U_v2}.}
\label{Fig:CU}
\end{figure}

Some unitary operations can be decomposed using less gates that the general decomposition introduced above. For instance, a unitary operation $U$ with $\beta=\alpha$ and $\delta=0$
\begin{equation}
U_{1}=\left(\begin{matrix}e^{i\alpha}\cos(\theta/2) & \sin(\theta/2)\\
-\sin(\theta/2) & e^{-i\alpha}\cos(\theta/2)
\end{matrix}\right),
\label{eq:U_v1}
\end{equation}
can be implemented with the upper circuit of Fig. \ref{Fig:CU}. Notice that adding a CNOT gate at the beginning of this circuit modifies the unitary gate implemented as
\begin{equation}
U_{2}=\left(\begin{matrix}\sin(\theta/2) & e^{i\alpha}\cos(\theta/2)\\
e^{-i\alpha}\cos(\theta/2) & -\sin(\theta/2) 
\end{matrix}\right),
\label{eq:U_v2}
\end{equation}
which circuit is shown in the bottom of Fig. \ref{Fig:CU}.

We can use these gates decomposition to obtain common controlled-unitary gates. A simple example is the controlled-Z gate, CZ, which corresponds with the unitary $U_2(\alpha=0,\theta=\pi)$. Applying the decomposition shown in Fig. \ref{Fig:CU}, we obtain the relation $\mathrm{CZ}=(I\otimes H)\mathrm{CNOT} (I\otimes H)$. Moreover, applying Hadamard gates in the second qubit, we obtain the reverse relation, $\mathrm{CNOT}=(I\otimes H)\mathrm{CZ}(I\otimes H)$.

Another example is the decomposition of controlled-Hadamard gate, CH. Again, Hadamard gate is of the type $U_2$ with $\theta=\pi/2$ and $\alpha=0$, so it is only necessary two $R_y$ gates and a CNOT gate to obtain this operation. It can be also implemented with $S$, $T$ and $H$ gates, as shows Fig. \ref{Fig:CH}.

\begin{figure}[t!]
\[
\Qcircuit @C=1em @R=1em @!R {
&\ctrl{1} & \qw & \push{\rule{0em}{0em} \rule{0em}{0em}} & & \qw & \ctrl{1} & \qw & \ctrl{1} & \qw & \qw & \qw \\
&\gate{R_{X}(\theta_{k})} & \qw & \push{\rule{0em}{0em}\equiv\rule{0em}{0em}}&  & \gate{R_{z}\left(-\frac{\pi}{2}\right)} & \targ & \gate{R_{y}\left(-\frac{\theta}{2}\right)} & \targ & \gate{R_{y}\left(\frac{\theta}{2}\right)} & \gate{R_{z}\left(\frac{\pi}{2}\right)}& \qw
}
\]
\caption{Controlled-$R_{X}$ gate decomposition in terms of the rotational gates.}
\label{Fig:RX}
\end{figure}
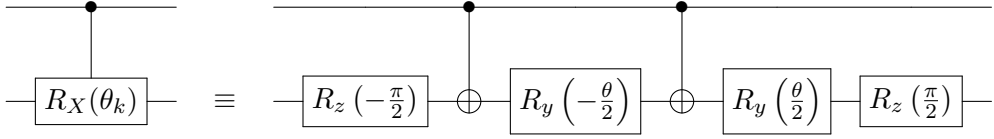

Finally, a last example that appears in Chapter \ref{Ch:Ising}, is the Controlled-$R_x$ gate, which decomposition is shown in Fig.\ref{Fig:RX}.

\begin{figure}[t!]
\[
\Qcircuit @C=0.5em @R=1.7em 
{
& &\ctrl{1} & \qw &\push{\rule{0em}{0em} \quad\quad \rule{0em}{0em}}& \qw & \ctrl{1} & \qw & \qw &\push{\rule{0em}{0em} \quad\quad \rule{0em}{0em}}&\qw & \qw & \qw & \ctrl{1} & \qw & \qw & \qw & \qw \\
& & \gate{H} & \qw & \push{\rule{0em}{0em}\equiv\quad\rule{0em}{0em}}& \gate{R_{y}\left(-\frac{\pi}{2}\right)} & \targ & \gate{R_{y}\left(\frac{\pi}{2}\right)} & \qw & \push{\rule{0em}{0em}\equiv\quad\rule{0em}{0em}}&\gate{S^{\dagger}} & \gate{H} & \gate{T^{\dagger}} & \targ & \gate{T} & \gate{H} & \gate{S} & \qw
}
\]
\caption{Controlled-Hadamard gate. 
}
\label{Fig:CH}
\end{figure}
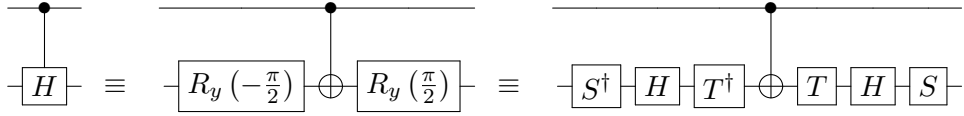

\section{Three qubit gates}

Three qubit gates are not as extensively used as one and two-qubit gates. The reason behind is basically experimentally, since the control of the interaction between three qubits is more difficult and challenging. However, there are some widely used three-qubit gates that are necessary to implement many quantum algorithms. In particular, in this section we introduce the Toffoli gate or CCNOT gate.

The CCNOT gate applies an $X$ gate on target qubit if the two controlled qubits are in the $|1\rangle$ state. Its matrix representation is

\begin{equation}
\mathrm{CCNOT} = 
\left(\begin{array}{cccccccc}
1 & 0 & 0 & 0 & 0 & 0 & 0 & 0 \\
0 & 1 & 0 & 0 & 0 & 0 & 0 & 0 \\
0 & 0 & 1 & 0 & 0 & 0 & 0 & 0 \\
0 & 0 & 0 & 1 & 0 & 0 & 0 & 0 \\
0 & 0 & 0 & 0 & 1 & 0 & 0 & 0 \\
0 & 0 & 0 & 0 & 0 & 1 & 0 & 0 \\
0 & 0 & 0 & 0 & 0 & 0 & 0 & 1 \\
0 & 0 & 0 & 0 & 0 & 0 & 1 & 0 
\end{array}
\right) \ .
\end{equation}

\begin{figure}[t!]
\[
\Qcircuit @C=.7em @R=0.2em @!R {
& \ctrl{1} & \qw & & & \qw &\qw &\qw & \ctrl{2} \qw & \qw & \qw & \qw & \ctrl{2} & \qw & \ctrl{1} & \gate{T} & \ctrl{1} & \qw\\
& \ctrl{1} & \qw & \push{\rule{.3em}{0em}=\rule{.3em}{0em}} & & \qw & \ctrl{1} & \qw & \qw & \qw & \ctrl{1} & \qw & \qw & \gate{T} & \targ & \gate{T^\dagger} & \targ  & \qw \\
& \targ & \qw & & & \gate{H} & \targ & \gate{T^\dagger} & \targ & \gate{T} & \targ & \gate{T^\dagger} & \targ & \gate{T} & \gate{H} & \qw & \qw & \qw 
}
\]
\caption{Toffoli gate implementation with one-qubit gates and CNOTs.}
\label{Fig:CCNOT}
\end{figure}
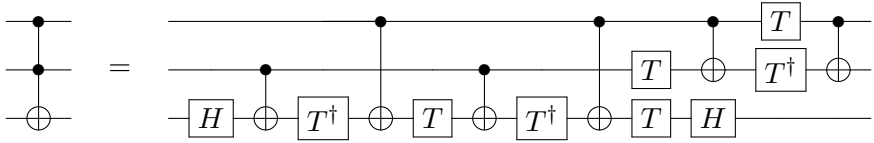

Due to its uses in many quantum algorithms but its difficult implementation, what is actually used is its decomposition in terms of one and two-qubit gates. One of the most used decompositions is shown in Fig. \ref{Fig:CCNOT}.

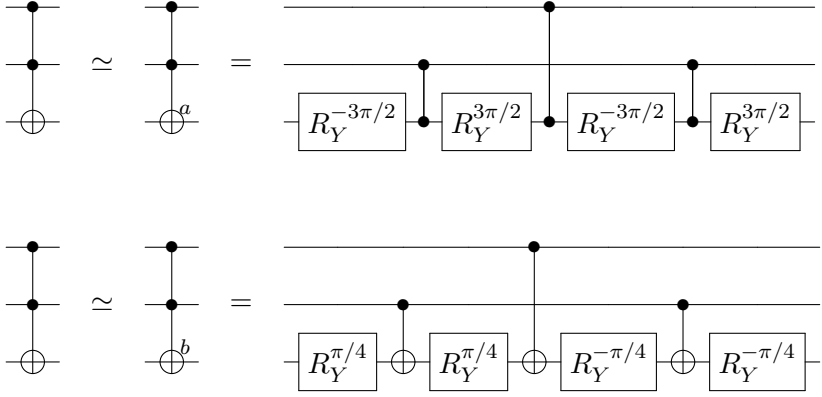
\begin{figure}[t!]
\[
\Qcircuit @C=0.5em @R=0em @!R {
& \ctrl{1} & \qw & & & \ctrl{1} & \qw & & &  \qw & \qw & \qw & \ctrl{2} & \qw & \qw & \qw & \qw\\
& \ctrl{1} & \qw & \push{\rule{.3em}{0em}\simeq\rule{.3em}{.0em}} & & \ctrl{1} & \qw & \push{\rule{.3em}{0em}=\rule{.3em}{.0em}} & & \qw & \ctrl{1} & \qw & \qw & \qw & \ctrl{1} & \qw & \qw &  \\
& \targ & \qw & & & \targ & \qw_{a} & & & \gate{R_{Y}^{-3\pi/2}} & \ctrl{-1} & \gate{R_{Y}^{3\pi/2}} & \ctrl{-2} & \gate{R_{Y}^{-3\pi/2}} & \ctrl{-1} & \gate{R_{Y}^{3\pi/2}} & \qw
} \]\\
\[
\Qcircuit @C=0.5em @R=0em @!R {
& \ctrl{1} & \qw & & & \ctrl{1} & \qw & & &  \qw & \qw & \qw & \ctrl{2} & \qw & \qw & \qw & \qw\\
& \ctrl{1} & \qw & \push{\rule{.3em}{0em}\simeq\rule{.3em}{.0em}} & & \ctrl{1} & \qw & \push{\rule{.3em}{0em}=\rule{.3em}{.0em}} & & \qw & \ctrl{1} & \qw & \qw & \qw & \ctrl{1} & \qw & \qw &  \\
& \targ & \qw & & & \targ & \qw_{b} & & & \gate{R_{Y}^{\pi/4}} & \targ & \gate{R_{Y}^{\pi/4}} & \targ & \gate{R_{Y}^{-\pi/4}} & \targ & \gate{R_{Y}^{-\pi/4}} & \qw
}
\]
\caption{Approximations of CCNOT gate. They introduce a change of sign in some states, in particular $\mathrm{CCNOT}_{a}|101\rangle =  -|101\rangle$ and $\mathrm{CCNOT}_{b}|100\rangle = -|100\rangle$.}
\label{Fig:Toffaprox}
\end{figure}

The exact decomposition of CCNOT gate involves six CNOT gates plus some one-qubit gates, so any circuit that needs from CCNOT gates will increase significantly its depth. We can try to reduce this problem by using an approximate CCNOT gates, as the ones shown in Fig. \ref{Fig:Toffaprox} \cite{Barenco95}. Their matrix representations are

\vfill
\newpage

\setstacktabbedgap{0.2ex}
\medmuskip=0mu
\begin{equation}
\mathrm{CCNOT}_{a} = \left(\begin{array}{cccccccc}
1 & 0 & 0 & 0 & 0 & 0 & 0 & 0 \\
0 & 1 & 0 & 0 & 0 & 0 & 0 & 0 \\
0 & 0 & 1 & 0 & 0 & 0 & 0 & 0 \\
0 & 0 & 0 & 1 & 0 & 0 & 0 & 0 \\
0 & 0 & 0 & 0 & 1 & 0 & 0 & 0 \\
0 & 0 & 0 & 0 & 0 & -1 & 0 & 0 \\
0 & 0 & 0 & 0 & 0 & 0 & 0 & 1 \\
0 & 0 & 0 & 0 & 0 & 0 & 1 & 0 
\end{array}\right) 
\ , \
\mathrm{CCNOT}_{b} = \left(\begin{array}{cccccccc}
1 & 0 & 0 & 0 & 0 & 0 & 0 & 0 \\
0 & 1 & 0 & 0 & 0 & 0 & 0 & 0 \\
0 & 0 & 1 & 0 & 0 & 0 & 0 & 0 \\
0 & 0 & 0 & 1 & 0 & 0 & 0 & 0 \\
0 & 0 & 0 & 0 & -1 & 0 & 0 & 0 \\
0 & 0 & 0 & 0 & 0 & 1 & 0 & 0 \\
0 & 0 & 0 & 0 & 0 & 0 & 0 & 1 \\
0 & 0 & 0 & 0 & 0 & 0 & 1 & 0 
\end{array}
\right) \ .
\end{equation}
\medmuskip=4mu plus 2mu minus 4mu 
These gates are equivalent to CCNOT gate except for one or phase. In particular, the state $|101\rangle$ carries a $-1$ in the case of $\mathrm{CCNOT}_{a}$ and similarly with the state $|100\rangle$ for $\mathrm{CCNOT}_{b}$.

\chapterimage{feynman_diagram}
\chapter{Feynman Rules \label{app:Feynman}}


\vspace{-1.5cm}
\begin{flushright}
\begin{minipage}{0.6\textwidth}
\textit{Nature uses only the longest threads to weave her patterns, so each small piece of her fabric reveals the organization of the entire tapestry.}
\begin{flushright}
--Richard P. Feynman, \\
``The character of physical law'', 1965.
\end{flushright}
\end{minipage}

\end{flushright}
\vspace{1cm}

In this appendix, we introduce some basic definitions and define the conventions used to compute the particle processes presented in Chapter \ref{Ch:MaxEnt}. We provide a finite set of Feynman rules, in particular we only show the necessary ones to do the computations of Chapter \ref{Ch:MaxEnt}. For detailed definitions, derivations and a more complete set of Feynman rules see for instance Ref. \cite{Peskin,QFTSM}.

\vspace{-0.1cm}

\section{Conventions and definitions}

\subsection{Dirac equation}

The Dirac equation describes all spin-$\frac{1}{2}$ massive particles. It is a relativistic wave equation which solutions are called Dirac spinors.

\begin{definition}[Dirac equation]
A free spin-$\frac{1}{2}$ particle of mass $m$ is described by the Dirac equation as
\begin{equation}
\bar{\psi}\left(i\partial_{\mu}\gamma^\mu-m\right)\psi=0,
\end{equation}
where $\bar{\psi}=\psi^\dagger\gamma^0$, $\vec{\sigma}=\left(\sx,\sy,\sz\right)$ and $\gamma^{\mu}$ matrices generate the Clifford algebra, i.e. $\{\gamma^\mu,\gamma^\nu\}=2g^{\mu\nu}$, where $g^{\mu\nu}$ is the Minkowsky metric with signature $+---$.
\label{def:Diraceq}
\end{definition}

Dirac equation can be written in terms of Weyl spinors $\psi_{R}$ and $\psi_{L}$, which are two component spinors that form the four component spinor $\psi=\left(\psi_{L} \ \psi_{R}\right)^{T}$. In matrix form, the Dirac equation becomes
\begin{equation}
\left(\begin{array}{cc}
-m & i\sigma\cdot\partial \\ i\bar{\sigma}\cdot\partial & -m
\end{array}\right)\left(\begin{array}{c}
\psi_{L}\\ \psi_{R}
\end{array}\right)=0,
\end{equation}
where $\sigma=(\mathbb{I},\vec{\sigma})$, $\bar{\sigma}=(\mathbb{I},-\vec{\sigma})$ and $\partial=\partial_{\mu}$. For massless fermions, the Dirac equation decouples into left and right spinors equations
\begin{align}
i\bar{\sigma}\cdot\partial\psi_{L}&=0, \\
i\sigma\cdot\partial\psi_{R}&=0.
\end{align}

There are different conventions for $\gamma^{\mu}$ matrices. In this work, we use the Weyl or chiral representation:
\begin{definition}[Weyl representation of gamma matrices]
\begin{equation}
\gamma^{0}\equiv\left(\begin{matrix} 0 & & \mathbb{I}\\ \mathbb{I} & & 0 \end{matrix}\right), \ 
\gamma^{i}\equiv\left(\begin{matrix} 0 & \sigma^i \\ -\sigma^{i} & 0 \end{matrix}\right), \ 
\gamma^{5}\equiv\left(\begin{matrix} -\mathbb{I} & & 0 \\ 0 & & \mathbb{I} \end{matrix}\right).
\end{equation}
\end{definition}

The fifth gamma matrix is defined as $\gamma^{5}=i\gamma^{0}\gamma^{1}\gamma^{2}\gamma^{3}$ and, although it is not part of the rest of gamma matrices as a generator of Clifford group, it is useful to define the \emph{chirality} operators
\begin{equation}
P_{R}=\frac{1+\gamma^{5}}{2}, \quad P_{L}=\frac{1-\gamma^{5}}{2}.
\end{equation}
These operators project right and left handed Weyl spinors, i.e. $P_{R}\psi=\left(0 \ \psi_{R}\right)^{T}$ and $P_{L}\psi=\left(\psi_{L} \ 0\right)^{T}$.

Thus, the solutions of Dirac equations are plane-wave equations of the form
\begin{equation}
\psi(x)=u(p,s)e^{-ipx}, \quad \psi(x)=v(p,s)e^{+ipx},
\end{equation}
where $u(p,s)$ and $v(p,s)$ are called \emph{Dirac spinors} and correspond to the particle and anti-particle solutions respectively. They should obey the Dirac equation for particles and anti-particles, i.e.
\begin{align}
\left(\gamma^\mu p_\mu -m\right)u(p,s)&=0,\\
\left(\gamma^\mu p_\mu +m\right)v(p,s)&=0.
\end{align}
Then, they can be written as
\begin{equation}
u(p,s)=\left(\begin{matrix} \sqrt{p\cdot\sigma}\xi^{s} \\ \sqrt{p\cdot\bar{\sigma}}\xi^{s}\end{matrix}\right), \qquad
v(p,s)=\left(\begin{matrix} \sqrt{p\cdot\sigma}\xi^{s} \\ -\sqrt{p\cdot\bar{\sigma}}\xi^{s}\end{matrix}\right),
\end{equation}
where $\xi^s$ 
are the two component spinors, eigenstates of helicity operator $\hat{h}$ that are defined in the next subsection.

\subsection{Spin, helicity and chirality}

\begin{definition}[Spin]
Spin is a vector quantity. For fermions, it is represented with the operator $\vec{S}=\vec{\sigma}/2$. However, we sometimes refer to it as a scalar, being one of the eigenvalues of $\vec{S}$ or the quantity $s$ of the eigenvalue $s(s+1)$ of the operator $\vec{S}^2$.
\end{definition}

If we want to use spin as a quantum number to describe spinors, we should choose one spin direction, for instance $S_{z}$. However, in general, spinors $u(p,s)$ and $v(p,s)$ are not eigenstates of $S_{z}$ (only if $\vec{p}=p_{z}\hat{z}$), so spin is not a particularly useful basis. In addition, $\left[\mathcal{H},\vec{S}\right]\neq 0$ in general, so what is actually a good quantum number is angular momentum, i.e. $\vec{J}=\vec{L}+\vec{S}$ where $\vec{L}$ is the orbital angular momentum, due to $\left[\mathcal{H},\vec{L}+\vec{S}\right]= 0$. 

Let us define another quantity that is more convenient to use with Dirac spinors.
\begin{definition}[Helicity]
Helicity is the projection of spin on the direction of momenta:
\begin{equation}
\hat{h} \ \xi= \frac{\vec{S}\cdot\vec{p}}{|\vec{p}|}\xi=\lambda \ \xi \ ,
\end{equation}
where $\lambda$ are the eigenvalues of $\hat{h}$ and $\xi$ are the helicity eigenstates.
\end{definition}

For fermions, we have $\lambda=\pm 1$. If the projection of the fermion spin points towards the direction of momentum, its value is $s=+1/2$ and therefore $\lambda=+1$ (\emph{right-handed helicity}). On the contrary, if it points in the opposite direction, $s=-1/2$ and $\lambda=-1$ (\emph{left-handed helicity}). The opposite convention holds for anti-fermions, since the spin for antiparticles is measured with the operator $-\vec{S}$. Thus, what dictates if a particle is right or left handed is the eigenstate of the helicity operator.

The two-component spinors $\xi$ are helicity eigenstates. For an arbitrary direction of momentum $\hat{p}=(\sin\theta\cos\phi,\sin\theta\sin\phi,\cos\theta)$,
\begin{align}
&\hat{p}\cdot\vec{\sigma}=\left(\begin{array}{cc}
\cos\theta & e^{-i\phi}\sin\theta \\
e^{i\phi}\sin\theta & -\cos\theta
\end{array}\right), \nonumber\\
&\hat{p}\cdot\vec{\sigma}\left(\begin{array}{c}
\xi^s_{1}\\ \xi^s_{2}
\end{array}\right)=\pm\left(\begin{array}{c}
\xi^s_{1}\\ \xi^s_{2}
\end{array}\right) \longrightarrow \frac{\xi^s_2}{\xi^s_{1}}=\frac{\pm 1-\cos\theta}{\sin\theta}e^{i\phi},
\end{align}
where $s=\uparrow,\downarrow$ and $\xi^s=\left(\xi^s_{1} \ \xi^s_{2}\right)$. Thus, the helicity eigenstates can be written as
\begin{equation}
\xi^\uparrow=\left(\begin{array}{c}
\cos(\theta/2) \\ e^{i\phi}\sin(\theta/2)
\end{array}\right), \qquad \xi^\downarrow=\left(\begin{array}{c}
-\sin(\theta/2) \\ e^{i\phi}\cos(\theta/2)
\end{array}\right) \ .
\end{equation}
Therefore,
\begin{equation}
\hat{h} \ \xi^\uparrow=+\xi^\uparrow \ , \qquad \hat{h} \ \xi^\downarrow=-\xi^\downarrow \ ,
\end{equation}
when applied to a $u(s,p)$ spinor and
\begin{equation}
\hat{h} \ \xi^\uparrow=-\xi^\uparrow \ , \qquad \hat{h} \ \xi^\downarrow=+\xi^\downarrow 
\end{equation}
when applied to a $v(s,p)$ spinor. For example, a right handed electron and a left handed positron are described respectively by $u(\uparrow,p)$ and $v(\uparrow,p)$ spinors.

For photons, a similar convention is used. A photon with momentum $\vec{k}$ have the circular polarization vectors
\begin{equation}
\vec{\epsilon}(\lambda,\vec{k})=-\frac{\lambda}{\sqrt{2}}\left(\cos\theta\cos\phi-i\lambda\sin\phi,\cos\theta\sin\phi+i\lambda\cos\phi,-\sin\theta\right),
\end{equation}
where $\lambda=\pm 1$ and correspond with right and left handed polarization vectors. 

To distinguish between fermion and photon helicities, we denote with $|R\rangle$ and $|L\rangle$ a right and left handed states for fermions and $|+\rangle$ and $|-\rangle$ the two circular photon polarizations.

As a final remark, for high energies or massless particles, helicity and chirality are equivalent. For that reason, many references, including the work presented in this thesis, use the term right and left handed particles to refer to helicity eigenstates.

\subsection{The \texorpdfstring{$S$}{} matrix}

A Hamiltonian that describes free particles and the interaction between them can be decomposed as
\begin{equation}
\mathcal{H}=\mathcal{H}_{0}+\mathcal{H}_{I},
\end{equation}
where $\mathcal{H}_{0}$ are the Hamiltonians of free particles and $\mathcal{H}_{I}$ the interaction Hamiltonian. Working in the interaction picture, the time evolution is dictated by the Schr\"odinger equation
\begin{equation}
i\frac{d}{dt}|\chi(t)\rangle=\mathcal{H}_{I}|\chi(t)\rangle.
\end{equation}
We define as initial state the state occurring at $t\rightarrow-\infty$ and we are interested in the state long after the interaction, i.e. at $t\rightarrow +\infty$. The transformation that drive us from the initial state to the final state is called $S$ \emph{matrix},
\begin{equation}
|\chi(+\infty)\rangle\equiv \mathcal{S}|i\rangle = \mathcal{S}|\chi(-\infty)\rangle.
\end{equation}
We will be interested in the projection on the possible final states, i.e.
\begin{equation}
\langle f|\chi(+\infty)\rangle=\langle f|\mathcal{S}|i\rangle\equiv \mathcal{S}_{fi} \ .
\end{equation}
Thus,
\begin{equation}
|\chi(t)\rangle=|i\rangle + (-i)\int_{-\infty}^{t}d\tau \mathcal{H}_{I}(\tau)|\chi(\tau)\rangle.
\end{equation}
We can solve the above integral perturbatively. If we stop at first order, we can define $\mathcal{S}\equiv \mathbb{I}+i\mathcal{T}$ and therefore obtain
\begin{equation}
\mathcal{S}_{fi}=i\mathcal{T}_{fi}\equiv (2\pi)^4\delta^4(p_{i}-p_{f})(i\mathcal{M}).
\end{equation}
Here, $i\mathcal{M}$ is called \emph{Feynman amplitude} and the rules to construct this amplitude are called \emph{Feynman rules}. By using the \emph{Feynman diagrams}, we can extract the correct form of $i\mathcal{M}$. Since we have computed $\mathcal{S}$ until first order, the corresponding amplitudes are called \emph{tree-level} amplitudes. If we expand the $S$ matrix until the second order, they are called \emph{one-loop} amplitudes, etc.

\section{Feynman rules}

Feynman rules are extracted from the Lagrangian of the theory. For instance, the description of free fermions are obtained from Dirac equation, of free photons from Maxwell equations, etc. The interaction terms are dictated from the interaction part of the Lagrangian that will depend on the order that we are considering in perturbation theory. For the tree-level processes that are computed in Chapter \ref{Ch:MaxEnt}, we only have simple vertices, which simplifies the amount of diagrams for a given process and its computation. 

Table \ref{Tab:FR_QED} introduces the tree-level Feynman rules for QED. We should add an extra rule that has not been included in the table: for a given process, diagrams that differ in the exchange of identical fermions or anti-fermions carry a relative $-$ sign.

\begin{table}[t!]
\centering
\begin{tabular}{c c c}
\toprule
\textbf{Process} & \textbf{Diagram} & \textbf{Rule} \\
\midrule
\raisebox{0.75cm}{Fermion-photon vertex} & \includegraphics[scale=0.7]{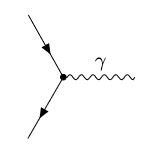} & \raisebox{0.75cm}{$-ieQ_{f}\gamma^\mu$}\\
\raisebox{0.17cm}{Incoming fermion} & \includegraphics[scale=0.7]{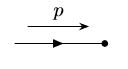} & \raisebox{0.17cm}{$u(p,s)$}\\
\raisebox{0.17cm}{Outgoing fermion} & \includegraphics[scale=0.7]{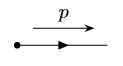} & \raisebox{0.17cm}{$\bar{u}(p,s)$}\\
\raisebox{0.17cm}{Incoming anti-fermion} & \includegraphics[scale=0.7]{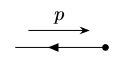} & \raisebox{0.17cm}{$\bar{v}(p,s)$}\\
\raisebox{0.17cm}{Outgoing anti-fermion} & \includegraphics[scale=0.7]{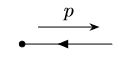} & \raisebox{0.17cm}{$v(p,s)$}\\
\raisebox{0.17cm}{Incoming photon} & \includegraphics[scale=0.7]{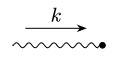} & \raisebox{0.17cm}{$\epsilon_{\mu}(k,\lambda)$}\\
\raisebox{0.17cm}{Outgoing photon} & \includegraphics[scale=0.7]{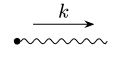} & \raisebox{0.17cm}{$\epsilon^{*}_{\mu}(k,\lambda)$}\\
\raisebox{0.17cm}{Fermion propagator} & \includegraphics[scale=0.7]{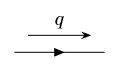} & \raisebox{0.17cm}{$\frac{i(\cancel{q}+m)}{q^2-m^2+i\varepsilon}$}\\
\raisebox{0.17cm}{Photon propagator} & \includegraphics[scale=0.7]{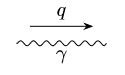} & \raisebox{0.17cm}{$-\frac{ig^{\mu\nu}}{q^2+i\varepsilon}$} \\
\bottomrule
\end{tabular}
\caption{Feynman rules for QED.}
\label{Tab:FR_QED}
\end{table}

Tables \ref{Tab:FR_weak} and \ref{Tab:FR_QCD} show some basic tree-level Feynman rules for weak and QCD interactions. The constants $f$ that appear in QCD rules are the structure constants of $SU(3)$, which are obtained from the corresponding generators called Gell-Man matrices.
\begin{definition}[Gell-Mann matrices]
\begin{equation}
\begin{array}{lll}
\lambda_{1}\equiv\left(\begin{array}{ccc}0&1&0\\ 1&0&0\\ 0&0&0 \end{array}\right),&
\lambda_{2}\equiv\left(\begin{array}{ccc}0&-i&0\\ i&0&0\\ 0&0&0 \end{array}\right),&
\lambda_{3}\equiv\left(\begin{array}{ccc}1&0&0\\ 0&-1&0\\ 0&0&0 \end{array}\right),\\
\lambda_{4}\equiv\left(\begin{array}{ccc}0&0&1\\ 0&0&0\\ 1&0&0 \end{array}\right),&
\lambda_{5}\equiv\left(\begin{array}{ccc}0&0&-i\\ 0&0&0\\ i&0&0 \end{array}\right),&
\lambda_{6}\equiv\left(\begin{array}{ccc}0&0&0\\ 0&0&1\\ 0&1&0 \end{array}\right),\\
\lambda_{7}\equiv\left(\begin{array}{ccc}0&0&0\\ 0&0&-i\\ 0&i&0 \end{array}\right),&
\lambda_{8}\equiv\frac{1}{\sqrt{3}}\left(\begin{array}{ccc}1&0&0\\ 0&1&0 \\ 0&0&-2 \end{array}\right). &
\end{array}
\label{eq:GM}
\end{equation}
These matrices satisfy the commutation relations $[\lambda_{i},\lambda_{j}]=2i\sum_{k}f^{ijk}\lambda_{k}$ with 
\begin{align}
f^{123}&=1, \nonumber\\ 
f^{147}&=f^{165}=f^{246}=f^{257}= f^{345}=f^{376}=1/2, \nonumber\\ f^{458}&=f^{678}=\sqrt{3}/2.
\end{align}
The structure constants $f^{ijk}$ are completely antisymmetric in the three indices.
\end{definition}

\begin{table}[t!]
\centering
\begin{tabular}{c c c}
\toprule
\textbf{Process} & \textbf{Diagram} & \textbf{Rule} \\
\midrule
\raisebox{0.75cm}{Charged vertex} & \includegraphics[scale=0.7]{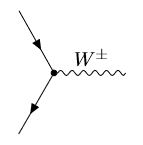} & \raisebox{0.75cm}{$i\frac{g}{\sqrt{2}}\gamma^\mu\frac{1-\gamma^{5}}{2}$}\\
\multirow{2}{*}{\raisebox{-0.4cm}{Neutral vertex}} & \multirow{2}{*}{\includegraphics[scale=0.7]{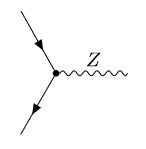}} & \raisebox{0.0cm}{$i\frac{g}{\cos\theta_{W}}\gamma^\mu\left(g_V^{f}-\gamma^{5}g_{A}^{f}\right)$}\\
 & & {\footnotesize $\begin{array}{l}g_{V}^{f}=\frac{T_{3}^{f}}{2} \\ g_{A}^{f}=\frac{T_{3}^{f}}{2}-Q_{f}\sin^2\theta_{W}\end{array}$} \\
\raisebox{0.17cm}{Incoming boson} & \includegraphics[scale=0.7]{incoming_boson} & \raisebox{0.17cm}{$\epsilon_{\mu}(k,\lambda)$}\\
\raisebox{0.17cm}{Outgoing boson} & \includegraphics[scale=0.7]{outgoing_boson} & \raisebox{0.17cm}{$\epsilon^{*}_{\mu}(k,\lambda)$}\\
\raisebox{0.25cm}{$Z$ propagator} & \includegraphics[scale=0.7]{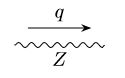} & \raisebox{0.25cm}{$-\frac{ig^{\mu\nu}}{q^2-M_{Z}^2+i\varepsilon}$} \\
\bottomrule
\end{tabular}
\caption{Some Feynman rules for weak interaction.}
\label{Tab:FR_weak}
\end{table}

\begin{table}[t!]
\centering
\begin{tabular}{c c c}
\toprule
\textbf{Process} & \textbf{Diagram} & \textbf{Rule} \\
\midrule
\raisebox{0.75cm}{Triple gluon vertex} & \includegraphics[scale=0.7]{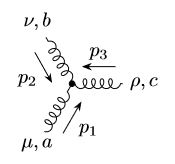} & \raisebox{0.75cm}{$\begin{array}{l} gf^{abc}\Big[g^{\mu\nu}\left(p_{1}-p_{2}\right)^\rho \\ + g^{\nu\rho}\left(p_{2}-p_{3}\right)^\mu \\ + g^{\rho\mu}\left(p_{3}-p_{1}\right)^\nu\Big] \end{array} $}\\
\raisebox{0.75cm}{Quartic gluon vertex} & \includegraphics[scale=0.7]{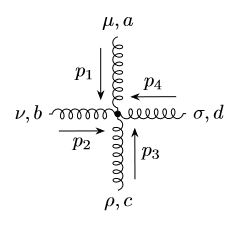} & \raisebox{0.75cm}{$\begin{array}{l}-ig^2\Big[f_{eab}f_{ecd}\left(g_{\mu\rho}g_{\nu\sigma} -g_{\mu\sigma}g_{\nu\rho}\right) \\ + f_{eac}f_{edb}\left(g_{\mu\sigma}g_{\rho\nu}-g_{\mu\rho}g_{\rho\sigma}\right) \\ + f_{ead}f_{ebc}\left(g_{\mu\nu}g_{\rho\sigma}-g_{\mu\rho}g_{\nu\sigma}\right)\Big]\end{array} $}\\
\raisebox{0.17cm}{Incoming gluon} & \includegraphics[scale=0.7]{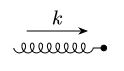} & \raisebox{0.17cm}{$\epsilon_{\mu}(k,\lambda)$}\\
\raisebox{0.17cm}{Outgoing gluon} & \includegraphics[scale=0.7]{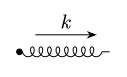} & \raisebox{0.17cm}{$\epsilon^{*}_{\mu}(k,\lambda)$}\\
\bottomrule
\end{tabular}
\caption{Some Feynman rules for QCD.}
\label{Tab:FR_QCD}
\end{table}

\vfill

\chapterimage{feynman_diagram}
\chapter{Entanglement at Tree-level QED \label{app:QED}}


\vspace{-1.5cm}
\begin{flushright}
\begin{minipage}{0.6\textwidth}
\textit{If and when all the laws governing physical phenomena are finally discovered 
(...) we will be able to say that physical science has reached its end, that no excitement is left in further explorations, and that all that remains to a physicist is either tedious work on minor details or the self-educational study and adoration of the magnificence of the completed system.}
\begin{flushright}
-- George Gamow, \\
``Any Physics Tomorrow'', 1949.
\end{flushright}
\end{minipage}
\end{flushright}
\vspace{1cm}

In this appendix, we provide all amplitudes of tree-level QED processes in terms of the helicities of the incoming and outgoing particles. We analyze explicitly the entanglement generation in each process as described in Chapter \ref{Ch:MaxEnt}.

\section{\texorpdfstring{$e^-\mu^-\rightarrow e^-\mu^-$}{}}


The matrix element for this process is
\begin{equation}
i\mathcal{M}_{t}=\bar{u}(s'_{2},q_{2})(-ie\gamma^{\mu})u(s_{2},p_{2})\frac{-ig_{\mu\nu}}{(p_{1}-q_{1})^{2}}\bar{u}(s'_{1},q_{1})(-ie\gamma_{\nu})u(s_{1},p_{1}),
\end{equation}
and the unpolarized amplitude
\begin{equation}
|\overline{\mathcal{M}}|^2=
\frac{8e^{4}}{t^{2}}\left(\left(\frac{s-(m^{2}+M^{2})}{2}\right)^{2}+\left(\frac{u-(m^{2}+M^{2})}{2}\right)^{2}+\frac{t(m^{2}+M^{2})}{2}\right),
\end{equation}
where $m$ is the electron mass, $M$ is the muon mass and
\begin{equation}
|\overline{\mathcal{M}}|^2 \equiv \overline{\sum_{i}}\sum_{f}|\mathcal{M}|^{2} \ .
\end{equation}
In the following sections, we will keep this definition for the unpolarized amplitude $|\overline{\mathcal{M}}|^2$.

The amplitudes for this process are:
\begin{align}
\mathcal{M}_{|RR\rangle\rightarrow|RR\rangle}&= \mathcal{M}_{|LL\rangle\rightarrow|LL\rangle} = e^2  \frac{\mu^2\left(3-\cos\theta\right)+ \sqrt{\left(1+\mu^2\right)\left(\mu^2+\lambda^2\right)}\left(1+\cos\theta\right)}{\mu^2\left(-1+\cos\theta\right)} \ , \nonumber\\
\mathcal{M}_{|RR\rangle\rightarrow|RL\rangle}&= -\mathcal{M}_{|LL\rangle\rightarrow|LR\rangle} = e^2  \frac{\sqrt{\mu^2+\lambda^2}}{\mu^2}\cot(\theta/2) \ , \nonumber\\
\mathcal{M}_{|RR\rangle\rightarrow|LR\rangle}&= -\mathcal{M}_{|LL\rangle\rightarrow|RL\rangle} = e^2  \frac{\lambda\sqrt{1+\mu^2}}{\mu^2}\cot(\theta/2) \ , \nonumber\\
\mathcal{M}_{|RR\rangle\rightarrow|LL\rangle}&=\mathcal{M}_{|LL\rangle\rightarrow|RR\rangle} = -e^{2} \frac{\lambda}{\mu^2} \ ,\nonumber\\
\mathcal{M}_{|RL\rangle\rightarrow|RR\rangle}&= -\mathcal{M}_{|LR\rangle\rightarrow|LL\rangle} = -e^{2} \frac{\sqrt{\mu^2+\lambda^2}}{\mu^2}\cot(\theta/2) \ , \nonumber\\
\mathcal{M}_{|RL\rangle\rightarrow|RL\rangle}&= \mathcal{M}_{|LR\rangle\rightarrow|LR\rangle} = -e^{2}  \frac{\mu^2+\sqrt{\left(1+\mu^2\right)\left(\mu^2+\lambda^2\right)}}{\mu^2}\cot^2(\theta/2) \ , \nonumber\\
\mathcal{M}_{|RL\rangle\rightarrow|LR\rangle}&= \mathcal{M}_{|LR\rangle\rightarrow|RL\rangle} = e^{2}  \frac{\lambda}{\mu^2} \ , \nonumber\\
\mathcal{M}_{|RL\rangle\rightarrow|LL\rangle}&= \mathcal{M}_{|LR\rangle\rightarrow|RR\rangle} = e^{2}  \frac{\lambda\sqrt{1+\mu^2}}{\mu^2}\cot(\theta/2) \ ,
\end{align}
where $\mu=|\vec{p}|/M$ and $\lambda=m/M$.

\subsubsection{Concurrence analysis}

The concurrence from the initial state $|RR\rangle$ or $|LL\rangle$ is the same,  
\begin{equation}
\Delta_{RR\rangle}=\Delta_{LL\rangle}= \frac{2\mu^2\lambda\left(3-\cos\theta\right)\sin^2(\theta/2)}{f(\mu,\lambda,\theta)},
\end{equation}
where
\begin{multline}
f(\mu,\lambda,\theta)=4\lambda^2+ 4\mu^2\left(1+\lambda^2+\sqrt{\left(1+\mu^2\right)\left(\mu^2+\lambda^2\right)} \left(3-\cos\theta\right)\right)\cos^2(\theta/2) \\
+\mu^4\left(11-4\cos\theta+\cos 2\theta\right) \ .
\end{multline}

Taking $|\vec{p}|\rightarrow\infty$, i.e. $\mu\rightarrow\infty$, we obtain the high energy limit 
\begin{equation}
\Delta_{RR}^{HE} = 0 +\mathcal{O}\left(\frac{1}{\mu^2}\right) \ .
\end{equation}
The low energy limit corresponds to $\mu\rightarrow 0$, that leads to
\begin{equation}
\Delta_{RR}^{LE} = 0 + \mathcal{O}\left(\mu^2\right).
\end{equation}
Then, it is not possible to achieve MaxEnt at those limits. However, a maximization of $\Delta_{RR}$ give us MaxEnt at $\theta=\pi$,
\begin{equation}
\Delta_{RR}^{\theta\rightarrow\pi} = \frac{1}{2}\frac{\mu^{2}\lambda}{2\mu^{4}+\lambda^{2}} - \mathcal{O}\left(\left(\theta-\pi\right)^{2}\right).
\end{equation}
For $\mu=\sqrt{\lambda/2}$ the zeroth order is $1/2$ and the next order negative, which means that it is a maximum. Although $\mu$ is not zero, as $\lambda=m/M\ll 1$ we can consider this limit as a low energy limit.

For an initial state $|RL\rangle$ or $|LR\rangle$ the concurrence is also the same, i.e. $\Delta_{RL}=\Delta_{LR}$ with
\begin{equation}
\Delta_{RL}=\frac{\mu^2\lambda\sin^2\theta}{4\left(\lambda^2\mu^2\left(1+\lambda^2+ \sqrt{\left(1+\mu^2\right)\left(\mu^2+\lambda^2\right)}\left(1+\cos\theta\right)\right) +2\mu^4\cos^4(\theta/2)\right)},
\end{equation}
and the corresponding high and low energy limits are
\begin{align}
\Delta_{RL}^{HE} &= 0 + \mathcal{O}\left(\frac{1}{\mu^{2}}\right), \\
\Delta_{RL}^{LE} &= 0 + \mathcal{O}\left(\mu^{2}\right),
\end{align}
so it is not possible to generate MaxEnt at high neither at low energy limits. We also performed a maximization of the above concurrences and did not find any angle or values of $\mu$ and $\lambda$ for those concurrence is 1/2.

\subsubsection{Generated states}

If the initial electron and muon are in the product state $|RR\rangle$, the final state become
\begin{multline}
|\psi\rangle_{RR} =-\frac{1}{\sqrt{\mathcal{N}}}\sum_{\{+,-\}} \Bigg( +|\Psi^{\pm}\rangle\left(\lambda\sqrt{1+\mu^{2}}\pm\sqrt{\mu^{2}+\lambda^{2}}\right) \sin\theta \\
+|\Phi^{\pm}\rangle\left(2\mu^{2}+ \sqrt{\left(1+\mu^{2}\right)\left(\mu^{2}+\lambda^{2}\right)}\left(1+\cos\theta\right) +\left(\mu^{2}\pm\lambda\right)\left(1-\cos\theta\right)\right)\Bigg),
\end{multline}
where $\mathcal{N}$ is the corresponding norm. This state have the following high and low energy limits:
\begin{align}
\lim_{\mu\rightarrow\infty}|\psi\rangle_{RR}&=-\frac{1}{\sqrt{2}}\left(|\Phi^{+}\rangle+|\Phi^{-}\rangle\right) = -|RR\rangle,\\
\lim_{\mu\rightarrow 0}|\psi\rangle_{RR}&=-\frac{1}{\sqrt{2\left(1+\sin^2\theta\right)}}\left(|\Phi^{+}\rangle+|\Phi^{-}\rangle\cos\theta-|\Psi^{+}\rangle\sin\theta\right)\nonumber\\
&=-\frac{\sqrt{2}}{1+\sin^2\theta}\left(c|R\rangle-s|L\rangle\right)\left(c|R\rangle-s|L\rangle\right),\\
\lim_{\substack{\theta\rightarrow\pi\\ \mu\rightarrow\sqrt{\lambda/2}}}|\psi\rangle_{RR}&=-|\Phi^{+}\rangle,
\end{align}
where $c\equiv\cos(\theta/2)$ and $s\equiv\sin(\theta/2)$. As expected from concurrence analysis, only one MaxEnt state is generated, the $|\Phi^{+}\rangle$, at $\theta=\pi$ and $\mu=\sqrt{\lambda/2}$.

If the initial electron and muon are in the product state $|RL\rangle$, the final state become
\begin{multline}
|\psi\rangle_{RL}=-\frac{1}{\sqrt{\mathcal{N}}}\sum_{\{+,-\}}\Big(|\Phi^{\pm}\rangle\left(\sqrt{\mu^{2}+\lambda^{2}}\mp\lambda\sqrt{1+\mu^{2}}\right)\sin\theta \\
+|\Psi^{\pm}\rangle\left(\mp\lambda\left(1-\cos\theta\right)+ \left(\mu^2+ \sqrt{\left(1+\mu^{2}\right)\left(\mu^{2}+\lambda^{2}\right)}\right) \left(1+\cos\theta\right)\right)\Big) \ ,
\end{multline}
and the high and low energy limits are
\begin{align}
\lim_{\mu\rightarrow\infty}|\psi\rangle_{RL}&=-\frac{1}{\sqrt{2}}\left(|\Psi^{+}\rangle+|\Psi^{-}\rangle\right) = -|RL\rangle, \\
\lim_{\mu\rightarrow 0}|\psi\rangle_{RL}&=-\frac{1}{\sqrt{2\left(1+\sin^2\theta\right)}}\left(|\Psi^{+}\rangle\cos\theta+|\Psi^{-}\rangle+|\Phi^{-}\rangle\sin\theta\right) \nonumber\\
& =-\frac{\sqrt{2}}{1+\sin^2\theta}\left(c|R\rangle-s|L\rangle\right)\left(s|R\rangle+c|L\rangle\right).
\label{eq:pmLE_state_mott}\\
\end{align}
Both limits lead to a product state.

\section{\texorpdfstring{$e^-e^+\rightarrow \mu^-\mu^+$}{}}

The matrix element of this process is
\begin{equation}
i\mathcal{M}_{s}=\bar{v}(s_{2},p_{2})(-ie\gamma^{\mu})u(s_{1},p_{1}) \frac{-ig_{\mu\nu}}{(p_{1}+p_{2})^{2}} \bar{u}(s_{1}',q_{1})(-ie\gamma_{\nu})v(s_{2}',q_{2}),
\end{equation}
and the unpolarized amplitude
\begin{equation}
|\overline{\mathcal{M}}|^2=\frac{8e^{4}}{s^{2}}\left(\left(\frac{t-(m^{2}+M^{2})}{2}\right)^{2}+\left(\frac{u-(m^{2}+M^{2})}{2}\right)^{2}+\frac{s(m^{2}+M^{2})}{2}\right), \nonumber\\
\end{equation}
where $m$ is the electron mass and $M$ is the muon mass as in the previous process.

The helicity amplitudes of this process are:
\begin{align}
\mathcal{M}_{|RR\rangle\rightarrow|\substack{RR\\ LL}\rangle} &= \mathcal{M}_{|LL\rangle\rightarrow|\substack{RR\\ LL}\rangle}=-e^{2}\frac{\lambda}{\mu^2+\lambda^2}\cos\theta \ , \nonumber\\
\mathcal{M}_{|RR\rangle\rightarrow|\substack{RL\\ LR}\rangle} &= \mathcal{M}_{|LL\rangle\rightarrow|\substack{RL\\ LR}\rangle} = \mp e^{2}\frac{\lambda}{\sqrt{\mu^2+\lambda^2}}\sin\theta \ , \nonumber\\
\mathcal{M}_{|RL\rangle\rightarrow|\substack{RR\\ LL}\rangle} &=-\mathcal{M}_{|LR\rangle\rightarrow|\substack{RR\\ LL}\rangle} = e^{2}\frac{1}{\sqrt{\mu^2+\lambda^2}}\sin\theta \ , \nonumber\\
\mathcal{M}_{|RL\rangle\rightarrow|\substack{RL\\ LR}\rangle} &= -e^{2}\left(1\pm \cos\theta\right) \ , \nonumber\\
\mathcal{M}_{|LR\rangle\rightarrow|\substack{LR\\ RL}\rangle} &= -e^{2}\left(1\mp \cos\theta \right) \ ,
\label{eq:emuannih_amplitudes}
\end{align}
where, again, $\mu=|\vec{p}|/M$ and $\lambda=m/M$. 

\subsubsection{Concurrence analysis}

The concurrences of the final states are
\begin{align}
\Delta_{RR}&=\Delta_{LL}=\frac{1}{2} \ ,\\
\Delta_{RL}&=\Delta_{LR}=\frac{(\mu^{2}+\lambda^{2}-1)\sin^2\theta}{2\left((\mu^{2}+\lambda^{2})(1 + \cos^{2}\theta) + \sin^2\theta\right)} \ .
\label{eq:Deltapm_emuannh}
\end{align}
Then, from an initial $|RR\rangle$ or $|LL\rangle$ state it is always possible to generate a MaxEnt state at any energy except for very high energies, i.e. $\mu\rightarrow\infty$, since $|RR\rangle\rightarrow \cdots $ and $|LL\rangle\rightarrow \cdots $ amplitudes of Eq. \eqref{eq:emuannih_amplitudes} vanish at that limit. From an initial $|RL\rangle$ or $|LR\rangle$ state, the entanglement generation depends on the energy and COM angle.

Let's analyze the $\Delta_{RL}$ with more detail. The limit for high energies corresponds to $\mu\rightarrow\infty$,
\begin{equation}
\Delta_{RL}^{HE}= \frac{\sin^2\theta}{3+\cos 2\theta}-\frac{4\sin^2\theta}{(3+\cos 2\theta)^2}\frac{1}{\mu^2} + \mathcal{O}\left(\frac{1}{\mu}\right)^{4} .
\end{equation}
The first non-zero order give us MaxEnt for $\theta=\pi/2$. Let's perform again an expansion around this angle:
\begin{equation}
\Delta_{RL}^{HE}=\frac{1}{2}-\left(\theta-\frac{\pi}{2}\right)^2-\frac{1}{\mu^{2}} + 3\frac{1}{\mu^{2}}\left(\theta-\frac{\pi}{2}\right)^2.
\end{equation}
As energy decreases, MaxEnt is lost, although it is a fourth order effect.

For Low energies, the limit corresponds to the muon production threshold, i.e. $\mu\rightarrow\sqrt{1-\lambda^2}$. Defining $\epsilon=\mu-\sqrt{1-\lambda^2}$ as the expansion parameter,
\begin{equation}
\Delta_{RL}^{LE}\simeq\epsilon\left(\frac{1}{2}\sqrt{1-\lambda^2}+\frac{1}{4}\epsilon\left(-2+3\lambda^2-(1-\lambda^2)\cos 2\theta\right)\right)\sin^2\theta \ .
\end{equation}
Only for $\lambda\rightarrow 0$ MaxEnt could be achieved if $\theta=\pi/2$, although it is suppressed by an $\epsilon$ factor.

\subsubsection{Generated States}

From an initial state $|RR\rangle$, the final state become
\begin{equation}
|\psi\rangle_{RR}
=\frac{1}{\mathcal{N}}\left(|\Phi^{+}\rangle \cos\theta +|\Psi^{-}\rangle\sqrt{\mu^2+\lambda^2}\sin\theta\right) \ .
\end{equation}
This state is always maximally entangled as any state with the form $a|\Phi^{+}\rangle+b|\Psi^{-}\rangle$ is. For high energy, the $|\Psi^{-}\rangle$ dominates $\forall\theta$ whereas for $\mu=\sqrt{1+\lambda^2}$ the generated state is $|\Phi^{+}\rangle\cos\theta+|\Psi^-\rangle\sin\theta$.

From Eq. \eqref{eq:Deltapm_emuannh} we only obtain a MaxEnt state at high energies and $\theta=\pi/2$. Indeed, the general state obtained is
\begin{equation}
|\psi\rangle_{RL}=\frac{1}{\mathcal{N}}\left(\sqrt{\mu^2 + \lambda^2}(|\Psi^{+}\rangle+|\Psi^{-}\rangle\cos\theta) - |\Phi^{+}\rangle\sin\theta\right),
\end{equation}
where for high energies the first term dominates leading to a state $|\Psi^{+}\rangle$ for $\theta=\pi/2$. On the other hand, for low energies the above state become
\begin{equation}
|\psi\rangle_{RL}^{LE}=\frac{1}{\sqrt{2}}\left(|\Psi^{+}\rangle +|\Psi^{-}\rangle\cos\theta - |\Phi^{-}\rangle\sin\theta\right),
\end{equation}
where each particle can be described separately in a similar way as the state of Eq. \eqref{eq:pmLE_state_mott}.

\section{\texorpdfstring{$e^-e^-\rightarrow e^-e^-$}{}}

This process has two channels
\begin{align}
i\mathcal{M}_{t}&=\bar{u}(s'_{1},q_{1})(-ie\gamma^{\mu})u(s_{1},p_{1})\frac{-ig_{\mu\nu}}{(p_{1}-q_{1})^{2}}\bar{u}(s'_{2},q_{2})(-ie\gamma_{\nu})u(s_{2},p_{2}) \ , \\
i\mathcal{M}_{u}&=(-1)\bar{u}(s'_{1},q_{1})(-ie\gamma^{\mu})u(s_{2},p_{2})\frac{-ig_{\mu\nu}}{(p_{2}-q_{1})^{2}}\bar{u}(s'_{2},q_{2})(-ie\gamma_{\nu})u(s_{1},p_{1}) \ .
\end{align}

Thus the unpolarized amplitude become
\medmuskip=0mu
\begin{equation}
|\overline{\mathcal{M}}|^2= 2e^{4}\left(\frac{s^{2}+u^{2}+8(t-m^{2})m^{2}}{t^{2}}+\frac{s^{2}+t^{2}+8(u-m^{2})m^{2}}{u^{2}}+\frac{2(s-2m^2)(s-6m^2)}{ut}\right).
\label{eq:Moller_totalA}
\end{equation}
\medmuskip=4mu plus 2mu minus 4mu 

The amplitudes for this process are:
\begin{align}
\mathcal{M}_{|RR\rangle\rightarrow|RR\rangle} &=& \mathcal{M}_{|LL\rangle\rightarrow|LL\rangle} &=& - &e^{2}\left(\frac{3+\cos 2\theta}{\mu^2}+8\right)\csc^{2}\theta \ , \nonumber\\
\mathcal{M}_{|RR\rangle\rightarrow|\substack{RL\\ LR}\rangle} &=&\mathcal{M}_{|LL\rangle\rightarrow|\substack{RL\\ LR}\rangle} &=&\mp &2e^{2}\frac{\sqrt{1+\mu^2}}{\mu^2}\cot\theta \ , \nonumber\\
\mathcal{M}_{|RR\rangle\rightarrow|LL\rangle} &=& \mathcal{M}_{|LL\rangle\rightarrow|RR\rangle} &=& &e^{2}\frac{2}{\mu^2} \ , \nonumber\\
\mathcal{M}_{|RL\rangle\rightarrow|\substack{RR\\ LL}\rangle} &=&-\mathcal{M}_{|LR\rangle\rightarrow|\substack{RR\\ LL}\rangle} &=&  &2e^{2}\frac{\sqrt{1+\mu^2}}{\mu^2}\cot\theta \ , \nonumber\\
\mathcal{M}_{|RL\rangle\rightarrow|RL\rangle} &=& \mathcal{M}_{|LR\rangle\rightarrow|LR\rangle} &=& -&e^{2}\left(2\cot^{2}(\theta/2)+\frac{1}{\mu^2}\cos\theta\csc^{2}(\theta/2)\right) \ , \nonumber\\
\mathcal{M}_{|RL\rangle\rightarrow|LR\rangle} &=& \mathcal{M}_{|LR\rangle\rightarrow|RL\rangle}  &=& &e^{2}\left(2\tan^{2}(\theta/2)-\frac{1}{\mu^2}\cos\theta\sec^{2}(\theta/2)\right) \ ,
\label{eq:Moller_exact}
\end{align}
where $\mu=|\vec{p}|/m$.

\subsubsection{Concurrence analysis}

The concurrences of the final states are:
\begin{align}
\Delta_{RR}&= \frac{2(2+\mu^2(7-\cos 2\theta))\sin^2\theta}{64\mu^4 +\mu^2\left(49+16\cos 2\theta-\cos 4\theta\right)+4\left(3+\cos 2\theta\right)} \ , \label{eq:DMollerpp}\\
\Delta_{RL}&= \Big|\frac{2\mu^2(1+\cos 2\theta-\mu^2(1-\cos 2\theta)\sin^2\theta}{\mu^4\left(35+28\cos 2\theta+\cos 4\theta\right)+\mu^2\left(31+32\cos 2\theta +\cos 4\theta\right)+8\left(1+\cos 2\theta\right)}\Big| \ , \nonumber\\
\label{eq:DMollerpm}
\end{align}
with $\Delta_{LL}=\Delta_{RR}$ and $\Delta_{LR}=\Delta_{RL}$.

In the high energy limit, i.e. $\mu\rightarrow\infty$,
\begin{align}
\Delta_{RR}^{HE}&=\frac{\left(7 -\cos 2\theta\right)\sin^2\theta}{32}\frac{1}{\mu^2} + \mathcal{O}\left(\frac{1}{\mu^4}\right) \ , \\
\Delta_{RL}^{HE}&=\frac{4\sin^4\theta}{35+28\cos 2\theta+\cos 4 \theta} - 
\mathcal{O}\left(\frac{1}{\mu^2}\right) \ .
\end{align}
Thus, for an initial state $|RR\rangle$ it is not possible to generate MaxEnt whereas it is possible for an initial state $|RL\rangle$ if $\theta=\pi/2$.

The low energy limit corresponds with $\mu\rightarrow 0$ since this process does not need an energy threshold (no particles are created),
\begin{align}
\Delta_{RR}^{LE}&= \frac{\sin^2\theta}{3 +\cos 2\theta}-\mathcal{O}(\mu^2) \ ,\\ 
\Delta_{RL}^{LE}&= 0 + \mathcal{O}(\mu^2) \ .
\end{align}
In this limit it is possible to generate MaxEnt for $\theta=\pi/2$ if the initial state is $|RR\rangle$ and, apparently, it is not for $|RL\rangle$. 

To analyse if MaxEnt is lost at next orders, let's perform an expansion around the solution $\theta=\pi/2$:
\begin{align}
\Delta_{RR}^{HE}&\simeq \left(\frac{1}{4}-\frac{5}{16}\left(\theta-\frac{\pi}{2}\right)^2\right)\frac{1}{\mu^2},\\
\Delta_{RR}^{LE}&\simeq \frac{1}{2}-\left(1+\mu^2\right)\left(\theta-\frac{\pi}{2}\right)^2,\\
\Delta_{RL}^{HE}&\simeq \frac{1}{2} - 4\left(1 + \frac{1}{\mu^2}\right)\left(\theta-\frac{\pi}{2}\right)^2,\\
\Delta_{RL}^{LE}&\simeq \frac{1}{2}-4\left(\theta-\frac{\pi}{2}\right)^2-\frac{1}{\mu^2}\left(4+\frac{1}{\mu^2}\left(\theta-\frac{\pi}{2}\right)^2\right).
\end{align}
Then, MaxEnt at $\theta=\pi/2$, when it is generated, is a global maximum. Moreover it is actually generated for an initial state $|RL\rangle$: the $\mu\rightarrow 0$ expansion covered this result. The concurrence \eqref{eq:DMollerpm} shows the indeterminate form $0/0$ when $\mu\rightarrow 0$ and $\theta\rightarrow\pi/2$: the nominator goes like $\mu^2$ while the denominator goes like $1+\cos 2\theta$ when $\mu\rightarrow 0$. As $\mu=0$ could only be a limit while one can fix the angle at $\theta=\pi/2$, the correct analysis consist on first taking the limit $\theta\rightarrow\pi/2$ and after that studying the $\mu$ dependence. As in this case the limit $\lim_{\theta\rightarrow\pi/2}\Delta_{RL}=1/2$, there is always MaxEnt $\forall\mu$.

\subsubsection{Generated States}

Starting from an initial state $|RR\rangle$ the final state become:
\begin{equation}
|\psi\rangle_{RR}=\frac{1}{\sqrt{\mathcal{N}}}\left(2|\Phi^{+}\rangle\left(1+2\mu^2\right)+|\Phi^-\rangle\left(1+\cos 2\theta+\mu^2\right)-|\Psi^+\rangle\sqrt{1+\mu^2}\sin 2\theta\right).
\end{equation}
In the limit $\mu\rightarrow\infty$ the generated state tends to $|\psi\rangle_{RR}^{HE}=\left(|\Phi^+\rangle+|\Phi^-\rangle\right)/\sqrt{2} = |RR\rangle$, which is a product state. However, in the low energy limit $\mu\rightarrow 0$ the generated state become 
\begin{equation}
|\psi\rangle_{RR}^{LE}=\frac{1}{\mathcal{N}}\left(2|\Phi^+\rangle + 
|\Phi^-\rangle (1 + \cos 2\theta) -|\Psi^+\rangle \sin 2\theta\right),
\end{equation}
which at $\theta=\pi/2$ becomes maximally entangled, $|\psi\rangle_{RR}^{LE}=|\Phi^+\rangle$.

For an initial state $|RL\rangle$ the final generated state is:
\begin{multline}
|\psi\rangle_{RL}=\frac{1}{\mathcal{N}}\bigg(|\Phi^-\rangle\sqrt{1+\mu^2}\sin 2\theta + |\Psi^+\rangle\left(1+\cos 2\theta +\mu^2\left(3+\cos 2\theta\right)\right) \\
+2|\Psi^-\rangle\left(1+2\mu^2\right)\cos\theta\bigg) \ .
\end{multline}
One can easily check that at $\theta=\pi/2$ $|\psi\rangle_{RL}(\theta=\pi/2)=|\Psi^+\rangle$, so it is maximally entangled for all $\mu$.

\section{\texorpdfstring{$e^-e^+\rightarrow e^-e^+$}{}}

This process has also two channels:
\begin{align}
i\mathcal{M}_{s}&=\bar{v}(s_{2},p_{2})(-ie\gamma^{\mu})u(s_{1},p_{1}) \frac{-ig_{\mu\nu}}{(p_{1}+p_{2})^{2}} \bar{u}(s_{1}',q_{1})(-ie\gamma_{\nu})v(s_{2}',q_{2}),\\
i\mathcal{M}_{t}&=(-1)\bar{v}(s_{2},p_{2})(-ie\gamma^{\mu})v(s_{2}',q_{2}) \frac{-ig_{\mu\nu}}{(p_{2}-q_{2})^{2}} \bar{u}(s_{1}',q_{1})(-ie\gamma_{\nu})u(s_{1},p_{1}).
\end{align}
The unpolarized amplitude is
\begin{equation}
|\overline{\mathcal{M}}|^{2}= 2e^{4}\left(\frac{s^{2}+u^{2}+8(t-m^{2})m^{2}}{t^{2}}+\frac{t^{2}+u^{2}+8(s-m^{2})m^{2}}{s^{2}}+\frac{2(u-2m^2)(u-6m^2)}{st}\right).
\label{eq:Bhabha_totalA}
\end{equation}

The helicity amplitudes are
\begin{align}
\mathcal{M}_{|RR\rangle\rightarrow|RR\rangle}&= \mathcal{M}_{|LL\rangle\rightarrow|LL\rangle} = e^{2}\frac{2\left(1+\cos\theta\right)+\mu^{2} \left(11+\cos 2\theta\right)+8\mu^{4}}{4\mu^{2}\left(1+\mu^{2}\right)}\csc^{2}(\theta/2) \ , \nonumber\\
\mathcal{M}_{|RR\rangle\rightarrow|RL\rangle}&= \mathcal{M}_{|LL\rangle\rightarrow|LR\rangle} = \pm e^{2}\frac{1+\mu^2\cos\theta}{\mu^2\sqrt{1+\mu^2}}\cot(\theta/2) \ , \nonumber\\
\mathcal{M}_{|RR\rangle\rightarrow|LL\rangle}&= \mathcal{M}_{|LL\rangle\rightarrow|RR\rangle} = -e^{2}\frac{1+\mu^2\left(1+\cos\theta\right)}{\mu^2\left(1+\mu^2\right)} \ , \nonumber\\
\mathcal{M}_{|RL\rangle\rightarrow|\substack{RR\\ LL}\rangle}&= -\mathcal{M}_{|LR\rangle\rightarrow|\substack{RR\\ LL}\rangle} = -e^{2}\frac{1+\mu^2\cos\theta}{\mu^2\sqrt{1+\mu^2}}\cot(\theta/2) \ , \nonumber\\
\mathcal{M}_{|RL\rangle\rightarrow|RL\rangle}&= \mathcal{M}_{|LR\rangle\rightarrow|LR\rangle} = e^{2}\frac{1+\mu^2\left(1+\cos\theta\right)}{\mu^2}\cot^2(\theta/2) \ , \nonumber\\
\mathcal{M}_{|RL\rangle\rightarrow|LR\rangle}&= \mathcal{M}_{|LR\rangle\rightarrow|RL\rangle} = -e^{2}\left(1-\cos\theta-\frac{1}{\mu^2}\right) \ ,
\label{eq:Bhabha_exact}
\end{align}
where $\mu=|\vec{p}|/m$.

\subsubsection{Concurrence analysis}

The concurrence for an initial state $|RR\rangle$ or $|LL\rangle$ is
\begin{equation}
\Delta_{RR}=\Delta_{LL} = \Bigg|\frac{1}{2}+\frac{8(1+3 \mu^2 + 2 \mu^4)^2}{f(\mu,\theta)}\Bigg| \ , 
\end{equation}
where
\begin{multline}
f(\mu,\theta)=64\mu^8+\mu^6\left(177+16\cos 2\theta-\cos 4\theta\right)+\mu^2\left(52+48\cos\theta-4\cos 2\theta\right) \\
+4\mu^4\left(39+9\cos\theta+5\cos 2\theta-\cos 3\theta\right)+16 \ .
\end{multline}

The high and low energy limits for an initial state $|RR\rangle$ become
\begin{align}
\Delta_{RR}^{HE}&= 0- \mathcal{O}\left(\frac{1}{\mu^2}\right), \\
\Delta_{RR}^{LE}&= 0 + \mathcal{O}(\mu^2),
\end{align}
so in these limits it is not possible to generate MaxEnt. However, we can solve the equation $\Delta_{RR}=1/2$ to see if there is any possible solution for other energies. The result is
\begin{equation}
\Delta_{RR}\left(\theta=\pi,\mu=\frac{1}{2}\sqrt{-3+\sqrt{17}}\right)=\frac{1}{2} \ . 
\end{equation}
MaxEnt is allowed at that point but one can check it is not a global maximum.

The concurrence for an initial state $|RL\rangle$ is:
\begin{equation}
\Delta_{RL}=\Bigg|\frac{4\mu^2\left(\mu^4\sin^2\theta+ \mu^2\left(1-2\cos\theta\right)-1\right)\sin^2\theta}{g(\mu,\theta)}\Bigg| \ , 
\end{equation}
where
\begin{multline}
g(\mu,\theta)= \mu^6\left(35+28\cos 2\theta+\cos 4\theta\right)+ 4\mu^2\left(3+16\cos\theta+\cos 2\theta\right) \\
+4\mu^4\left(9+15\cos\theta+7\cos 2\theta+\cos 3\theta\right)+16 \ , 
\end{multline}
and $\Delta_{LR}=\Delta_{RL}$. The corresponding expansion at high and low energy limits
\begin{align}
\Delta_{RL}^{HE}&=\frac{4 \sin^4\theta}{35+28\cos 2\theta +\cos 4\theta} + \mathcal{O}\left(\frac{1}{\mu^2}\right), \\ 
\Delta_{RL}^{LE}&=0 - \mathcal{O}(\mu^2).
\end{align}
At high energies, MaxEnt is achieved for $\theta=\pi/2$. After analyze this point, we obtain:
\begin{equation}
\Delta_{RL}^{HE}\simeq \frac{1}{2}-4\left(\theta-\frac{\pi}{2}\right)^2 + \frac{4}{\mu^2}\left(\theta-\frac{\pi}{2}\right),
\end{equation}
so $\theta=\pi/2$ is not a global maximum. We can solve the equation $\Delta_{RL}=\frac{1}{2}$ as we did in the previous case looking for other possible solutions. The result is
\begin{equation}
\mu_{MaxEnt}=\sqrt{-\frac{1}{2\cos\theta}} \quad , \quad \frac{\pi}{2}\leq\theta<\pi \ .
\end{equation}
As energy decreases, the angle where MaxEnt is found increases. The lowest energy that allows MaxEnt is $\mu\rightarrow\left(1/\sqrt{2}\right)^{+}$ (notice that at $\theta=\pi$, which is the extremal value $\mu=1/\sqrt{2}$, the amplitudes are zero except $|RL\rangle\rightarrow|LR\rangle$). 

\subsubsection{Generated states}

The generated state when the initial particles are in the helicity configuration $|RR\rangle$ is
\begin{multline}
|\psi\rangle_{RR}= \frac{1}{\mathcal{N}}\Big(|\Phi^+\rangle\left(2\cos\theta+ \mu^2\left(5+\cos 2\theta\right)+4\mu^4\right)+ 2|\Phi^-\rangle\left(1+3\mu^2+2\mu^4\right) \\
+ 2|\Psi^-\rangle\sqrt{1+\mu^2}\left(1+\mu^2\cos\theta\right)\sin\theta\Big) \ .
\end{multline}
The corresponding energy limits of this state are:
\begin{align}
\lim_{\mu\rightarrow\infty}|\psi\rangle_{RR} &= \frac{1}{\sqrt{2}}\left(|\Phi^+\rangle+|\Phi^-\rangle\right)=|RR\rangle, \\
\lim_{\mu\rightarrow 0}|\psi\rangle_{RR} &= \frac{1}{\sqrt{2\left(1+\sin^2\theta\right)}}\left(|\Phi^+\rangle\cos\theta+ |\Phi^-\rangle+|\Psi^-\rangle\sin\theta\right), \nonumber\\
&=\frac{1}{\sqrt{2\left(1+\sin^2\theta\right)}}\left(c|R\rangle-s|L\rangle\right)\left(c|R\rangle+s|L\rangle\right),\\
\lim_{\substack{\theta\rightarrow\pi\\\mu\rightarrow\mu_{0}}}|\psi\rangle_{RR} &= |\Phi^-\rangle,
\end{align}
where $\mu_{0}=\frac{1}{2}\left(-3+\sqrt{7}\right)^{1/2}$. As expected, we have only MaxEnt in the last case.

If the initial state is $|RL\rangle$ the corresponding state generated is
\begin{multline}
|\psi\rangle_{RL}=\frac{1}{\mathcal{N}}\Big(-2|\Phi^+\rangle\left(1+\mu^2\cos\theta\right)\sin\theta +2|\Psi^+\rangle\sqrt{1+\mu^2}\left(1+2\mu^2\cos\theta\right) \\
+|\Psi^-\rangle\sqrt{1+\mu^2}\left(2+\cos\theta+\mu^2\left(3+\cos 2\theta\right)\right)\Big) \ ,
\end{multline}
from which we can obtain a MaxEnt state for $\mu_{MaxEnt}=\sqrt{-\sec\theta/2}$,
\begin{equation}
\lim_{\mu\rightarrow \mu_{MaxEnt}^{+}}|\psi\rangle_{RL} = \frac{1}{\mathcal{N}}\left(2|\Phi^+\rangle+|\Psi^-\rangle\sqrt{4-2\sec\theta}\tan\theta\right). 
\label{eq:state_mu+}
\end{equation}
Written in this form, it is not evident, but if we perform, for instance, a $R_{y}(\phi)$ transformation with $\phi=-2\tan^{-1}(\cot\theta/\sqrt{1-\sec\theta/2})$ on the second particle, we obtain the $|\Psi^{-}\rangle$ state.

\section{\texorpdfstring{$e^-e^+\rightarrow \gamma\gamma$}{}}

The amplitudes for this process are
\begin{align}
i\mathcal{M}_{t}&=-ie^{2}\epsilon^{*}_{\mu}(\lambda,q)\epsilon^{*}_{\nu}(\lambda',q') \bar{v}(s',p')\gamma^{\nu}\frac{i}{\cancel{t}-m}\gamma^{\mu}u(s,p), \nonumber\\
i\mathcal{M}_{t}&=-ie^{2}\epsilon^{*}_{\mu}(\lambda,q)\epsilon^{*}_{\nu}(\lambda',q') \bar{v}(s',p')\gamma^{\mu}\frac{i}{\cancel{u}-m}\gamma^{\nu}u(s,p).
\end{align}
These can be written in a more compact way using $1/(\cancel{k}-m)=(\cancel{k}+m)/(k^{2}-m^{2})$:
\begin{equation}
\mathcal{M}=\epsilon^{*}_{\mu}(\lambda,q)\epsilon^{*}_{\nu}(\lambda',q')\mathcal{M}^{\mu\nu},
\end{equation}
where $\mathcal{M}^{\mu\nu}=\mathcal{M}^{\mu\nu}_{t}+\mathcal{M}^{\mu\nu}_{u}$ and
\begin{align}
\mathcal{M}^{\mu\nu}_{t}&=-\frac{e^{2}}{t^{2}-m^{2}}\bar{v}(s',p')\gamma^{\nu}(\cancel{t}+m)\gamma^{\mu}u(s,p), \\
\mathcal{M}^{\mu\nu}_{u}&=-\frac{e^{2}}{u^{2}-m^{2}}\bar{v}(s',p')\gamma^{\mu}(\cancel{u}+m)\gamma^{\nu}u(s,p).
\end{align}

The total unpolarized amplitude is
\begin{equation}
|\overline{\mathcal{M}}|^2 =-2e^4\left(\frac{p\cdot k'}{p\cdot k}+\frac{p\cdot k}{p\cdot k'}+2m^2\left(\frac{1}{p\cdot k}+\frac{1}{p\cdot k'}\right)-m^4\left(\frac{1}{p\cdot k}+\frac{1}{p\cdot k'}\right)^2\right).
\end{equation}

The amplitudes for all helicity and polarization configurations are 
\begin{align}
\mathcal{M}_{|RR\rangle\rightarrow|\pm\pm\rangle} &=& \mathcal{M}_{|LL\rangle\rightarrow|\pm\pm\rangle} &=& \pm &e^2\frac{4\left(\mu^2\mp\sqrt{1+\mu^2}\right)}{\mu^2\left(1-\cos 2\theta\right)+2} \ , \nonumber\\
\mathcal{M}_{|RR\rangle\rightarrow|\pm\mp\rangle} &=& \mathcal{M}_{|LL\rangle\rightarrow|\pm\mp\rangle} &=& -&e^2\frac{4\mu^2}{\mu^2\left(1-\cos 2\theta\right)+2}\sin^2\theta \ , \nonumber\\
\mathcal{M}_{|RL\rangle\rightarrow|\pm\pm\rangle} &=& \mathcal{M}_{|LR\rangle\rightarrow|\pm\pm\rangle} &=& &0 \ , \nonumber\\
\mathcal{M}_{|RL\rangle\rightarrow|\pm\mp\rangle} &=& -\mathcal{M}_{|LR\rangle\rightarrow|\mp\pm\rangle} &=& \pm &e^2\frac{2\mu\sqrt{1+\mu^2}}{\mu^2\sin^2\theta+1}\sin\theta\left(1\pm\cos\theta\right) \ ,
\end{align}
where $\mu=|\vec{p}|/m$, $|+\rangle$ and $|-\rangle$ are the right and left circular polarizations of the photon. The signs $\pm$ and $\mp$ correspond to the right and left final photon polarizations respectively.

\subsubsection{Concurrence analysis}

The concurrences for this process are
\begin{align}
\Delta_{RR}&= \frac{1}{2}-\frac{8\mu^2}{8+\mu^2(19-4\cos 2\theta + \cos 4\theta)}, \\
\Delta_{RL\rangle}&= \frac{\sin^2\theta}{3 + \cos 2\theta},
\end{align}
with $\Delta_{LL}=\Delta_{RR}$ and $\Delta_{LR}=\Delta_{RL}$.
For $\Delta_{RR}$, the high and low energy limits are
\begin{align}
\Delta_{RR}^{HE}&= \frac{4\sin^4\theta}{ 19-4\cos 2\theta +\cos 4\theta} + \mathcal{O}\left(\frac{1}{\mu^2}\right) \ , \\
\Delta_{RR}^{LE}&= \frac{1}{2}-\mathcal{O}(\mu^2) \ . 
\end{align}
The maximum value of the zero order in the high energy limit is $1/6$ for $\theta=\pi/2$, so it is not possible to generate maximal entanglement. On the contrary, at low energies MaxEnt is achieved independently of the angle.

For $\Delta_{RL}$ there is always MaxEnt for all energies except $\mu=0$ if we take $\theta=\pi/2$. As this is an exact result, this angle is a global maximum.

\subsubsection{Generated states}

The generated state from an initial $|RR\rangle$ is
\begin{equation}
|\psi\rangle_{RR}=\frac{1}{\mathcal{N}}\left(|\Phi^{+}\rangle\mu-|\Phi^{-}\rangle\sqrt{1+\mu^2}-|\Psi^{+}\rangle\mu\sin^2\theta\right).
\end{equation}
For high energies, this state has the form $|\Psi^{+}\rangle\sin^2\theta +2|--S\rangle$ which cannot be maximally entangled for any $\theta$. But for low energies, the second term dominates leading to a maximally entangled state $-|\Phi^{-}\rangle$ independently of the COM angle.

If the initial state is $|RL\rangle$, the final particles state become
\begin{equation}
|\psi\rangle_{RL}=\frac{1}{\sqrt{2\cos 2\theta}}\left(|\Psi^{-}\rangle-|\Psi^{+}\rangle\cos\theta\right) \ .
\end{equation}

The only possible solution to generate a maximally entangled state is taking $\theta=\pi/2$, which leads to the $|\Psi^{-}\rangle$ as a final state independently of the energy regime.

\section{\texorpdfstring{$e^-\gamma\rightarrow e^-\gamma$}{}}

The two channels of this process are
\begin{align}
i\mathcal{M}_{s}&=-ie^{2}\bar{u}(s',p')\gamma^{\mu}\epsilon_{\mu}^{*}(\lambda',k')\frac{\cancel{p}+\cancel{k}+m}{s^{2}-m^{2}}\gamma^{\nu}\epsilon_{\nu}(\lambda,k)u(s,p), \nonumber\\
i\mathcal{M}_{u}&=-ie^{2}\bar{u}(s',p')\gamma^{\nu}\epsilon_{\nu}^{*}(\lambda,k)\frac{\cancel{p}-\cancel{k'}+m}{u^{2}-m^{2}}\gamma^{\mu}\epsilon_{\mu}^{*}(\lambda',k')u(s,p).
\end{align}
Using $s^{2}-m^{2}=(p+k)^{2}-m^{2}=2p\cdot k$, we can simplify the matrix element:
\begin{equation}
i\mathcal{M}=-ie^{2}\epsilon_{\mu}^{*}(\lambda',k')\epsilon_{\nu}(\lambda,k)\bar{u}(s',p')\left(\frac{\gamma^{\mu}\cancel{k}\gamma^{\nu}+2\gamma^{\mu}p^{\nu}}{2p\cdot k}+ \frac{-\gamma^{\nu}\cancel{k'}\gamma^{\mu}+2\gamma^{\nu}p^{\mu}}{-2p\cdot k'}\right)u(s,p).
\end{equation}

The total unpolarized amplitude is \cite{Peskin}:
\begin{equation}
|\overline{\mathcal{M}}|^2=2e^4\left(\frac{p\cdot k'}{p\cdot k}+\frac{p\cdot k}{p\cdot k'}+2m^2\left(\frac{1}{p\cdot k}-\frac{1}{p\cdot k'}\right)+m^4\left(\frac{1}{p\cdot k}-\frac{1}{p\cdot k'}\right)^2\right).
\end{equation}

The exact amplitudes for this process are
\begin{align}
\mathcal{M}_{|R+\rangle\rightarrow|R+\rangle} &= \mathcal{M}_{|L-\rangle\rightarrow|L-\rangle} = 2e^2\frac{\mu+\sqrt{1+\mu^2}}{\mu\cos\theta+\sqrt{1+\mu^2}}\cos^3(\theta/2) \ , \nonumber\\
\mathcal{M}_{|R+\rangle\rightarrow|R-\rangle} &= -\mathcal{M}_{|L-\rangle\rightarrow|L+\rangle} =  e^2\frac{1+\cos\theta}{\mu\cos\theta+\sqrt{1+\mu^2}}\sin(\theta/2)\ , \nonumber\\
\mathcal{M}_{|R+\rangle\rightarrow|L+\rangle} &= \mathcal{M}_{|L-\rangle\rightarrow|R-\rangle}  = e^2\frac{\left(\mu-\sqrt{1+\mu^2}\right)\cos(\theta/2)}{\mu\cos\theta+\sqrt{1+\mu^2}}(-1+\cos\theta) \ , \nonumber\\
\mathcal{M}_{|R+\rangle\rightarrow|L-\rangle} &= -\mathcal{M}_{|L-\rangle\rightarrow|R+\rangle} = 2e^2\frac{1}{\mu\cos\theta+\sqrt{1+\mu^2}}\sin^3(\theta/2)\ , \nonumber \\
\mathcal{M}_{|R-\rangle\rightarrow|R+\rangle} &= -\mathcal{M}_{|L+\rangle\rightarrow|R+\rangle} = -\frac{e^2}{2}\frac{1}{\mu\cos\theta+\sqrt{1+\mu^2}}\csc(\theta/2)\sin^2\theta \ , \nonumber\\
\mathcal{M}_{|R-\rangle\rightarrow|R-\rangle} &= \mathcal{M}_{|L+\rangle\rightarrow|L+\rangle} =e^2\frac{1+\cos\theta+4\mu\left(\mu+\sqrt{1+\mu^2}\right)}{\left(\mu+\sqrt{1+\mu^2}\right)\left(\sqrt{1+\mu^2}+\mu\cos\theta\right)}\cos(\theta/2)\ , \nonumber\\
\mathcal{M}_{|R-\rangle\rightarrow|L+\rangle} &= -\mathcal{M}_{|L+\rangle\rightarrow|R-\rangle} = -\frac{2\sin(\theta/2)}{\mu+\sqrt{1+\mu^2}}+\frac{\cos(\theta/2)\sin\theta}{\sqrt{1+\mu^2}+\mu\cos\theta}\ , \nonumber\\
\mathcal{M}_{|R-\rangle\rightarrow|L-\rangle} &= \mathcal{M}_{|L+\rangle\rightarrow|L-\rangle} = e^2\frac{1}{\left(\mu+\sqrt{1+\mu^2}\right)\left(\sqrt{1+\mu^2}+\mu\cos\theta\right)}\sin(\theta/2) \sin\theta  \ ,
\end{align}
where $\mu=|\vec{p}|/m$.

\subsubsection{Concurrence analysis}
The concurrences for this process are
\begin{align}
\Delta_{R+}&=\Delta_{L-} = \frac{\mu|\sin^3\theta|}{f(\mu,\theta)}, \\
f(\mu,\theta)&= 2\mu^2\left(3+\cos 2\theta\right)\cos^2(\theta/2)+2\mu\sqrt{1+\mu^2}\left(2\cos\theta\cos 2\theta+1\right)+3+\cos 2\theta,
\end{align}
and
\begin{equation}
\Delta_{R-}=\Delta_{L+} \frac{8\mu\sin^4(\theta/2)|\sin\theta|}{g(\mu,\theta)} \ ,
\end{equation}
where
\begin{multline}
g(\mu,\theta)= 32\mu^4\left(1+\cos\theta\right)+\mu^2\left(34+25\cos\theta+6\cos 2\theta-\cos 3\theta\right) \\
+4\mu\sqrt{1+\mu^2}\left(8\mu^2\left(1+\cos\theta\right) +1+3\cos\theta\right)+2\left(3+\cos 2\theta\right) \ . 
\end{multline}


For an initial state $|R+\rangle$ or $|L-\rangle$ the corresponding high and low energy limits are
\begin{align}
\Delta_{R+}^{HE}&=\frac{\tan^3(\theta/2)}{2\mu} +\mathcal{O}\left(\frac{1}{\mu^{3}}\right), \\
\Delta_{R+}^{LE}&= \frac{\sin^3\theta}{3 + \cos 2\theta}\mu - \frac{2 (1 + 2\cos\theta + \cos 2\theta) \sin^3\theta}{3 + \cos 2 \theta)^2}\mu^2+\mathcal{O}(\mu^3),
\end{align}
and for an initial state $|R-\rangle$ or $|L+\rangle$,
\begin{align}
\Delta_{R-}^{HE}&= 0 +\mathcal{O}\left(\frac{1}{\mu_{m}}\right)^{3},\\
\Delta_{R-}^{LE}&=  \frac{4\sin^4(\theta/2) \sin\theta}{
 3 + \cos 2\theta}\mu_{m}-\frac{(16 (5\cos(\theta/2 + 3\cos (3\theta/2))\sin^5(\theta/2)}{(3 + \cos 2\theta)^2}\mu_{m}^2 +\mathcal{O}(\mu_{m}^3),\nonumber\\
\\
\Delta_{R-}^{\theta\rightarrow\pi}&= \frac{\mu}{1+2\mu\left(\mu-\sqrt{1+\mu^2}\right)}\left(\theta-\pi\right) + \mathcal{O}\left(\theta-\pi\right)^3, \\
\Delta_{R-}^{\theta\rightarrow\pi}&= -\frac{1}{4}\mu\left(\theta-\pi\right)^3 + \mathcal{O}\left(\theta-\pi\right)^5.
\end{align}
Clearly, it is not possible to achieve the MaxEnt in any case.

\subsubsection{Generated states}

For an initial $|R+\rangle$ particles, the final state become
\begin{multline}
|\psi\rangle_{R+}=\frac{1}{\mathcal{N}}\sum_{\{+,-\}}\Big(2|\Phi^{\pm}\rangle\left(\left(\mu+\sqrt{1+\mu^2}\right)\cos^3(\theta/2) \mp\sin^3(\theta/2)\right) \\
-|\Psi^{\pm}\rangle\left(1+\cos\theta\pm\left(\mu-\sqrt{1+\mu^2}\right) \sin\theta\right)\sin(\theta/2)\Big)  \ , 
\end{multline}
and the corresponding limits are
\begin{align}
\lim_{\mu\rightarrow\infty}|\psi\rangle_{R+}&=\frac{1}{\sqrt{2}}\left(|\Phi^+\rangle+|\Phi^-\rangle\right), \\
\lim_{\mu\rightarrow 0}|\psi\rangle_{R+}&= \frac{1}{\mathcal{N}}\sum_{\{+,-\}}\Big(2|\Phi^\pm\rangle\left(\cos^3(\theta/2)\mp\sin^3(\theta/2)\right) \nonumber\\
&-|\Psi^\pm\rangle\left(1\pm\cos\theta-\sin\theta\right)\sin(\theta/2)\Big), \\
\lim_{\theta\rightarrow\pi}|\psi\rangle_{R+}&=-\frac{1}{\sqrt{2}}\left(|\Phi^+\rangle-|\Phi^-\rangle\right).
\end{align}
In all cases the result is a product state.\\

For an initial state $|RL\rangle$, the final one become:
\begin{align}
|\psi\rangle_{R-}&= \frac{1}{\mathcal{N}}\sum_{\{+,-\}}\Big( |\Phi^\pm\rangle\left(\left(\mu+\sqrt{1+\mu^2}\right)\cos(\theta/2)\pm \sin(\theta/2)\right)\sin\theta +\nonumber\\
&+\frac{1}{2}|\Psi^\pm\rangle\bigg(\left(3+8\mu\left(\mu+\sqrt{1+\mu^2}\right)\right)\cos(\theta/2) +\cos(3\theta/2)\mp \nonumber\\
&\mp 4\left(\mu-\sqrt{1+\mu^2}\right)\sin^3(\theta/2)\bigg)\Big),
\end{align}
which have the limits
\begin{align}
\lim_{\mu\rightarrow\infty}|\psi\rangle_{R-}&=\frac{1}{\sqrt{2}}\left(|\Psi^+\rangle+|\Psi^-\rangle \right),\\
\lim_{\mu\rightarrow 0}|\psi\rangle_{R-}&=\frac{1}{\mathcal{N}} \sum_{\{+,-\}}\Big(|\Phi^\pm\rangle\left(\cos(\theta/2)\pm\sin(\theta/2)\right)\sin\theta \nonumber\\
& + |\Psi^\pm\rangle\left(\cos^3(\theta/2)\pm\sin^3(\theta/2)\right)\Big), \\
\lim_{\theta\rightarrow\pi}|\psi\rangle_{R-}&=\frac{1}{\sqrt{2}}\left(|\Psi^+\rangle-|\Psi^-\rangle\right).
\end{align}
Again, all limits are product states.

\chapterimage{Cajon_sastre}
\chapter{Odds and Ends\label{app:OddsEnds}}


\vspace{-1.5cm}
\begin{flushright}
\begin{minipage}{0.6\textwidth}
\textit{One never notices what has been done; one can only see what remains to be done.}
\begin{flushright}
--Marie Sk\l odowska Curie\\
Letter to her brother, 1894.
\end{flushright}
\end{minipage}

\end{flushright}
\vspace{1cm}

\section{Novel Bell Inequalities}

In this section, we explain in detail how to obtain the optimal settings that lead to a maximal violation of a Bell inequality following the method explained in Ref. \cite{Collins02bis,Acin02,Acin04}. We also provide some properties of the settings found in the Bell inequalities discussed in Chapter \ref{Ch:Bell_Ineq}.

\subsection{Maximal violation of Bell inequalities \label{sec:MaxViolation}}

The method explained in detail in Ref. \cite{Acin02} is as follows. First, we take a general state $|\Phi\rangle=\sum_{i=0}^{d^n-1}\alpha_{i}|e_{i}\rangle$ where $|e_{i}\rangle$ are the $d^n$ computational basis states and $\sum_{i=0}^{d^n-1}|\alpha_{i}|^2=1$. Second, each party, applies a unitary operation on each subsystem consisting on a phase shift followed by a Fourier transform, i.e.
\begin{equation}
U_{FT}U(\vec{\varphi})|\Phi\rangle_{N},
\end{equation}
where $N$ is each party, e.g $A$ and $B$ if it is a bipartite Bell inequality, and
\begin{align}
F_{d}&=\sum_{j,k=0}^{d-1}e^{\frac{2\pi i}{d}jk}|j\rangle\langle k|, \label{eq:FT2}\\
U(\vec{\varphi})&=\sum_{j=0}^{d-1}e^{i\varphi(j)}|j\rangle\langle j|,
\end{align}
where $\vec{\varphi}=(\varphi(0),\varphi(1),\cdots,\varphi(d-1))$ for each settings. After this transformation, one can show that
\begin{equation}
p(a,b,\cdots)=|\langle ab\cdots|F_{d}U(\vec{\varphi_{a}})\otimes F_{d}U(\vec{\varphi_{b}})\otimes\cdots|\Phi\rangle|^2.
\end{equation}
Finally, we rewrite the Bell inequality and maximize its violation to obtain the free parameters $\vec{\varphi}$. Notice that the first phase angle of each setting can be set to 0 without loss of generality and similarly $\vec{\varphi_{a}}=0$. Thus, the optimization procedure involves $n\times s\times(d-1)-1$ free parameters. 

This procedure can be translated into operators instead of probabilities. It can be shown that the above method is equivalent to compute the operators corresponding to each setting as
\begin{equation}
s = (F_{d}U(\vec{\varphi}))^\dagger\Pi F_{d}U(\vec{\varphi}),
\end{equation}
where $s=a,a',b,b',\cdots$ and $\Pi$ is the projector on the states $|\Phi\rangle$. In particular, for qubit inequalities, $\Pi=\sigma_{z}$ and for qutrits inequalities, $\Pi=\lambda_{3}$, since eigenstates of $\sigma_{z}$ and $\lambda_{3}$ are the computational basis states.

The optimal settings written in Chapter \ref{Ch:Bell_Ineq} are not the same as the ones obtained with the above method. We look for a final unitary operation that leads to a more compact expression for the optimal settings, i.e. $\tilde{s}=U^{\dagger}s U$.

\subsection{Mutually Unbiased Bases}

Some of the optimal settings that maximally violate Bell inequalities have special properies. In particular, we found that several Bell inequalities studied in Chapter \ref{Ch:Bell_Ineq} are maximally violated by mutually unbiased bases (MUB).

\begin{definition}[Mutually Unbiased Bases]
Given two sets of orthonormal bases $\{|\phi_{1}\rangle,\cdots,|\phi_{d}\rangle\}$ and $\{|\psi_{1}\rangle,\cdots,|\psi_{d}\rangle\}$ in Hilbert space $\mathbb{C}^{d}$, they are mutually unbiased if
\begin{equation}
|\langle\phi_{i}|\psi_{k}\rangle|^2=\frac{1}{d}, \qquad \forall j,k\in\{1,\cdots,d\}.
\end{equation}
\end{definition}

Notice that the definition do not depend on which basis elements are being taken. Information obtained from a projective measurement associated to one basis set is completely independent to the information obtained from the other set. Measurement outcomes obtained with respect to one basis occur with equal probability if the state belongs to the other set.

We can also understand MUB as an attempt to extend uncertainty principle to finite Hilbert spaces, although the generalization of MUB to infinite Hilbert spaces is still an open question.

If the local dimension $d$ is a power of a prime number, i.e. $d=p^n$ for $p$ prime and $n\in\mathbb{N}$, then there exist a maximal set of $d+1$ MUB \cite{Ivonovic81,Wootters89}. However, the maximum number of MUB for $d$ no prime remains unanswered. An example of MUB for $d=2$ are the Pauli operators basis defined in App. \ref{app:quantum_gates}:
\begin{equation}
\langle\psi_{i}^{\pm}|\psi_{j}^{\pm}\rangle=\frac{1}{2}, \qquad \mathrm{for} \ i\neq j \ and \ i,j=x,y,z.
\end{equation}
For $i=j$, $\langle\psi_{i}^{k}|\psi_{i}^{l}\rangle=\delta_{kl}$ for $k,l=\pm$, as expected from an orthonormal basis. 

We say that a set of normal operators is MUB if their eigenvectors bases are MUB. So Pauli operators are MUB operators. On the contrary, the set of Gell-Mann matrices of Eq. \eqref{eq:GM}, are not MUB, but some specific combinations are. For instance the optimal settings for the three qutrits Bell inequality of Eq. \eqref{eq:BI_3qutrits}, $A=\lambda_{3}$ and $A'=\frac{1}{3}\left(\lambda_{2}+\lambda_{4}+\lambda_{6}\right)$ are MUB.

\subsubsection{Weyl-Heisenberg construction of MUB}

Although there are several methods to find MUB basis or operators (see for instance Ref. \cite{Durt10}), we explain here the Weyl-Heisenberg construction.

The \emph{shift} and \emph{clock} matrices first defined by John Sylvester \cite{Sylvester},
\begin{align}
X&=\sum_{j=0}^{d-1}|j+1\rangle\langle j| =
\left(\begin{array}{cccccc}0& 0& 0&\cdots &0 & 1 \\
1&0&0&\cdots&0&0\\
0&1&0&\cdots&0&0\\
0&0&1&\cdots&0&0 \\
\vdots&\vdots&\vdots&\ddots&\vdots&\vdots \\
0&0&0&\cdots&1&0
\end{array}\right),\\
Z&=\sum_{j=0}^{d-1}\omega^{j}|j\rangle\langle j| =
\left(\begin{array}{cccccc}1& 0& 0&\cdots &0 & 0 \\
0&\omega&0&\cdots&0&0\\
0&0&\omega^2&\cdots&0&0\\
\vdots&\vdots&\vdots&\ddots&\vdots&\vdots \\
0&0&0&\cdots&0&\omega^{d-1}
\end{array}\right),
\end{align}
where $\omega=e^{2\pi i/d}$, are the generators of the Weyl-Heisenberg group. An orthonormal basis is given by
\begin{equation}
X^{j}Z^{k}=\sum_{i=0}^{d-1}\omega^{ki}|i+j\rangle\langle i|.
\end{equation}
Specific combinations of these basis elements form MUB \cite{MUB}:
\begin{align}
X, \ Z, \ XZ^i  \ \mathrm{for} \ i=1,\cdots d-1  \ \mathrm{are} \ \mathrm{MUB} \ , \label{eq:MUB_basisd}\\
X^{i}, \ Z^{j}, \ (XZ)^{k}  \ \mathrm{for} \ i,j,k=1,\cdots d-1  \ \mathrm{are} \ \mathrm{MUB} \ . \label{eq:MUB_basisd2}
\end{align}

So now is clear that in three and four qutrits inequalities of Eq. \eqref{eq:3qutrits} and Eq. \eqref{eq:B423} the optimal settings are mutually unbiased ($A=X$ and $A'=Z$) while in the two and six qutrit cases are not, since $A=X$ and $A'$ is a combination that includes $X$. The $\mathcal{B}_{223}$ inequality of Eq. \eqref{eq:B223A} is maximally violated with MUB settings that fulfilled the relation \eqref{eq:MUB_basisd2} for $i=j=1$ and $k=2$.

\subsection{Multiplets of Optimal Settings}

We have introduced in Chapter \ref{Ch:Bell_Ineq} the multiplets of optimal settings (MOS) which denotes any set of matrices that maximize the two-qutrit, six-qutrit and all two-qudit inequalities. 

If we set one of the settings of these inequalities to $X$, then, in order to obtain maximal violation, the other takes the form
\begin{equation}
MOS = e^{i\phi} \left( \begin{array}{ccccccc}
0 & 0 & 0 & \cdots & 0 & 0 & 1 \\
-1 & 0 & 0 & \cdots & 0 & 0 & 0 \\
0 & -1 & 0 & \cdots & 0 & 0 & 0 \\
\vdots & \vdots & \vdots & \ddots & \vdots & \vdots & \vdots\\
0 & 0 & 0 & \cdots & 0  & -1 & 0
\end{array}\right),
\label{eq:MOS}
\end{equation}
where $\phi$ is a global phase that depends on the specific form of the inequality.

At the moment, the properties that have been found for these kind of operators are that both commutator and anticommutator of any pair of MOS are nilpotent matrices, i.e. matrices $M$ such that $M^{k}=0$ for some integer $k$.

\section{Hyperdeterminant in Spin Chains}

In this section we present the exact diagonalization of $n=4$ spin chains and the computation of hyperdeterminant and $S$ and $T$ invariants.

\subsection{Ising model eigenstates}

The diagonalization of Ising Hamiltonian of Eq. \eqref{eq:HIsing} for a $n=4$ spin chain is written in Tab. \ref{Tab:Ising}. Energy levels are labeled from the ground state to the $15^{th}$ excited state for $0<\lambda<2/\sqrt{3}$. 

The coefficients $\alpha$, $\beta$ and $\gamma$ appearing in Tab. \ref{Tab:Ising} are
\begin{align}
\alpha_{0\pm} &= \frac{1}{\lambda}\left(2 \lambda^3 \pm\sqrt{2}\lambda^2\sqrt{\lambda' +\sqrt{\lambda''}}\mp\sqrt{2}\sqrt{\lambda' +\sqrt{\lambda''}} \left(1 - \sqrt{\lambda''}\right) - \lambda\left(1 -2\sqrt{\lambda''}\right)\right),  \nonumber \\ 
\alpha_{2\pm}&= \frac{1}{\lambda}\left(2 \lambda^3 \pm\sqrt{2}\lambda^2\sqrt{\lambda' -\sqrt{\lambda''}}\mp\sqrt{2}\sqrt{\lambda' -\sqrt{\lambda''}}\left(1 + \sqrt{\lambda''}\right) - \lambda\left(1 +2\sqrt{\lambda''}\right)\right),\nonumber\\
\beta_{0\pm}&= \lambda\pm\frac{1}{\sqrt{2}}\sqrt{\lambda' +\sqrt{\lambda''}}, \nonumber \\ 
\beta_{2\pm}&=  \lambda\pm\frac{1}{\sqrt{2}}\sqrt{\lambda' -\sqrt{\lambda''}},
\nonumber \\ 
\gamma_{0\pm}&= 1 \pm \frac{\sqrt{2} \lambda}{\sqrt{\lambda'} +\sqrt{\lambda''}},  \nonumber \\ 
\gamma_{2\pm}&=  1 \pm \frac{\sqrt{2} \lambda}{\sqrt{\lambda'} -\sqrt{\lambda''}},
\label{eq:Isingvar}
\end{align}
with $\lambda'=1+\lambda^2$ and $\lambda''=1+\lambda^4$.

\begin{table}[t!]
\centering
\begin{tabular}{@{\extracolsep{\fill}}ccc}
\toprule
\textbf{Level} & \textbf{Energy} & \textbf{Eigenstate (up to normalization)} \\
\midrule
$|\Psi_{0}\rangle$ & $\scriptstyle -2\sqrt{2}\sqrt{\lambda'+\sqrt{\lambda''}}$ & $\alpha_{0+}|0000\rangle+ 2\beta_{0+}|\Psi^{+}\rangle_{13}|\Psi^{+}\rangle_{24}+\gamma_{0+}(|0101\rangle+|1010\rangle)+|1111\rangle$ \\
$|\Psi_{1}\rangle$ & $\scriptstyle -2\left(\sqrt{\lambda'}+1\right)$ & $\left(\lambda+\sqrt{\lambda'}\right)\left(|00\rangle|\Psi^{+}\rangle+|\Psi^{+}\rangle|00\rangle\right) +|11\rangle|\Psi^{+}\rangle+|\Psi^{+}\rangle|11\rangle$ \\
$|\Psi_{2}\rangle$ & $\scriptstyle -2\sqrt{2}\sqrt{\lambda'-\sqrt{\lambda''}}$ & $\alpha_{2+}|0000\rangle+ 2\beta_{2+}|\Psi^{+}\rangle_{13}|\Psi^{+}\rangle_{24} +\gamma_{2+}(|0101\rangle+|1010\rangle)+|1111\rangle$ \\
$|\Psi_{3}\rangle$ & $\scriptstyle -2\lambda$ & $|\Psi^{-}\rangle_{13}|00\rangle_{24}$ \\
$|\Psi_{4}\rangle$ & $\scriptstyle -2\lambda$ & $|00\rangle_{13}|\Psi^{-}\rangle_{24}$ \\
$|\Psi_{5}\rangle$ & $\scriptstyle -2\left(\sqrt{\lambda'}-1\right)$ & $-\left(\lambda+\sqrt{\lambda'}\right)\left(|00\rangle|\Psi^{-}\rangle + |\Psi^{-}\rangle|00\rangle\right)-\left(|11\rangle|\Psi^{-}\rangle + |\Psi^{-}\rangle|11\rangle\right)$ \\
$|\Psi_{6}\rangle$ & $\scriptstyle 0$ & $|0011\rangle-|1100\rangle$ \\
$|\Psi_{7}\rangle$ & $\scriptstyle 0$ & $|01\rangle_{13}|\Psi^{-}\rangle_{24}$ \\
$|\Psi_{8}\rangle$ & $\scriptstyle 0$ & $|\Psi^{-}\rangle_{13}|01\rangle_{24}$ \\
$|\Psi_{9}\rangle$ & $\scriptstyle 0$ & $|0101\rangle-|1010\rangle$ \\
$|\Psi_{10}\rangle$ & $\scriptstyle 2\left(\sqrt{\lambda'}-1\right)$ & $\left(\lambda-\sqrt{\lambda'}\right)\left(|00\rangle|\Psi^{+}\rangle+|\Psi^{+}\rangle|00\rangle\right)+|11\rangle|\Psi^{+}\rangle+|\Psi^{+}\rangle|11\rangle$ \\
$|\Psi_{11}\rangle$ & $\scriptstyle 2\lambda$ & $|11\rangle_{13}|\Psi^{-}\rangle_{24}$\\
$|\Psi_{12}\rangle$ & $\scriptstyle 2\lambda$ & $|\Psi^{-}\rangle_{13}|11\rangle_{24}$ \\
$|\Psi_{13}\rangle$ & $\scriptstyle 2\sqrt{2}\sqrt{\lambda'-\sqrt{\lambda''}}$ & $\alpha_{2-}|0000\rangle+ 2\beta_{2-}|\Psi^{+}\rangle_{13}|\Psi^{+}\rangle_{24}+\gamma_{2-}(|0101\rangle+|1010\rangle)+|1111\rangle$ \\
$|\Psi_{14}\rangle$ & $\scriptstyle 2\left(\sqrt{\lambda'}+1\right)$ & $-\left(\lambda+\sqrt{\lambda'}\right)\left(|00\rangle|\Psi^{-}\rangle + |\Psi^{-}\rangle|00\rangle\right)-\left(|11\rangle|\Psi^{-}\rangle + |\Psi^{-}\rangle|11\rangle\right)$ \\
$|\Psi_{15}\rangle$ & $\scriptstyle 2\sqrt{2}\sqrt{\lambda'+\sqrt{\lambda''}}$ & $\alpha_{0-}|0000\rangle+ 2\beta_{0-}|\Psi^{+}\rangle_{13}|\Psi^{+}\rangle_{24}+\gamma_{0-}(|0101\rangle+|1010\rangle)+|1111\rangle$\\
\bottomrule
\end{tabular}
\caption{Energies and eigenstates of $n=4$ Ising Hamiltonian from Eq. \eqref{eq:HIsing}. Coefficients $\alpha_{0\pm}$, $\alpha_{2\pm}$, $\beta_{0\pm}$, $\beta_{2\pm}$, $\gamma_{0\pm}$ and $\gamma_{2\pm}$ are those from Eq. \eqref{eq:Isingvar}. Values for $\lambda'$ and $\lambda''$ are $1+\lambda^2$ and $1+\lambda^4$ respectively. To simplify the notation, we introduced the states $|\Psi^{\pm}\rangle=(|01\rangle\pm|10\rangle)/\sqrt{2}$. Subscripts 13 and 24 refer to spins that are being described by the state, leaving blank when these spins are 12 and 34 respectively.}
\label{Tab:Ising}
\end{table}

The analysis of $S$ and $T$ invariants together with $\hdet_{4}$ explained in Chapter \ref{Ch:HDet} is extended below to all eigenstates of this model. States with $S=T=0$ can be factorized into two subsystems:
\begin{equation}
\begin{array}{lll}
 |\Psi_{3}\rangle = |\Psi^{-}\rangle_{13}|00\rangle_{24} \ ,  &  \quad |\Psi_{7}\rangle = |01\rangle_{13}|\Psi^{-}\rangle_{24} \ , & \quad |\Psi_{11}\rangle = |11\rangle_{13}|\Psi^{-}\rangle_{24} \ ,
\nonumber \\
 |\Psi_{4}\rangle = |00\rangle_{13}|\Psi^{-}\rangle_{24} \ , & \quad |\Psi_{8}\rangle = |\Psi^{-}\rangle_{13}|01\rangle_{24} \ , & \quad |\Psi_{12}\rangle = |\Psi^{-}\rangle_{13}|11\rangle_{24} \ .
\end{array}
\end{equation}

States with energies $\pm 2(\sqrt{\lambda'}\pm 1)$ -- that is $|\Psi_{1}\rangle$, $|\Psi_{5}\rangle$, $|\Psi_{10}\rangle$ and $|\Psi_{14}\rangle$ -- are a superposition of two $W$ states or a local transformation of a $W$ state. As a consequence, $\mathrm{HDet}_{4}$ is zero but not $S$ and $T$.

States $|\Psi_{6}\rangle$ and $|\Psi_{9}\rangle$ have  $\hdet_{4} = 0$, $S \neq 0$ and $T \neq 0$. These states entangle  maximally  two spins in one direction with the other spins  in the opposite direction.

Finally, there are four states with $\hdet_{4}$ different from zero: $|\Psi_{0}\rangle$, $|\Psi_{2}\rangle$, $|\Psi_{13}\rangle$ and $|\Psi_{15}\rangle$. Their analysis is explained in detail in Chapter \ref{Ch:HDet}.

\subsection{\texorpdfstring{$XXZ$}{} model eigenstates\label{app:XXZ}}

The $XXZ$ spin chain with 4 sites can be solved analytically as the Ising model. 
Table \ref{Tab:XXZstates} collects all energies and eigenstates of this model. 
 
For $\Delta<-1$, the ground state is doubly degenerate with energy $4\Delta$: it describes a ferromagnetic phase where all spins are aligned. For $\Delta>-1$ its energy is $-2\left(\Delta+\sqrt{8+\Delta^2}\right)$ and the ground state is a resonating valence bound, as explained in Chapter \ref{Ch:HDet}. 

\begin{table}[t!]
\centering
\begin{tabular}{@{}cc}
\toprule
\textbf{Energy} & \textbf{Eigenstate (up to normalization)} \\
\midrule
-4 & $|\Psi^{-}\rangle|11\rangle+|11\rangle|\Psi^{-}\rangle$ \\
-4 & $|\Psi^{-}\rangle|00\rangle+|00\rangle|\Psi^{-}\rangle$ \\
4 & $|\Psi^{+}\rangle|11\rangle+|11\rangle|\Psi^{+}\rangle$ \\
4 & $|\Psi^{+}\rangle|00\rangle+|00\rangle|\Psi^{+}\rangle$ \\
0 & $|\Psi^{-}\rangle_{13}|11\rangle_{24}$  \\
0 & $|11\rangle_{13}|\Psi^{-}\rangle_{24}$  \\
0 & $|\Psi^{-}\rangle_{13}|10\rangle_{24}$  \\
0 & $|10\rangle_{13}|\Psi^{-}\rangle_{24}$  \\
0 & $|\Psi^{-}\rangle_{13}|00\rangle_{24}$  \\
0 & $|00\rangle_{13}|\Psi^{-}\rangle_{24}$  \\
0 & $|0011\rangle -|1100\rangle$  \\
$-4\Delta$ & $|0101\rangle -|1010\rangle$  \\
$4\Delta$ & $|0000\rangle$  \\
$4\Delta$ & $|1111\rangle$  \\
$-2\left(\Delta-\sqrt{8+\Delta^2}\right)$ & $|\Psi^{+}\rangle_{13}|\Psi^{+}\rangle_{24}-\frac{1}{2}\left(\Delta-\sqrt{8+\Delta^2}\right)\left(|0101\rangle+|1010\rangle\right)$ \\
$-2\left(\Delta+\sqrt{8+\Delta^2}\right)$ & $|\Psi^{+}\rangle_{13}|\Psi^{+}\rangle_{24}-\frac{1}{2}\left(\Delta+\sqrt{8+\Delta^2}\right)\left(|0101\rangle+|1010\rangle\right)$ \\
\bottomrule
\end{tabular}
\caption{Energies and eigenstates of $n=4$ spin chain with a $XXZ$ interaction. To compact the notation, it has been used the states $|\Psi^{\pm}\rangle=(|01\rangle\pm|10\rangle)/\sqrt{2}$.}
\label{Tab:XXZstates}
\end{table}

All eigenstates have zero $\hdet_{4}$, and only four states have $S$ and $T$ invariants non zero. Two of these states correspond with the two configurations that maximally entangled two spins up with two spins down:
\begin{align}
|\Psi_{6}\rangle&=\frac{1}{\sqrt{2}}\left(|0011\rangle-|1100\rangle\right)=\frac{1}{\sqrt{2}}\left(|\upuparrows\rangle|\downdownarrows\rangle-|\downdownarrows\rangle|\upuparrows\rangle\right), \\
|\Psi_{9}\rangle&=\frac{1}{\sqrt{2}}\left(|0101\rangle-|1010\rangle\right)=\frac{1}{\sqrt{2}}\left(|\upuparrows\rangle_{13}|\downdownarrows\rangle_{24}-|\downdownarrows\rangle_{13}|\upuparrows\rangle_{24}\right),
\end{align}
where $|\upuparrows\rangle=|00\rangle$ and $|\downdownarrows\rangle=|11\rangle$. The other two states are the ones with energies $-2\left(\Delta+\sqrt{8+\Delta^2}\right)$ and $-2\left(\Delta-\sqrt{8+\Delta^2}\right)$ and correspond, respectively, with the ground state and $15^{\mathrm{th}}$ excited state for $-1<\Delta<1$. States that can be factorized into two subsystems have $S$ and $T$ zero, as happens with Ising model eigenstates. Finally, states with energy $\pm 4$ are $W$-type and, consequently, have $S$ and $T$ zero. States with energy 4 have the typical form of a $W$ state and states with energy $-4$ correspond to the local operation $\sz_1\sz_3|W\rangle$, where $\sz_{i}$ is the Pauli matrix operation over $i$-th qubit.

\vfill
\section{Absolute Maximal Entanglement in Quantum Computation}

\subsection{C\texorpdfstring{$_3$}{}--adder gate construction}\label{app:C3}

To construct the C$_3$--adder gate with qubits we should find a sequence of gates that perform the following operations:
\begin{equation}
\begin{array}{lll}
\mathrm{C}_{3}|00\rangle |00\rangle= |00\rangle |00\rangle, & \mathrm{C}_{3}|01\rangle |00\rangle= |01\rangle |01\rangle, & \mathrm{C}_{3}|10\rangle |00\rangle= |10\rangle |10\rangle \ , \\
\mathrm{C}_{3}|00\rangle |01\rangle= |00\rangle |01\rangle, & \mathrm{C}_{3}|01\rangle |01\rangle= |01\rangle |10\rangle, & \mathrm{C}_{3}|10\rangle |01\rangle= |10\rangle |00\rangle \ ,\\
\mathrm{C}_{3}|00\rangle |10\rangle= |00\rangle |10\rangle, & \mathrm{C}_{3}|01\rangle |10\rangle= |01\rangle |00\rangle, & \mathrm{C}_{3}|10\rangle |10\rangle= |10\rangle |01\rangle \ .
\end{array}
\label{eq:C3}
\end{equation}

As a result, besides from CNOT gates, we will need from CCNOT gates. Three-qubit gates are difficult to implement experimentally, so we should decompose them in terms of one and two-qubit gates. The exact decomposition of CCNOT gate is shown in App. \ref{app:quantum_gates}, which is a circuit of 12 gates of depth. However, we can use instead an approximate decomposition which differ from the previous for some phase shifts of the quantum states other than zero \cite{Barenco95}. In particular, we can use the approximate CCNOT gates shown in Fig. \ref{Fig:Toffaprox}. The only changes that those gates introduce respect the exact CCNOT gate are
\begin{align}
\mathrm{CCNOT}_{a}|101\rangle &= -|101\rangle, \\
\mathrm{CCNOT}_{b}|100\rangle &= -|100\rangle.
\end{align}
This is translated into the use of controlled-Z gate in the first approximation to obtain the desired result after applying the gate sequence to construct the C$_3$--adder. The sign introduced in the CCNOT$_{b}$ gate is canceled after this sequence, so the circuit remains equal as exact CCNOT gates were used.

We can keep saving more gates. Notice that the firsts two C$_{3}$--adders of the AME circuit of Fig. \ref{Fig:AME43} are implemented on qutrits in the state $|\bar{0}\rangle$. Let's write it explicitly. After the Fourier transform on qutrit 1, the circuit applies the C$_{3}$--adder on qutrit 3:
\begin{equation}
(\bar{\mathrm{C}}_{3})_{_{13}}\left[\frac{1}{\sqrt{3}}\left(|\bar{0}\rangle+|\bar{1}\rangle+|\bar{2}\rangle\right)_{1}\otimes|\bar{0}\rangle_{3} \right]= \frac{1}{\sqrt{3}}\left(|\bar{0}\bar{0}\rangle+|\bar{1}\bar{1}\rangle+|\bar{2}\bar{2}\rangle\right)_{13},
\end{equation}
where the subindex 13 stands for the qutrits affected from this operation. In qubits form
\begin{equation}
(\mathrm{C}_{3})_{_{13}}\left[\frac{1}{\sqrt{3}}\left(|00\rangle+|01\rangle+|10\rangle\right)_{1}\otimes|00\rangle_{3} \right]= \frac{1}{\sqrt{3}}\left(|00\rangle|00\rangle+|01\rangle|01\rangle+|10\rangle|10\rangle\right)_{13}.
\end{equation}
Then, the above operation consists uniquely in two CNOT gates between even and odd qubits. Similarly, the next C$_{3}$--adder acting on qutrit 4 can be implemented in the same way:
\begin{equation}
(\bar{\mathrm{C}}_{3})_{_{14}}\otimes\mathbb{I}_{_{3}}\left[\frac{1}{\sqrt{3}}\left(|\bar{0}\bar{0}\rangle+|\bar{1}\bar{1}\rangle+|\bar{2}\bar{2}\rangle\right)_{13}\otimes|\bar{0}\rangle_{4}\right]= \frac{1}{\sqrt{3}}\left(|\bar{0}\bar{0}\bar{0}\rangle+|\bar{1}\bar{1}\bar{1}\rangle+|\bar{2}\bar{2}\bar{2}\rangle\right)_{134},
\end{equation}
which in the qubit form becomes
\begin{multline}
(\mathrm{C}_{3})_{_{14}}\otimes\mathbb{I}_{_{3}}\left[\frac{1}{\sqrt{3}}\left(|00\rangle|00\rangle+|01\rangle|01\rangle+|10\rangle|10\rangle\right)_{13}\otimes|00\rangle_{4} \right] \\
= \frac{1}{\sqrt{3}}\left(|00\rangle|00\rangle|00\rangle+|01\rangle|01\rangle|01\rangle+|10\rangle|10\rangle|10\rangle\right)_{134}.
\end{multline}
Again, the above state can be obtained from the previous using two CNOT gates, between even and odd qubits. This enormous simplification cannot be extended to the other C$_{3}$--adder gates, as all elements of the basis appear once we implement the $F_{3}$ gate on qutrit 2.

\subsection{Karamata's inequality and its applications to bipartite figures of merit \label{app:majorization}}

In this section, we provide a proof of Karamata's inequality and apply it to deduce the inequalities of Eq. \eqref{eq:majS} and Eq. \eqref{eq:majP}.

\begin{theorem}[Karamata's inequality]
Let \textbf{a} and \textbf{b} be two finite sequences of real numbers from an interval $(\alpha,\beta)$. If $\mathbf{a}\succ\mathbf{b}$ and if $f:(\alpha,\beta)\rightarrow\mathbb{R}$ is a convex function, then the inequality
\begin{equation}
\sum_{i=1}^{n}f(a_{i})\geq\sum_{i=1}^{n}f(b_{i})
\end{equation}
holds.
\end{theorem}

To prove this theorem, we need to use the definition of a convex function:
\begin{definition}[Convex function of one variable]
$f(x)$ is a convex function if for $x_{1}\neq x_{2}$ in some interval $(\alpha,\beta)$, the slope
\begin{equation}
\frac{f(x_{1})-f_(x_{2})}{x_{1}-x_{2}}
\end{equation}
is monotonically non-decreasing in $x_{1}$ $\forall x_{2}$ or vice versa.
\end{definition}

Next, we define the following functions applied to the elements of \textbf{a} and \textbf{b}:
\begin{definition}[Slope function]
\begin{equation}
c_{i}\equiv\frac{f(b_{i})-f(a_{i})}{b_{i}-a_{i}}.
\label{eq:slope}
\end{equation}
\end{definition}
Since $\mathbf{a}\succ\mathbf{b}$, this function is decreasing, i.e. $c_{i+1}\leq c_{i}$.
\begin{definition}[Partial sums]
\begin{equation}
A_{k}\equiv\sum_{i=1}^{k}a_{i}, \qquad B_{k}\equiv\sum_{i=1}^{k}b_{i},
\label{eq:partialsum}
\end{equation}
with $A_{0}=B_{0}=0$.
\end{definition}
Notice that majorization implies $A_{k}\geq B_{k}$ and $A_{n}=B_{n}$.

With the above definitions, we are ready to proof Karamata's inequality. An equivalent way to formulate this inequality is to prove that
\begin{equation}
\sum_{i=1}^{n}f(a_{i})-\sum_{i=1}^{n}f(b_{i})\geq 0.
\end{equation}
\begin{align}
\sum_{i=1}^{n}\left(f(a_{i})-f(b_{i})\right)&=\sum_{i=1}^{n} c_{i}(a_{i}-b_{i})=\sum_{i=1}^{n}(A_{i}-A_{i-1}-(B_{i}-B_{i-1}))\nonumber\\
&= \sum_{i=1}^{n}c_{i}(A_{i}-B_{i})+ \sum_{i=1}^{n}c_{i}(A_{i-1}-B_{i-1}) \nonumber\\
&= \sum_{i=1}^{n}c_{i}(A_{i}-B_{i}) - \sum_{i=0}^{n-1}c_{i+1}(A_{i}-B_{i}) \nonumber \\
&=\sum_{i=1}^{n-1}(c_{i}-c_{i+1})(A_{i}-B_{i}) + c_{1}(A_{0}-B_{0}) + c_{n}(A_{n}-B_{n}) \nonumber\\
&= \sum_{i=1}^{n-1}(c_{i}-c_{i+1})(A_{i}-B_{i}) \geq 0 \ . \hspace{4cm} \blacksquare 
\end{align}

We can use this inequality to obtain inequality relations in terms of Von Neumann entropy and purity. These figures of merit can be written in terms of eigenvalues of the reduced density matrix: $S=-\sum_{i}\lambda_{i}\log\lambda_{i}$ and $\gamma=\sum_{i}\lambda_{i}^2$, as has been explained in Chapter \ref{Ch:HDet}. Both functions of $\lambda_{i}$ are convex and if we assume majorization in eigenvalues each time we apply a CZ gate, i.e. $\sum_{i}\lambda_{i}^{s}\geq\sum_{i}\lambda_{i}^{s+1}$, then
\begin{equation}
\sum_{i=1}^{n}\lambda_{i}^{s}\log\lambda_{i}^{s}\geq \sum_{i=1}^{n}\lambda_{i}^{s+1}\log\lambda_{i}^{s+1} \Rightarrow S^{s}\leq S^{s+1},
\end{equation}
for the entropy and
\begin{equation}
\sum_{i=1}^{n}(\lambda_{i}^{s})^2\geq\sum_{i=1}^{n}(\lambda_{i}^{s+1})^2 \Rightarrow \gamma^{s}\geq \gamma^{s+1},
\end{equation}
for the purity.

\end{appendix}

\part*{References}

\chapterimage{library3}

\addcontentsline{toc}{chapter}{\textcolor{clr}{Bibliography}} 

\chapter*{Bibliography}

\begin{flushright}
\begin{minipage}{0.6\textwidth}
\textit{M\'as sabe el diablo por viejo que por diablo.}
\begin{flushright}
--Spanish proverb.
\end{flushright}
\end{minipage}
\end{flushright}
\vspace{1cm}

\let\cleardoublepage\clearpage
\phantomsection
\begingroup
\let\clearpage\relax
\let\cleardoublepage\clearpage
\vspace{-10cm}


%

\endgroup




\end{document}